\pdfoutput=1

\documentclass[11pt]{article}%
\usepackage[sc]{mathpazo}
\usepackage[T1]{fontenc}
\usepackage[utf8]{inputenc}%

\usepackage{amsmath}
\usepackage{amsfonts}
\usepackage{amssymb}
\usepackage{graphicx}
\usepackage[margin=.85in]{geometry}

\usepackage{natbib}
\usepackage{xcolor}
\usepackage[title]{appendix}
\usepackage{booktabs}

\usepackage{pdflscape}
\usepackage{afterpage}

\usepackage{textcomp}

\definecolor{webgreen}{rgb}{0,.5,0}
\definecolor{webbrown}{rgb}{.6,0,0}
\definecolor{webpurple}{rgb}{0.7,0,0.7}

\usepackage{amsthm}
\newtheorem{proposition}{Proposition}%
\newtheorem{remark}{Remark}
\theoremstyle{definition}
\newtheorem{assumption}{Assumption}

\usepackage[onehalfspacing]{setspace}

\usepackage{hyperref}
\hypersetup{%
  breaklinks = true,%
  colorlinks = true,%
  anchorcolor = webbrown,%
  citecolor = webbrown,%
  filecolor = webbrown,%
  linkcolor = webbrown,%
  menucolor = webbrown,%
  urlcolor= webbrown,%
  citebordercolor= 1 0 0,%
  menubordercolor=1 0 0,%
  urlbordercolor=1 0 0,%
  runbordercolor=1 0 0,%
  pdftitle=Shift-Share Designs: Theory and Inference,%
  pdfauthor=Rodrigo Adão \& Michal Kolesár \& Eduardo Morales}
\usepackage{hypcap}

\newcounter{subassumption}[assumption]
\renewcommand{\thesubassumption}{({\roman{subassumption}})}
\makeatletter
\renewcommand{\p@subassumption}{\theassumption}%
\makeatother
\newcommand{\aitem}{\refstepcounter{subassumption}\thesubassumption~\ignorespaces}

\usepackage{cleveref}
\crefname{subassumption}{assumption}{assumptions}
\crefname{assumption}{assumption}{assumptions}
\crefname{remark}{remark}{remarks}
\crefname{proposition}{proposition}{propositions}

\usepackage{caption}
\usepackage{subcaption}

\newcommand{\N}{\ensuremath{N}}          %
\newcommand{\JN}{N}          %
\usepackage[cal=dutchcal]{mathalfa}      %
\newcommand{\Xs}{\ensuremath\mathcal{X}} %
\newcommand{\xs}{\ensuremath\mathcal{x}} %
\newcommand{\Zs}{\ensuremath\mathcal{Z}} %

\DeclareMathOperator{\var}{var} %

\usepackage{mathtools}
\DeclarePairedDelimiter\indicatorfence{\{}{\}}
\newcommand\1{\operatorname{\mathbb{I}}\indicatorfence}
\DeclarePairedDelimiter\abs{\lvert}{\rvert} %

\allowdisplaybreaks

\usepackage{xr}
\crefname{sappsec}{Online Appendix}{Online Appendices}
\crefname{sappsubsec}{Online Appendix}{Online Appendices}
\crefname{sappsubsubsec}{Online Appendix}{Online Appendices}

\begin{document}

\title{SHIFT-SHARE DESIGNS: THEORY AND INFERENCE\protect\footnote{We thank the
    editor, four anonymous referees, Kirill Borusyak, Peter Egger, Gordon
    Hanson, Bo Honoré, seminar participants at Carleton University, CEMFI,
    Chicago Fed, EESP-FGV, Fed Board, GTDW, Johns Hopkins University, Princeton
    University, PUC-Rio, Stanford University, UC Irvine, Unil, University of
    Michigan, University of Nottingham, University of Virginia, University of
    Warwick, and Yale University, and conference participants at the CEPR-CURE
    conference, CEPR-ERWIT conference, EIIT-FREIT, Globalization \& Inequality
    BFI conference, NBER-ITI winter meeting, Sciences Po Summer Workshop, and
    Princeton-IES Workshop, for very useful comments and suggestions. We thank
    Juan Manuel Castro Vincenzi and Xiang Zhang for excellent research
    assistance. We thank David Autor, David Dorn and Gordon Hanson for sharing
    their code and data. All errors are our own.}}%
\author{{\Large Rodrigo Adão}
  \\{\Large Michal Kolesár\thanks{Corresponding author. 287 Julis Romo
      Rabinowitz Building, Princeton University, Princeton NJ 08544. Phone:
      (609) 258-6726, Fax: (609) 258-6419. Email:
      \href{mailto:mkolesar@princeton.edu}{mkolesar@princeton.edu}}}\\
  {\Large Eduardo Morales} }

\maketitle
\begin{abstract}
  \setstretch{1.1}%
  We study inference in shift-share regression designs, such as when a regional outcome is regressed on a weighted average of sectoral shocks, using regional sector shares as weights. We conduct a placebo exercise in which we estimate the effect of a shift-share regressor constructed with randomly generated sectoral shocks on actual labor market outcomes across U.S. Commuting Zones. Tests based on commonly used standard errors with 5\% nominal significance level reject the null of no effect in up to 55\% of the placebo samples. We use a stylized economic model to show that this overrejection problem arises because regression residuals are correlated across regions with similar sectoral shares, independently of their geographic location. We derive novel inference methods that are valid under arbitrary cross-regional correlation in the regression residuals. We show using popular applications of shift-share designs that our methods may lead to substantially wider confidence intervals in practice.\\[1ex]
\end{abstract}

\thispagestyle{empty}
\setcounter{page}{0}
\clearpage

\section{Introduction}\label{sec:introduction}

We study how to perform inference in shift-share designs: regression
specifications in which one studies the impact of a set of shocks, or
``shifters'', on units differentially exposed to them, with the exposure
measured by a set of weights, or ``shares''. Specifically, shift-share
regressions have the form
\begin{equation}\label{eq:bartikreg}
  Y_{i} = \beta X_{i} + Z_{i}'\delta + \epsilon_{i},
  \qquad\text{where}\quad
  X_{i}=\sum_{s=1}^{S}w_{is}\Xs_{s},\quad w_{is}\geq 0\;\text{for all $s$},
  \quad\text{and}\quad\sum_{s=1}^{S}w_{is} \le 1.
\end{equation}
For example, in an investigation of the impact of sectoral demand shifters on
regional employment changes, $Y_{i}$ is the change in employment in region $i$,
the shifter $\Xs_{s}$ is a measure of the change in demand for the good produced
by sector $s$, and the share $w_{is}$ may be measured as the initial share of
region $i$'s employment in sector $s$. Other observed characteristics of region
$i$ are captured by the vector $Z_{i}$, which includes the intercept, and
$\epsilon_{i}$ is the regression residual. Shift-share specifications are
increasingly common in many contexts (see, e.g., \citet{bartik1991benefits},
\citet{blanchard1992regional}, \citet{card2001immigrant}, or
\citet*{autordornhanson2013}). However, their formal properties are relatively
understudied.

Our starting point is the observation that usual standard error formulas may
substantially understate the true variability of OLS estimators of $\beta$ in
\cref{eq:bartikreg}. We illustrate the importance of this issue through a
placebo exercise. As outcomes, we use 2000--2007 changes in employment rates and
average wages for 722 Commuting Zones in the United States. We build a
shift-share regressor by combining actual sectoral employment shares in 1990
with randomly drawn sector-level shifters for 396 4-digit SIC manufacturing
sectors. The placebo samples thus differ exclusively in the randomly drawn
sectoral shifters. For each sample, we compute the OLS estimate of $\beta$ in
\cref{eq:bartikreg} and test if its true value is zero. Since the shifters are
randomly generated, their true effect is indeed zero. Valid 5\% significance
level tests should therefore reject the null of no effect in at most 5\% of the
placebo samples. We find, however, that usual standard errors---clustering on
state as well as heteroskedasticity-robust errors---are much smaller than the
standard deviation of the OLS estimator and, as a result, lead to severe
overrejection. Depending on the labor market outcome used, the rejection rate
for 5\% level tests can be as high as 55\% for heteroskedasticity-robust
standard errors and 45\% for standard errors clustered on state, and it is never
below 16\%.

To explain the source of this overrejection problem, we introduce a stylized
economic model featuring multiple regions, each of which produces output in
multiple sectors. The key ingredients of our model are a sector- and
region-specific labor demand and a regional labor supply. We assume that labor
demand in each sector-region pair has a sector-specific elasticity with respect
to wages and an intercept that aggregates several sector-specific components
(e.g.\ sectoral productivities and demand shifters for the corresponding
sectoral good). Labor supply in each region is upward-sloping and has a
region-specific intercept that may aggregate group-specific labor supply
shifters (e.g.\ push factors that raise immigration from different countries of
origin). Up to a first-order approximation, the impact of sector-level shocks on
labor market outcomes takes the form of a shift-share specification similar to
that in \cref{eq:bartikreg}.

A key insight of our model is that the regression residual $\epsilon_{i}$ in
\cref{eq:bartikreg} will generally account for shift-share components that
aggregate all unobserved sector-level shocks using the same shares $w_{is}$ that
enter the construction of the regressor $X_{i}$, as well as shift-share
components that aggregate unobserved group-specific labor supply shifters using
exposures $\tilde{w}_{ig}$ of region $i$ to group-$g$ specific shocks. Thus, the
residual may incorporate multiple shift-share terms with shares correlated with
those defining the shift-share regressor $X_{i}$. Consequently, whenever two
regions have similar shares, they will not only have similar exposure to the
shifters $\Xs_{s}$, but will also tend to have similar values of the residuals
$\epsilon_{i}$. While traditional inference methods allow for some forms of
dependence between the residuals, such as spatial dependence within a state,
they do not directly address the possible dependence between residuals generated
by unobserved shift-share components. This is why, in our placebo exercise,
traditional inference methods underestimate the variance of the OLS estimator of
$\beta$, creating the overrejection problem.

We then establish the large-sample properties of the OLS estimator of $\beta$ in
\cref{eq:bartikreg} under repeated sampling of the shifters $\Xs_s$,
conditioning on the realized shares $w_{is}$, controls $Z_{i}$, and residuals
$\epsilon_i$. This sampling approach is motivated by our economic
model: we are interested in what would have happened to outcomes if the
sector-level shocks $\Xs_s$ had taken different values, holding everything
else constant. Our framework allows for heterogeneous effects of the shifters:
one unit increase in $\Xs_{s}$ causes the outcome in region $i$ to increase by
$w_{is}\beta_{is}$, where $\beta_{is}$ is an unknown parameter.

Our key assumption is that, conditional on the controls and the shares, the
shifters are as good as randomly assigned and independent across sectors. An
advantage of this assumption is that it allows us to do inference conditionally
on $\epsilon_{i}$; as a result, we can allow for \textit{any} correlation
structure of the regression residuals across
regions.\footnote{\label{fn:spatial}This is similar to the insight in
  \citet{barrios2012clustering}, who consider cross-section regressions
  estimated at an individual level when the variable of interest varies only
  across groups of individuals. They show that, as long as the regressor of
  interest is as good as randomly assigned and independent across the groups,
  standard errors clustered on groups are valid under any correlation structure
  of the residuals.} In contrast, if, instead of assuming independence of the
shifters across sectors, we modeled the correlation structure in the residual,
as in the spatial econometrics literature \citep[e.g.][]{conley_gmm_1999} or in
the interactive fixed effects literature
\citep[e.g.][]{bai2009,gobillonmagnac2016}, the resulting inference would be
sensitive to the validity of the modeling assumptions. We show that the
regression estimand $\beta$ in \cref{eq:bartikreg} corresponds to a weighted
average of the heterogeneous parameters $\beta_{is}$ and derive novel confidence
intervals that are valid in samples with many regions and sectors.
We also derive an analogous formula when $X_{i}$ is used as an instrument in an
instrumental variables regression, which follows directly from the fact that the
associated first-stage and reduced-form regressions take the form in
\cref{eq:bartikreg}.

To gain intuition for our formula, it is useful to consider the special case in
which each region is fully specialized in one sector (i.e.\ for every $i$,
$w_{is}=1$ for some sector $s$). In this case, our procedure is identical to
using the usual clustered standard error formula, but with clusters defined as
groups of regions specialized in the same sector. This is in line with the rule
of thumb that one should ``cluster'' at the level of variation of the regressor
of interest. In the general case, our standard error formula essentially forms
sectoral clusters, the variance of which depends on the variance of a weighted
sum of the regression residuals $\epsilon_{i}$, with weights that correspond to
the shares $w_{is}$.

We extend our baseline results in three ways. We provide versions of our
standard errors that only require the shifters to be independent across
``clusters'' of sectors, allowing for arbitrary correlation among sectors
belonging to the same ``cluster.'' We also show how to apply our framework to
panel data settings in which we have multiple observations of each region over
time. Finally, we cover applications in which the shifter is unobserved, but can
be estimated using observable local shocks.

We illustrate the finite-sample properties of our novel inference procedure in
the same placebo exercise that we use to show the bias of the usual standard
error formulas. Our new formulas give a good approximation to the variability of
the OLS estimator across the placebo samples; consequently, they yield rejection
rates that are close to the nominal significance level. As predicted by the
theory, our standard error formula remains accurate under alternative
distributions of both the shifters and the regression residuals. When the number
of sectors is small or there is a sector that is significantly larger than the
rest, our method overrejects, although the overrejection is milder in comparison
with the usual standard error formulas. If the shifters are not independent
across sectors, we show that it is important to properly account for their
correlation structure.

In the final part of the paper, we illustrate the implications of our new
inference procedure for two popular applications of shift-share regressions.
First, we study the effect of changes in sector-level Chinese import
competition on labor market outcomes across U.S. Commuting Zones, as in
\citet*{autordornhanson2013}. Second, we use changes in sector-level national
employment to estimate the regional inverse labor supply elasticity, as in
\citet{bartik1991benefits}.\footnote{Additionally, in \begin{NoHyper}\Cref{appsec:applicationBP}\end{NoHyper}, we use changes in the stock of immigrants
from various origin countries to investigate the impact of immigration on employment and wages, following \citet{altonji1991effects}
and \citet{card2001immigrant}.} Our new confidence intervals for the effects of Chinese
competition on local labor markets increase by 23\%--66\% relative to those
implied by state-clustered or heteroskedasticity-robust standard errors,
although these effects remain statistically significant.
In contrast, our confidence intervals for the inverse labor supply elasticity estimated using the
procedure in \citet{bartik1991benefits} are very similar to those
constructed using standard approaches.

Shift-share designs have been applied to estimate the effect of a wide range of
shocks. For example, in seminal papers, \citet{bartik1991benefits} and
\citet{blanchard1992regional} use shift-share designs to analyze the impact on
local labor markets of shifters measured as changes in national sectoral
employment. More recently, shift-share strategies have been applied to
investigate the local labor market impact of various shocks, including
international trade competition \citep*{topalova2007trade,topalova2010factor,
  kovak2013regional,autordornhanson2013, dixcarneirokovak2017, pierce2017trade},
credit supply \citep*{Greenstone2015credit}, technological change
\citep{acemoglu2017robots,acemoglurestrepo2018}, and industry reallocation
\citep{chodorow2018reallocation}. Shift-share regressors have been used as well
to estimate the impact of immigration on labor markets, as in
\citet{card2001immigrant} and many other papers following his approach; see
reviews in \citet{lewisperi2015} and \citet*{dustmann2016different}.
Furthermore, recent papers use shift-share strategies to estimate how
firms respond to changes in outsourcing costs and foreign demand
\citep{hummels2014wage, aghion2018impact}.\footnote{Shift-share regressors have
  also been used to study the impact of sectoral shocks on political preferences
  \citep{autor2017elections, che2017tradechinaelections,colantonestanig2018},
  marriage patterns \citep*{autordornhanson2018}, crime levels
  \citep*{dixcarneiro2017crime}, and innovation
  \citep{acemoglulinn2004,autordornhanson2017innovation}. In addition to using
  shift-share designs to estimate the overall impact of a shifter of interest,
  other work has used them as part of a more general structural estimation
  approach; see \citet{diamond2016}, \citet{adao2016},
  \citet*{galle2017slicing}, \citet{bursteinhansontianvogel2018},
  \citet{bartelme2018tradecosts}. \citet{baum2015causal} review additional
  applications in the context of urban economics.}

Our paper is related to two other papers studying the statistical properties of
shift-share instrumental variables. First, \citet*{goldsmith2018bartik} consider
using the full vector of shares $(w_{i1}, \dotsc, w_{iS})$ as an instrument for
endogenous treatment. They conclude that this approach requires the entire
vector of shares to be as good as randomly assigned conditional on the shifters.
Second, \citet*{borusyakhulljaravel2018shiftshare}, focusing on the use of a
shift-share regressor as an instrument, show it is a valid instrument if the set
of shifters is as good as randomly assigned conditional on the shares, and
discuss consistency of the instrumental variables estimator in this context. We
follow \citet*{borusyakhulljaravel2018shiftshare} by modeling the shifters as
randomly assigned, since this approach follows naturally from our economic
model. Using this assumption, we point out the
potential bias of standard inference procedures when applied to shift-share
designs, and provide a novel inference procedure that is valid in this context.

While our paper focuses on the statistical properties of the OLS estimator of
$\beta$ in \cref{eq:bartikreg}, there exists a prior literature that has focused
on studying the validity of different economic interpretations that one may
attach to the estimand $\beta$. For example, this prior literature has studied
how this interpretation may be affected by the presence of cross-regional
general equilibrium effects \citep*{beraja2016aggregate,adao2018spatial}, slow
adjustment of labor market outcomes to the shifters $\Xs_{s}$
\citep*{jaeger2018shift}, and heterogeneous effects of the shifters across
sectors and regions \citep*{montereddingrossi2018}.

The rest of this paper is organized as follows. \Cref{sec:overr-usual-stand}
presents a placebo exercise illustrating the properties of the usual inference
procedures. \Cref{sec:model} introduces a stylized economic model and maps its
implications into a potential outcome framework. \Cref{sec:asympt-prop-bart}
establishes the asymptotic properties of the OLS estimator of $\beta$ in
\cref{eq:bartikreg}, as well as the properties of an instrumental variables
estimator that uses a shift-share variable as an instrument.
\Cref{sec:extensions} discusses extensions of our baseline framework.
\Cref{sec:placebo-exercise} examines the performance of our novel inference
procedures in a series of placebo exercises. \Cref{sec:empapp} revisits two
prior applications of shift-share designs, and \Cref{sec:conclusion} concludes. Proofs and additional results are collected
in an Online Appendix. %

\section{Overrejection of usual standard errors: placebo evidence}\label{sec:overr-usual-stand}

In this section, we implement a placebo exercise to evaluate the finite-sample
performance of the two inference methods most commonly applied in shift-share
regression designs: (a) Eicker-Hubert-White---or
heteroskedasticity-robust---standard errors, and (b) standard errors clustered
on groups of regions geographically close to each other. In our placebo, we
regress observed changes in U.S. regional labor market outcomes on a shift-share
regressor that is constructed by combining actual data on initial sectoral
employment shares for each region with randomly generated sector-level shocks.
We describe the setup in \Cref{sec:data} and discuss the results in
\Cref{sec:placeboresults}.

\subsection{Setup and Data}\label{sec:data}

We generate $30,000$ placebo samples indexed by $m$. Each of them contains
$N = 722$ regions and $S = 396$ sectors. We identify each region $i$ with a
U.S.\ Commuting Zone (CZ) and each sector $s$ with a 4-digit SIC manufacturing
industry.

Using the notation from \cref{eq:bartikreg}, the shares
$\{w_{is}\}_{i=1,s=1}^{N, S}$, and the outcomes $\{Y_{i}\}_{i=1}^{N}$ are
identical in each placebo sample. The shares correspond to
employment shares in 1990, and the outcomes correspond to changes in employment
rates and average wages for different subsets of the population between 2000 and
2007. Our source of data on employment shares is the County Business Patterns,
and our measures of changes in employment rates and average wages are based on
data from the Census Integrated Public Use Micro Samples in 2000 and the
American Community Survey for 2006 through 2008. Given these data sources, we
construct our variables following the procedure described in the Online Appendix
of \citet*{autordornhanson2013}.

The placebo samples differ exclusively in the shifters
$\{\Xs_{s}^{m}\}_{s=1}^{\N}$, which are drawn i.i.d.\ from a normal distribution
with zero mean and variance equal to five in each placebo sample $m$. Since the
shifters are independent of both the outcomes and the shares, the parameter
$\beta$ is zero; this is true irrespective of the dependence structure between
the outcomes and the shares.

For each placebo sample $m$, given the observed outcome $Y_{i}$, the generated
shift-share regressor $X^{m}_{i}$ and a vector of controls $Z_{i}$ including
only an intercept, we compute the OLS estimate of $\beta$, the
heteroskedasticity-robust standard error (which we label \emph{Robust}), and the
standard error that clusters CZs in the same state (labeled \emph{Cluster}).

\subsection{Results}\label{sec:placeboresults}

\Cref{tab:simple_placebo} presents the median and standard deviation of the
empirical distribution of the OLS estimates of $\beta$ across the 30,000 placebo
samples, along with the median standard error estimates,
and rejection rates for 5\% significance level tests of the null hypothesis
$H_{0}\colon \beta=0$. We present these statistics for several outcome variables, which are listed in the leftmost column.

Column (1) of \Cref{tab:simple_placebo} shows that, up to simulation error, the
average of the OLS estimates is zero for all outcomes. Column (2) reports the
standard deviation of the estimated coefficients. This dispersion is the target
of the estimators of the standard error of the OLS
estimator.\footnote{\begin{NoHyper}\Cref{fig:EmpDist} in
    \Cref{sec:empirical_distribution_placebo}\end{NoHyper} reports the empirical
  distribution of the OLS estimates when the dependent variable is the change in
  each CZ's employment rate. Its distribution resembles a normal distribution
  centered around $\beta=0$.} Columns (3) and (4) report the median standard
error estimates for the \emph{Robust} and \emph{Cluster} procedures,
respectively, and show that both standard error estimators are downward biased.
On average across all outcomes, the median magnitudes of the
heteroskedasticity-robust and state-clustered standard errors are, respectively,
55\% and 46\% lower than the standard deviation.

\begin{table}[!tp]
\centering
\caption{Standard errors and rejection rate of the hypothesis
  $H_{0}\colon \beta=0$ at 5\% significance level.}\label{tab:simple_placebo}
\begin{tabular}{@{}l rc cc cc@{}}
\toprule
&\multicolumn{2}{c}{Estimate} & \multicolumn{2}{c}{Median std.\ error} &
\multicolumn{2}{c}{Rejection rate}\\ [1pt]
\cmidrule(rl){2-3} \cmidrule(rl){4-5}\cmidrule(rl){6-7}
&Mean & Std.\ dev.\ & Robust & Cluster & Robust & Cluster \\
& \multicolumn{1}{c}{(1)} & \multicolumn{1}{c}{(2)} & \multicolumn{1}{c}{(3)} & \multicolumn{1}{c}{(4)}
& \multicolumn{1}{c}{(5)} & \multicolumn{1}{c}{(6)} \\
\midrule

\multicolumn{7}{@{}l}{\textbf{Panel A\@: Change in the share of working-age population}} \\ [2pt]

Employed        &       $-0.01$ &       $2.00$  &       $0.73$  &       $0.92$  &       $48.5$\%        &       $38.1$\%        \\
Employed in manufacturing       &       $-0.01$ &       $1.88$  &       $0.60$  &       $0.76$  &       $55.7$\%        &       $44.8$\%        \\
Employed in non-manufacturing   &       $0.00$  &       $0.94$  &       $0.58$  &       $0.67$  &       $23.2$\%        &       $17.6$\%        \\ [6pt]

\multicolumn{7}{@{}l}{\textbf{Panel B\@: Change in average log weekly wage}} \\ [2pt]
Employed        &       $-0.03$ &       $2.66$  &       $1.01$  &       $1.33$  &       $47.3$\%        &       $34.2$\%        \\
Employed in manufacturing       &       $-0.03$ &       $2.92$  &       $1.68$  &       $2.11$  &       $26.7$\%        &       $16.8$\%        \\
Employed in non-manufacturing   &       $-0.02$ &       $2.64$  &       $1.05$  &       $1.33$  &       $45.4$\%        &       $33.7$\%        \\ [2pt]

  \bottomrule
  \multicolumn{7}{p{6.1in}}{\scriptsize{Notes: For the outcome variable
  indicated in the leftmost column, this table indicates the mean and standard deviation of the OLS estimates of $\beta$ in~\cref{eq:bartikreg} across the placebo samples (columns (1) and (2)),
  the median standard error estimates (columns (3) and (4)), and the percentage of placebo samples for which we reject the null
  hypothesis $H_{0}\colon\beta=0$ using a 5\% significance level test (columns (5) and (6)).
  \textit{Robust} is the Eicker-Huber-White standard error, and
  \textit{Cluster} is the standard error that clusters CZs in the same state. Results are based on 30,000 placebo samples.}}
\end{tabular}
\end{table}

The downward bias in the \emph{Robust} and \emph{Cluster} standard errors
translates into a severe overrejection of the null hypothesis
$H_{0}\colon\beta=0$. Since the true value of $\beta$ equals $0$ by
construction, a correctly behaved test with significance level 5\% should have a
5\% rejection rate. Columns (5) and (6) in \Cref{tab:simple_placebo} show that
traditional standard error estimators yield much higher rejection rates. For
example, when the outcome variable is the CZ's employment rate, the rejection
rate is 48.5\% and 38.1\% when \emph{Robust} and \emph{Cluster} standard errors
are used, respectively. These rejection rates are very similar when the
dependent variable is instead the change in the average log weekly wage.

These results are quantitatively important. To see this, consider the following
thought-experiment. Suppose we were to provide the $30,000$ simulated samples to
$30,000$ researchers without disclosing the origin of the data to them. Instead,
we would tell them that the shifters correspond to changes in a sectoral shock
of interest---for instance, trade flows, tariffs, or national employment. If the researchers set out to test the null that the impact of this shock is zero using standard inference procedures at a 5\%
significance level, then over a third of them would conclude that our computer
generated shocks had a statistically significant effect on the evolution of
employment rates between 2000 and 2007.

The following remark summarizes the results of our placebo exercise.

\begin{remark}\label{rem:overrejection}
  In shift-share regressions, traditional inference methods may suffer from a
  severe overrejection problem, and yield confidence intervals that are too
  short.
\end{remark}

To understand the source of this overrejection problem, note that the standard
error estimators reported in \Cref{tab:simple_placebo} assume that the
regression residuals are either independent across all regions (for
\emph{Robust}), or between geographically defined groups of regions (for
\emph{Cluster}). Given that shift-share regressors are correlated across regions
with similar employment shares $\{w_{is}\}_{s=1}^{S}$, these methods generally
lead to a downward bias in the standard error estimate whenever regions with
similar employment shares $\{w_{is}\}_{s=1}^{S}$ also have similar regression
residuals. In the next section, we show how such correlations between regression
residuals may arise.

\section{Stylized economic model}\label{sec:model}

This section presents a stylized economic model mapping labor demand and labor
supply shocks to labor market outcomes for a set of regional economies. The aim
of the model is twofold. First, it illustrates the economic mechanisms behind
the overrejection problem documented in \Cref{sec:placeboresults}. Second, it
provides guidance on how to estimate: (i) the impact of
sector-specific labor demand shifters on regional labor market outcomes; and (ii) the
regional inverse labor supply elasticity. We describe the model fundamentals in
\Cref{sec:environment_main}, discuss its main implications in \Cref{sec:impacteconshocks}, and map these implications to a
potential outcome framework in \Cref{sec:transition_econometrics}.

\subsection{Environment}\label{sec:environment_main}

We consider an economy with multiple sectors $s=1, \dotsc, S$ and multiple
regions $i=1, \dotsc, \JN$. We assume that the labor demand in sector $s$ and
region $i$, $L_{is}$, is given by
\begin{equation}\label{labor_demand_maintext}
  \log L_{is} = - \sigma_s \log \omega_{i} + \log D_{is}, \qquad\sigma_{s}>0,
\end{equation}
where $\omega_i$ is the wage rate in region $i$, $\sigma_s$ is the labor demand
elasticity in sector $s$, and $D_{is}$ is a region- and sector-specific labor
demand shifter. This shifter may account for multiple sectoral
components. Specifically, we decompose $D_{is}$ into a sectoral shifter of
interest $\chi_{s}$, other shifters that vary by sector $\mu_s$, and a residual
region- and sector-specific shifter $\eta_{is}$:
\begin{equation}\label{technology_maintext}
  \log D_{is} = \rho_s \log \chi_s + \log \mu_s + \log \eta_{is}.
\end{equation}

We assume that the labor supply in region $i$ is given by
\begin{equation}\label{labor_supply_maintext}
  \log L_i = \phi \log \omega_i + \log v_i, \qquad\phi>0,
\end{equation}
where $\phi$ is the labor supply elasticity, and $v_i$ is a region-specific
labor supply shifter. We allow this shifter to have a shift-share structure that yields region-specific aggregates of group-specific labor supply
shocks. In particular, indexing labor groups by $g=1, \dots, G$, we decompose
\begin{equation}\label{labor_supply_shift_share}
\log v_i = \sum_{g=1}^G \tilde{w}_{ig} \log \nu_g + \log \nu_i,
\end{equation}
where $\nu_g $ is a group-specific labor supply shifter, $\tilde{w}_{ig}$
measures the exposure of region $i$ to group $g$ labor supply shifter,
and $\nu_i$ captures region-specific factors affecting labor supply. The
variable $\nu_g$ captures factors that affect the supply of labor of group
$g$ in all regions in the population of interest. Workers may be classified
into groups according to their education level, gender, or country of
origin.

We assume that workers cannot move across regions but are freely mobile
across sectors. Thus, labor markets clear if
\begin{equation}\label{market_clearing_maintext}
L_i = \sum_{s=1}^S L_{is}, \qquad i=1,\dotsc, \JN.
\end{equation}

\subsection{Labor market equilibrium}\label{sec:impacteconshocks}

We assume that, in each period, the model described by
\cref{labor_demand_maintext,technology_maintext,labor_supply_maintext,labor_supply_shift_share,market_clearing_maintext}
characterizes the labor market equilibrium in every region, and
that, across periods, changes in the labor market outcomes
$\{\omega_{i}, L_{i}\}_{i=1}^{\JN}$ are due to changes in either the labor demand
shifters, $\{\chi_s, \mu_{s}\}_{s=1}^S$ and $\{\eta_{is}\}_{i=1,s=1}^{\JN, S}$, or
the labor supply shifters, $\{\nu_g\}_{g=1}^{G}$ and $\{\nu_{i}\}_{i=1}^{\JN}$.

We use $\hat{z} = \log(z^{t}/z^{0})$ to denote log-changes in a variable $z$
between a period $t=0$ and some other period $t$. We assume that the realized changes between any two periods in all labor demand and supply
shifters are draws from a joint distribution $F(\cdot)$:
\begin{equation}\label{eq:jointdistshocks}
\left(\{\hat{\chi}_s, \hat{\mu}_s\}_{s=1}^{S}, \{\hat{\eta}_{is}\}_{i=1,s=1}^{\JN, S}, \{\hat{\nu}_g\}_{g=1}^{G} \{\hat{\nu}_i\}_{i=1}^{\JN} \right)\sim F(\cdot).
\end{equation}

Up to a first-order approximation around the initial equilibrium,
\cref{labor_demand_maintext,technology_maintext,labor_supply_maintext,labor_supply_shift_share,market_clearing_maintext}
imply that the changes in employment and wages in region $i$ are given by
\begin{align}
\hat{L}_i&= \sum_{s=1}^{S} l_{is}^0(\theta_{is} \hat{\chi}_s + \lambda_i \hat{\mu}_s + \lambda_i \hat{\eta}_{is}) + \left(1 - \lambda_i\right)(\sum_{g=1}^G \tilde{w}_{ig} \hat{\nu}_g + \hat{\nu}_i),\label{eq:emp_change}\\
\hat{\omega}_i&= \phi^{-1} \sum_{s=1}^{S} l_{is}^0 (\theta_{is} \hat{\chi}_s + \lambda_i \hat{\mu}_s + \lambda_i \hat{\eta}_{is})-\phi^{-1}\lambda_i(\sum_{g=1}^G \tilde{w}_{ig} \hat{\nu}_g + \hat{\nu}_i),\label{eq:wage_change}
\end{align}
where $l_{is}^{0}=L_{is}^{0}/L_{i}^{0}$ is the initial employment share of
sector $s$ in region $i$,
$\lambda_i =\phi \left[\phi + \sum_{s=1}^{S} l_{is}^0 \sigma_s\right]^{-1}$, and
$\theta_{is} = \rho_s \lambda_i$.

Consider first the model's implications for the impact on
regional labor market outcomes of changes in
sector-specific labor demand. We focus here on the impact of the demand shocks
$\{\hat{\chi}_{s}, \hat{\mu}_{s}\}_{s=1}^{S}$ on the change in the employment rate $\hat{L}_i$;
however, given the symmetry between \cref{eq:emp_change,eq:wage_change}, the
model's implications for the impact of these shocks on the change in the wage
level $\hat{\omega}_i$ are analogous.

According to \cref{eq:emp_change}, the change in the employment rate in region $i$ depends on
two shift-share components that aggregate the impact of the sector-specific labor
demand shocks. In both components, the ``share'' term is the initial
employment share $l_{is}^{0}$;  the ``shift'' term corresponds in each of them to
one of the two sector-specific labor demand shocks, $\hat{\chi}_{s}$ or $\hat{\mu}_s$. Furthermore, $\hat{L}_i$ also depends on additional shift-share terms that aggregate the impact of group-specific labor supply shocks. In this case, the ``share'' term is the
region's exposure to each group-specific shock,
$\tilde{w}_{ig}$. Conditional on a sector $s$ and a labor group $g$, the
shares $\{l_{is}^0\}_{i=1}^{\JN}$ and $\{\tilde{w}_{ig}\}_{i=1}^{\JN}$ may be
correlated. Settings in which the outcome of interest depends on multiple
shift-share terms with potentially correlated shares is central to understanding
the placebo results presented in \Cref{sec:overr-usual-stand}.

Another implication of \cref{eq:emp_change} is that, even conditional on the
initial employment share $l_{is}^0$, the impact of sectoral
labor demand shocks on regional employment may be heterogeneous across
sectors and regions; e.g., the impact of $\hat{\chi}_{s}$ on $\hat{L}_i$ depends
not only on $l_{is}^0$ but also on $\theta_{is}$, which may vary across $i$ and
$s$. While datasets usually
contain information on the initial employment shares for every sector and region
$\{l_{is}^0\}_{i=1,s=1}^{\JN, S}$, the parameters
$\{\theta_{is}\}_{i=1,s=1}^{\JN, S}$ are not
generally known.

We summarize the discussion in the last two paragraphs in the following remark:

\begin{remark}\label{rem:shiftshare}
  In our model, the equilibrium equations for the change in regional labor market
  outcomes combines multiple shift-share terms, and the shifter effects depend
  on unknown parameters that may be heterogeneous.
\end{remark}

\begin{NoHyper}\Cref{app:general_model,sec:econ-model}\end{NoHyper} show that there are multiple microfoundations consistent with the insights summarized in \Cref{rem:shiftshare}. Alternative microfoundations may differ in the mapping between the labor demand and supply elasticities, $\sigma_{s}$ and $\phi$, and structural parameters, or in the interpretation of the different terms entering the labor demand shifter $D_{is}$ in \cref{technology_maintext}.\footnote{In \begin{NoHyper}\Cref{app:general_model}\end{NoHyper}, we derive \cref{eq:emp_change,eq:wage_change} from a multisector gravity model with endogenous labor supply that follows closely that in \citet*{adao2018spatial}. In \begin{NoHyper}\Cref{appsec:specific}\end{NoHyper}, we show that \Cref{rem:shiftshare} is consistent with a \citet{jones1971specific} model featuring sector-specific production inputs, as in \citet{kovak2013regional}. In \begin{NoHyper}\Cref{appsec:specific_pref}\end{NoHyper}, we show that it is also consistent with a \citet{roy1951distribution} model featuring workers with heterogeneous preferences for employment across sectors, as in \citet*{galle2017slicing}, \citet{lee2017inequality}
  and \citet*{burstein2018inequality}.}
In addition, \begin{NoHyper}\Cref{appsec:migration}\end{NoHyper} shows that
similar insights arise in a model that allows for migration across regions. In this
case, the change in regional employment depends not only on the
region's own shift-share terms included in \cref{eq:emp_change}, but also on a
component, common to all regions, that combines the shift-share terms
corresponding to all $N$ regions. In this environment,
$l_{is}^0 \theta_{is}$ is the partial effect of the shifter $\hat{\chi}_{s}$ on
$\hat{L}_i$ conditional on a fixed effect that absorbs cross-regional spillovers
created by migration.

Turning to the estimation of the inverse labor supply elasticity, \cref{labor_supply_maintext,labor_supply_shift_share} imply that
\begin{equation}\label{eq: invlaborsupeq}
  \hat{\omega}_{i}=\tilde{\phi}\hat{L}_{i}-\tilde{\phi}
  (\sum_{g=1}^G \tilde{w}_{ig}
  \hat{\nu}_g + \hat{\nu}_i)\quad\text{with}\quad\tilde{\phi}=\phi^{-1}.
\end{equation}
It follows from \cref{eq:emp_change} that the change in region $i$'s employment
rate, $\hat{L}_{i}$, also depends on the term
$\sum_{g=1}^G \tilde{w}_{ig} \hat{\nu}_g + \hat{\nu}_i$. Thus, the two terms on
the right-hand side of \cref{eq: invlaborsupeq} are correlated with each other,
creating an endogeneity problem. The instrumental variables solution to this
problem relies on the observation that using
\cref{eq:emp_change,eq:wage_change}, one can write the inverse labor supply
elasticity as the ratio of the impact of a sector-specific labor demand shock
(e.g. $\hat{\chi}_{s}$) on wages to that on employment:
\begin{equation*}
\tilde{\phi} = \frac{\partial \hat{\omega}_i}{\partial \hat{\chi}_s}\bigg/\frac{\partial \hat{L}_i}{\partial \hat{\chi}_s}.
\end{equation*}

In \Cref{sec:asympt-prop-bart,sec:extensions}, we use the model described here
to provide an economic interpretation for the econometric assumptions we impose
when discussing identification and estimation in shift-share designs. These
assumptions imply restrictions on the distribution of labor supply and demand
shocks $F(\cdot)$ introduced in \cref{eq:jointdistshocks}. In \Cref{sec:empapp},
we return to this economic model when interpreting empirical estimates of the
impact of sector-specific labor demand shifters on regional labor market
outcomes (\Cref{sec:ADH}); and the regional inverse labor supply elasticity
(\Cref{sec:estlaborsupply}).

\subsection{From economic model's equilibrium conditions to a potential outcome framework}\label{sec:transition_econometrics}

We build on the results in \Cref{sec:impacteconshocks} to propose a general
framework for the estimation of the impact of shifters on outcomes measured at a
different unit of observation. For concreteness, we refer to the level at which
shifters vary as sectors and to the level at which the outcome varies as
regions.

To make precise what we mean by ``the effect of shifters on an outcome'', we use
the potential outcomes notation, writing $Y_{i}(\xs_{1}, \dotsc, x_{S})$ to
denote the potential (counterfactual) outcome that would occur in region $i$ if
the shocks to the $S$ sectors were exogenously set to $\{\xs_{s}\}_{s=1}^{S}$.
Consistently with \cref{eq:emp_change,eq:wage_change}, we assume that the
potential outcomes are linear in the shocks,

\begin{equation}\label{eq:potential-outcomes}
  Y_{i}(\xs_{1}, \dotsc, \xs_{S})=Y_{i}(0)+\sum_{i=1}^{S}w_{is}\xs_{s}\beta_{is},
  \qquad\text{where}\qquad w_{is}\geq0\;\text{for all $s$},
  \qquad\sum_{s=1}^{S}w_{is}\leq 1,
\end{equation}
and $Y_{i}(0)=Y_{i}(0,\dotsc,0)$ denotes the potential outcome in region $i$
when all shocks $\{\xs_{s}\}_{s=1}^{S}$ are set to zero. Thus, increasing
$\xs_{s}$ by one unit, holding the shocks to the other sectors constant, leads
to an increase in region $i$'s outcome of $w_{is}\beta_{is}$ units. This is the
treatment effect of $\xs_{s}$ on $Y_{i}(\xs_{1}, \dotsc, \xs_{S})$. The actual
(observed) outcome is given by $Y_{i}=Y_{i}(\Xs_{1}, \dotsc, \Xs_{S})$, which
depends on the realization of the shifters, $(\Xs_{1}, \dotsc, \Xs_{S})$.

If the shifters of interest are the sectoral labor demand shocks $\{\hat{\chi}_{s}\}_{s=1}^{S}$,
and the outcome of interest is the employment change $\hat{L}_{i}$, we can map
\cref{eq:emp_change} into \cref{eq:potential-outcomes} by defining
\begin{equation}\label{eq:potential_outcome0_wage}
  Y_{i}=\hat{L}_{i}, \; w_{is}=l_{is}^0,\;
  \xs_{s}=\hat{\chi}_s, \; \beta_{is}=\theta_{is}, \;
  Y_{i}(0)= \lambda_i\sum_{s=1}^{S}
  w_{is}(\hat{\mu}_s + \hat{\eta}_{is}) +
  (1 - \lambda_i)
  (\sum_{g=1}^G \tilde{w}_{ig} \hat{\nu}_g + \hat{\upsilon}_i).
\end{equation}
Observe that $Y_{i}(0)$ aggregates all shifters other than the sectoral shocks
of interest $\{\hat{\chi}_{s}\}_{s=1}^{S}$.\footnote{Given the mapping in
  \cref{eq:potential_outcome0_wage}, the expression in
  \cref{eq:potential-outcomes} captures the first-order impact of the labor
  demand shocks $\{\hat{\chi}_{s}\}_{s=1}^{S}$ on changes in the employment
  rate. We focus on this first-order impact because it helps connecting our analysis to
  linear specifications used extensively in the shift-share
  literature. See \begin{NoHyper}\Cref{app:approximation_error}\end{NoHyper} for a
  discussion of the approximation error arising from the linear specification
  imposed in \cref{eq:emp_change}.}

We are interested in the properties of the OLS estimator $\hat{\beta}$ of the
coefficient on the shift-share regressor $X_{i}=\sum_{s=1}^{S}w_{is}\Xs_{s}$ in
a regression of $Y_{i}$ onto $X_{i}$.\footnote{\label{fn:observable-shifters}We
  assume for now that the shifters $\{\Xs_{s}\}_{s=1}^{S}$ are directly
  observable. In \Cref{sec:mismeasuredIV}, we consider the case in which we only
  observe noisy estimates of these shifters.} To focus on the key conceptual
issues, we abstract away from any additional covariates or controls for now, and
assume that $\Xs_{s}$ and $Y_{i}$ have been demeaned, so that we can omit the
intercept in a regression of $Y_{i}$ on $X_{i}$ (see
\Cref{sec:adding-covariates} for the case with controls). In this simplified setting, the OLS estimator of the coefficient on $X_{i}$ is given by
\begin{equation}\label{eq:OLS-no-covariates}
  \hat{\beta}=\frac{\sum_{i=1}^{\N}X_{i}Y_{i}}{\sum_{i=1}^{\N}X_{i}^{2}},
\end{equation}
and we can write the regression equation as
\begin{equation}\label{eq:bartikreg2}
  Y_{i} = \beta X_{i} + \epsilon_{i},
  \qquad\text{where}\quad
  X_{i}=\sum_{s=1}^{S}w_{is}\Xs_{s}.
\end{equation}

The definition of the estimand $\beta$ in \cref{eq:bartikreg2} and the
properties of the estimator $\hat{\beta}$ will depend on: (a) what is the
population of interest; and (b) how we think about repeated sampling. For
(a), we define the population of interest to be the observed set of $N$ regions,
as opposed to focusing on a large superpopulation of regions from which the $N$
observed regions are drawn. Consequently, we are interested in the parameters
$\{\beta_{is}\}_{i=1,s=1}^{N, S}$ and the treatment effects
$\{w_{is}\beta_{is}\}_{i=1,s=1}^{N, S}$ themselves, rather than the
distributions from which they are drawn, which would be the case if we were
interested in a superpopulation of regions.\footnote{\label{fn:finitepop}Treating the set of
  observed regions as the population of interest is common in applications of
  the shift-share approach. For example, the abstract of
  \citet*{autordornhanson2013} reads: ``We analyze the effect of rising Chinese
  import competition between 1990 and 2007 \textit{on U.S. local labor
    markets}''. Similarly, the abstract of \citet{dixcarneirokovak2017} reads:
  ``We study the evolution of trade liberalization's effects \textit{on
    Brazilian local labor markets}'' (emphases added).} For
(b), given our interest on estimating the \emph{ceteris paribus} impact of a
specific set of shocks $(\Xs_{1}, \dotsc, \Xs_{S})$, we consider repeated
sampling of these shocks, while holding the shares $\{w_{is}\}_{i=1,s=1}^{N, S}$,
the parameters $\{\beta_{is}\}_{i=1,s=1}^{N, S}$, and the potential outcomes
$\{Y_{i}(0)\}_{i=1}^{N}$ fixed.

Given these assumptions, the estimand $\beta$ is defined as the population
analog of \cref{eq:OLS-no-covariates} under repeated sampling of the shocks
$\Xs_{s}$,
\begin{equation}\label{eq:beta-estimand}
\beta=\frac{\sum_{i=1}^{\N}E[{X}_{i}Y_{i}\mid \mathcal{F}_{0}]}{\sum_{i=1}^{\N}E[{X}_{i}^{2}\mid\mathcal{F}_{0}]}, \qquad\text{with}\qquad\mathcal{F}_{0}=\{Y_{i}(0), \beta_{is}, w_{is} \}_{i=1,s=1}^{N, S},
\end{equation}
and, given \cref{eq:potential-outcomes,eq:bartikreg2}, the regression error
$\epsilon_{i}$ is then defined as the residual
\begin{align}\label{eq:epsilon}
\epsilon_{i}=Y_{i}-X_{i}\beta = Y_{i}(0)+\sum_{i=1}^{S}w_{is}\Xs_{s}(\beta_{is}-\beta).
\end{align}

Thus, the statistical properties of the regression residual $\epsilon_{i}$
depend on the properties of the potential outcome $Y_{i}(0)$, the shifters
$\{\Xs_{s}\}_{s=1}^{S}$, the shares $\{w_{is}\}_{s=1}^{S}$, and the difference
between the parameters $\{\beta_{is}\}_{s=1}^{S}$ and the estimand $\beta$.
Importantly, as illustrated in \cref{eq:potential_outcome0_wage}, the potential
outcome $Y_{i}(0)$ will generally incorporate terms that have a shift-share
structure with shares that are either identical to (e.g.\ the term
$\sum_{s=1}^{S} w_{is}\hat{\mu}_s$) or different from but potentially correlated
with (e.g.\ the term $\sum_{g=1}^G \tilde{w}_{ig} \hat{\nu}_g$) the shares
$\{w_{is}\}_{s=1}^{S}$ that define the shift-share regressor $X_{i}$. It then
follows from \cref{eq:epsilon} that the residuals $\epsilon_{i}$ and
$\epsilon_{i'}$ will generally be correlated for any pair of regions $i$ and
$i'$ with similar values of the shift-share regressor.

We summarize this discussion in the following remark.

\begin{remark}\label{rem:corr_residual}
  Correct inference for the coefficient on a shift-share regressor requires
  taking into account potential cross-regional correlation in residuals across
  observations with similar values of the shift-share covariate of interest. One
  possible source of such correlation is the presence in these residuals of
  shift-share components with shares identical to or correlated with those
  entering the covariate of interest.
\end{remark}

\Cref{rem:corr_residual} has important implications for estimating the sampling
variability of $\hat{\beta}$. In particular, traditional inference procedures do
not account for correlation in $\epsilon_i$ among regions with similar shares
and, therefore, tend to underestimate the variability of $\hat{\beta}$. As we
formalize in the next section, this is the main reason for the overrejection
problem described in \Cref{sec:overr-usual-stand}.

\section{Asymptotic properties of shift-share regressions}\label{sec:asympt-prop-bart}

In this section, we formulate the statistical assumptions that we impose on the
data generating process (DGP), use them to derive asymptotic results, and
provide an economic interpretation of these assumptions using the model
introduced in \Cref{sec:model}. In \Cref{sec:no-covariates}, we consider
the case in which there is a single shift-share regressor and no controls. We
account for controls in \Cref{sec:adding-covariates}. In \Cref{sec:ivreg}, we
consider using the shift-share variable as an instrument for a regional
treatment variable. All proofs and technical details are collected in
\begin{NoHyper}\Cref{sec:proofs-econometrics}\end{NoHyper}.

We follow the notation from
\cref{eq:bartikreg} by writing sector-level variables (such as the shifter
$\Xs_{s}$) in script font style and region-level aggregates (such as $X_{i}$) in
normal style. We use standard matrix and vector notation. In particular, for a
(column) $L$-vector $A_{i}$ that varies at the regional level, $A$ denotes the
$\N\times L$ matrix with the $i$th row given by $A_{i}'$. For an $L$-vector
$\mathcal{A}_{s}$ that varies at the sectoral level, $\mathcal{A}$ denotes the
$S\times L$ matrix with the $s$th row given by $\mathcal{A}_{s}'$. If $L=1$,
then ${A}$ and $\mathcal{A}$ are an $\N$-vector and an $S$-vector, respectively.
Let $W$ denote the $\N\times S$ matrix of shares, so that its $(i, s)$ element
is given by $w_{is}$, and let $B$ denote the $\N\times S$ matrix with $(i, s)$
element given by $\beta_{is}$.

\subsection{Simple case without controls}\label{sec:no-covariates}

We focus here on the statistical properties of the OLS estimator $\hat{\beta}$
defined in \cref{eq:OLS-no-covariates}.

\subsubsection*{Assumptions}

We consider large-sample properties of $\hat{\beta}$ as the number sectors goes
to infinity, $S\to\infty$. The assumptions below imply that $N\to\infty$ as
$S\to\infty$. To assess how large $S$ needs to be in order that these
asymptotics provide a good approximation to the finite sample distribution of
$\hat{\beta}$, we conduct a series of placebo simulations in
\Cref{sec:placebo-exercise}. We describe here the main substantive assumptions,
and collect technical regularity conditions in
\begin{NoHyper}\Cref{sec:assumptions}\end{NoHyper}. As in
\cref{eq:beta-estimand}, let $\mathcal{F}_{0}=(Y(0), B, W)$.

\begin{assumption}[Identification]\label{assumption:DGP-noZ}
  \aitem\label{sec:DGPlinear} The observed outcome is given by
  $Y_{i}=Y_{i}(\Xs_{1}, \dotsc, \Xs_{S})$, such that
  \cref{eq:potential-outcomes} holds; %
  \aitem\label{item:random-assignment}~The shifters are as good as randomly
  assigned conditional on $\mathcal{F}_{0}$ in the sense that, for all
  $s=1,\dotsc, S$,
  \begin{equation}
    \label{eq:random-assignment}
    E[\Xs_{s}\mid\mathcal{F}_{0}]=0.
  \end{equation}
\end{assumption}

\Cref{sec:DGPlinear} requires that the potential outcomes are linear in the
shifters $\{\Xs_{s}\}_{s=1}^{S}$. As discussed in
\Cref{sec:transition_econometrics}, one can generate such linear specification
from a first-order approximation of the impact of the shifters
$(\Xs_{1}, \dotsc, \Xs_{S})$ on the outcome $Y_{i}$. This approximation may be
subject to error. In
\begin{NoHyper}\Cref{sec:assumptions}\end{NoHyper}, we generalize
\cref{eq:potential-outcomes} to allow for a linearization error and derive
restrictions on this error under which our inference procedures remain valid.

\Cref{item:random-assignment} imposes that the sectoral shifters
$\Xs$ are mean independent of the shares $W$,
potential outcomes $Y(0)$, and parameters $B$; the assumption that the shifters
are mean zero is a normalization to allow us to drop the intercept; we relax it
in~\Cref{sec:adding-covariates}. This random assignment assumption is a key
assumption for identifying the causal impact of a shift-share covariate; a
version of this assumption has been previously proposed by
\citet*{borusyakhulljaravel2018shiftshare}.

If we are interested in studying the effect of labor demand shifters in the
context of the model in \Cref{sec:model} (i.e.
$\Xs_{s}=\hat{\chi}_{s}$), \Cref{item:random-assignment} will hold if the
shifters $\{\hat{\chi}_{s}\}_{s=1}^{S}$ are mean independent of the other labor
demand shifters, $\{\hat{\mu}_{s}\}_{s=1}^{S}$ and
$\{\hat{\eta}_{is}\}_{i=1,s=1}^{\JN, S}$, and of the labor supply shifters,
$\{\hat{\nu}_{g}\}_{g=1}^{G}$ and $\{\hat{\nu}_{i}\}_{i=1}^{N}$. The plausibility of this
restriction depends on the specific empirical
application. For example, if all $N$ regions in the sample are regions within a
small open economy, $\hat{\chi}_{s}$ denotes changes in
international prices in sector $s$, and $\hat{\mu}_{s}$ denotes
changes in the tariffs that this small open economy charges on its sector $s$
imports; then, \Cref{item:random-assignment} requires these
changes in tariffs to be independent of the changes in tariffs in any country that is large enough for their tariff changes to affect international prices
(see \begin{NoHyper}\Cref{appsec:sector_region_employment_small}\end{NoHyper}
for additional details).

\begin{assumption}[Consistency and Inference]\label{assumption:inference}
  \aitem\label{item:indepX} The shifters $(\Xs_{1}, \dotsc, \Xs_{S})$ are
  independent conditional on $\mathcal{F}_{0}$; %
  \aitem\label{item:sector-size} $\max_{s}n_{s}/\sum_{t=1}^{S}n_{t}\to 0$, where
  $n_{s}=\sum_{s=1}^{S}w_{is}$ denotes the total share of sector $s$;
  \aitem\label{item:sector-size-normality}
  $\max_{s} n_{s}^{2}/\sum_{t=1}^{S}n_{t}^{2}\to 0$.
\end{assumption}

\Cref{item:indepX} requires the shifters to be independent. It adapts to our
setting the assumption underlying randomization-style inference in randomized
controlled trials that the treatment assignment is independent across entities
\citep[see][for a review]{imbens_causal_2015}. An independence or a weak
dependence assumption of this type is generally necessary in order to do
inference.\footnote{For example, for inference on average treatment effects,
  which is commonly the goal when running a regression, one typically assumes
  that the sample is a random sample from the population of interest and, thus,
  that the treatment variable is independent across the individuals in the
  sample.} One could alternatively impose assumptions on the correlation
structure of the regression residuals, either by imposing a particular structure
on them, as in the literature on interactive fixed effects
\citep[e.g.][]{gobillonmagnac2016}, or by imposing a distance metric on the
observations, as in the spatial econometrics literature
\citep[e.g.][]{conley_gmm_1999}. However, as the economic model in
\Cref{sec:model} shows, the structure of the residuals may be very
complex. %
The residuals may include potentially correlated region-specific terms as well
as several shift-share terms, which may or may not use the same shares as the
covariate of interest $X_{i}$. It is thus difficult to conceptualize which exact
restriction on their joint distribution one should impose.

By instead imposing restrictions on the distribution of the vector of shifters
$(\Xs_{1}, \dotsc, \Xs_{S})$ conditional on $\mathcal{F}_{0}=(Y(0), B, W)$,
\Cref{item:indepX} ensures that the standard errors we derive remain valid under
\emph{any} dependence structure between the shares $w_{is}$ across sectors and
regions, and under \emph{any} correlation structure of the potential outcomes
$Y_{i}(0)$ or, equivalently, of the regression errors $\epsilon_{i}$, across
regions.\footnote{\label{fn:conditional}Since our inference is valid conditional
  on $\{\epsilon_{i}\}_{i=1}^{N}$, it accounts for any correlation structure
  they may have, including spatial, or, in applications with multiple periods,
  temporal correlations. See \Cref{sec:panel} for settings with multiple
  periods.} We thus do not have to worry about correctly specifying this
correlation structure, as one would under the alternative approaches mentioned
above. Our approach allows (but does not require) the residual to have a
shift-share structure; it similarly allows all $\{w_{is}\}_{i=1,s=1}^{N, S}$ to
be equilibrium objects responding to the same economic shocks, and thus be
correlated across regions and sectors.\footnote{This conceptualization of all
  the shares $w_{is}$ as equilibrium objects that respond (at least partly) to
  the same set of shocks is consistent with the model in \Cref{sec:model}. As
  shown in \cref{eq:potential_outcome0_wage}, each share $w_{is}$ corresponds to
  the share of workers in region $i$ employed in sector $s$ in an initial
  equilibrium, $l^{0}_{is}$. Furthermore, each of these initial employment
  shares will be a function of the same sector-specific demand shocks and
  group-specific labor supply shocks; consequently $l^{0}_{is}$ will generally
  be correlated with $l^{0}_{i's'}$ even for $i\neq i'$ and $s\neq s'$.} In
\Cref{sec:clustersectors}, we relax \Cref{item:indepX} and allow for a non-zero
correlation in the shifters $(\Xs_{1}, \dotsc, \Xs_{S})$ within clusters of
sectors; we only require that the shifters are independent across the clusters.
Additionally, in the context of the empirical application in
\Cref{sec:ADH}, we discuss how to perform inference in a setting in which all
shifters of interest are generated by a common shock that has heterogeneous
effects across sectors.

In the economic model in \Cref{sec:model}, if $\Xs_{s}=\hat{\chi}_{s}$ and we
interpret these shocks as, for example, sector-specific productivity shocks,
\Cref{item:indepX} requires that there is no common component driving the
changes in sectoral productivities. Our approach does not
require the shifters $\{\Xs_{s}\}_{s=1}^{S}$ to be identically distributed; we allow, for
example, the variance of the shock to differ across sectors.

\Cref{item:sector-size,item:sector-size-normality} are our main regularity
conditions.\footnote{\label{fn:rotemberg}In the context of a shift-share
  instrumental variables regression, \citet*{goldsmith2018bartik} discuss
  similar conditions stated in terms of Rotemberg weights. This is convenient
  under the baseline assumption considered in \citet*{goldsmith2018bartik} that
  the vector of shares $(w_{i1}, \dotsc, w_{iS})$ is exogenous, because the
  Rotemberg weights determine the asymptotic bias of the estimator under local
  failures of this exogeneity condition. Since we do not assume exogeneity of
  the shares, this interpretation is not available under our setup.}
\Cref{item:sector-size} is needed for consistency: it requires that the size of
each sector, $n_{s}$, is asymptotically negligible. This assumption is analogous
to the standard consistency condition in the clustering literature that the
largest cluster be asymptotically negligible. To see the connection, consider
the special case with ``concentrated sectors'', in which each region $i$
specializes in one sector $s(i)$; i.e. $w_{is}=1$ if $s=s(i)$ and $w_{is}=0$
otherwise, and $n_{s}$ is thus the number of regions that specialize in sector
$s$. In this case, $X_{i}=\Xs_{s(i)}$, so that, if \cref{eq:random-assignment}
holds, $\hat{\beta}$ is equivalent to an OLS estimator in a randomized
controlled trial in which the treatment varies at a cluster level; here the
$s$th cluster consists of regions that specialize in sector $s$. The condition
$\max_{s}n_{s}/\sum_{t=1}^{S}n_{t}\to 0$ then reduces to the assumption that the
largest cluster be asymptotically negligible. \Cref{item:sector-size-normality}
is needed for asymptotic normality---it ensures that the Lindeberg condition
holds. It strengthens~\Cref{item:sector-size} slightly by requiring that the
contribution of each sector to the asymptotic variance is asymptotically
negligible; otherwise the estimator will not generally be asymptotically normal,
even if it is consistent.%

In terms of the economic model introduced in \Cref{sec:model},
\Cref{item:sector-size,item:sector-size-normality} require that no sector
dominates the rest in terms of initial employment at the national level; i.e.
$\sum_{i=1}^{N}l^{0}_{is}$ is not too large for any sector.
\Cref{sec:placebo-baseline} shows that this assumption is reasonable for the
U.S. if the $S$ sectors used to construct the treatment of interest $X_{i}$
correspond to the 396 4-digit manufacturing sectors (see \Cref{sec:data}). In
\Cref{sec:altenative_placebo}, we illustrate the consequences of the failure of
this assumption due to the inclusion of a large aggregate sector, the
non-manufacturing sector, in $X_{i}$.

\subsubsection*{Asymptotic theory}

We now establish that the OLS estimator in \cref{eq:OLS-no-covariates} is
consistent and asymptotically normal.

\begin{proposition}\label{theorem:consistency-noZ}
  Suppose \Cref{assumption:DGP-noZ}, \Cref{item:indepX,item:sector-size}, and
  \begin{NoHyper}\Cref{sitem:beta-bounded,sitem:detX,sitem:2nu} in
    \Cref{sec:assumptions}\end{NoHyper} hold. Then
  \begin{equation}\label{eq:beta-noZ}
    \beta=\frac{\sum_{i=1}^{\N}\sum_{s=1}^{S}\pi_{is}\beta_{is}}{
      \sum_{i=1}^{\N}\sum_{s=1}^{S}\pi_{is}}, \qquad \text{and}\qquad \hat{\beta}=\beta+o_{p}(1),
  \end{equation}
  where $\pi_{is}=w_{is}^{2}\var(\Xs_{s}\mid \mathcal{F}_{0})$.
\end{proposition}

This \namecref{theorem:consistency-noZ} gives two results. First, it shows that
the estimand $\beta$ in \cref{eq:beta-estimand} can be expressed as a weighted
average of the region- and sector-specific parameters
$\{\beta_{is}\}_{i=1,s=1}^{N, S}$, with the weight $\pi_{is}$ increasing in the
share $w_{is}$ and in the conditional variance of the shifter
$\var(\Xs_{s}\mid \mathcal{F}_{0})$. Second, it states that the OLS estimator
$\hat{\beta}$ converges to this estimand as $S\to\infty$. The special case with
concentrated sectors is again useful in interpreting
\Cref{theorem:consistency-noZ}. In this case,
$\sum_{s=1}^{S}\pi_{is}\beta_{is}=\var(\Xs_{s(i)}\mid \mathcal{F}_{0})\beta_{is(i)}$ and,
therefore, the first result in \Cref{theorem:consistency-noZ} reduces to the
standard result from the randomized controlled trials literature with
cluster-level randomization (with each ``cluster'' defined as all regions
specialized in the same sector) that the weights are proportional to the
variance of the shock.

The estimand $\beta$ \emph{does not} in general equal a weighted average of the
heterogeneous treatment effects. As discussed in
\Cref{sec:transition_econometrics}, the effect on the outcome in region $i$ of
increasing the value of the sector $s$ shock in one unit is equal to
$w_{is}\beta_{is}$; weighting this effect using a set of region- and
sector-specific weights $\{\xi_{is}\}_{i=1,s=1}^{N, S}$, yields the weighted
average treatment effect
\begin{equation*}
  \tau_{\xi}=\frac{\sum_{i=1}^{N}\sum_{s=1}^{S}\xi_{is}w_{is}\beta_{is}}{\sum_{i=1}^{N}\sum_{s=1}^{S}\xi_{is}}.
\end{equation*}
Alternatively, the total effect of increasing the shifters simultaneously in
every sector by one unit is $\sum_{s=1}^{S}w_{is}\beta_{is}$; weighting it using
a set of region-specific weights $\{\zeta_{i}\}_{i=1}^{\N}$ yields the weighted
total treatment effect
$\tau^{T}_{\zeta}=\sum_{i=1}^{\N}\zeta_{i}\sum_{s=1}^{S}w_{is}\beta_{is}/\sum_{i=1}^{\N}\zeta_{i}$.
If $\beta_{is}$ is constant across $i$ and $s$, then
$\beta=\tau^{T}_{\zeta}$, provided $\sum_{s=1}^{S}w_{is}=1$ in every region $i$; otherwise, we can
consistently estimate $\tau^{T}_{\zeta}$ by
$\hat{\beta}\cdot
\sum_{i=1}^{\N}\zeta_{i}\sum_{s=1}^{S}w_{is}/\sum_{i=1}^{\N}\zeta_{i}$.
Similarly, if $\beta_{is}$ is constant across $i$ and $s$, $\tau_{\xi}$ is consistently estimated by
$\hat{\beta}\cdot\sum_{i=1}^{N}\sum_{s=1}^{S}\xi_{is}w_{is} /
\sum_{i=1}^{N}\sum_{s=1}^{S}\xi_{is}$. On the other hand, if $\beta_{is}$ varies
across regions and sectors, then it is not clear in general how to exploit
knowledge of the estimand $\beta$ defined in \cref{eq:beta-noZ} to learn
something about $\tau_{\xi}$ or $\tau^{T}_{\zeta}$. A special case in which it
is possible to consistently estimate $\tau_{\xi}$ even if $\beta_{is}$ varies
across $i$ or $s$ arises when $\Xs_{s}$ is homoskedastic,
$\var(\Xs_{s}\mid \mathcal{F}_{0})=\sigma^{2}$, and $\xi_{is}=w_{is}$; in this
case, a consistent estimate of $\tau_{\xi}$ is given by
$\hat{\beta}\sum_{i=1}^{\N}\sum_{s=1}^{S}w_{is}^{2}/\sum_{i=1}^{\N}\sum_{s=1}^{S}w_{is}$.\footnote{In
  general, one can consistently estimate $\tau_{\xi}$ or $\tau^{T}_{\zeta}$ by
  imposing a mapping between $\beta_{is}$ and structural parameters, and
  obtaining consistent estimates of these structural parameters. However, since
  this mapping will vary across models, the consistency of such estimator will
  not be robust to alternative modeling assumptions, even if all these
  assumptions predict an equilibrium relationship like that in
  \cref{eq:emp_change}; e.g.\ see
  \begin{NoHyper}\Cref{app:general_model,,appsec:specific,appsec:specific_pref}\end{NoHyper}
  for examples of this mapping in different models.}

\begin{proposition}\label{theorem:normality-noZ}
  Suppose \Cref{assumption:DGP-noZ,assumption:inference}, and
  \begin{NoHyper}\Cref{assumption:regularity} in \Cref{sec:assumptions}\end{NoHyper}
  hold. Suppose also that
  \begin{equation*}
    \mathcal{V}_{\N}=\frac{1}{\sum_{s=1}^{S}n_{s}^{2}}\var\left(\sum_{i=1}^{\N}X_{i}\epsilon_{i}\mid
      \mathcal{F}_{0}\right)
  \end{equation*}
  converges in probability to a non-random limit. Then
  \begin{equation*}
    \frac{\N}{\sqrt{\sum_{s=1}^{S}n_{s}^{2}}}(\hat{\beta}-\beta)
    =\mathcal{\N}\left(0,\frac{\mathcal{V}_{\N}}{\left(\frac{1}{\N}\sum_{i=1}^{\N}X_{i}^{2}\right)^{2}}\right)+o_{p}(1).
  \end{equation*}
\end{proposition}

This proposition shows that $\hat{\beta}$ is asymptotically normal, with a rate
of convergence equal to $\N(\sum_{s=1}^{S}n_{s}^{2})^{-1/2}$. If all sector
sizes $n_{s}$ are of the order $\N/S$, the rate of convergence equals
$\sqrt{S}$. However, if the sizes are unequal, the rate may be slower.

According to \Cref{theorem:normality-noZ}, the asymptotic variance formula has
the usual ``sandwich'' form. Since $X_{i}$ is observed, to construct a
consistent standard error estimate, it suffices to construct a consistent
estimate of $\mathcal{V}_{\N}$, the middle part of the sandwich. To motivate our
standard error formula, suppose that $\beta_{is}$ is constant across $i$ and
$s$, $\beta_{is}=\beta$. Then it follows from \cref{eq:random-assignment}
and \Cref{item:indepX} that
\begin{equation}\label{eq:Vn-constantbeta-noZ}
  \mathcal{V}_{\N}=\frac{\sum_{s=1}^{S}\var(\Xs_{s}\mid \mathcal{F}_{0})
    R_{s}^{2}}{\sum_{s=1}^{S}n_{s}^{2}}, \qquad R_{s}=\sum_{i=1}^{\N}w_{is}\epsilon_{i}.
\end{equation}
Replacing $\var(\Xs_{s}\mid \mathcal{F}_{0})$ by $\Xs_{s}^{2}$, and
$\epsilon_{i}$ by the regression residual
$\hat{\epsilon}_{i}=Y_{i}-X_{i}\hat{\beta}$, we obtain the estimate
\begin{align}\label{eq:se-constantbeta-noZ}
  \widehat{V}_{AKM}(\hat{\beta})=\frac{\hat{\mathcal{V}}_{AKM}(\hat{\beta})}{\left(\sum_{i=1}^{\N}X_{i}^{2}\right)^2}, \qquad
  \hat{\mathcal{V}}_{AKM}(\hat{\beta})=\sum_{s=1}^{S}\Xs_{s}^{2}
  \hat{R}_{s}^{2}, \qquad
  \hat{R}_{s}=\sum_{i=1}^{\N}w_{is}\hat{\epsilon}_{i}.
\end{align}
When $\beta_{is}=\beta$, we show formally that this variance estimate leads to
valid inference under regularity conditions in~\Cref{sec:adding-covariates}. In
\begin{NoHyper}\Cref{sec:infer-under-heter}\end{NoHyper}
we show that this variance estimate remains valid
under heterogeneous $\beta_{is}$ under further regularity conditions.

To gain intuition for the variance estimate in \cref{eq:se-constantbeta-noZ},
consider the case with concentrated sectors. Then the numerator in
\cref{eq:se-constantbeta-noZ} becomes
$\sum_{s=1}^{S}\Xs_{s}^{2}\hat{R}_{s}^{2}=\sum_{s=1}^{S}(\sum_{i=1}^{\N}\1{s(i)=s}X_{i}\hat{\epsilon}_{i})^{2}$,
so that \cref{eq:se-constantbeta-noZ} reduces to the cluster-robust variance
estimate that clusters on the sector that each region is specialized. This is consistent with the rule of thumb that one should ``cluster'' at the level of variation of
the regressor of interest. More generally, the variance estimate essentially
forms sectoral clusters with variance that depends on the variance of
$\hat{R}_{s}$, a weighted sum of the regression residuals
$\{\hat{\epsilon}_{i}\}_{i=1}^{N}$, with weights that correspond to the shares
$\{w_{is}\}_{i=1}^{N}$. An important advantage of
$\widehat{V}_{AKM}(\hat{\beta})$ is that it allows for an arbitrary structure of
cross-regional correlation in residuals:

\begin{remark}\label{remark:fixed-epsilon}
  In the expression for $\mathcal{V}_{\N}$ in \cref{eq:Vn-constantbeta-noZ}, the
  expectation is only taken over $\{\Xs_{s}\}_{s=1}^{S}$---we do not take any expectation over
  the shares $\{w_{is}\}_{i=1,s=1}^{N, S}$ or the residuals
  $\{\epsilon_{i}\}_{i=1}^{N}$. This is because our inference is conditional on
  the realized values of the shares and on the potential outcomes
  $\{Y_{i}(0)\}_{i=1}^{N}$. In terms of the regression in \cref{eq:bartikreg2},
  this means that we consider properties of $\hat{\beta}$ under repeated
  sampling of $X_{i}=\sum_{s=1}^{S}w_{is}\Xs_{s}$ conditional on the shares
  $\{w_{is}\}_{i=1,s=1}^{N, S}$ and on the residuals
  $\{\epsilon_{i}\}_{i=1}^{N}$ (as opposed to, say, considering properties of
  $\hat{\beta}$ under repeated sampling of the residuals conditional on
  $\{X_{i}\}_{i=1}^{N}$). As a result, our inference method allows for arbitrary
  dependence between the residuals $\{\epsilon_{i}\}_{i=1}^{N}$.
\end{remark}

To understand the source of the overrejection problem discussed in
\Cref{sec:overr-usual-stand}, let us compare the variance estimate
$\widehat{V}_{AKM}(\beta)$ with the cluster-robust variance estimate when the
residuals $\hat{\epsilon}_{i}$ are computed at the true $\beta$ (so that
$\hat{\epsilon}_{i}=\epsilon_{i}$). These variance estimates differ in the
middle sandwich, with the cluster-robust estimate replacing
$\hat{\mathcal{V}}_{AKM}({\beta})$ in \cref{eq:se-constantbeta-noZ} with
$\hat{\mathcal{V}}_{CL}({\beta})=\sum_{i=1}^{N}\sum_{j=1}^{N}\1{c(i)=c(j)}X_{i}X_{j}{\epsilon}_{i}{\epsilon}_{j}$,
where $c(i)$ denotes the cluster that region $i$ belongs to (the comparison with
heteroskedasticity-robust standard errors obtains as a special case if $c(i)=i$,
so that each region belongs to its own cluster). Assuming for simplicity that
the conditional variance of $\Xs_{s}$ does not depend on $Y(0)$, it follows by
simple algebra that the expectation of the difference between these terms is
given by
\begin{equation}\label{eq:Vakm_Vcl}
  E[\hat{\mathcal{V}}_{AKM}(\beta)-\hat{\mathcal{V}}_{CL}(\beta)\mid W]=\sum_{s=1}^{S}\var(\Xs_{s}\mid W)\sum_{i=1}^{N}\sum_{j=1}^{N}\1{c(i)\neq c(j)}
  w_{is}w_{js}E[\epsilon_{i}\epsilon_{j}\mid W].
\end{equation}
This expression is non-negative so long as the correlation between the residuals
is non-negative. The magnitude of the difference will be large if regions
located in different clusters (so that $c(i)\neq c(j)$) that have similar shares
(i.e.\ large values of $\sum_{s=1}^{S}w_{is}w_{js}$) also tend to have similar
residuals (i.e.\ large values of $E[\epsilon_{i}\epsilon_{j}\mid W]$). For
illustration, consider a simplified version of the model described in
\Cref{sec:model} in which: (a) $\sigma_{s}\geq 0$ for all $s$ and $\phi\geq 0$,
so that $0\leq \lambda_{i}\leq 1$; (b) region-specific labor demand and supply
shocks $\{\hat{\eta}_{is}\}_{s=1}^{S}$ and $\hat{\nu}_{i}$ are independent
across regions; and (c) all labor demand and supply shocks are independent of
each other. Then, it follows from \cref{eq:potential_outcome0_wage,eq:epsilon}
that, for any $i\neq j$,
\begin{equation}\label{eq:Eepseps}
E[\epsilon_{i}\epsilon_{j}\mid W, \tilde{W}]
=
  \lambda_{i}\lambda_{j}\sum_{s=1}^{S}
  w_{is}w_{js}E[\hat{\mu}_{s}^{2}\mid W, \tilde{W}]
 +
  (1 - \lambda_{i})(1 - \lambda_{j})
  \sum_{g=1}^G \tilde{w}_{jg}\tilde{w}_{ig} E[\hat{\nu}_{g}^{2}\mid
  W, \tilde{W}]\geq 0,
\end{equation}
which by the law of iterated expectations implies that
$E[\hat{\mathcal{V}}_{AKM}(\beta)-\hat{\mathcal{V}}_{CL}(\beta)\mid W]\geq 0$. This expression illustrates that regions with similar shares will tend
to have similar residuals in two cases. First, if the variance of the unobserved
shifter $\hat{\mu}_{s}$ is large, so that
$E[\hat{\mu}_{s}^{2}\mid W, \tilde{W}]$ is large. In other words, standard
inference methods lead to overrejection if the residual contains important
shift-share terms that affect the outcome of interest through the same shares
$\{w_{is}\}_{s=1}^{S}$ as those defining the covariate of interest $X_{i}$.
Second, if the variance of the unobserved shifter $\hat{\nu}_{g}$ is
large, so that $E[\hat{\nu}_{g}^{2} \mid W, \tilde{W}]$ is large, and the shares
$\tilde{w}_{ig}$ through which these shifters affect the outcome variable have a
correlation structure that is similar to that of $w_{is}$ (so that
$\sum_{g=1}^{G}\tilde{w}_{ig}\tilde{w}_{jg}$ is large whenever
$\sum_{s=1}^{S}w_{is}w_{js}$ is large). Thus, standard inference
methods may overreject even when the unobserved shifters contained in the
residual vary along a different dimension than the shift-share covariate of
interest.

\subsection{General case with controls}\label{sec:adding-covariates}

We now study the properties of the OLS estimator $\hat{\beta}$ of the
coefficient on $X_{i}$ in a regression of $Y_{i}$ onto $X_{i}$ and a $K$-vector
of controls $Z_{i}$. To this end, let $Z$ denote the $\N\times K$ matrix with
$i$-th row given by $Z_{i}'=(Z_{i1}, \dotsc, Z_{iK})$, and let
$\ddot{X}=X-Z(Z'Z)^{-1}Z'X$ denote an $N$-vector with $i$-th element equal
to the regressor $X_{i}$ with the controls $Z_{i}$ partialled out (i.e.\ the
residual from regressing $X_{i}$ onto $Z_{i}$). Then, by the Frisch–Waugh–Lovell
theorem, $\hat{\beta}$ can be written as
\begin{equation}\label{eq:OLS-adding-covariates}
  \hat{\beta}=\frac{\sum_{i=1}^{\N}\ddot{X}_{i}Y_{i}}{\sum_{i=1}^{\N}\ddot{X}_{i}^{2}}=
\frac{\ddot{X}'Y}{\ddot{X}'\ddot{X}}.
\end{equation}

The controls may play two roles. First, they may be included to increase the
precision of $\hat{\beta}$. Second, and more importantly, they may be included
because one may worry that the shifters $\{\Xs_{s}\}_{s=1}^{S}$ are correlated
with the potential outcomes $\{Y_{i}(0)\}_{i=1}^{N}$, violating~\Cref{item:random-assignment}. To formalize
how $Z_{i}$, a regional variable, may be a control variable for the shifters,
which vary at a sectoral level, we project $Z_{i}$ onto the sectoral space using
the same shares as those defining the shift-share regressor $X_{i}$,
\begin{equation}\label{eq:Z-proxy}
Z_{i}=\sum_{s=1}^{S}w_{is}\Zs_{s}+U_{i}.
\end{equation}
We think of $\{\Zs_{s}\}_{s=1}^{S}$ as latent sector-level shocks that may have
an independent effect on the outcome $Y$ and may also be correlated with the
shifters $\{\Xs_{s}\}_{s=1}^{S}$, with $U_{i}$, the residual in this projection,
mean-independent of the shifters. If the $k$th control $Z_{ik}$ is included for
precision, then the sector-level shocks $\{\Zs_{sk}\}_{s=1}^{S}$ and, thus,
$Z_{ik}$, are uncorrelated with $X_{i}$. If $Z_{ik}$ is included because one
worries that otherwise $X_{i}$ may not be as good as randomly assigned, we
interpret $Z_{ik}$ as a proxy for the confounding sector-level shocks
$\{\Zs_{sk}\}_{s=1}^{S}$, and think of $U_{ik}$ as a measurement error in this
proxy.

To make this concrete, consider the model in \Cref{sec:model}, with the equivalences in \cref{eq:potential_outcome0_wage}. Then we may include
$Z_{ik}=\sum_{s=1}^{S}l^{0}_{is}\hat{\mu}_{s}$ as a control. Here the measurement
error in \cref{eq:Z-proxy} is zero, and $\Zs_{sk}=\hat{\mu}_{s}$. If the shifters
$\{\hat{\chi}_{s}\}_{s=1}^{S}$ are correlated with the demand shocks $\{\hat{\mu}_{s}\}_{s=1}^{S}$,
then not including this control will generate omitted variable bias.
Alternatively, we may include $Z_{ik}=\sum_{s=1}^{S}w_{is} \hat{\eta}_{is}$ as a
control. Here $\Zs_{sk}=0$, and $U_{ik}=Z_{ik}$ is a regional aggregation of
idiosyncratic region- and sector-specific labor-demand shocks that are
independent of $\Xs_{s}$. In this case, if the shifters
$\{\hat{\chi}_{s}\}_{s=1}^{S}$ are independent of the demand shocks $\{\eta_{is}\}_{i=1,s=1}^{N, S}$, then including the control will help increase
the precision of $\hat{\beta}$, but it is not necessary for consistency.

\subsubsection*{Assumptions}

For clarity of exposition, we focus here on the main substantive assumptions
and relegate technical regularity conditions to
\begin{NoHyper}\Cref{sec:assumptions}\end{NoHyper}. Let
$\mathcal{F}_{0}=(Y(0), W, B, \Zs, U)$; without controls, this set of variables
reduces to $(Y(0), B, W)$, as in \Cref{sec:no-covariates}. Here, $\Zs$ denotes
the $S\times K$ matrix with $s$th row given by $\Zs_{s}'$, and $U$ denotes the
$N\times K$ matrix with $i$-th element given by $U_{i}'$.

We maintain \Cref{assumption:inference} with
$\mathcal{F}_{0}=(Y(0), W, B, \Zs, U)$. The inclusion of controls allows us to
weaken \Cref{assumption:DGP-noZ} and instead impose the following identification
assumption:
\begin{assumption}[Identification with controls]\label{assumption:DGP-z}
  \aitem\label{item:DGP-L} The observed outcome satisfies
  $Y_{i}=Y_{i}(\Xs_{1}, \dotsc, \Xs_{S})$, such that \cref{eq:potential-outcomes}
  holds, and the controls $Z_{i}$ satisfy \cref{eq:Z-proxy}; %
  \aitem\label{item:samplingZ} The shifters are as good as randomly assigned in
  the sense that, for every $s$,
  \begin{equation}\label{eq:random-assignment-Zcond}
    E[\Xs_{s}\mid \mathcal{F}_{0}]=E[\Xs_{s}\mid \Zs_{s}],
  \end{equation}
  and the right-hand side is linear in $\Zs_{s}$,
  \begin{equation}\label{eq:linear-controls}
    E[\Xs_{s} \mid \Zs_{s}]=\Zs_{s}'\gamma;
  \end{equation}
  \aitem\label{item:ugamma-cons} For elements $k$ such that $\gamma_{k}\neq 0$,
  $\N^{-1}\sum_{i=1}^{N}E[U_{ik}^{2}]\to 0$; %
  \aitem\label{item:ugamma-norm} For elements $k$ such that $\gamma_{k}\neq 0$,
  $(\sum_{s=1}^{S}n_{s}^{2})^{-1/2}\sum_{i=1}^{N}E[U_{ik}^{2}]\to 0$.
\end{assumption}

\Cref{item:samplingZ} weakens \Cref{item:random-assignment} by only requiring
the shifters to be as good as randomly assigned conditional on $\Zs$, in the
sense that \cref{eq:random-assignment-Zcond} holds. To
interpret this restriction, consider a projection of the
regional potential outcomes onto the sectoral space. For simplicity, consider
the case with constant effects, $\beta_{is}=\beta$ for all $i$ and $s$, and project $Y_{i}(0)$ onto the
shares $(w_{i1}, \dotsc, w_{iS})$, so that we may write
$Y_{i}(0)=\sum_{s=1}^{S}w_{is}\mathcal{Y}_{s}(0)+\kappa_{i}$. Then,
\cref{eq:random-assignment-Zcond} holds if (i) $\mathcal{Y}_{s}(0)$ is spanned
by the vector of controls $\Zs_{s}$; and (ii) $\{\Xs_{s}\}_{s=1}^{S}$ is mean-independent of
the projection residuals $\{\kappa_{i}\}_{i=1}^{N}$.

As an example, consider again the model in \Cref{sec:model}, with the outcomes
$Y_{i}$ generated by \cref{eq:potential_outcome0_wage}. Then
\cref{eq:random-assignment-Zcond} holds, for example, if we set
$\mathcal{Z}_{s}=\mathcal{Y}_{s}(0)=\hat{\mu}_{s}$ and if, conditional on the
sector-specific labor demand shocks $\{\hat{\mu}_{s}\}_{s=1}^{S}$, the shifters
of interest $\{\hat{\chi}_{s}\}_{s=1}^{S}$ are mean independent of the sector-
and region-specific labor demand shocks $\{\hat{\eta}_{is}\}_{i=1,s=1}^{N, S}$
and of the labor supply shocks $\{\hat{\nu}_{g}\}_{g=1}^{G}$ and $\{\hat{\nu}_{i}\}_{i=1}^{N}$. Suppose, for instance, the shocks of interest
$\{\hat{\chi}_{s}\}_{s=1}^{S}$ are changes in tariffs
\citep[e.g.][]{kovak2013regional} and that other potential labor demand shocks
are those induced by automation and robots \citep[e.g][]{acemoglu2017robots}.
Splitting the impact of automation into nationwide sector-specific effects, as
captured by $\{\hat{\mu}_{s}\}_{s=1}^{S}$, and sector- and region-specific
deviations from the nationwide effects, as captured by
$\{\hat{\eta}_{is}\}_{i=1,s=1}^{N, S}$, \cref{eq:random-assignment-Zcond} allows
the political entity responsible for setting the tariffs to do so influenced by
the nationwide sector-specific effects of automation, but not by any
region-specific deviation from those national effects. In contrast,
\Cref{item:random-assignment} would require that the tariffs are also
independent of the nationwide effects of automation.

Under \cref{eq:random-assignment-Zcond}, one generally needs to include the
controls non-parametrically; by imposing \cref{eq:linear-controls}, we ensure
that it suffices to include the controls as additional covariates in a linear
regression. If the shifters $\Xs_{s}$ are not mean zero (in the sense that the
regression intercept on the right-hand side of \cref{eq:linear-controls} is
non-zero), \cref{eq:linear-controls} requires that we include a constant
$\Zs_{sk}=1$ as one of the controls. If the shares sum to one,
$\sum_{s=1}^{S}w_{is}=1$, this amounts to including an intercept $Z_{ik}=1$ as a
control in the regression. Importantly, if the shares do not sum to one, this
amounts to including $\sum_{s=1}^{S}w_{is}$ as a control \citep*[see][for a more
extensive discussion of this point]{borusyakhulljaravel2018shiftshare}. For
instance, if the shares $w_{is}$ correspond to labor shares in different
manufacturing sectors, one needs to include the size of the manufacturing sector
$\sum_{s=1}^{S}w_{is}$ in each region as a control.

Given \Cref{item:samplingZ}, if we observed $\{\Zs_{s}\}_{s=1}^{S}$ directly, we could include
the vector $Z_{i}^{*}=\sum_{s=1}^{S}w_{is}\Zs_{s}$ directly as control. However, the definition
of each regional control $Z_{i}$ in \cref{eq:Z-proxy} allows for $Z_{i}^{*}$ to
be observed with measurement error $U_{i}$. If $\gamma_{k}=0$, such as when
$Z_{ik}$ is included for precision, then this measurement error in $Z_{ik}^{*}$
does not matter; if $\gamma_{k}\neq0$, this measurement error will in general induce a bias in
$\hat{\beta}$. This is analogous to the classic linear regression result that
measurement error in a control variable generally leads to a bias in the estimate of the
coefficient on the variable of interest. \Cref{item:ugamma-cons} ensures that
any such bias disappears in large samples by imposing that the variance of the
measurement error for controls that matter (i.e.\ those with $\gamma_{k}\neq 0$)
converges to zero as $S\to\infty$. This ensures consistency of $\hat{\beta}$.
For asymptotic normality, we need to strengthen this condition in
\Cref{item:ugamma-norm} by requiring that the variance of the measurement error
converges to zero sufficiently fast. \Cref{item:ugamma-norm} holds, for instance,
if $U_{i}=S^{-1}\sum_{s=1}^{S}\psi_{is}$, where $\psi_{is}$ is an idiosyncratic
measurement error that is independent across $s$. In intuitive terms, this
condition guarantees that $Z_{i}$ is a sufficiently good proxy for the
confounding latent shocks $\{\Zs_{s}\}_{s=1}^{S}$.

\subsubsection*{Asymptotic theory}

The following result generalizes \Cref{theorem:consistency-noZ}:

\begin{proposition}\label{theorem:consistency-Z}
  Suppose \Cref{item:indepX,item:sector-size} and
  \begin{NoHyper}\Cref{sitem:beta-bounded,sitem:detX,sitem:2nu} in \Cref{sec:assumptions}\end{NoHyper}
  hold with $\mathcal{F}_{0}=(\mathcal{Z}, U, Y(0), B, W)$. Suppose also that
  \Cref{item:DGP-L,item:samplingZ,item:ugamma-cons} and
  \begin{NoHyper}\Cref{sitem:detZZ,sitem:UZ-moments} in
    \Cref{sec:assumptions}\end{NoHyper} hold. Then
  \begin{equation}\label{eq:beta-Z}
    \beta=\frac{\sum_{i=1}^{\N}\sum_{s=1}^{S}\pi_{is}\beta_{is}}{
      \sum_{i=1}^{\N}\sum_{s=1}^{S}\pi_{is}}, \qquad\text{and}\qquad\hat{\beta}=\beta+o_{p}(1),
  \end{equation}
  where $\pi_{is}=w_{is}^{2}\var(\Xs_{s}\mid \mathcal{F}_{0})$.
\end{proposition}
The only difference in the characterization of the probability limit relative to
\Cref{theorem:consistency-noZ} is that the weights $\pi_{is}$ now reflect
the variance of $\Xs_{s}$ that also conditions on the controls.

To state the asymptotic normality result, define
$\delta=E[Z'Z]^{-1}E[Z'(Y-X\beta)]$, so that we can define
the regression residual in~\cref{eq:bartikreg} as
$\epsilon_{i}=Y_{i}-X_{i}\beta-Z_{i}'\delta$.
\begin{proposition}\label{theorem:normality-Z}
  Suppose \Cref{assumption:inference,assumption:DGP-z} and
  \begin{NoHyper}\Cref{assumption:regularity,assumption:regularity-Z} in
    \Cref{sec:assumptions}\end{NoHyper} hold with
  $\mathcal{F}_{0}=(\mathcal{Z}, U, Y(0), B, W)$. Suppose, in
  addition, that
  \begin{equation*}
    \mathcal{V}_{\N}=\frac{1}{\sum_{s=1}^{S}n_{s}^{2}}\var\left(\sum_{i=1}^{N}(X_{i}-Z_{i}'\gamma)\epsilon_{i}\mid
      \mathcal{F}_{0} \right)
  \end{equation*}
  converges in probability to a non-random limit. Then
  \begin{equation*}
    \frac{\N}{\sqrt{\sum_{s=1}^{S}n_{s}^{2}}}(\hat{\beta}-\beta)
    =\mathcal{\N}\left(0,\frac{\mathcal{V}_{\N}}{\left(\frac{1}{\N}\sum_{i=1}^{N}\ddot{X}_{i}^{2}\right)^{2}}\right)+o_{p}(1).
  \end{equation*}
\end{proposition}

Relative to \Cref{theorem:normality-noZ}, the main difference is that $X_{i}$ in
the definition of $\mathcal{V}_{\N}$ is replaced by $X_{i}-Z_{i}'\gamma$, and
that $X_{i}$ is replaced by $\ddot{X}_{i}$ in the outer part of the
``sandwich.'' To motivate our standard
error formula, suppose that $\beta_{is}=\beta$ for all $i$ and $s$. Under $\beta_{is}=\beta$, it follows from
\cref{eq:random-assignment-Zcond} and \Cref{item:indepX} that
\begin{equation*}
  \mathcal{V}_{\N}=\frac{\sum_{s=1}^{S}\var(\tilde{\Xs}_{s}\mid \mathcal{F}_{0})
    R_{s}^{2}}{\sum_{s=1}^{S}n_{s}^{2}}, \qquad R_{s}=\sum_{i=1}^{\N}w_{is}\epsilon_{i}, \qquad
  \tilde{\Xs}_{s}=\Xs_{s}-\Zs_{s}'\gamma.
\end{equation*}
A plug-in estimate of $R_{s}$ can be constructed by replacing $\epsilon_{i}$
with the estimated regression residuals
$\hat{\epsilon}_{i}=Y_{i}-X_{i}\hat{\beta}-Z_{i}\hat{\delta}$, where
$\hat{\delta}=(Z'Z)^{-1}Z'(Y-X\hat{\beta})$ is an OLS estimate of $\delta$. We
can estimate the variance $\var(\tilde{\Xs}_{s}\mid \mathcal{F}_{0})$ by
$\widehat{\Xs}^{2}$, where
\begin{equation}\label{eq:hatX}
  \widehat{{\Xs}}=(W'W)^{-1}W'\ddot{X}
\end{equation}
 projects the estimate $\ddot{X}$ of $X-Z'\gamma$ onto
the sectoral space by regressing it onto the shares $W$.
To carry out the regression in~\cref{eq:hatX}, $W$ must be full rank; this
requires that there are more regions than sectors, $\N\geq S$. These steps lead to the
standard error estimate
\begin{equation}\label{eq:se-AKM}
  \widehat{se}(\hat{\beta})=\frac{\sqrt{\sum_{s=1}^{S}\widehat{{\Xs}}_{s}^{2}
    \hat{R}_{s}^{2}}}{\sum_{i=1}^{\N}\ddot{X}_{i}^{2}}, \qquad \hat{R}_{s}=\sum_{i=1}^{\N}w_{is}\hat{\epsilon}_{i}.
\end{equation}

The next remark summarizes the steps needed for the construction of the standard error $\widehat{se}(\hat{\beta})$:
\begin{remark}\label{remark:akm-construction}
To construct the standard error
  estimate in \cref{eq:se-AKM}:
  \begin{enumerate}
  \item Obtain the estimates $\hat{\beta}$ and $\hat{\delta}$ by regressing
    $Y_{i}$ onto $X_{i}=\sum_{s=1}^{S}w_{is}\Xs_{s}$ and the controls $Z_{i}$. The
    estimate $\hat{\epsilon}_{i}$ corresponds to the estimated regression
    residuals.
  \item\label{item:akm-backout-Xt} Construct $\ddot{X}_{i}$, the residuals from regressing $X_{i}$ onto
    $Z_{i}$. Compute $\widehat{\Xs}_{s}$, the regression coefficients from
    regressing $\ddot{X}$ onto $W$.
    \item Plug the estimates $\hat{\epsilon}_{i}$, $\ddot{X}_{i}$, and
$\widehat{\Xs}_{s}$ into the standard error formula in \cref{eq:se-AKM}.
  \end{enumerate}
\end{remark}

To gain intuition for the procedure in \Cref{remark:akm-construction}, it is
useful to consider again the case with concentrated sectors. Suppose that
$U_{i}=0$ for all $i$, so that the regression of $Y_{i}$ onto $X_{i}$ and
$Z_{i}$ is identical to the regression of $Y_{i}$ onto $\Xs_{s(i)}$ and
$\Zs_{s(i)}$. Then the standard error formula in \cref{eq:se-AKM} reduces to the
usual cluster-robust standard error, with clustering on $s(i)$.

The cluster-robust standard error is generally biased due to estimation noise in
estimating ${\epsilon}_{i}$, which can lead to undercoverage, especially in
cases with few clusters (see \citealp{cameron_practitioners_2014} for a survey).
Since the standard error in \cref{eq:se-AKM} can be viewed as generalizing the
cluster-robust formula, similar concerns arise in our setting. We thus
consider a modification $\widehat{se}_{\beta_{0}}(\hat{\beta})$ of
$\widehat{se}(\hat{\beta})$ that imposes the null hypothesis when estimating the
regression residuals to reduce the estimation noise in estimating
${\epsilon}_{i}$.\footnote{Alternatively, one could construct a bias-corrected
  variance estimate; see, for example, \citet{BeMc02} for an example of this
  approach in the context of cluster-robust inference.} To
calculate the standard error $\widehat{se}_{\beta_{0}}(\hat{\beta})$ for testing
the hypothesis $H_{0}\colon \beta=\beta_{0}$ against a two-sided alternative at
significance level $\alpha$, one replaces $\hat{\epsilon}_{i}$ with
$\hat{\epsilon}_{\beta_{0}, i}$, the residual from regressing
$Y_{i}-X_{i}\beta_{0}$ onto $Z_{i}$ ($\hat{\epsilon}_{\beta_{0}, i}$ is
an estimate of the residuals with the null imposed). The null is rejected if the
absolute value of the $t$-statistic
$(\hat{\beta}-\beta_{0})/\widehat{se}_{\beta_{0}}(\hat{\beta})$ exceeds
$z_{1-\alpha/2}$, the $1-\alpha/2$ quantile of a standard normal distribution
(1.96 for $\alpha=0.05$). To construct a confidence interval (CI) with coverage
$1-\alpha$, one collects all hypotheses $\beta_{0}$ that are not rejected. The endpoints of this CI are a solution to a
quadratic equation, and are thus available in closed form---one does not
have to numerically search for all the hypotheses that are not rejected. The
next remark summarizes this procedure.
\begin{remark}[Confidence interval with null imposed]\label{remark:akm0-construction}
  To test the hypothesis $H_{0}\colon \beta=\beta_{0}$ with significance level $\alpha$ or, equivalently, to
  check whether $\beta_{0}$ lies in the confidence interval with confidence level $1-\alpha$:
  \begin{enumerate}
  \item Obtain the estimate $\hat{\beta}$ by regressing $Y_{i}$ onto
    $X_{i}=\sum_{s=1}^{S}w_{is}\Xs_{s}$ and the controls $Z_{i}$. Obtain the
    restricted regression residuals
    $\hat{\epsilon}_{\beta_{0}, i}$ as the residuals from regressing
    $Y_{i}-X_{i}\beta_{0}$ onto $Z_{i}$.
  \item Construct $\ddot{X}_{i}$, the residuals from regressing $X_{i}$ onto
    $Z_{i}$. Compute $\widehat{\Xs}_{s}$, the regression coefficients from
    regressing $\ddot{X}$ onto $W$ (this step is identical to
    step~\ref{item:akm-backout-Xt} in \Cref{remark:akm-construction}).
  \item Compute the standard error as
  \begin{equation}\label{eq:se-AKM0}
    \widehat{se}_{\beta_{0}}(\hat{\beta})=\frac{\sqrt{\sum_{s=1}^{S}\widehat{{\Xs}}_{s}^{2}
        \hat{R}_{\beta_{0}, s}^{2}}}{\sum_{i=1}^{\N}\ddot{X}_{i}^{2}}, \qquad \hat{R}_{\beta_{0}, s}
    =\sum_{i=1}^{\N}w_{is}\hat{\epsilon}_{\beta_{0}, i}.
  \end{equation}
  \item Reject the null if
  $\abs{(\hat{\beta}-\beta_{0})/\widehat{se}_{\beta_{0}}(\hat{\beta})}>z_{1-\alpha/2}$.
  A confidence set with coverage $1-\alpha$ is given by all nulls that
  are not rejected,
  $CI_{1-\alpha}=\{\beta_{0}\colon
  \abs{(\hat{\beta}-\beta_{0})/\widehat{se}_{\beta_{0}}(\hat{\beta})}<z_{1-\alpha/2}\}$.
  This set is an interval with endpoints given by
  \begin{equation}\label{eq:ci-AKM0}
    \hat{\beta}-A\pm \sqrt{A^{2}
      +\frac{\widehat{se}(\hat{\beta})^{2}}{Q/(\ddot{X}'\ddot{X})^{2}}}, \qquad
    A=\frac{
      \sum_{s=1}^{S}\widehat{{\Xs}}_{s}^{2}\hat{R}_{s}\sum_{i=1}^{N}w_{is}\ddot{X}_{i}}{Q},
  \end{equation}
  where
  $Q=
  (\ddot{X}'\ddot{X})^{2}/z_{1-\alpha/2}^{2}-\sum_{s=1}^{S}\widehat{{\Xs}}_{s}^{2}(\sum_{i}w_{is}\ddot{X}_{i})^{2}$
  and $\widehat{se}(\hat{\beta})$ and $\hat{R}_{s}$ are given in \cref{eq:se-AKM}.
\end{enumerate}
\end{remark}

\begin{proposition}\label{theorem:se-consistency}
  Suppose that the assumptions of \Cref{theorem:normality-Z} hold, and that
  $\beta_{is}=\beta$. Suppose also that $\N\geq S$, $W$ is full rank, and that
  either $\max_{s}\sum_{i=1}^{N}\abs{((W'W)^{-1}W')_{si}}$ is bounded and
  $\max_{i}E[({U}_{i}'\gamma)^{4}\mid W]\to 0$, or else that $U_{i}=0$ for
  $i=1,\dotsc, N$. Define $\widehat{\Xs}$ as in \cref{eq:hatX}, and let
  $\hat{R}_{s}=\sum_{i=1}^{\N}w_{is}\tilde{\epsilon}_{i}$, where
  $\tilde{\epsilon}_{i}=Y_{i}-X_{i}\tilde{\beta}-Z_{i}'\tilde{\delta}$, and
  $\tilde{\beta}$ and $\tilde{\delta}$ are consistent estimators of $\delta$ and
  $\beta$. Then
  \begin{equation}\label{eq:se-consistency}
    \frac{\sum_{s=1}^{S}\widehat{\Xs}_{s}^{2}\hat{R}_{s}^{2}}{\sum_{s=1}^{S}n_{s}^{2}}=\mathcal{V}_{\N}+o_{p}(1).
  \end{equation}
\end{proposition}
Since in both $\hat{\epsilon}_{i}$ and $\hat{\epsilon}_{\beta_{0}, i}$ are
consistent estimates of the residuals, this proposition shows that the
procedures in \Cref{remark:akm-construction,remark:akm0-construction} both yield
asymptotically valid confidence intervals. The additional assumptions of
\Cref{theorem:se-consistency} ensure that the estimation error in
$\widehat{\Xs}_{s}$ that arises from having to back out the sector-level shocks
$\Zs_{s}$ from the controls $Z_{i}$ is not too large. If the sectors are
concentrated, then $((W'W)^{-1}W')_{si}=\1{s(i)=s}/n_{s}$, so that
$\max_{s}\sum_{i=1}^{N}\abs{((W'W)^{-1}W')_{si}}=1$, and the assumption always
holds. We show in
\begin{NoHyper}\Cref{sec:infer-under-heter}\end{NoHyper}
that the procedures in
\Cref{remark:akm-construction,remark:akm0-construction} continue to yield valid
inference if $\beta_{is}$ is heterogeneous across regions and sectors, as long
as further regularity conditions hold.

Although both standard errors $\widehat{se}_{\beta_{0}}(\hat{\beta})$ and
$\widehat{se}(\hat{\beta})$ are consistent (and one could further show that the
resulting confidence intervals are asymptotically equivalent), they will in
general differ in finite samples. In particular, it can be seen from
\cref{eq:ci-AKM0} that the confidence interval with the null imposed is not
symmetric around $\hat{\beta}$, but its center is shifted by $A$.\footnote{This
  is analogous to the differences in likelihood models between confidence
  intervals based on the Lagrange multiplier test (which imposes the null and is
  not symmetric around the maximum likelihood estimate) and the Wald test (which
  does not impose the null and yields the usual confidence interval).} As we
show in \Cref{sec:placebo-exercise}, this recentering tends to improve the
finite-sample coverage properties of the confidence interval. On the other hand,
the confidence interval described in \Cref{remark:akm0-construction} tends to be
longer on average than that in \Cref{remark:akm-construction}.

\subsection{Instrumental variables regression}\label{sec:ivreg}

We now turn to the problem of estimating the effect of a regional treatment
variable $Y_{2i}$ on a regional outcome $Y_{1i}$ using the shift-share
variable $X_{i}=\sum_{s=1}^{S}w_{is}\Xs_{s}$ as an instrumental variable (IV). To set
up the problem precisely, we again use the potential outcome framework. In
particular, we assume that
\begin{equation}\label{eq:iv-outcome}
  Y_{1i}(y_{2})=Y_{1i}(0)+y_{2}\alpha,
\end{equation}
where $\alpha$, our parameter of interest, measures the causal effect of
$Y_{2i}$ onto $Y_{1i}$. We assume for simplicity that this causal effect is
linear and constant across regions.\footnote{\label{fn:late}If we weaken the
  assumption of constant treatment effects and instead assume
  $Y_{1i}(y_{2})=Y_{1i}(0)+y_{2}\alpha_{i}$, then it follows by a mild extension
  of the results in \begin{NoHyper}\Cref{sec:proofs-iv}\end{NoHyper} that our
  methods would deliver inference on the estimand
  $\sum_{i=1}^{\N}\pi_{i}\alpha_{i}/\sum_{i=1}^{\N}\pi_{i}$, with
  $\pi_{i}=\sum_{s=1}^{S}w_{is}^{2}\var(\Xs_{s}\mid\mathcal{F}_{0})\beta_{is}$,
  where $\mathcal{F}_{0}=(\Zs, U, Y_{1}(0), Y_{2}(0), B, \alpha, W)$, and
  $\beta_{is}$ is defined in \cref{eq:FS-eq}.} In analogy
with~\cref{eq:potential-outcomes}, we denote the region-$i$ treatment level that
would occur if the region received shocks $(\xs_{1}, \dotsc, \xs_{S})$ as
\begin{equation}\label{eq:FS-eq}
  Y_{2i}(\xs_{1}, \dotsc, \xs_{S})=Y_{2i}(0)+\sum_{s=1}^{S}w_{is}x_{s}\beta_{is}.
\end{equation}
The observed outcome and treatment variables are given by $Y_{1i}=Y_{1i}(Y_{2i})$ and
$Y_{2i}=Y_{2i}(\Xs_{1}, \dotsc, \Xs_{S})$, respectively.

The framework in \cref{eq:iv-outcome,eq:FS-eq} maps directly to the problem of estimating the regional inverse
labor supply elasticity. In particular, in the context of the model in
\Cref{sec:model}, \cref{eq:emp_change,eq: invlaborsupeq} map directly into \cref{eq:iv-outcome,eq:FS-eq} if we define
\begin{equation}\label{LaborSupplyMapping}
  Y_{1i} = \hat{\omega}_{i}, \; Y_{2i} =\hat{L}_{i}, \;
  \alpha = \tilde{\phi}, \; Y_{1i}(0) = - \tilde{\phi}(\sum_{g=1}^G \tilde{w}_{ig} \hat{\nu}_g + \hat{\nu}_i),\;
  w_{is}=l_{is}^{0}, \; \Xs_{s}=\hat{\chi}_s, \; \beta_{is}=\theta_{is},
\end{equation}
and $Y_{2i}(0)$ is given by the expression for $Y_{i}(0)$ in
\cref{eq:potential_outcome0_wage}.\footnote{In some applications of shift-share
  IVs, the shifters $\{\Xs_{s}\}_{s=1}^{S}$ are
  unobserved and have to be estimated. We assume here that $\Xs_{s}$ is directly
  measurable for every sector $s$, and study the case with estimated shifters in
  \Cref{sec:mismeasuredIV}.} As this mapping illustrates, the potential outcome
$Y_{1i}(0)$ will generally have a shift-share structure, with the shifters being group-specific labor supply shocks (e.g.\ growth in the number of
workers by education group). Consequently, the regression residual in the structural equation will generally have a shift-share structure. Similarly, as
\cref{eq:potential_outcome0_wage} illustrates, the potential outcome $Y_{2i}(0)$
will also generally include several shift-share components, with the shifters being
either sector-specific labor demand shocks or the same group-specific
labor supply shocks appearing in $Y_{1i}(0)$. Thus, the regression residual in the
first-stage regression of $Y_{2i}$ onto $X_{i}$ will also generally have a
shift-share structure.

Our estimate of $\alpha$ is given by an
IV regression of $Y_{1i}$ onto $Y_{2i}$ and a $K$-vector of controls $Z_{i}$,
with $X_{i}$ used as an instrument for $Y_{2i}$. This IV estimate can be written as
\begin{equation}\label{eq:IVestimator}
  \hat{\alpha}=\frac{\sum_{i=1}^{\N}\ddot{X}_{i}Y_{1i}}{
   \sum_{i=1}^{\N}\ddot{X}_{i}Y_{2i}},
\end{equation}
where, as in~\Cref{sec:adding-covariates}, $\ddot{X}_{i}$ denotes the
residual from regressing $X_{i}$ onto $Z_{i}$.

\subsubsection*{Assumptions}

\Cref{assumption:DGP-IV} is a generalization of
\Cref{assumption:DGP-z}. Let $\mathcal{F}_{0}=(\Zs, U, Y_{1}(0), Y_{2}(0), B, W)$.
\begin{assumption}[IV Identification]\label{assumption:DGP-IV}
  \aitem\label{item:consistencyZ-IV} The observed outcome and treatment variables satisfy
  $Y_{1i}=Y_{1i}(Y_{2i})$ and $Y_{2i}=Y_{2i}(\Xs_{1}, \dotsc, \Xs_{s})$ such
  that~\cref{eq:iv-outcome,eq:FS-eq} hold, and the controls $Z_{i}$
  satisfy~\cref{eq:Z-proxy}; %
  \aitem\label{item:random-assignment-IV} The shifters are exogenous in the
  sense that, for every $s$,
  \begin{equation}\label{eq:random-assignment-IV}
    E[\Xs_{s}\mid \mathcal{F}_{0}]=E[\Xs_{s}\mid \Zs_{s}],
  \end{equation}
  and the right-hand side satisfies \cref{eq:linear-controls}; %
  \aitem\label{item:ugamma-cons-IV} \Cref{item:ugamma-cons,item:ugamma-norm}
  hold; %
  \aitem\label{item:iv-relevance}
  $\sum_{i=1}^{N}\sum_{s=1}^{S}w^{2}_{is}\cdot\var(\Xs_{s}\mid \mathcal{F}_{0})\beta_{is}\neq 0$.
\end{assumption}

\Cref{item:random-assignment-IV} adapts the standard instrument exogeneity
condition (see, e.g., Condition 1 in \citealp{imbens_identification_1994}) to
our setting. Our approach follows \citet*{borusyakhulljaravel2018shiftshare},
who impose a similar identification condition. To illustrate the restrictions
that \Cref{item:random-assignment-IV} may impose, consider again the problem of
estimating the inverse labor supply elasticity within the context of the model
in \Cref{sec:model}, with the mapping between this model and the potential
outcomes in \cref{eq:iv-outcome,eq:FS-eq} given in
\cref{eq:potential_outcome0_wage,LaborSupplyMapping}. If the controls
$\{\Zs_{s}\}_{s=1}^{S}$ correspond to the shocks $\{\hat{\mu}_{s}\}_{s=1}^{S}$,
then \cref{eq:random-assignment-IV} requires that, conditional on
$\{\hat{\mu}_{s}\}_{s=1}^{S}$, the labor demand shocks
$\{\hat{\chi}_s\}_{s=1}^{S}$ used to construct our IV are mean-independent of
the idiosyncratic labor demand shocks $\{\hat{\eta}_{is}\}_{i=1,s=1}^{N, S}$ and
of the labor supply shifters $\{\hat{\nu}_{i}\}_{i=1}^{N}$ and
$\{\hat{\nu}_{g}\}_{g=1}^{G}$.\footnote{If, instead of \cref{eq:FS-eq}, we
  defined the first stage as simply the projection of $Y_{2i}$ onto the
  shift-share instrument, we could further relax this condition and only require
  $\{\hat{\chi}_s\}_{s=1}^{S}$ to be mean-independent of the labor supply
  shifters. An advantage of the current setup is that it allows us to derive
  primitive conditions for the consistency of the estimates of the first-stage
  regression and, thus, of the IV estimator.} For example, if
$\{\hat{\chi}_s\}_{s=1}^{S}$ are sectoral productivity shocks, then these
productivity shocks need to be independent of shocks to individuals' willingness
to work in different groups and regions. \Cref{item:iv-relevance} requires that
the coefficient on the instrument in the first-stage equation, which can be
written as
$\beta=\sum_{i=1}^{N}\sum_{s=1}^{S}w^{2}_{is}\var(\Xs_{s}\mid
\mathcal{F}_{0})\beta_{is}/\sum_{i=1}^{N}\sum_{s=1}^{S}w^{2}_{is}\var(\Xs_{s}\mid
\mathcal{F}_{0})$, is non-zero---this is the standard IV relevance assumption.
For consistency and inference, in an analogy to the OLS case, we assume that
\Cref{assumption:inference} holds with
$\mathcal{F}_{0}=(\Zs, U, Y_{1}(0), Y_{2}(0), B, W)$.

In a recent paper, \citet*{goldsmith2018bartik} explore a different approach to
identification and inference on the treatment effect $\alpha$. Focusing here for
simplicity on the case without controls, in place of
\Cref{item:random-assignment-IV}, they assume that the shares $\{w_{is}\}_{s=1}^{S}$ are as good as randomly assigned conditional on the
shifters $\{\Xs_{s}\}_{s=1}^{S}$; so that they are mean-independent of the potential outcomes
$Y_{1}(0)$ and $Y_{2}(0)$ conditional on $\Xs$. As \citet*{goldsmith2018bartik}
show, under this alternative assumption, one can replace the shift-share
instrument $X_{i} = \sum_{s=1}^{S} w_{is} \Xs_s$ by the full vector of shares
$(w_{i1}, \dotsc, w_{iS})$ in the first-stage equation. For estimation and
inference, this alternative approach requires that, conditionally on the
shifters, either the shares $(w_{i1}, \dotsc, w_{iS})$ or else the structural
residuals be independent across regions or clusters of regions.

For estimating the inverse labor supply elasticity in the context of the model
in \Cref{sec:model}, \cref{LaborSupplyMapping} illustrates that this alternative
identification assumption requires that, conditional on
$\{\hat{\chi}_{s}\}_{s=1}^{S}$, the region-specific employment shares in the
initial equilibrium $\{l^{0}_{is}\}_{s=1}^{S}$ are mean-independent of both the
region-specific exposure shares $\{\tilde{w}_{ig}\}_{g=1}^{G}$, and the
region-specific labor supply shock $\nu_{i}$. This assumption is violated if
regions more exposed to labor demand shocks in a sector $s$ (e.g.\ to changes in
tariffs in the food sector) are also more exposed to labor supply shocks
affecting workers of a group $g$ \citep[e.g. currency crisis in Mexico affecting
the number of Mexican migrants;
see][]{monras2018Mexico}.\footnote{\label{fn:manyinvalid}To allow for a
  shift-share component in the structural residual, \citet*{goldsmith2018bartik}
  view the shares $(w_{i1}, \dots, w_{iS})$ as ``invalid'' instruments, since,
  in this case, $E[\epsilon_{i}w_{{is}}\mid \Xs]\neq 0$, where $\epsilon_{i}$
  denotes the structural error. \citet*{goldsmith2018bartik} show that if these
  shares are used to construct a single shift-share instrument $X_{i}$, the bias
  in the IV estimator coming from the correlation between any $w_{is}$ and the
  structural residual averages out under certain conditions as $S\to\infty$, as
  in the many invalid instrument setting studied in \citet{kolesar15invalid}.
  Under the current setup, in contrast, \cref{eq:random-assignment-IV} implies
  that $X_{i}$ is a valid instrument for any fixed $S$. Leveraging exogeneity of
  $\Xs_{s}$ is a key difference between our approach and that in
  \citet{kolesar15invalid} and \citet*{goldsmith2018bartik}. It allows us to do
  inference without imposing a particular correlation structure on the residuals
  $\epsilon_{i}$, and it allows us to achieve identification without requiring
  $S\to\infty$; the latter is only needed for consistency and inference.}

In terms of inference, since the structural residuals will not be independent across regions unless
they contain no shift-share component (which, according to the economic model in \Cref{sec:model}, is unlikely), the approach in \citet*{goldsmith2018bartik} generally requires that the
shares are independent across (clusters of) regions. This assumption is, from
the perspective of the model in \Cref{sec:model}, conceptually very
different from assuming independence of the shifters $\Xs_{s}$ across sectors.
Since the shifters $\Xs_{s}=\hat{\chi}_{s}$ are exogenous, the latter only involves
assumptions on model fundamentals by restricting the distribution
in~\cref{eq:jointdistshocks}. In contrast, each share $w_{is}=l_{is}^{0}$
corresponds to the employment allocation across sectors in a region $i$ in an
initial equilibrium, so that the former involves imposing restrictions on an
endogenous outcome of the model. Furthermore, since all the shares
$\{w_{is}\}_{i=1,s=1}^{N, S}$ depend on the same set of sector-specific labor
demand shifters $\{(\chi_{s}, \mu_{s})\}_{s=1}^{S}$, they will generally be
correlated across regions.\footnote{For instance, if $\sigma_{s}=\sigma$ for all
  $s$, then $l_{is}^0 = D_{is}^0 / (\sum_{t=1}^{S}D_{it}^0)$, where
  $D^{0}_{is}$ is the labor demand shifter of sector $s$ in region $i$ in the
  initial equilibrium. According to \cref{technology_maintext}, for any $s$, all shifters $\{D^{0}_{is}\}_{i=1}^{N}$ depend on the same sector-level demand shocks,
  $\{(\chi_{s}, \mu_{s})\}_{s=1}^{S}$ and, thus, the labor shares $l_{is}^0$
  will generally be correlated across all regions for any given sector.}

Which identification and inference approach is more attractive depends on the
context of each particular empirical application. While the economic model in
\Cref{sec:model} motivates the approach we pursue here, this does not mean that
our approach is generally more attractive. In other empirical applications (e.g.\ when the shares are exogenous variables from the perspective
of an economic framework), the approach of \citet*{goldsmith2018bartik} may be
more appropriate.

\subsubsection*{Asymptotic theory}

It follows by adapting the arguments in the proof of \Cref{theorem:normality-Z}
that, if \Cref{assumption:DGP-IV} holds, and \Cref{assumption:inference} holds
with $\mathcal{F}_{0}=(\Zs, U, Y_{1}(0), Y_{2}(0), B, W)$, then, under mild
technical regularity conditions (see
\begin{NoHyper}\Cref{sec:proofs-iv}\end{NoHyper} for details and proof),
\begin{equation}\label{eq:se-IV}
  \frac{\N}{\sqrt{\sum_{s=1}^{S}n_{s}^{2}}}\left(
    \hat{\alpha}-\alpha
  \right)=\mathcal{N}\left(0,\frac{
      \mathcal{V}_{N}}{(\frac{1}{N}\sum_{i=1}^{N}\ddot{X}_{i}Y_{2i})^{2}}
  \right)+o_{p}(1), \;\mathcal{V}_{N}=\frac{\sum_{s=1}^{S}\var(\tilde{\Xs}_{s}\mid
  \mathcal{F}_{0}) R_{s
  }^{2}}{\sum_{s=1}^{S}n_{s}^{2}},
  \; R_{s}=\sum_{i=1}^{N}w_{is}\epsilon_{i},
\end{equation}
where $\epsilon_{i}=Y_{1i}-Y_{2i}\alpha-Z_{i}'\delta$ is the residual in the
structural equation, with $\delta=E[Z'Z]^{-1}E[Z'(Y_{1}-Y_{2}\alpha)]$. This
suggests the standard error estimate
\begin{equation}\label{eq:se-AKM-IV}
  \widehat{se}(\hat{\alpha}) =\frac{\sqrt{\sum_{s=1}^{S}\widehat{{\Xs}}_{s}^{2}
      \hat{R}_{s}^{2}}}{\abs{\sum_{i=1}^{\N}\ddot{X}_{i}Y_{2i}}}
  =\frac{\sqrt{\sum_{s=1}^{S}\widehat{{\Xs}}_{s}^{2}
      \hat{R}_{s}^{2}}}{{\sum_{i=1}^{\N}\ddot{X}_{i}^{2}}\abs{\hat{\beta}}},
  \qquad \hat{R}_{s}=\sum_{i=1}^{\N}w_{is}\hat{\epsilon}_{i},
\end{equation}
where $\widehat{\Xs}_{s}$ is constructed as in \Cref{remark:akm-construction},
$\hat{\epsilon}=Y_{1}-Y_{2}\hat{\alpha}-Z'(Z'Z)^{-1}Z'(Y_{1}-Y_{2}\hat{\alpha})$
is the estimated residual of the structural equation, and
$\hat{\beta}=\sum_{i=1}^{N}\ddot{X}_{i}Y_{2i}/\sum_{i=1}^{N}\ddot{X}_{i}^{2}$ is the
first-stage coefficient.

The difference between the IV standard error formula in \cref{eq:se-AKM-IV} and
the OLS version in \cref{eq:se-AKM} is analogous to the difference between IV standard errors
and OLS heteroskedasticity-robust standard errors for the corresponding
reduced-form specification: the residual $\hat{\epsilon}_{i}$ corresponds to
the residual in the structural equation, and the denominator is scaled by the
first-stage coefficient. To obtain the IV analog of the standard error estimator
under the null $H_{0}\colon \alpha=\alpha_{0}$, we use the formula in
\cref{eq:se-AKM-IV} except that, instead of $\hat{\epsilon}_{i}$, we use the
structural residual computed under the null,
$\hat{\epsilon}_{\alpha_{0}}=(I-Z'(Z'Z)^{-1}Z')(Y_{1}-Y_{2}\alpha_{0})$. The resulting confidence interval
is a generalization of the \citet{andersonrubin1949} confidence interval (which
assumes that the structural errors are independent). For this reason, this
confidence interval will remain valid even if the shift-share instrument is
weak.

\section{Extensions}\label{sec:extensions}

We now discuss three extensions to the basic setup. In
\Cref{sec:clustersectors}, we relax the assumption that the shifters
$\{\Xs_{s}\}_{s=1}^{S}$ are independent, allowing them to be correlated
within clusters of sectors. \Cref{sec:panel} generalizes our results to settings in which we have multiple observations for each region. \Cref{sec:mismeasuredIV} considers the case in which the shifters are
not directly observed, and have to be estimated.

\subsection{Clusters of sectors}\label{sec:clustersectors}

Suppose that the sectors can be grouped into larger units, which we refer to as
``clusters'', with $c(s)\in\{1,\dotsc, C\}$ denoting the cluster that sector $s$
belongs to; e.g., if each $s$ corresponds to a four-digit industry code, $c(s)$
may correspond to a three-digit code. With this structure, we replace
\Cref{item:indepX} with the weaker assumption that, conditional on
$\mathcal{F}_{0}$, the shocks $\Xs_{s}$ and $\Xs_{k}$ are independent if
$c(s)\neq c(k)$, and we replace \Cref{item:sector-size-normality} with the
assumption that, as $C\to\infty$, the largest cluster makes an asymptotically
negligible contribution to the asymptotic variance; i.e.
$\max_{c}\tilde{n}_{c}^{2}/\sum_{d=1}^{C}\tilde{n}_{d}^{2}\to 0$, where
$\tilde{n}_{c}=\sum_{s=1}^{S}\1{c(s)=c}n_{s}$ is the total share of cluster $c$.

Under this setup, by generalizing the arguments in \Cref{sec:adding-covariates},
one can show that, as $C\to\infty$,
\begin{equation*}
  \frac{\N}{\sqrt{\sum_{c=1}^{C}\tilde{n}_{c}^{2}}}(\hat{\beta}-\beta)
  =\mathcal{N}\left(0,\frac{\mathcal{V}_{\N}}{\left(\frac{1}{\N}
        \sum_{i=1}^{N}\ddot{X}_{i}^{2}\right)^{2}}\right)+o_{p}(1),
\end{equation*}
and, assuming that $\beta_{is}=\beta$ for every region and sector, the term $\mathcal{V}_{\N}$ is now given by
\begin{equation*}
  \mathcal{V}_{\N}=\frac{\sum_{c=1}^{C}\sum_{s=1,t=1}^{S, S}\1{c(s)=c(t)=c}
    E[\tilde{\Xs}_{s}\tilde{\Xs}_{t}\mid W, \Zs]R_{s}R_{t}}{
    \sum_{c=1}^{C}\tilde{n}_{c}^{2}}, \qquad R_{s}=\sum_{i=1}^{\N}
  w_{is}\epsilon_{i},
  \quad \tilde{\Xs}_{s}=\Xs_{s}-\Zs_{s}'\gamma.
\end{equation*}
As a result, we replace the standard error estimate in \cref{eq:se-AKM} with a
version that clusters $\widehat{\Xs}_{s}\hat{R}_{s}$,
\begin{equation}\label{eq:se-AKM-cluster}
  \widehat{se}(\hat{\beta})=\frac{\sqrt{\sum_{c=1}^{C}
      \sum_{s, t}\1{c(s)=c(t)=c}\widehat{\Xs}_{s}\hat{R}_{s}\widehat{\Xs}_{t}\hat{R}_{t}
}}{\sum_{i=1}^{\N}\ddot{X}_{i}^{2}}, \qquad \hat{R}_{s}=\sum_{i=1}^{\N}w_{is}\hat{\epsilon}_{i},
\end{equation}
where $\widehat{\Xs}_{s}$ is defined as in \Cref{remark:akm-construction}.
Confidence intervals with the null imposed can be constructed as in
\Cref{remark:akm0-construction}, replacing $\hat{\epsilon}_{i}$ with
$\hat{\epsilon}_{\beta_{0}, i}$ in \cref{eq:se-AKM-cluster}. In
the IV setting considered in \Cref{sec:ivreg}, the standard error for
$\hat{\alpha}$ is analogous to that in \cref{eq:se-AKM-cluster}, except that
$\hat{\epsilon}_{i}$ denotes the residual in the structural equation, and we
divide the expression by the absolute value of the first-stage coefficient,
$\sum_{i=1}^{N}\ddot{X}_{i}Y_{2i}/\sum_{i=1}^{N}\ddot{X}_{i}^{2}$.

\subsection{Panel data}\label{sec:panel}

Consider a setting with $j=1, \dotsc, J$ regions, $k=1, \dotsc, K$ sectors, and
$t=1, \dotsc, T$ periods. For each period $t$, we have data on shifters
$\{\Xs_{kt}\}_{k=1}^{K}$, outcomes $\{Y_{jt}\}_{j=1}^{J}$, and shares $\{w_{jkt}\}_{j=1,k=1}^{J, K}$. This setup maps into the
potential outcome framework in \cref{eq:potential-outcomes} if we identify a
``sector'' with a sector-period pair $s=(k, t)$, and a ``region'' with a
region-period pair $i=(j, t)$, so that we can index outcomes and shifters as
$Y_{i}=Y_{jt}$ and $\Xs_{s}=\Xs_{kt}$, with the shares given by
\begin{equation}\label{eq:panel_share}
  w_{is}=\begin{cases}
    w_{jkt} & \text{if $i=(j, t)$ and $s=(k, t)$,} \\
    0 & \text{if $i=(j, t)$, $s=(k, t')$, and $t\neq t'$.}
\end{cases}
\end{equation}
If the shifters $\Xs_{kt}$ are independent across time and sectors,
\Cref{theorem:consistency-Z,theorem:normality-Z} immediately give the
large-sample distribution of the OLS estimator. In general, however, it will be
important to allow the shifters $\Xs_{kt}$ to be correlated across time within
each sector $k$. In this case, one can use the clustered standard error derived
in \Cref{sec:clustersectors} by grouping observations over time for each sector
$k$ into a common cluster, so that $c(k, t)=c(k', t')$ if $k=k'$. We can then
apply the formula in \cref{eq:se-AKM-cluster} to allow for any arbitrary
time-series correlation in the sector-level shocks $\Xs_{kt}$ for
any given sector $k$. Regardless of whether the sector-period pairs $(k, t)$ are
clustered, as discussed in \Cref{remark:fixed-epsilon}, our standard error
formulas allow for arbitrary dependence patterns in the regression
residuals---in particular, they account for potential serial dependence in the
regression residuals.

If the shift-share regressor is used as an IV in a regression of an outcome
$Y_{1jt}$ onto a treatment $Y_{2jt}$, the mapping to
\cref{eq:iv-outcome,eq:FS-eq} is analogous, and one can use an IV version of the
formula in \cref{eq:se-AKM-cluster} for inference.

\subsection{IV with estimated shifters}\label{sec:mismeasuredIV}

We now consider a setting in which the sectoral shifters $\{\Xs_{s}\}_{s=1}^{S}$ that define
the shift-share IV studied in \Cref{sec:ivreg} are not directly observed. We follow the
setup in \Cref{sec:ivreg} but assume that, instead of observing $\Xs_{s}$
directly, we only observe a noisy measure of it,
\begin{equation}\label{eq:xis}
  X_{is}=\Xs_{s}+\psi_{is}
\end{equation}
for each sector-region pair. We consider IV regressions that use two
different estimates of $X_{i}=\sum_{s=1}^{S}w_{is}\Xs_{s}$. First, an estimate
that replaces $\Xs_{s}$ with an estimate
$\hat{\Xs}_{s}=\sum_{i=1}^{N}\check{w}_{is}X_{is}/\check{n}_{s}$, where
$\check{n}_{s}=\sum_{i=1}^{N}\check{w}_{is}$ and the weights $\check{w}_{is}$
are not necessarily related to $w_{is}$. The resulting estimate of $X_{i}$ is
\begin{equation}\label{eq:X-estimated}
  \hat{X}_{i}=\sum_{s=1}^{S}w_{is}\hat{\Xs}_{s}=
  \sum_{s=1}^{S}w_{is}\frac{1}{\check{n}_{s}}
    \sum_{j=1}^{\N}\check{w}_{js}X_{js},
\end{equation}
and it yields the IV estimate
$\tilde{\alpha}=\ddot{\hat{X}}'Y_{1}/\ddot{\hat{X}}'Y_{2}$, where
$\ddot{\hat{X}}=\hat{X}-Z(Z'Z)^{-1}Z'\hat{X}$ is the residual from regressing
$\hat{X}_{i}$ onto $Z_{i}$. Second, we consider the leave-one-out estimator

\begin{equation}\label{eq:X-leave-one-out}
  \hat{X}_{i,-}=\sum_{s=1}^{S}w_{is}\hat{\Xs}_{s, -i}=
  \sum_{s=1}^{S}w_{is}\frac{1}{\check{n}_{s, -i}}
    \sum_{j=1}^{\N}\1{j\neq i}\check{w}_{js}X_{js}, \qquad
    \check{n}_{s, -i}=\sum_{j=1}^{\N}\1{j\neq i}\check{w}_{js},
\end{equation}
where
$\hat{\Xs}_{s, -i}=\sum_{j=1}^{N}\1{j\neq i}\check{w}_{js}X_{js}/\check{n}_{s, -i}$ is
an estimate of $\Xs_{s}$ that excludes region $i$. A version of this estimator
has been used in \citet{autorduggan2003}. This leave-one-out estimator of the
shift-share instrument $X_{i}$ yields the IV estimate
$\hat{\alpha}_{-}=\ddot{\hat{X}}_{-}'Y_{1}/\ddot{\hat{X}}_{-}'Y_{2}$, where
$\ddot{\hat{X}}_{-}=\hat{X}_{-}-Z'(Z'Z)^{-1}Z'\hat{X}_{-}$.

While we assume that $\Xs_{s}$ satisfies the exogeneity restriction in
\Cref{item:random-assignment-IV} for every $s$, we allow the measurement errors
$\psi_{i}=(\psi_{i1}, \dotsc, \psi_{iS})'$ to be potentially correlated with the
potential outcomes $Y_{1i}(0)$ and $Y_{2i}(0)$ in the same region $i$. We
assume, however, that $\psi_{i}$ is independent of the errors $\psi_{j}$ and of
the potential outcomes $Y_{1j}(0)$ and $Y_{2j}(0)$ for any region $j\neq i$ (see
\begin{NoHyper}\Cref{sec:proofs-iv}\end{NoHyper} for a formal
statement). In \begin{NoHyper}\Cref{app:missIVandModel}\end{NoHyper}, we use the
model in \Cref{sec:model} to discuss these assumptions in the context of
estimating the inverse labor supply elasticity.\footnote{\label{fn:leaveoneout}Specifically, we show
  in \begin{NoHyper}\Cref{app:missIVandModel}\end{NoHyper} that, if $X_{is}$ corresponds to employment growth rates, then $\psi_{i}$
  will generally not be independent of $(\psi_{j}, Y_{1j}(0), Y_{2j}(0))$ in
  others regions $j\neq i$, unless one makes restrictive assumptions about the
  demand elasticities $\sigma_{s}$, such as $\sigma_{s}=0$. We also construct alternative shift-share IVs that satisfy this independence assumption under weaker restrictions on $\sigma_{s}$, but require adjusting the shifter used in estimation.}

The potential correlation between $\psi_{i}$ and the potential outcomes in
region $i$ implies that the estimation error in $\hat{X}_{i}$, which is a
function on $\psi_{i}$, may be correlated with the residual in the structural
equation. Thus, including the $i$th observation in the construction of
$\hat{X}_{i}$ induces an own-observation bias in the IV estimator
$\tilde{\alpha}$ of $\alpha$. See \citet*{goldsmith2018bartik} and
\citet*{borusyakhulljaravel2018shiftshare} for a discussion. This bias is
analogous to the bias of the two-stage least squares estimator in settings with
many instruments (e.g.\
\citealp*{bekker_alternative_1994,angrist_jackknife_1999}), such as when one
uses group indicators as instruments.\footnote{See, e.g.,
  \citet*{maestas2013,DoSo15,AiDo15}, or \citet{Silver16}.} We show
in \begin{NoHyper}\Cref{sec:proofs-iv}\end{NoHyper} that the magnitude of the
bias is of the order
$\frac{1}{\N}
\sum_{i=1}^{N}\sum_{s=1}^{S}\frac{w_{is}\check{w}_{is}}{\check{n}_{s}}\leq
S/\N$, so that consistency of $\tilde{\alpha}$ generally requires the number of
sectors to grow more slowly than the number of regions. Furthermore, to ensure
that the asymptotic bias in $\tilde{\alpha}$ does not induce undercoverage of
the resulting confidence intervals, one generally requires $S^{3/2}/N\to 0$.

The estimator $\hat{\alpha}_{-}$, which can be thought of as a shift-share analog of
the jackknife IV estimator studied in \citet*{angrist_jackknife_1999}, remains
consistent, as shown in \citet*{borusyakhulljaravel2018shiftshare}
and in \begin{NoHyper}\Cref{sec:proofs-iv}\end{NoHyper}. We also show in
this appendix that, under regularity conditions, its asymptotic distribution is
given by
\begin{equation}\label{eq:se-IV-psi}
  \frac{\N}{\sqrt{\sum_{s=1}^{S}n_{s}^{2}}}(\hat{\alpha}_{-}-\alpha)
  =\mathcal{N}\left(0,\frac{\mathcal{V}_{\N}+\mathcal{W}_{\N}}{
      \left(\frac{1}{\N}\sum_{i=1}^{N}\ddot{X}_{i}Y_{2i}\right)^{2}}\right)+o_{p}(1),
\end{equation}
with $\mathcal{V}_{\N}$ defined as in \cref{eq:se-IV}, and
\begin{equation*}
  \mathcal{W}_{\N}
  =\frac{1}{\sum_{s=1}^{S}n_{s}^{2}}(\sum_{j=1}^{N}(\sum_{i=1}^{N}\mathcal{S}_{ij})^{2}+\sum_{i=1}^{N}\sum_{j=1}^{N}\mathcal{S}_{ij}\mathcal{S}_{ji}), \qquad
  \mathcal{S}_{ij}=\sum_{s=1}^{S}\1{i\neq
    j}\frac{w_{is}\check{w}_{js}\psi_{js}\epsilon_{i}}{\check{n}_{s, -i}}.
\end{equation*}
The term $\mathcal{W}_{\N}$ accounts for the additional uncertainty stemming
from the fact that the shift-share IV is estimated. It is analogous to
the many-instrument term in the jackknife IV estimator under many instrument
asymptotics (see \citealp{chao_asymptotic_2012}). Using simulations, we show in
\begin{NoHyper}\Cref{app:placebo_leave_one_out}\end{NoHyper} several designs in which, while
correcting for the own-observation bias by using
$\hat{\alpha}_{-}$ instead of $\tilde{\alpha}$ is quantitatively important,
accounting for the additional variance term $\mathcal{W}_{N}$ is less important.

\section{Performance of new methods: placebo evidence}\label{sec:placebo-exercise}

In \Cref{sec:placebo-baseline}, we revisit the placebo exercise
in~\Cref{sec:overr-usual-stand} to examine the finite-sample properties of the
inference procedures described in \Cref{remark:akm-construction,remark:akm0-construction}. In
\Cref{sec:altenative_placebo}, we show that our baseline placebo results are
robust to several changes in the placebo design.

\subsection{Baseline specification}\label{sec:placebo-baseline}

We first consider the performance of the standard error estimator in
\cref{eq:se-AKM} (which we label \emph{AKM}), and the standard error and
confidence interval in~\cref{eq:se-AKM0,eq:ci-AKM0} (with label \emph{AKM0}) in
the baseline placebo design described in~\Cref{sec:overr-usual-stand}.\footnote{We fix the matrix $Z$ to be a column of ones when implementing the formulas in \cref{eq:se-AKM,eq:ci-AKM0}.}$^{,}$\footnote{In \begin{NoHyper}\Cref{app:placebo_additional_results}\end{NoHyper}, we explore the sensitivity of our results to using counties (instead of CZs) as the regional unit of analysis, and occupations (instead of sectors) as the unit at which the shifter is defined.}

For the \emph{AKM} and \emph{AKM0} inference procedures,
\Cref{tab:simple_placebo_akm} presents median standard error estimates
and rejection rates for 5\% significance level tests of the null hypothesis
$H_{0}\colon\beta=0$. In the case of \emph{AKM0}, since the standard error
depends on the null being tested, the table reports the median ``effective
standard error'', defined as the length of the 95\% confidence interval divided
by $2\times 1.96$. %

The results in \Cref{tab:simple_placebo_akm} show that the inference procedures
introduced in \Cref{sec:asympt-prop-bart} perform well. The median \textit{AKM}
standard error is slightly lower than the standard deviation of $\hat{\beta}$,
by about 5\% on average across all outcomes. The median \textit{AKM0} effective
standard error is slightly larger than the standard deviation of $\hat{\beta}$,
by about 11\% on average. The implied rejection rates are close to the 5\%
nominal rate: the \textit{AKM} procedure has rejection rates between 7.5\% and
9.1\% and the \textit{AKM0} rejection rates are always between 4.3\% and 4.5\%.
As discussed in \Cref{sec:adding-covariates}, the \textit{AKM} and \textit{AKM0}
confidence intervals are asymptotically equivalent. The differences in rejection
rates between the \textit{AKM} and \textit{AKM0} inference procedures are thus
due to differences in finite-sample performance. As noted in other contexts
(see, e.g., \citealp{lazarus2018har}), imposing the null can lead to improved
finite-sample size control. The better size control of the \textit{AKM0}
procedure is consistent with these results. %

\begin{table}[!t]
\centering
\caption{Median standard errors and rejection rates for
  $H_{0}\colon \beta=0$ at 5\% significance level.}\label{tab:simple_placebo_akm} %
\begin{tabular}{@{}l rc cc cc@{}}
\toprule
&\multicolumn{2}{c}{Estimate} & \multicolumn{2}{c}{Median eff.\ s.e.} &
\multicolumn{2}{c}{Rejection rate}\\ [1pt]
\cmidrule(rl){2-3} \cmidrule(rl){4-5}\cmidrule(rl){6-7}
&Mean & Std.\ dev & AKM & AKM0 & AKM & AKM0 \\
& \multicolumn{1}{c}{(1)} & \multicolumn{1}{c}{(2)} & \multicolumn{1}{c}{(3)} & \multicolumn{1}{c}{(4)}
& \multicolumn{1}{c}{(5)} & \multicolumn{1}{c}{(6)} \\
\midrule
\multicolumn{7}{@{}l}{\textbf{Panel A\@: Change in the share of working-age population}} \\ [2pt]

Employed        &       $-0.01$ &       $2.00$  &       $1.90$  &       $2.21$  &       $7.8$\% &       $4.5$\% \\
Employed in manufacturing       &       $-0.01$ &       $1.88$  &       $1.77$  &       $2.06$  &       $8.0$\% &       $4.3$\% \\
Employed in non-manufacturing   &       $0.00$  &       $0.94$  &       $0.89$  &       $1.04$  &       $8.2$\% &       $4.5$\% \\ [6pt]

\multicolumn{7}{@{}l}{\textbf{Panel B\@: Change in average log weekly wage}} \\ [2pt]
Employed        &       $-0.03$ &       $2.66$  &       $2.57$  &       $2.99$  &       $7.5$\% &       $4.3$\% \\
Employed in manufacturing       &       $-0.03$ &       $2.92$  &       $2.74$  &       $3.18$  &       $9.1$\% &       $4.5$\% \\
Employed in non-manufacturing   &       $-0.02$ &       $2.64$  &       $2.55$  &       $2.96$  &       $7.8$\% &       $4.5$\% \\ [2pt]

  \bottomrule
  \multicolumn{7}{p{5.8in}}{\scriptsize{Notes: For the outcome variable
  indicated in the leftmost column, this table indicates the mean and standard deviation of the OLS estimates of $\beta$ in~\cref{eq:bartikreg} across the placebo samples (columns (1) and (2)),
  the median effective standard error estimates (columns (3) and (4)), and the percentage of placebo samples for which we reject the null
  hypothesis $H_{0}\colon\beta=0$ using a 5\% significance level test (columns (5) and (6)).
  \textit{AKM} is the standard error
  in \Cref{remark:akm-construction}; and \textit{AKM0} is the confidence interval
  in \Cref{remark:akm0-construction}. The median effective standard error is equal to the median length of the corresponding 95\% confidence interval divided by $2\times1.96$. Results are based on 30,000 placebo samples.}}
\end{tabular}
\end{table}

\subsection{Alternative placebo specifications}\label{sec:altenative_placebo}

In \Cref{sec:asympt-prop-bart}, we show theoretically that the \textit{AKM} and \textit{AKM0} inference procedures are valid in large
samples only if: (a) the number of sectors goes to infinity; (b) all sectors are
asymptotically ``small''; (c) the sectoral shocks are independent across
sectors. Given these conditions, these inference
procedures remain valid under (d) any distribution of the sectoral shifters; and (e) arbitrary correlation structure
of the regression residuals. In this section, we
evaluate the sensitivity of these inference procedures
to requirements (a) to (c) above, and illustrate points (d) and (e) by
documenting the robustness of these procedures to alternative
distributions of the shifters and the residuals. In all cases, we
also report \textit{Robust} and \textit{Cluster} standard errors estimates and
rejection rates. We focus on the change in the share of working-age population employed as the outcome variable of interest.

We first evaluate how the performance of different inference procedures depends on the number of sectors. Panel A of
\Cref{tab:departures_placebo} shows that the overrejection problem affecting
standard inference procedures worsens when the number of sectors decreases: the
rejection rates of 5\% significance level tests based on \textit{Robust} and
\textit{Cluster} standard errors reach 70.6\% and 56.1\%, respectively, when we
construct the shift-share covariate using 20 2-digit SIC sectors (instead of the
396 4-digit SIC sectors we use in the baseline placebo). In line with the
findings of the literature on clustered standard errors with few clusters, the
rejection rates of hypothesis tests that rely on \textit{AKM} standard errors
also increase to 12\%, but rejection rates for hypothesis tests that apply the
\textit{AKM0} inference procedure remain very close to the nominal 5\%
significance level.

Panels B to D of \Cref{tab:departures_placebo} examine the robustness of the
results in \Cref{tab:simple_placebo,tab:simple_placebo_akm} to alternative
distributions of the shifters. In Panel B, as in our baseline placebo exercise,
the shifters are drawn i.i.d.\ from a normal distribution, but we change the
variance to both a lower ($\sigma^{2}=0.5$) and a higher value ($\sigma^{2}=10$)
than in the baseline ($\sigma^{2}=5$). In Panel C, we draw the shifters from a
log-normal distribution re-centered to have mean zero and scaled to have the
same variance as in the baseline. Panel D investigates the robustness of our
results to heteroskedasticity in the sector-level shocks. We set variance of the
shock in each sector $s$, to $\sigma^{2}_{s}=5 + \lambda(n_{s} - S/N)$. Thus,
the cross-sectional average of the variance of the sector-level shocks is the
same as in the baseline (which corresponds to setting $\lambda=0$), but this
variance now varies across sectors. Comparison of the results in Panels B to D
of \Cref{tab:departures_placebo} to those in
\Cref{tab:simple_placebo,tab:simple_placebo_akm} suggests that our baseline
results are not sensitive to specific details of the distribution of
sector-level shifters. This is consistent with the claim (d) above.

\begin{table}[!t]
\centering
\caption{Alternative number of sectors, shifter distributions and residuals' correlation patterns}\label{tab:departures_placebo}
\tabcolsep=4pt
\begin{tabular}{@{}l rc cccc rrrr @{}}
  \toprule
  &\multicolumn{2}{c}{Estimate} & \multicolumn{4}{c}{Median eff.\ s.e.} &
\multicolumn{4}{c}{Rejection rate}\\ [1pt]
\cmidrule(rl){2-3} \cmidrule(rl){4-7}\cmidrule(rl){8-11}
  &Mean & Std.\ dev & Robust & Cluster & AKM & AKM0 & Robust & Cluster & AKM & AKM0 \\
  & \multicolumn{1}{c}{(1)} & \multicolumn{1}{c}{(2)} & \multicolumn{1}{c}{(3)} & \multicolumn{1}{c}{(4)}
& \multicolumn{1}{c}{(5)} & \multicolumn{1}{c}{(6)} & \multicolumn{1}{c}{(7)} & \multicolumn{1}{c}{(8)} & \multicolumn{1}{c}{(9)}
& \multicolumn{1}{c}{(10)} \\
  \midrule

\multicolumn{10}{@{}l}{\textbf{Panel A\@: Sensitivity to the number of sectors}} \\ [2pt]

2-digit ($S = 20$)      &       $-0.01$ &       $3.19$  &       $0.65$  &       $0.96$  &       $2.84$  &       $6.06$  &       $70.6$\%        &       $56.1$\%        &       $12.0$\%        &       $5.8$\% \\
3-digit ($S=136$)       &       $0.00$  &       $2.25$  &       $0.73$  &       $0.94$  &       $2.18$  &       $2.72$  &       $54.2$\%        &       $42.5$\%        &       $7.5$\% &       $4.5$\% \\ [6pt]

\multicolumn{10}{@{}l}{\textbf{Panel B\@: Sensitivity to the variance of the shifters}} \\ [2pt]
$\sigma^{2}=0.5$    &       $-0.04$ &       $6.33$  &       $2.33$  &       $2.91$  &       $6.04$  &       $7.02$  &       $48.5$\%        &       $38.0$\%        &       $7.9$\% &       $4.5$\% \\
$\sigma^{2}=10$     &       $0.00$  &       $1.41$  &       $0.52$  &       $0.65$  &       $1.35$  &       $1.57$  &       $48.1$\%        &       $37.8$\%        &       $7.5$\% &       $4.5$\% \\ [6pt]

\multicolumn{10}{@{}l}{\textbf{Panel C\@: Log-normal shifters}} \\ [2pt]
$\sigma^{2}=5$      &       $0.27$  &       $2.26$  &       $0.86$  &       $1.05$  &       $2.17$  &       $3.7$   &       $44.6$\%        &       $35.3$\%        &       $7.7$\% &       $5.2$\% \\ [6pt]

\multicolumn{10}{@{}l}{\textbf{Panel D\@: Heteroskedastic shifters}} \\ [2pt]
$\lambda=3$&        $-0.01$&   1.63&    0.55&    0.72&    1.51&    2.14& 52.1\%&    40.1\%&    8.7\%&    4.0\%\\
$\lambda=7$&       0.01 &   1.38  &  0.44  &  0.58  &  1.23  &  2.01  &  53.7\%  &  41.1\%  &  9.5\%  &  4.2\%\\[6pt]
\multicolumn{10}{@{}l}{\textbf{Panel E\@: Simulated state-level shocks in regression residual}} \\ [2pt]
        &       $0.00$  &       $2.11$  &       $0.86$  &       $1.11$  &       $1.99$  &       $2.32$  &       $42.8$\%        &       $30.4$\%        &       $7.9$\% &       $4.6$\% \\ [6pt]

\multicolumn{10}{@{}l}{\textbf{Panel F\@: Simulated `large' sector shifter in regression residual}} \\ [2pt]
        &       $-0.01$ &       $2.01$  &       $0.74$  &       $0.92$  &       $1.90$  &       $2.21$  &       $48.4$\%        &       $37.8$\%        &       $7.9$\% &       $4.6$\% \\ [6pt]

\multicolumn{10}{@{}l}{\textbf{Panel G\@: Including a `large' sector in shift-share regressor}} \\ [2pt]
        &       $-0.02$ &       $4.25$  &       $0.59$  &       $0.76$  &       $1.18$  &       $1.34$  &       $92.0$\%        &       $89.6$\%        &       $77.2$\%        &       $76.3$\%        \\ [2pt]

  \bottomrule
  \multicolumn{11}{p{6.6in}}{\scriptsize{Notes: All estimates in this table use
  the change in the share of the working-age population employed in each CZ as
  the outcome variable $Y_{i}$ in~\cref{eq:bartikreg}. This table indicates the
  mean and standard deviation of the OLS estimates of $\beta$
  in~\cref{eq:bartikreg} across the placebo samples (columns (1) and (2)), the
  median effective standard error estimates (columns (3) to (6)), and the
  percentage of placebo samples for which we reject the null hypothesis
  $H_{0}\colon\beta=0$ using a 5\% significance level test (columns (7) to
  (10)). \textit{Robust} is the Eicker-Huber-White standard error;
  \textit{Cluster} is the standard error that clusters CZs in the same state;
  \textit{AKM} is the standard error in \Cref{remark:akm-construction};
  \textit{AKM0} is the confidence interval in \Cref{remark:akm0-construction}.
  For each inference procedure, the median effective standard error is equal to
  the median length of the corresponding 95\% confidence interval divided by
  $2\times1.96$. Results are based on 30,000 placebo samples. This table
  presents results for placebo simulations that depart from the baseline; the
  results should thus be compared to those in
  \Cref{tab:simple_placebo,tab:simple_placebo_akm}. In Panel A, we reduce the
  number of sectors relative to the baseline. In Panel B, we change the variance
  of the distribution from which all shifters are drawn. In Panel C, we assume
  that the distribution from which all shifters are drawn is log-normal
  (re-centered at zero) with variance equal to five. In Panel D, we allow the
  variance of the shock in each sector to be heteroskedastic, $\sigma^{2}_{s}=5+\lambda(n_{s}-S/N)$.  In Panel E, we simulate state-level shocks and include them in our regression residual. In Panels F and G, we simulate a shifter for the non-manufacturing sector and include it in our regression residual and in our shift-share regressor, respectively.}}
\end{tabular}
\end{table}

Panels E and F of \Cref{tab:departures_placebo} explore the robustness of our
baseline results to different patterns of correlation in the regression
residuals. In the baseline placebo, since $\beta=0$, the regression residuals
inherit the correlation patterns in the outcome variable. Here, we modify these
patterns by adding a random shock $\eta^{m}_{i}$ in each placebo sample $m$ to
the outcome $Y_{i}$. Panel E explores the impact of increasing the correlation
between the regression residuals of CZs that belong to the same state.
Specifically, we generate a random variable $\tilde{\eta}_{k}^m$ for each state
$k$ and simulation $m$ such that $\tilde{\eta}_{k}^m \sim \mathcal{N}(0,6)$. We
then set $\eta_{i}^m = \tilde{\eta}_{k(i)}^m$ where $k(i)$ is the state of CZ
$i$. Since we have now increased the relative importance of the correlation
pattern accounted for by \emph{Cluster} standard errors, the resulting
overrejection decreases from 38.3\% to 30.4\%. In line with claim (e) above, the
rejection rates of the \emph{AKM} and \emph{AKM0} inference procedures are not
affected. In Panel F, we evaluate the robustness of our results to adding a
shock to the non-manufacturing sector that is included in the regression
residual. Specifically, in each simulation $m$, we set
$\eta_i^m = (1-\sum_{s=1}^{S}w_{is})\hat{\eta}^m_{S}$ with
$\hat{\eta}^m_{S} \sim \mathcal{N}(0,5)$, where $\sum_{s=1}^{S}w_{is}$ is the
1990 aggregate employment share of the 396 4-digit SIC manufacturing sectors
included in the definition of the shift-share regressor of interest. The results
in Panel F of \Cref{tab:departures_placebo} show that adding this component to
the regression residual does not affect the rejection rates.

Lastly, Panel G in \Cref{tab:departures_placebo} explores the consequences of
adding the non-manufacturing sector to the shift-share regressor. In Panel F,
the shock to the non-manufacturing sector is part of the regression residual; in
Panel G, we use this shock, in combination with the shocks to all manufacturing
sectors, to construct the shift-share regressor. Across CZs, the average initial
employment share in the non-manufacturing sector is 77.5\%; i.e.
$N^{-1}\sum_{i=1}^{N}(1-\sum_{s=1}^{S}w_{is})=77.5\%$. Including such a large
sector in the shift-share regressor violates
\Cref{item:sector-size,item:sector-size-normality}. As a result, the
\textit{AKM} and \textit{AKM0} inference procedures overreject severely;
standard inference procedures fare even worse, with rejection rates reaching up
to 92\%. The results in Panels F and G suggest that, provided that the shifters
are independent across sectors, it is better to exclude large sectors from the
shift-share regressor of interest, and thus let the shocks associated with them
enter the regression residual. One should, however, bear in mind that, if
$\beta_{is}$ in \cref{eq:potential-outcomes} varies across sectors, excluding
large sectors from the shift-share regressor will change the estimand $\beta$
(see \Cref{theorem:consistency-Z}).

In the placebo simulations described in
\Cref{tab:simple_placebo,tab:simple_placebo_akm,tab:departures_placebo}, we have drawn the
shifters independently from a mean-zero distribution. In \Cref{tab:correlation_cross_section_placebo}, we allow for
non-zero correlation in the shifters within ``clusters'' of sectors.\footnote{In
\begin{NoHyper}\Cref{sec:control_residual_sector}\end{NoHyper}, we study the
impact of drawing the shifters from a distribution with non-zero mean. We show
that, in line with the discussion in \Cref{sec:adding-covariates}, it is
important to control for the region-specific sum of shares
$\sum_{s=1}^{S}w_{is}$.} Specifically, we report results from placebo exercises
in which the shifters are drawn from the joint distribution
$(\Xs_{1}^m, \dotsc, \Xs_{S}^m) \sim \mathcal{N}\left(0,\Sigma\right)$, where
$\Sigma$ is an $S\times S$ covariance matrix with elements
$\Sigma_{sk}=(1-\rho)\sigma\1{s=k}+\rho\sigma\1{c(s)=c(k)}$ and $c(s)$ indicates
the ``cluster'' that industry $s$ belongs to. In panels A, B, and C, these
clusters correspond to the 3-, 2-, and 1-digit SIC sector that the 4-digit SIC
sector $s$ belongs to, respectively.

\afterpage{\begin{landscape}
\begin{table}[ht]
\centering
\caption{Correlation in sectoral shocks}\label{tab:correlation_cross_section_placebo}
\begin{tabular}{@{}l rc cccccc rrrrrr @{}}
\toprule
        &\multicolumn{2}{c}{Estimate} & \multicolumn{6}{c}{Median eff.\ s.e.} & \multicolumn{6}{c}{Rejection rate}\\ [1pt]
\cmidrule(rl){2-3} \cmidrule(rl){4-9}\cmidrule(rl){10-15}
  &&&&& \multicolumn{2}{c}{Independent} & \multicolumn{2}{c}{2-digit SIC} &&& \multicolumn{2}{c}{Independent} & \multicolumn{2}{c}{2-digit SIC} \\
   \cmidrule(rl){6-7} \cmidrule(rl){8-9} \cmidrule(rl){12-13} \cmidrule(rl){14-15}
  &Mean & \multicolumn{1}{c}{Std.\ dev}  & Robust & Cluster &  AKM  & AKM0 & AKM  &  AKM0 & Robust & Cluster &  AKM  & AKM0 & AKM  & AKM0 \\
  & \multicolumn{1}{c}{(1)} & \multicolumn{1}{c}{(2)} & \multicolumn{1}{c}{(3)} & \multicolumn{1}{c}{(4)}
& \multicolumn{1}{c}{(5)} & \multicolumn{1}{c}{(6)} & \multicolumn{1}{c}{(7)} & \multicolumn{1}{c}{(8)} & \multicolumn{1}{c}{(9)}
& \multicolumn{1}{c}{(10)} & \multicolumn{1}{c}{(11)} & \multicolumn{1}{c}{(12)} & \multicolumn{1}{c}{(13)} & \multicolumn{1}{c}{(14)} \\
  \midrule

\multicolumn{15}{@{}l}{\textbf{Panel A\@: Simulated shifters with correlation within 3-digit SIC sectors}} \\ [2pt]
$\rho = 0.00$   &       $-0.01$ &       $2.00$  &       $0.74$  &       $0.92$  &       $1.91$  &       $2.22$  &       $1.86$  &       $2.64$  &       $48.6$\%        &       $37.8$\%        &       $7.9$\% &       $4.6$\% &       $8.9$\% &       $4.7$\% \\
$\rho = 0.50$   &       $0.02$  &       $2.14$  &       $0.77$  &       $1.07$  &       $2.03$  &       $2.16$  &       $2.24$  &       $2.87$  &       $49.2$\%        &       $32.3$\%        &       $6.4$\% &       $7.1$\% &       $4.4$\% &       $4.7$\% \\
$\rho = 1.00$   &       $0.01$  &       $2.27$  &       $0.76$  &       $1.08$  &       $1.99$  &       $2.10$  &       $2.38$  &       $3.14$  &       $52.2$\%        &       $35.0$\%        &       $8.6$\% &       $9.7$\% &       $4.7$\% &       $4.8$\% \\ [6pt]

\multicolumn{15}{@{}l}{\textbf{Panel B\@: Simulated shifters with correlation within 2-digit SIC sectors}} \\ [2pt]
$\rho = 0.00$   &       $-0.01$ &       $1.99$  &       $0.73$  &       $0.92$  &       $1.90$  &       $2.22$  &       $1.86$  &       $2.65$  &       $48.2$\%        &       $37.7$\%        &       $7.6$\% &       $4.5$\% &       $8.8$\% &       $4.5$\% \\
$\rho = 0.50$   &       $-0.01$ &       $2.73$  &       $0.73$  &       $1.13$  &       $1.82$  &       $1.89$  &       $2.92$  &       $4.07$  &       $62.3$\%        &       $43.2$\%        &       $20.5$\%        &       $23.5$\%        &       $5.8$\% &       $5.0$\% \\
$\rho = 1.00$   &       $0.01$  &       $3.20$  &       $0.69$  &       $1.18$  &       $1.67$  &       $1.63$  &       $3.38$  &       $6.16$  &       $68.4$\%        &       $48.1$\%        &       $31.2$\%        &       $35.7$\%        &       $6.3$\% &       $5.2$\% \\ [6pt]

\multicolumn{15}{@{}l}{\textbf{Panel C\@: Simulated shifters with correlation within 1-digit SIC sectors}} \\ [2pt]
$\rho = 0.00$   &       $0.01$  &       $1.98$  &       $0.73$  &       $0.92$  &       $1.90$  &       $2.21$  &       $1.85$  &       $2.65$  &       $48.5$\%        &       $37.7$\%        &       $7.3$\% &       $4.4$\% &       $8.5$\% &       $4.4$\% \\
$\rho = 0.50$   &       $0.02$  &       $4.95$  &       $0.74$  &       $1.41$  &       $1.88$  &       $1.59$  &       $2.42$  &       $2.36$  &       $84.2$\%        &       $72.3$\%        &       $58.3$\%        &       $64.7$\%        &       $42.7$\%        &       $53.2$\%        \\
$\rho = 1.00$   &       $0.42$  &       $59.63$ &       $0.71$  &       $1.74$  &       $1.86$  &       $1.02$  &       $2.94$  &       $1.67$  &       $90.2$\%        &       $78.3$\%        &       $75.2$\%        &       $82.7$\%        &       $52.1$\%        &       $65.6$\%        \\ [2pt]
  \bottomrule
  \multicolumn{15}{p{8.9in}}{\scriptsize{Notes: All estimates in this table use the change in the share of the working-age population employed in each CZ as the outcome variable $Y_{i}$ in~\cref{eq:bartikreg}. This table indicates the mean and standard deviation of the OLS estimates of $\beta$ in~\cref{eq:bartikreg} across the placebo samples (columns (1) and (2)), the median effective standard error estimates (columns (3) to (8)), and the percentage of placebo samples for which we reject the null hypothesis $H_{0}\colon\beta=0$ using a 5\% significance level test (columns (9) to (14)). \textit{Robust} is the Eicker-Huber-White standard error; \textit{Cluster} is the standard error that clusters CZs in the same state; in columns (5) and (11), \textit{AKM} is the standard error in \Cref{remark:akm-construction}; in columns (7), and (13), \textit{AKM} is the standard error in \cref{eq:se-AKM-cluster} for 2-digit SIC sector clusters; in columns (6) and (12), \textit{AKM0} is the confidence interval in \Cref{remark:akm0-construction}; in columns (8) and (14), \textit{AKM0} adjusts the confidence interval in \Cref{remark:akm0-construction} as indicated in \Cref{sec:clustersectors}. For each inference procedure, the median effective standard error is equal to the median length of the corresponding 95\% confidence interval divided by $2\times1.96$. Results are based on 30,000 placebo samples.}}
\end{tabular}
\end{table}
\end{landscape}}

Panel A of \Cref{tab:correlation_cross_section_placebo} shows that introducing
correlation within 3-digit SIC sectors has a moderate effect on the rejection
rates of both the traditional methods and versions of the \textit{AKM} and
\textit{AKM0} methods that assume that the sectoral shocks are independent. Rejection rates close
to 5\% are obtained with versions of the \textit{AKM} and \textit{AKM0}
inference procedures that cluster the shifters at a 2-digit SIC level (see
\Cref{sec:clustersectors}). As shown in Panel B, the overrejection problem
affecting both traditional inference procedures and versions of the \textit{AKM}
and \textit{AKM0} procedures that assume independence of shifters is more severe
when the shifters are correlated at the 2-digit level. However, the last two
columns show that, in this case,
the versions of \textit{AKM} and \textit{AKM0} that cluster the sectoral shocks at
the 2-digit level achieve rejection rates close to the nominal level. Finally,
Panel C shows that the overrejection problem is much more severe in the presence
of high correlation in shifters within the two 1-digit aggregate
sectors, and this problem is not solved by clustering at the
2-digit level.

The last panel in \Cref{tab:correlation_cross_section_placebo} illustrates the
inferential problems that arise in empirical applications of shift-share designs
when all shifters are correlated with each other. Such correlations also arise,
for example, when all shifters are generated (at least in part) by a common
shock with potentially heterogeneous effects across sectors.\footnote{There is
  an extensive empirical literature documenting the importance of common factors
  driving changes in sector-specific variables such as sectoral industrial
  production, employment and value added \citep*[see,
  e.g.,][]{altonji_comovement_1990,shea_2002,foerster_aggshocks_2011}.} As
simulations presented in \begin{NoHyper}\Cref{tab:simulation_china_factor} in
  \Cref{sec:plac-exerc-altern}\end{NoHyper} illustrate, if there is a common
component affecting all shifters, it is important to first estimate this common
component and to control for it in the shift-share regression of interest.
Otherwise, hypothesis tests based on standard inference procedures as well as on
the \textit{AKM} and \textit{AKM0} inference procedures may suffer from an
overrejection problem.

We summarize the conclusions from
\Cref{tab:departures_placebo,tab:correlation_cross_section_placebo}
in the following remark.

\begin{remark}\label{remark:overerejection}
  In shift-share regressions, overrejection of the usual inference procedures is
  more severe when there is a small number sectors. In this case, the methods we
  provide attenuate the overrejection problem, but may still overreject when the
  number of sectors is very small. Our methods perform well under different
  distributions of shifters and regression residuals, but they lead to an
  overrejection problem when the shift-share covariate aggregates over a large
  sector. Finally, when the shifters are not independent across sectors, it is
  important to properly account for their correlation structure.
\end{remark}

In \begin{NoHyper}\Cref{sec:placebo_confounding,app:panel_placebo,app:approximation_error,app: placebo_other_share,sec:heter-treatm-effects}\end{NoHyper} we present results from additional placebo simulations in which we investigate the consequences of: (a) the violation of the assumption that the shifters of interest are as good as randomly assigned; (b) the presence of serial correlation in both the shifters of interest and the regression residuals, in panel data settings; (c) the true potential outcome function being nonlinear, implying that the linearly additive potential outcome framework in \cref{eq:potential-outcomes} is misspecified; (d) the presence in the regression residuals of shift-share components with shares correlated in different degrees with those entering the shift-share covariate of interest; and, (e) the presence of treatment heterogeneity across regions and sectors.

\section{Empirical applications}\label{sec:empapp}

We now apply the \textit{AKM} and \textit{AKM0} inference procedures to two
empirical applications. First, the effect of Chinese competition on U.S. local
labor markets, as in \citet*{autordornhanson2013}. Second, the estimation of the
local inverse elasticity of labor supply, as in \citet{bartik1991benefits}.
Additionally, in \begin{NoHyper}\Cref{appsec:applicationBP}\end{NoHyper}, we
apply the \textit{AKM} and \textit{AKM0} inference procedures to the study of
the impact of immigration on labor market outcomes of U.S. natives.

\subsection{Effect of Chinese exports on U.S. labor market outcomes}\label{sec:ADH}

\citet*[henceforth ADH]{autordornhanson2013}, explore the impact of exports from
China on labor market outcomes across U.S. CZs. Specifically, ADH present IV
estimates for a specification that fits within the panel data setting described
in \Cref{sec:panel}, with each region $j=1,\dots,722$ denoting a CZ, each sector
$k=1,\dots,396$ denoting a 4-digit SIC industry, and each period $t=1,2$
denoting either 1990--2000 changes or 2000--2007 changes. As in
\Cref{sec:panel}, we index here the intersection of a region $j$ and a period
$t$ by $i$, and the intersection of a sector $k$ and a period $t$ by $s$. In
ADH, the outcome $Y_{1i}$ is a ten-year equivalent change in a labor-market
outcome, the endogenous treatment is
$Y_{2i} = \sum_{s=1}^{S} \bar{w}_{is} \Xs_s^{US}$, where $\Xs_{s}^{US}$ is the
change in U.S. imports from China normalized by the start-of-period total U.S.
employment in the sector, and $\bar{w}_{is}$ is the start-of-period employment
share of a sector in a CZ\@. ADH use the shift-share IV
$X_{i} = \sum_{s=1}^{S} w_{is} \Xs_s$, where $\Xs_s$ denotes imports from China
by high-income countries other than the U.S. normalized by a ten-year-lag of the start-of-period
total U.S.\ employment in the sector, and $w_{is}$ is the ten-year-lag of the
employment share $\bar{w}_{is}$. To measure these variables, we
use the data sources described in \Cref{sec:data}. In all regression
specifications, we include a vector of controls $Z_{i}$ corresponding to the
largest set of controls used in ADH.\footnote{\label{fn:adh_controls}See column
  (6) of Table 3 in ADH\@. The vector $Z_{i}$ aims to control for labor supply
  shocks and labor demand shocks other than the changes in imports from China,
  and it includes the start-of-period percentage of employment in manufacturing.
  The discussion in \Cref{sec:adding-covariates} implies that one should
  instead control for the ten-year-lagged of the start-of-period employment share in
  manufacturing, to match the shares that enter the definition of the
  shift-share IV\@. However, to facilitate the comparison with the original results
  in ADH, we use their vector of controls. As shown in
  \citet*{borusyakhulljaravel2018shiftshare}, controlling for the
  ten-year-lagged manufacturing employment shares does not substantively affect
  the estimates.}

\Cref{TabADH} reports 95\% CIs computed using different methodologies for the
specifications in Tables 5 to 7 in ADH\@. Panels A, B, and C present the IV,
reduced-form and first-stage estimates, respectively. Following
\citet{autor2014trade}, the \textit{AKM} and \textit{AKM0} CIs cluster the
shifters $\{\Xs_{s}\}_{s=1}^{S}$ by 3-digit SIC industry; thus, the
\textit{AKM} and \textit{AKM0} CIs we report are robust to serial correlation in
the shifters as well as to cross-sectoral correlation in the shifters within
3-digit SIC industries. \begin{NoHyper}\Cref{TabADH_RF,TabADH_2SLS} in
  \Cref{appsec:applicationADH}\end{NoHyper} report \textit{AKM} and
\textit{AKM0} CIs for alternative definitions of clusters.

In \begin{NoHyper}\Cref{appsec:applicationADH}\end{NoHyper}, we present placebo
simulations that depart from our baseline placebo design in ways that explore
specific features of the empirical setting studied in this section.
In \begin{NoHyper}\Cref{tab:ADHshifters_placebo}\end{NoHyper}, we draw the
shifters from the empirical distribution of shifters used to construct the ADH
IV (instead of drawing them from a normal distribution); the resulting rejection
rates are very similar to those in the baseline simulation.
In \begin{NoHyper}\Cref{tab:simulation_china_factor}\end{NoHyper}, we draw
shifters that have a common component with factor structure; since the resulting
correlation structure cannot be captured by clustering, we show that it is
important in this case to include an estimate of the common factor component as
an additional control.\footnote{For placebo simulation evidence under our
  baseline assumption that the shifters are independent across 3-digit clusters,
  using data for outcomes $Y_{1i}$ and shares $w_{is}$ identical to that used in
  this section, see \begin{NoHyper}\Cref{app:panel_placebo}\end{NoHyper}.}

\begin{table}[!t]
\centering
\caption{Effect of Chinese exports on U.S. commuting
  zones---\citet*{autordornhanson2013}}\label{TabADH}
\tabcolsep=0.09cm
\begin{tabular}{@{}lccc cccc@{}}
\toprule
& \multicolumn{3}{c}{{Change in the employment share}} &&
\multicolumn{3}{c}{{Change in avg.\ log weekly wage}}          \\
  \cmidrule(rl){2-4} \cmidrule(rl){6-8}
& All & Manuf. & Non-Manuf. && All & Manuf. & Non-Manuf. \\
 & (1) & (2) & (3) && (4) & (5) & (6) \\
\midrule
\multicolumn{8}{@{}l}{\textbf{Panel A\@: 2SLS Regression}} \\ [2pt]
$\hat{\beta}$  &$-0.77$         & $-0.60$         & $-0.18$ &&       $-0.76$ &       $0.15$  &       $-0.76$ \\
Robust         &$[-1.10,-0.45]$ & $[-0.78,-0.41]$ & $[-0.47,0.12]$  &&       $[-1.23,-0.29]$ &       $[-0.81,1.11]$  &       $[-1.27,-0.25]$ \\
Cluster        &$[-1.12,-0.42]$ & $[-0.79,-0.40]$ & $[-0.45,0.10]$  &&       $[-1.26,-0.26]$ &       $[-0.81,1.11]$  &       $[-1.28,-0.24]$ \\
AKM            &$[-1.25,-0.30]$ & $[-0.84,-0.35]$ & $[-0.54,0.18]$  &&       $[-1.37,-0.15]$ &       $[-0.81,1.11]$  &       $[-1.42,-0.10]$ \\
AKM0           &$[-1.69,-0.39]$ & $[-1.01,-0.36]$ & $[-0.84,0.14]$  &&       $[-1.77,-0.17]$ &       $[-1.49,1.05]$  &       $[-1.97,-0.19]$ \\ [6pt]

\multicolumn{8}{@{}l}{\textbf{Panel B\@: OLS Reduced-Form Regression}} \\ [2pt]

 $\hat{\beta}$  & $-0.49$ &       $-0.38$ &       $-0.11$ && $-0.48$         &       $0.10$  &       $-0.48$ \\
Robust          & $[-0.71,-0.27]$ &       $[-0.48,-0.28]$ & $[-0.31,0.08]$  &&       $[-0.80,-0.16]$ &       $[-0.50,0.69]$  &       $[-0.83,-0.13]$ \\
 Cluster        & $[-0.64,-0.34]$ &       $[-0.45,-0.30]$ & $[-0.27,0.05]$  &&       $[-0.78,-0.18]$ &       $[-0.51,0.70]$  &       $[-0.81,-0.15]$ \\
 AKM            & $[-0.81,-0.17]$ &       $[-0.52,-0.23]$ & $[-0.35,0.12]$  &&       $[-0.88,-0.07]$ &       $[-0.50,0.69]$  &       $[-0.93,-0.03]$ \\
 AKM0           & $[-1.24,-0.24]$ &       $[-0.67,-0.25]$ & $[-0.64,0.08]$  &&       $[-1.27,-0.10]$ &       $[-1.16,0.61]$  &       $[-1.47,-0.11]$ \\ [6pt]

\multicolumn{8}{@{}l}{\textbf{Panel C\@: 2SLS First-Stage}} \\ [2pt]

 $\hat{\beta}$  &       \multicolumn{6}{c}{     $0.63$  }                                                                       \\
Robust  &       \multicolumn{6}{c}{     $[0.46,0.80]$   }                                                                       \\
 Cluster        &       \multicolumn{6}{c}{     $[0.45,0.81]$   }                                                                       \\
 AKM    &       \multicolumn{6}{c}{     $[0.53,0.73]$   }                                                                       \\
 AKM0           &       \multicolumn{6}{c}{     $[0.54,0.84]$   }                                                                       \\ [2pt]
\bottomrule

\multicolumn{8}{p{6.7in}}{\scriptsize{Notes:  $N = 1,444$ (722 CZs $\times$ 2
  time periods). Observations are weighted by the start of period CZ share of national
  population. All regressions include the full vector of baseline controls in ADH\@; i.e.\ those in column (6) of Table 3 in \citet*{autordornhanson2013}. 95\% confidence intervals are reported in square brackets.
\textit{Robust} is the Eicker-Huber-White standard error; \textit{Cluster} is
  the standard error that clusters of CZs in the same state; \textit{AKM} is the
  standard error in \cref{eq:se-AKM-cluster} with 3-digit SIC clusters;
  \textit{AKM0} is the confidence interval with 3-digit SIC
  clusters described in the last sentence of~\Cref{sec:clustersectors}.}}

\end{tabular}
\end{table}

In \Cref{TabADH}, state-clustered CIs are very similar to the
heteroskedasticity-robust ones. In contrast, our proposed CIs are wider than
those implied by state-clustered standard errors. For the IV estimates reported
in Panel A, the average increase across all outcomes in the length of the 95\%
CI is 24\% with the \textit{AKM} procedure and 65\% with the \textit{AKM0}
procedure. When the outcome is the change in the manufacturing employment rate,
the length of the 95\% CI increases by 26\% with the \textit{AKM} procedure and
by 65\% with the \textit{AKM0} procedure. In light of the lack of impact of
state-clustering on the 95\% CI, the wider intervals implied by our inference
procedures indicate that cross-region residual correlation is driven by
similarity in sectoral compositions rather than by geographic proximity.

Panel B of \Cref{TabADH} reports CIs for the reduced-form specification. In this
case, the increase in the CI length is slightly larger than for the IV estimates: across outcomes, it
increases on average by 54\% for \textit{AKM} and 130\% for \textit{AKM0}. The
smaller relative increase in the CI length for the IV estimate
relative to its increase for the reduced-form estimate is a consequence of the
fact that all inference procedures yield similar CIs for the first-stage
estimate, as reported in Panel C.

As discussed in \Cref{sec:placebo-exercise}, the differences between
\textit{AKM} (or \textit{AKM0}) CIs and state-clustered CIs are related to the
importance of shift-share components in the regression residual. The results in
Panel C suggest that, once we account for changes in sectoral imports from China
to other high-income countries, there is not much sectoral variation left in the
first-stage regression residual; i.e., there are no other sectoral variables
that are important to explain changes in sectoral imports from China to the
U.S.\footnote{This is analogous to what we would observe in a regression in
  which the regressor of interest varies at the state level, and we control for
  all state-specific covariates affecting the outcome variable: state-clustered
  standard errors would be similar to heteroskedasticity-robust standard errors,
  since there is little within-state correlation left in the residuals.} To
investigate this claim, \begin{NoHyper}\Cref{tab:ADH_FS_placebo} in
  \Cref{appsec:applicationADH}\end{NoHyper} reports the rejection rates implied
by a placebo exercise designed to match the first-stage specification reported
in Panel C of \Cref{TabADH}. The placebo results show that, while traditional
methods still suffer from severe overrejection when no controls are included,
the overrejection is attenuated once we include as controls the shift-share IV
and the control vector $Z_{i}$ we use in \Cref{TabADH}, indicating that these
variables soak up much of the cross-CZ correlation in the treatment variable
used in ADH\@.

Overall, \Cref{TabADH} shows that, despite the wider confidence intervals
obtained with our procedures, the qualitative conclusions in ADH remain valid at
usual significance levels. However, the increased width of the 95\% CI shows
that the uncertainty regarding the magnitude of the impact of Chinese import
exposure on U.S. labor markets is greater than that implied by usual
inference procedures. In particular, the \emph{AKM0} CI is much wider than that
based on state-clustered standard errors; furthermore due to its asymmetry
around the point estimate, using the \emph{AKM0} CI, we cannot rule out impacts
of the China shock that are two to three times larger than the point estimates
of these effects.\footnote{It follows from \Cref{remark:akm0-construction} (see
  the expression for the quantity $A$) that the asymmetry in the \textit{AKM0}
  CI comes from the correlation between the regression residuals $\hat{R}_{s}$
  and the shifters cubed. In large samples, this correlation is zero and the
  \emph{AKM} and \emph{AM0} CIs are asymptotically equivalent. The differences
  between both CIs in \Cref{TabADH} thus reflect differences in their
  finite-sample properties. This notwithstanding, the placebo exercise presented
  in \begin{NoHyper}\Cref{app:panel_placebo}\end{NoHyper} shows that both inference procedures yield close
  to correct rejection rates in a sample analogous to that used in ADH.}

\subsection{Estimation of inverse labor supply elasticity}\label{sec:estlaborsupply}

In our second application, we estimate the inverse labor supply elasticity.
Specifically, using the notation of \Cref{sec:model}, we estimate the parameter $\tilde{\phi}$ in the equation
\begin{align}\label{eq:labor_sup_empirical}
\hat{\omega}_{i}=\tilde{\phi}\hat{L}_{i}+\delta Z_{i}+\epsilon_{i}, \qquad\tilde{\phi}=\phi^{-1},
\end{align}
where $\hat{L}_{i}$ denotes the log change in the employment rate in CZ $i$,
$\hat{\omega}_{i}$ denotes the log change in wages, $Z_{i}$ is a vector of
controls, and $\epsilon_{i}$ is a regression residual. We use the same sample,
data sources, and vector of controls $Z_{i}$ as in
\Cref{sec:ADH}.\footnote{\begin{NoHyper}\Cref{tab:laborsupply_controls} in
    \Cref{app:estlaborsupply} \end{NoHyper} investigates the robustness of our
  results to alternative sets of controls.}

The model in \Cref{sec:model} has implications for the properties of different
strategies for estimating the inverse labor supply elasticity $\tilde{\phi}$. By
\cref{eq: invlaborsupeq}, the residual $\epsilon_{i}$ in
\cref{eq:labor_sup_empirical} accounts for changes in labor supply shocks,
$\sum_{g=1}^G \tilde{w}_{ig} \hat{\nu}_g + \hat{\nu}_i$, not controlled for by
the vector $Z_{i}$. Second, it follows from \cref{eq:emp_change} that, up to a
first-order approximation around an initial equilibrium, changes in regional
employment rates, $\hat{L}_{i}$, can be written as a function of both
shift-share aggregators of sectoral labor demand shocks and the same labor
supply shocks potentially entering $\epsilon_{i}$ in \cref{eq: invlaborsupeq},
$\sum_{g=1}^G \tilde{w}_{ig} \hat{\nu}_g + \hat{\nu}_i$. Thus, $\hat{L}_i$ and
$\epsilon_{i}$ will generally be correlated and the OLS estimator of
$\tilde{\phi}$ in \cref{eq:labor_sup_empirical} will be biased. However, as
discussed in \Cref{sec:ivreg}, the model in \Cref{sec:model} also implies that
we can instrument for $\hat{L}_i$ using shift-share aggregators of sectoral
labor demand shocks that are independent of the unobserved labor supply shocks
(see \begin{NoHyper}\Cref{app:estlaborsupply}\end{NoHyper} for more details).

\begin{table}[!t]
\centering
\caption{Estimation of inverse labor supply elasticity}\label{tab:laborsupply}
\begin{tabular}{@{}lccc@{}}
\toprule
& First-Stage & Reduced-Form & 2SLS \\

Dependent variable: & $\hat{L}_i$ & $\hat{\omega}_i$ & $\hat{\omega}_i$ \\
 & (1) & (2) & (3) \\

\midrule

\multicolumn{4}{@{}l}{\textbf{Panel A\@: Bartik IV---Not leave-one-out estimator}} \\ [2pt]
$\hat{\beta}$   & $0.90$          &  $0.73$          & $0.80$   \\
Robust          & $[0.70,1.10]$    &  $[0.54,0.91]$   & $[0.64,0.97]$   \\
Cluster         & $[0.64,1.16]$   &  $[0.47,0.98]$   & $[0.60,1.01]$    \\
AKM             & $[0.65,1.16]$   &  $[0.49,0.96]$   & $[0.62,0.98]$   \\
AKM0            & $[0.61,1.17]$   &  $[0.44,0.96]$   & $[0.59,1.02]$   \\ [6pt]

\multicolumn{4}{@{}l}{\textbf{Panel B\@: Bartik IV---Leave-one-out estimator}} \\ [2pt]

 $\hat{\beta}$  &       $0.87$  &       $0.71$  &       $0.82$  \\
Robust  &       $[0.68,1.06]$   &       $[0.53,0.89]$   &       $[0.65,0.98]$   \\
Cluster &       $[0.62,1.12]$   &       $[0.46,0.96]$   &       $[0.60,1.03]$    \\
AKM (\textit{leave-one-out})    &       $[0.59,1.15]$   &       $[0.47,0.94]$   &       $[0.61,1.02]$   \\
AKM0  (\textit{leave-one-out})  &       $[0.53,1.15]$   &       $[0.42,0.94]$   &       $[0.59,1.09]$   \\ [6pt]

\multicolumn{4}{@{}l}{\textbf{Panel C\@: ADH IV}} \\ [2pt]
 $\hat{\beta}$  &       $-0.72$ &       $-0.48$ &       $0.67$  \\
Robust  &       $[-1.04,-0.39]$ &       $[-0.80,-0.16]$ &       $[0.36,0.98]$   \\
Cluster &       $[-0.93,-0.50]$ &       $[-0.78,-0.18]$ &       $[0.35,0.99]$   \\
AKM     &       $[-1.19,-0.24]$ &       $[-0.88,-0.07]$ &       $[0.27,1.07]$   \\
AKM0            &       $[-1.83,-0.35]$ &       $[-1.27,-0.10]$  &       $[0.18,1.14]$   \\ [2pt]

\bottomrule

  \multicolumn{4}{p{4.7in}}{\scriptsize{Notes:  $N = 1,444$ (722 CZs $\times$ 2
  time periods). The variable $\hat{L}_i$ denotes the log-change in the employment rate in CZ $i$. The variable $\hat{\omega}_i$ denotes the log change in mean weekly earnings. Observations are weighted by the start of period CZ share of national population. All regressions include the full vector of baseline controls in ADH\@; i.e.\ those in column (6) of Table 3 in \citet*{autordornhanson2013}. 95\% confidence intervals in square brackets. \textit{Robust} is the Eicker-Huber-White standard error; \textit{Cluster} is the standard error that clusters of CZs in the same state; \textit{AKM} is the standard error in \cref{eq:se-AKM-cluster} with 3-digit SIC clusters; \textit{AKM0} is the confidence interval with 3-digit SIC clusters described in the last sentence of~\Cref{sec:clustersectors}; \textit{AKM (leave-one-out)} is the standard error in \Cref{sec:mismeasuredIV} with 3-digit SIC clusters; \textit{AKM0 (leave-one-out)} is the confidence interval with 3-digit SIC clusters described in \Cref{sec:mismeasuredIV}.}}

\end{tabular}
\end{table}

In this section, we use three different shift-share IVs to estimate
$\tilde{\phi}$ in \cref{eq:labor_sup_empirical}. For each of them,
\Cref{tab:laborsupply} presents the reduced-form, first-stage and 2SLS
estimates. First, in Panel A, we use the instrumental variable in
\citet{bartik1991benefits}; i.e. $\hat{X}_{i}=\sum_{i=1}^{N}w_{is}\hat{L}_{s}$,
where $\hat{L}_{s}$ denotes the nation-wide employment growth in sector $s$.
Second, in Panel B, we use the leave-one-out version of this instrument; i.e.
$\hat{X}_{i}=\sum_{i=1}^{N}w_{is}\hat{L}_{s, -i}$, where $\hat{L}_{s, -i}$
denotes the employment growth in sector $s$ over all CZs excluding CZ
$i$.\footnote{\label{fn:duggan}The leave-one-out version of the instrument in
  \citet{bartik1991benefits} was originally proposed by \citet{autorduggan2003}.
  In Online Appendix \begin{NoHyper}\ref{app:missIVandModel}\end{NoHyper}, we
  clarify the assumptions under which the model in \Cref{sec:model} is
  consistent with the validity of the leave-one-version of the Bartik
  IV\@. \begin{NoHyper}\Cref{app:placebo_leave_one_out}\end{NoHyper} presents
  placebo exercises attesting that the \textit{AKM} and \textit{AKM0} CIs
  reported in this section have appropriate coverage in the context of this
  empirical application.} Third, in Panel C, we use the IV used in
\citet*{autordornhanson2013}, which we denote as ADH IV and describe in detail
in \Cref{sec:ADH}.\footnote{The effect of these IVs on the changes in the
  employment rate may be heterogeneous across regions and sectors
  (see~\cref{eq:emp_change}). This does not affect the validity of our inference
  procedures since, as discussed in \Cref{sec:ivreg}, we allow for heterogeneous
  effects in the first-stage regression.} As in \Cref{sec:ADH}, we report
versions of the \textit{AKM} and \textit{AKM0} CIs with shifters clustered at
the 3-digit SIC industry for all periods.

Column (3) of \Cref{tab:laborsupply} shows that the estimates of the inverse
labor supply elasticity are similar no matter which IV we use: 0.80 when using
the original Bartik IV, 0.82 when using the leave-one-out version of this
estimator, and 0.67 when using the ADH
IV\@.\footnote{\label{fn:loo-vs-bartik}One explanation for the similarity
  between the leave-one-out and the original Bartik IV is that, as discussed in
  \Cref{sec:mismeasuredIV}, the bias of the original Bartik IV is of the order
  $\frac{1}{\N}
  \sum_{i=1}^{N}\sum_{s=1}^{S}\frac{w_{is}\check{w}_{is}}{\check{n}_{s}}$. This
  quantity equals $0.004$ in this application, indicating that the
  own-observation bias is likely to be small.} In both Panel A and Panel B, the
\textit{AKM} and \textit{AKM0} CIs are very similar to the state-clustered CI\@.
In Panel C, the \textit{AKM} and \textit{AKM0} CIs are only moderately wider
than those obtained with state-clustered standard errors.

Columns (1) and (2) of \Cref{tab:laborsupply} show the first-stage and
reduced-form estimates, respectively. In Panel A and Panel B, the \textit{AKM0}
CIs are similar to the state-clustered CIs; in contrast, in Panel C,
the first-stage and reduced-form \textit{AKM0} CIs more twice as wide, and more
than three times as wide as the state-clustered CI, respectively.
Thus, the first-stage and reduced-form \textit{AKM} and \textit{AKM0} CIs differ
more from the state-clustered CI when the ADH IV is used than when the Bartik IV
is used. A possible explanation for this finding is that the shift-share
component of the first-stage and reduced-form regression residuals is much
smaller in the latter than in the former case. The Bartik IV absorbs the bulk of
the shift-share covariates that affect the change in the employment rate and
wages across CZs. In contrast, the ADH IV is just one of the possibly various
shift-share terms affecting the change in the outcome and endogenous treatment
of interest. With the remaining shift-share entering the regression residual, it
becomes quantitatively important to use our inference procedures to obtain CIs
with the right coverage.

\section{Concluding remarks}\label{sec:conclusion}

This paper studies inference in shift-share designs. We show that standard
economic models predict that changes in regional outcomes depend on observed and
unobserved sector-level shocks through several shift-share terms. Our model thus
implies that the residual in shift-share regressions is likely to be correlated
across regions with similar sectoral composition, independently of their
geographic location, due to the presence of unobserved shift-share terms. Such
correlations are not accounted for by inference procedures typically used in
shift-share regressions, such as when standard errors are clustered on
geographic units. To illustrate the importance of this shortcoming, we conduct
a placebo exercise in which we study the effect of randomly generated
sector-level shocks on actual changes in labor market outcomes across CZs in the
United States. We find that traditional inference procedures severely overreject
the null hypothesis of no effect. We derive two novel inference procedures that
yield correct rejection rates.

It has become standard practice to report cluster-robust standard errors in
regression analysis whenever the variable of interest varies at a more aggregate
level than the unit of observation. This practice guards against potential
correlation in the residuals that arises whenever these residuals contain
unobserved shocks that also vary at the same level as the variable of interest.
In the same way, we recommend that researchers report confidence intervals in
shift-share designs that allow for a shift-share structure in the residuals,
such as one of the two confidence intervals that we propose.

\vspace{1em}
\noindent%
\textsc{University of Chicago Booth School of Business}\\
\textsc{Princeton University}\\
\textsc{Princeton University}
\bibliographystyle{aer}
\bibliography{Refs}

\end{document}


\title{Online supplement to:\\ ``Shift-Share Designs: Theory and Inference''}%
\author{{\Large Rodrigo Adão\thanks{University of Chicago Booth
      School of Business. Email:
      \href{mailto:rodrigo.adao@chicagobooth.edu}{rodrigo.adao@chicagobooth.edu}}} \and {\Large Michal
    Kolesár\thanks{Princeton University. Email:
      \href{mailto:mkolesar@princeton.edu}{mkolesar@princeton.edu}}} \and
  {\Large Eduardo Morales\thanks{Princeton University. Email:
      \href{mailto:ecmorales@princeton.edu}{ecmorales@princeton.edu}}} }

\maketitle
\clearpage
\allowdisplaybreaks

\tableofcontents
\newpage

\begin{appendices}

\crefalias{section}{sappsec}
\crefalias{subsection}{sappsubsec}
\crefalias{subsubsection}{sappsubsubsec}

\numberwithin{equation}{section}%
\numberwithin{table}{section}%
\numberwithin{figure}{section}%
\numberwithin{assumption}{section}%
\numberwithin{lemma}{section}%
\numberwithin{proposition}{section}%

\section{Proofs and additional theoretical results}\label{sec:proofs-econometrics}

\Cref{sec:ols-proofs} gives proofs and additional details for the results in
\begin{NoHyper}\Cref{sec:no-covariates,sec:adding-covariates}\end{NoHyper}.
\Cref{sec:proofs-iv} gives proofs and additional details for the results in
\begin{NoHyper}\Cref{sec:ivreg,sec:mismeasuredIV}\end{NoHyper}.

\subsection{Proofs and additional details for OLS regression}\label{sec:ols-proofs}

Since \begin{NoHyper}\Cref{theorem:consistency-noZ,theorem:normality-noZ} are special cases or
\Cref{theorem:consistency-Z,theorem:normality-Z}, we only prove
\Cref{theorem:consistency-Z,theorem:normality-Z,,theorem:se-consistency}\end{NoHyper}.
We give the proofs under a slightly more general setup that allows for a
linearization error in the potential outcome equation. We introduce this more
general setup in \Cref{sec:assumptions}, where we also collect the assumptions
that we impose on the DGP\@. We collect some auxiliary Lemmata used in the
proofs in \Cref{sec:auxilliary-results}, and we prove these propositions in
\Cref{sec:proof-texorpdfstr-z,,sec:proof-texorpdfstr-z,sec:proof-texorpdfstr-co}.
\Cref{sec:infer-under-heter} discusses inference when the effects $\beta_{is}$
are heterogeneous.

Throughout the \namecref{sec:proofs-econometrics}, we assume that
$\sum_{s=1}^{S}w_{is}\leq 1$ for all $i$. Thus, $\sum_{s=1}^{S}n_{s}\leq N$,
where $n_{s}=\sum_{i=1}^{N}w_{is}$ denotes the size of sector $s$. We use the
notation $A_{S}\preceq B_{S}$ to denote $A_{S}=O(B_{S})$, i.e.\ there exists a
constant $C$ independent of $S$ such that $A_{S}\leq C B_{S}$. Let
$\mathcal{F}_{0}$ denote the $\sigma$-field generated by $(\Zs, U, Y(0), B, W)$
(for the case with no covariates, $\mathcal{F}_{0}$ denotes the $\sigma$-field
generated by $(Y(0), B, W)$). Define
$\overbar{w}_{st}=\sum_{i=1}^{\N}w_{is}w_{it}$,
$\tilde{\Xs}_{s}=\Xs_{s}-\Zs_{s}'\gamma$, and
$\sigma_{s}^{2}=\var(\Xs_{s}\mid\mathcal{F}_{0})$. Finally, let
$r_{\N}=(\sum_{s}n_{s}^{2})^{-1}$, and let $E_{W}$ denote expectation conditional on
$W$.

\subsubsection{General setup and assumptions}\label{sec:assumptions}

We first list and discuss the regularity conditions needed for the results in
\begin{NoHyper}\Cref{sec:no-covariates}\end{NoHyper}. We then generalize the setup
from~\begin{NoHyper}\Cref{sec:adding-covariates} by allowing for a linearization
  error in the potential outcome
  equation~\eqref{eq:potential-outcomes}\end{NoHyper}. Unless stated otherwise,
all limits are taken as $S\to\infty$. We leave the dependence of the number of
regions $\N=\N_{S}$ on $S$ implicit.

For the results in \begin{NoHyper}\Cref{sec:no-covariates}\end{NoHyper}, we
assume that the observed data $({Y}, X, {W})$ is generated by the variables
$({Y}(0), B, {W}, \Xs)$, which we model as a triangular array, so that the
distribution of the data may change with the sample size.\footnote{In other
  words, to allow the distribution of the data to change with the sample size
  $S$, we implicitly index the data by $S$. Making this index explicit, for each
  $S$, the data is thus given by the array
  $\{(Y_{iS}(0), \beta_{isS}, w_{isS}, \Xs_{sS}) \colon i=1,\dotsc, N_{S},
  s=1,\dotsc, S\}$.} The additional regularity conditions we impose on these
variables, in addition to
\begin{NoHyper}\Cref{assumption:DGP-noZ,assumption:inference}\end{NoHyper}
as
follows:

\begin{assumption}\label{assumption:regularity}
  \aitem\label{sitem:beta-bounded} The support of $\beta_{is}$ is bounded; %
  \aitem\label{sitem:detX}
  $\frac{1}{\N}\sum_{i=1}^{\N}\sum_{s=1}^{S}\var(\Xs_{s}\mid \mathcal{F}_{0})
  w_{is}^{2}$ converges in probability to a strictly positive non-random
  limit; %
  \aitem\label{sitem:2nu} For some $\nu>0$,
  $E[\abs{\Xs_{s}}^{2+\nu}\mid \mathcal{F}_{0}]$ exists and is uniformly
  bounded, and conditional on $W$, the second moments of ${Y}_{i}(0)$ exist, and
  are bounded uniformly over $i$; %
  \aitem\label{sitem:4nu} For some $\nu>0$,
  $E[\abs{\Xs_{s}}^{4+\nu}\mid \mathcal{F}_{0}]$ is uniformly bounded, and
  conditional on $W$, the fourth moments of $Y_{i}(0)$ exist, and are bounded
  uniformly over $i$.
\end{assumption}
The bounded support condition on $\beta_{is}$ in \Cref{sitem:beta-bounded} is
made to keep the proofs simple and can be relaxed. \Cref{sitem:detX} is a
standard regularity condition ensuring that the shocks $\Xs$ have sufficient
variation so that the denominator of $\hat{\beta}$, scaled by $\N$, does not
converge to zero. This requires that there is at least one ``non-negligible''
sector in most regions in the sense that its share $w_{is}$ is bounded away from
zero. This implies that $\sum_{s=1}^{S}n_{s}/N$ is also bounded away from zero.
\Cref{sitem:2nu} imposes some mild assumptions on the existence of moments of
$\Xs$ and $Y_{i}(0)$. \Cref{sitem:4nu}, which is only needed for asymptotic
normality, strengthens this condition.

For the results in~\begin{NoHyper}\Cref{sec:adding-covariates}\end{NoHyper}, we generalize the setup in the
main text by allowing for a linearization error in the expression for potential
outcomes,
\begin{equation}\label{eq:potential-outcomes-alt}
  Y_{i}(\xs_{1}, \dotsc, \xs_{S})=Y_{i}(0)+\sum_{s=1}^{S}w_{is}\xs_{s}\beta_{is}
  + L_{i}(\xs_{1}, \dotsc, \xs_{S}),
  \qquad \sum_{s=1}^{S}w_{is}\leq 1,
\end{equation}
and we weaken \begin{NoHyper}\Cref{item:DGP-L}\end{NoHyper}
by replacing it with the assumption that the
observed outcome is given by $Y_{i}=Y_{i}(\Xs_{1}, \dotsc, \Xs_{S})$, such that
\cref{eq:potential-outcomes-alt} holds with
$L_{i}(\Xs_{1}, \dotsc, \Xs_{S})=L_{i}$.

We assume that the observed data $(Y, \Xs, Z, W)$ is generated by the triangular
array of variables $({Y}(0), B, W, U, \Xs, \Zs, L)$. Let
$\check{\delta}=(Z'Z)^{-1}Z'(Y-X\beta)$ denote the regression coefficient in a
regression of ${Y}-{X} \beta$ on ${Z}$, that is, the regression coefficient on
$Z_{i}$ in a regression in which $\hat{\beta}$ is restricted to equal to the
true value $\beta$.

\begin{assumption}\label{assumption:regularity-Z}
  \aitem\label{sitem:UZ-moments} $\N^{-1}\sum_{i=1}^{\N}E[L_{i}^{2}]^{1/2}\to 0$,
  and conditional on $W$, the second moments of $U_{i}$ and $\Zs_{s}$ exist and
  are bounded uniformly over $i$ and $s$; %
  \aitem\label{sitem:detZZ} $Z'Z/\N$ converges in probability to a positive
  definite non-random limit; %
  \aitem\label{sitem:UZ-fourth}
  $(\sum_{s}n_{s}^{2})^{-1/2} \sum_{i=1}^{\N}E[L_{i}^{2}]^{1/2}\to 0$,
  $\max_{i}E[L_{i}^{4}\mid W]\to 0$, and conditional on $W$, the fourth moments
  of $\Zs_{s}$, and ${U}_{i}$ exist and are bounded uniformly over $s$ and
  $i$; %
  \aitem\label{sitem:delta} $\check{\delta}-\delta=O_{p}(q_{s})$ for some
  sequence $q_{S}\to 0$; %
  \aitem\label{sitem:deltagamma}
  $q_{S}^{2}\N/\sum_{s}n_{s}^{2} \cdot\sum_{i}E[(U_{i}'\gamma)^{2}]\to 0$ and
  $\gamma'U'\epsilon=o_{p}((\sum_{s}n_{s}^{2})^{1/2})$.
\end{assumption}

\Cref{sitem:UZ-moments} imposes some mild moment restrictions on the controls
$Z_{i}$. It also requires that on average, the variance of the linearization
error $L_{i}$ vanishes with sample size. This ensures that the linearization
error does not impact the consistency of $\hat{\beta}$. \Cref{sitem:detZZ}
ensures that the controls are not collinear.

\Cref{sitem:UZ-fourth,sitem:delta,sitem:deltagamma} are only needed for
asymptotic normality. \Cref{sitem:UZ-fourth} strengthens the moment conditions
in \Cref{sitem:UZ-moments}. It also imposes a stricter condition on the
linearization error: it requires that, on average over $N$, the standard
deviation of $L_{i}$ is of smaller order than $(\sum_{s}n_{s}^{2})^{1/2}/N$, the
rate of convergence of $\hat{\beta}$. A sufficient condition is that
$L_{i}=o_{p}(S^{-1/2})$. This ensures that the linearization error is of smaller
order than the variance of the estimator, so that the distribution of
$\hat{\beta}$ does not suffer from asymptotic bias. This formalizes the
assumption that the linearization error is ``small''. The condition that
$\max_{i}E[L_{i}^{4}\mid W]\to 0$ is only needed for showing consistency of the
standard error estimator; it is not needed for asymptotic normality.
\Cref{sitem:delta} requires that $\check{\delta}$ is consistent, which ensures
that the error in estimation of $\delta$ does not affect the asymptotic
distribution of $\hat{\beta}$. Finally, \Cref{sitem:deltagamma} imposes
conditions on $U_{i}'\gamma$, the measurement error for controls that matter,
which ensure that measurement error in the controls that matter does not impact
the asymptotic distribution of $\hat{\beta}$. They are stated as high-level
conditions to cover a range of different cases, and depend on the rate of
convergence $q_{S}$ of $\check{\delta}$. In typical cases, the rate will be
$q_{S}=(\sum_{s}n_{s}^{2})^{1/2}/N$, the same as that of $\hat{\beta}$, and the
condition
$q_{S}^{2}\N/\sum_{s}n_{s}^{2} \cdot\sum_{i}E[(U_{i}'\gamma)^{2}]\to 0$ is
implied
by~\begin{NoHyper}\Cref{item:ugamma-cons}\end{NoHyper}. %
Let $U_{1i}$ denote the subset of elements of $U_{i}$ for which $\gamma_{k}\neq
0$, and let $U_{2i}$ denote the remaining elements. If $U_{i1}$ is mean zero and
independent across $i$ conditional on the remaining variables
($(Y(0), W, B, \Zs, \Xs, U_{2})$), so that these elements are pure measurement error,
then the second condition is implied by~\begin{NoHyper}\Cref{item:ugamma-norm}\end{NoHyper}.

\subsubsection{Auxiliary results}\label{sec:auxilliary-results}
\begin{lemma}\label{theorem:z-moments}
  $\{\mathcal{A}_{S1}, \dotsc, \mathcal{A}_{SS}\}_{S= 1}^{\infty}$ be a triangular
  array of random variables. Fix $\eta\geq 1$, and let
  ${A}_{Si}=\sum_{s=1}^{S}w_{is}\mathcal{A}_{Ss}$, $i=1\dotsc, \N_{S}$. Suppose
  $E[\abs{\mathcal{A}_{Ss}}^{\eta}\mid W]$ exists and is uniformly bounded. Then
  $E[\abs{{A}_{Si}}^{\eta}\mid W]$ exists and is bounded uniformly over $S$ and
  $i$.
\end{lemma}
\begin{proof}
  The result follows by triangle inequality for $\eta=1$. Suppose therefore that
  $\eta>1$. By Hölder's inequality,
  \begin{multline*}
    E[\abs{{A}_{Si}}^{\eta}\mid W]
    =E\left[\abs*{\sum_{s=1}^{S}w_{is}^{\frac{\eta-1}{\eta}}
        w_{is}^{\frac{1}{\eta}}\mathcal{A}_{Ss}}^{\eta}\mid W\right]\leq
    \left(\sum_{s=1}^{S}w_{is}\right)^{\eta-1}\sum_{s=1}^{S}w_{is}E[\abs{\mathcal{A}_{Ss}}^{\eta}\mid W]\\
    \leq \max_{s}E[\abs{\mathcal{A}_{Ss}}^{\eta}\mid W]\cdot (\textstyle\sum_{s=1}^{S}w_{is})^{\eta}\leq
    \max_{s}E[\abs{\mathcal{A}_{Ss}}^{\eta}\mid W],
  \end{multline*}
  which yields the result.
\end{proof}
\begin{lemma}\label{theorem:wA-bound}
  $\{A_{S1}, \dotsc, A_{S\N_{S}}\}_{S=1}^{\infty}$ be a triangular array of random
  variables. Suppose $E[A_{Si}^{2}\mid W]$ exists and is uniformly bounded. Then
  $\sum_{s=1}^{S}E\big[(\sum_{i=1}^{\N}w_{is}A_{Si})^{2}\mid W\big]\preceq \sum_{s}n_{s}^{2}$.
\end{lemma}
\begin{proof}
  By Cauchy-Schwarz inequality,
  \begin{multline*}
    \sum_{s=1}^{S}E\left[\left(\sum_{i=1}^{\N}w_{is}A_{Si}\right)^{2} \;\Big|\;
      W\right] \leq \sum_{s=1}^{S}\sum_{i=1}^{\N}\sum_{j=1}^{\N}
    w_{is}w_{js}E[A_{Si}^{2}\mid W]^{1/2} E[A_{Sj}^{2}\mid W]^{1/2}\\ \preceq
    \sum_{s=1}^{S}\sum_{i=1}^{\N}\sum_{j=1}^{\N} w_{is}w_{js}
    =\sum_{s=1}^{S}n_{s}^{2}.
\end{multline*}
\end{proof}
\begin{lemma}\label{theorem:Op1-variance}
  Let
  $\{A_{S1}, \dotsc, A_{S\N_{S}}, B_{S1}, \dotsc, B_{S\N_{S}}, \mathcal{A}_{S1}, \dotsc, \mathcal{A}_{SS}\}_{S=1}^{\infty}$
  be a triangular array of random variables. Suppose $E[A_{Si}^{4}\mid W]$,
  $E[B_{Si}^{4}\mid W]$, and $E[\mathcal{A}_{Ss}^{2}\mid W]$ exist and are
  uniformly bounded. Then
  $(\sum_{s}n_{s}^{2})^{-1}\cdot\sum_{i, j, s}w_{is}w_{js}A_{Si}B_{Sj}\mathcal{A}_{Ss}=O_{p}(1)$.
\end{lemma}
\begin{proof}
  Let
  $R_{S}=(\sum_{s}n_{s}^{2})^{-1}\sum_{i, j, s}w_{is}w_{js}A_{Si}B_{Sj}\mathcal{A}_{Ss}$.
  By the triangle and Cauchy-Schwarz inequalities,
  \begin{multline*}
    E[\abs{R_{S}}\mid W]\leq
    \frac{1}{\sum_{s}n_{s}^{2}}\sum_{i, j, s}w_{is}w_{js}E[\abs{A_{Si}B_{Sj}\mathcal{A}_{Ss}}\mid
    W]\\
    \leq \frac{1}{\sum_{s}n_{s}^{2}}\sum_{i, j, s}w_{is}w_{js}
    E[\abs{B_{Sj}}^{4}\mid W]^{1/4}E[\abs{A_{Si}}^{4}\mid
    W]^{1/4}E[\mathcal{A}_{Ss}^{2}\mid W]^{1/2}
    \preceq \frac{1}{\sum_{s}n_{s}^{2}}\sum_{i, j, s}w_{is}w_{js}=1.
  \end{multline*}
  The result then follows by Markov inequality.
\end{proof}

\subsubsection{Proof of Proposition~3}\label{sec:proof-texorpdfstr-z}

First we show that
\begin{equation}\label{eq:z-tildeX}
  Z'W\tilde{\Xs}= O_{p}(1/\sqrt{r_{\N}}).
\end{equation}
Conditional on $W$, the left-hand side has mean zero by
\begin{NoHyper}\Cref{item:samplingZ}\end{NoHyper},
and by \begin{NoHyper}\Cref{item:indepX}\end{NoHyper}, the variance of the $k$th row given by
\begin{equation*}
  \var\left(\sum_{i, s}w_{is}\tilde{\Xs}_{s}{Z}_{ik}\mid W\right)
  =
  \sum_{s}E_{W}\sigma^{2}_{s}\left(\sum_{i}w_{is}{Z}_{ik}\right)^{2}
  \preceq
  \sum_{s}E_{W}\left(\sum_{i}w_{is}{Z}_{ik}\right)^{2}.
\end{equation*}
By \Cref{theorem:z-moments}, \Cref{sitem:UZ-moments}, and the $C_{r}$-inequality,
$E_{W}[Z_{ik}^{2}] =E_{W}[(\sum_{s}w_{is}\Zs_{sk}+{U}_{ik})^{2}]$ is uniformly
bounded. Therefore, by \Cref{theorem:wA-bound}, the right-hand side is bounded
by $\sum_{s}n_{s}^{2}$, so the result follows by Markov inequality and dominated
convergence theorem.

Since $X=W\tilde{\Xs}+Z\gamma-U\gamma$, it follows from~\cref{eq:z-tildeX}
and~\Cref{sitem:detZZ} that
\begin{equation}\label{eq:gamma-consistent}
  \hat{\gamma}-\gamma
  =(Z'Z/\N)^{-1}Z'W\tilde{\Xs}/\N-(Z'Z/\N)^{-1}Z'U\gamma/\N
  =o_{p}(1),
\end{equation}
where $\hat{\gamma}=(Z'Z)^{-1}Z'X$, and the last equality follows since
$\sum_{s}n_{s}^{2}/N^{2}\leq \max_{s}n_{s}/N\to 0$ by
\begin{NoHyper}\Cref{item:sector-size}\end{NoHyper},
and since $Z'U\gamma/\N=o_{p}(1)$ by the Cauchy-Schwarz inequality
and~\begin{NoHyper}\Cref{item:ugamma-cons}\end{NoHyper}.

Next, we will show that
\begin{equation}\label{eq:dotxdotx}
  \ddot{X}'\ddot{X}/\N=
  \frac{1}{\N}\sum_{i, s}w_{is}^{2}\sigma^{2}_{s} +o_{p}(1).
\end{equation}
To this end, we have
\begin{equation*}
  \begin{split}
    \ddot{X}'\ddot{X}/\N&=
(W\tilde{\Xs}-U\gamma-Z(\hat{\gamma}-\gamma))'(W\tilde{\Xs}-U\gamma-Z(\hat{\gamma}-\gamma))/\N\\
    &=(W\tilde{\Xs})'(W\tilde{\Xs})/\N+o_{p}(1)\\
    &= \frac{1}{\N}\sum_{s}\overbar{w}_{ss}\sigma^{2}_{s}+
    \frac{2}{\N}\sum_{s<t}\overbar{w}_{st}\tilde{\Xs}_{s}\tilde{\Xs}_{t} +
    \frac{1}{\N}\sum_{s}\overbar{w}_{ss}(\tilde{\Xs}_{s}^{2}-\sigma^{2}_{s})
    +o_{p}(1).
  \end{split}
\end{equation*}
where the first line follows from the decomposition
\begin{equation}
  \label{eq:dotx}
  \ddot{X}=X-Z(Z'Z)^{-1}Z'X=X-Z\hat{\gamma}=W\tilde{\Xs}-U\gamma-Z(\hat{\gamma}-\gamma),
\end{equation}
the second line follows by the Cauchy-Schwarz inequality,
\begin{NoHyper}\Cref{item:ugamma-cons}\end{NoHyper}, and~\cref{eq:gamma-consistent}, and the third line
follows by expanding $(W\tilde{\Xs})'(W\tilde{\Xs})/\N$. Therefore, to
show~\cref{eq:dotxdotx}, it suffices to show that the second and third term in
the above expression are $o_{p}(1)$. Since the second term has mean zero
conditional on $W$, it suffices to show that its variance converges to zero. To
that end,
\begin{multline*}
  \var\left(\frac{2}{\N}\sum_{s< t}
    \tilde{\Xs}_{s}\tilde{\Xs}_{t}\overbar{w}_{st}\mid W\right) =
  \frac{4}{\N^{2}}\sum_{s<
    t}E_{W}[\sigma_{s}^{2}\sigma_{t}^{2}]\overbar{w}_{st}^{2} \preceq
  \frac{1}{\N^{2}}\sum_{s, t}\overbar{w}_{st}^{2} =\frac{1}{\N^{2}}\sum_{i, j,
    s, t}w_{is}w_{it}w_{js}w_{jt}\\
\leq \frac{1}{\N^{2}}\sum_{i, j,
    s, t}w_{is}w_{it}w_{js}\leq \frac{1}{\N^{2}}\sum_{i, j,
    s}w_{is}w_{js}=\frac{1}{\N^{2}}\sum_{s}n_{s}^{2}\leq \frac{\max_{t}n_{t}\sum_{s}n_{s}}{\N^{2}}\to
  0.
\end{multline*}
where the convergence to $0$ follows
by~\begin{NoHyper}\Cref{item:sector-size}\end{NoHyper}. By the inequality of von
Bahr and Esseen, \Cref{sitem:2nu}, and the inequality
$\overbar{w}_{ss}\leq n_{s}$,
\begin{multline}\label{eq:vonbahr-x2}
  E[\N^{-1}\abs*{\textstyle\sum_{s}(\tilde{\Xs}_{s}^{2}-\sigma^{2}_{s})\overbar{w}_{ss}}^{1+\nu/2}\mid \mathcal{F}_{0}]\leq
  \frac{2}{\N^{1+\nu/2}}
  \sum_{s}\overbar{w}_{ss}^{1+\nu/2} E[\abs*{\tilde{\Xs}_{s}^{2}-\sigma^{2}_{s}}^{1+\nu/2}\mid\mathcal{F}_{0}]\\
  \preceq \frac{1}{\N^{1+\nu/2}}
  \sum_{s}\overbar{w}_{ss}^{1+\nu/2} \leq
  (\max_{s}n_{s}/N)^{\nu/2},
\end{multline}
which converges to zero by~\begin{NoHyper}\Cref{item:sector-size}\end{NoHyper}. \Cref{eq:dotxdotx} then
follows by Markov inequality.

Next, we show that
\begin{equation}\label{eq:xy}
  \ddot{X}'Y/\N=\frac{1}{\N}\sum_{i, s}\sigma^{2}_{s}w_{is}^{2}\beta_{is}+o_{P}(1)
\end{equation}
Using \cref{eq:dotx}, we can write the left-hand side as
\begin{equation*}
  \begin{split}
    \ddot{X}'Y/\N&=\tilde{\Xs}'W'Y/\N-\gamma'U'Y/\N-Y'Z/N\cdot (\hat{\gamma}-\gamma)\\
    &=\tilde{\Xs}'W'Y/\N+o_{p}(1)\\
    &= \frac{1}{\N}\sum_{s, i}w_{is}\tilde{\Xs}_{s}L_{i}
    +\frac{1}{\N}\sum_{s, i}w_{is}^{2}(\tilde{\Xs}_{s}\Xs_{s}-\sigma_{s}^{2})\beta_{is}+\frac{1}{\N}\sum_{s, i}w_{is}\tilde{\Xs}_{s}Y_{i}(0)\\
    &\qquad +\frac{1}{\N}\sum_{s<t}\sum_{i}w_{is}w_{it}\tilde{\Xs}_{s}\Xs_{t}\beta_{it}
     +\frac{1}{\N}\sum_{s< t}\sum_{i}w_{is}w_{it}\tilde{\Xs}_{t}\Xs_{s}\beta_{is}+\frac{1}{\N}\sum_{s, i}w_{is}^{2}\sigma_{s}^{2}\beta_{is}+o_{p}(1)
  \end{split}
\end{equation*}
where the second line follows since by the $C_{r}$-inequality,
\Cref{theorem:z-moments}, \Cref{sitem:beta-bounded,sitem:UZ-moments,sitem:2nu},
$N^{-1}\sum_{i}E[Y_{i}^{2}]$ is bounded, so that $Y'Z/N=O_{p}(1)$ and
$\gamma'U'Y/\N=o_{p}(1)$ by Cauchy-Schwarz inequality
and~\begin{NoHyper}\Cref{item:ugamma-cons}\end{NoHyper}, and the third line follows by expanding
$\tilde{\Xs}'W'Y$. We therefore need to show that the first five terms in the
expression above are $o_{p}(1)$. By the Cauchy-Schwarz inequality, the
expectation of the absolute value of the first term is bounded by
\begin{equation*}
  \N^{-1}\sum_{i}E[L_{i}^{2}]^{1/2}(E \sum_{s}w_{is}^{2}\sigma^{2}_{s})^{1/2}
  \preceq \N^{-1}\sum_{i}E[L_{i}^{2}]^{1/2},
\end{equation*}
which converges to zero by \Cref{sitem:UZ-moments}.
Thus, the first term is $o_{p}(1)$ by Markov inequality and the dominated
convergence theorem. The second term is $o_{p}(1)$ by an argument analogous to
\cref{eq:vonbahr-x2}. The third to fifth terms are mean zero conditional on
$\mathcal{F}_{0}$, so it suffices to show that their variances conditional on
$W$ converge to zero. The variance of the third summand is bounded by
\begin{equation*}
  \var\left(\frac{1}{\N}\sum_{s}\tilde{\Xs}_{s}\sum_{i}w_{is}{Y}_{i}(0)\mid W\right)
  =\frac{1}{\N^{2}}\sum_{s}E_{W}\sigma_{s}^{2}\left(\sum_{i}w_{is}{Y}_{i}(0)\right)^{2}
  \preceq\frac{1}{\N^{2}}
  \sum_{s}E_{W}\left(\sum_{i}w_{is}{Y}_{i}(0)\right)^{2},
\end{equation*}
which converges to zero by~\Cref{theorem:wA-bound}. The variance of the fourth
term is bounded by
\begin{multline*}
  \var\left(\frac{1}{\N}\sum_{s<
    t}\sum_{i}w_{is}w_{it}\tilde{\Xs}_{s}\Xs_{t}\beta_{it}\mid W\right)
  =\frac{1}{\N^{2}}\sum_{s<
    t, t'}\sum_{i, i'}w_{is}w_{it}\sigma_{s}^{2}E_{W}[\Xs_{t}\Xs_{t'}]\beta_{it}
  w_{i's}w_{i't'}\beta_{i't'}\\
  \preceq\frac{1}{\N^{2}}\sum_{s, t, t', i, i'}w_{is}w_{it}w_{i's}w_{i't'}
  \leq\frac{1}{\N^{2}}\sum_{s}n_{s}^{2}\leq \max_{s}n_{s}/\N\to 0.
\end{multline*}
Variance of the fifth term converges to zero by analogous arguments.

Combining \cref{eq:dotxdotx} with~\cref{eq:xy} and~\Cref{sitem:detX} then yields
the result.

\subsubsection{Proof of Proposition 4}\label{sec:proof-texorpdfstr-zp}

Using~\cref{eq:dotx}, we have
\begin{equation*}
  \begin{split}
    r_{\N}^{1/2}(\ddot{X}'\ddot{X})(\hat{\beta}-\beta)&
    =r_{\N}^{1/2}X'(I-Z(Z'Z)^{-1}Z')(Z\delta+\epsilon)
    =r_{\N}^{1/2}X'(I-Z(Z'Z)^{-1}Z')\epsilon\\
    & =r_{\N}^{1/2}\tilde{\Xs}'W'\epsilon-r_{\N}^{1/2}\gamma'U'\epsilon
    -r_{\N}^{1/2}(\hat{\gamma}-\gamma)'Z'\epsilon.
  \end{split}
\end{equation*}
The third term can be written as
\begin{equation*}
  \begin{split}
    r_{\N}^{1/2}(\hat{\gamma}-\gamma)'Z'\epsilon &
    =r_{\N}^{1/2}\epsilon'Z(Z'Z)^{-1}(Z'W\tilde{\Xs}-Z'U\gamma)=r_{\N}^{1/2}(\check{\delta}-\delta)'(Z'W\tilde{\Xs}-Z'U\gamma)\\
    &=(\check{\delta}-\delta)'(O_{p}(1)-r_{\N}^{1/2}Z'U\gamma)\\
    &=o_{p}(1)-O_{p}(1)\cdot q_{S} r_{\N}^{1/2}Z'U\gamma=o_{p}(1),
  \end{split}
\end{equation*}
where the first line follows from the decomposition in
\cref{eq:gamma-consistent}, the second line follows from~\cref{eq:z-tildeX},
the third line follows by~\Cref{sitem:delta}, and the last equality follows since
by Cauchy-Schwarz inequality and \Cref{sitem:deltagamma},
$q_{S}r_{\N}^{1/2}E[\abs{Z_{k}'U\gamma}]\preceq\sqrt{q_{S}^{2}r_{\N}N\sum_{i}E(U_{i}'\gamma)^{2}}\to
0$.
Since $r_{\N}^{1/2}\gamma'U'\epsilon=o_{p}(1)$ by~\Cref{sitem:deltagamma}, and
since by \cref{eq:dotxdotx} and \Cref{sitem:detX},
$(\ddot{X}'\ddot{X}/\N)^{-1}=(1+o_{p}(1))\cdot (\N^{-1}\sum_{i,
  s}\pi_{is})^{-1}$, it follows that
\begin{equation*}
  \frac{\N}{(\sum_{s}n_{s}^{2})^{1/2}}(\hat{\beta}-\beta)
  =(1+o_{p}(1))\frac{1}{\N^{-1}\sum_{i, s}\pi_{is}}
  r_{\N}^{1/2}\sum_{s, i}\tilde{\Xs}_{s}w_{is}{\epsilon}_{i}+o_{p}(1).
\end{equation*}
Therefore, it suffices to show
\begin{equation}\label{eq:OLS-middle}
  r_{\N}^{1/2}\sum_{s, i}\tilde{\Xs}_{s}w_{is}{\epsilon}_{i}=\mathcal{N}(0,
  \operatorname{plim}\mathcal{V}_{\N})+o_{p}(1).
\end{equation}
Define ${V}_{i}={Y}_{i}(0)-{Z}_{i}'\delta
+\sum_{t}w_{it}\Zs_{t}'\gamma(\beta_{it}-\beta)$, and
\begin{align}\label{eq:as-bs}
  a_{s}&=\sum_{i}w_{is}{V}_{i},&
  b_{st}&=\sum_{i}w_{is}w_{it}(\beta_{it}-\beta).
\end{align}
Then we can write
${\epsilon}_{i}={V}_{i}+\sum_{t}w_{it}\tilde{\Xs}_{t}(\beta_{it}-\beta)+L_{i}$.
Since
\begin{equation*}
  E\abs{r_{\N}^{1/2}\sum_{i, s}\tilde{\Xs}_{s}w_{is}L_{i}}\leq r_{\N}^{1/2}
  \sum_{i}(\sum_{s}Ew^{2}_{is}\sigma^{2}_{s})^{1/2}E[L_{i}^{2}]^{1/2}\preceq
  r_{\N}^{1/2}
  \sum_{i}E[L_{i}^{2}]^{1/2}\to 0
\end{equation*}
by \Cref{sitem:UZ-fourth}, and since
$0=\sum_{i, s}\pi_{is}(\beta_{is}-\beta)= \sum_{s}\sigma^{2}_{s}b_{ss}$, we can
decompose
\begin{equation*}
  r_{\N}^{1/2}\sum_{s, i}\tilde{\Xs}_{s}w_{is}{\epsilon}_{i}
  =r_{\N}^{1/2}\sum_{s}\tilde{\Xs}_{s}\sum_{i}w_{is}
  \left({V}_{i} +\sum_{t}w_{it}\tilde{\Xs}_{t}(\beta_{it}-\beta)+L_{i}\right)
  =r_{\N}^{1/2}\sum_{s}\mathcal{Y}_{s}+o_{P}(1),
\end{equation*}
where
\begin{equation*}
  \mathcal{Y}_{s}= \tilde{\Xs}_{s}a_{s}
  +(\tilde{\Xs}_{s}^{2}-\sigma_{s}^{2})b_{ss}
  +\sum_{t=1}^{s-1}\tilde{\Xs}_{s}\tilde{\Xs}_{t}(b_{st}+b_{ts}).
\end{equation*}
Observe that $\mathcal{Y}_{s}$ is a martingale difference array with respect to
the filtration $\mathcal{F}_{s}=\sigma(\Xs_{1}, \dotsc, \Xs_{s}, \mathcal{F}_{0})$.

By the dominated convergence theorem and the martingale central limit theorem,
it suffices to show that
$r_{\N}^{1+\nu/4}\sum_{s=1}^{S}E_{W}[\mathcal{Y}_{s}^{2+\nu/2}]\to 0$ for some
$\nu>0$ so that the Lindeberg condition holds, and that the conditional
variance converges,
\begin{equation*}
  r_{\N}\sum_{s=1}^{S}
  E[\mathcal{Y}_{s}^{2} \mid\mathcal{F}_{s-1}]-\mathcal{V}_{\N}=o_{p}(1).
\end{equation*}
To verify the Lindeberg condition, by the $C_{r}$-inequality, it suffices to
show that
\begin{align*}
  r_{\N}^{2}\sum_{s}E_{W}[\tilde{\Xs}_{s}^{4}a_{s}^{4}]
  &\to 0,
  & r_{\N}^{1+\nu/4}\sum_{s}E_{W}[(\tilde{\Xs}_{s}^{2}-\sigma_{s}^{2})^{2+\nu/2}b_{ss}^{2+\nu/2}]&\to 0,\\
  r_{\N}^{2}\sum_{s}E_{W}\left(\sum_{t=1}^{s-1}\tilde{\Xs}_{s}\tilde{\Xs}_{t}b_{st}\right)^{4}
  &\to 0,
  & r_{\N}^{2}\sum_{s}E_{W}\left(\sum_{t=1}^{s-1}\tilde{\Xs}_{s}\tilde{\Xs}_{t}b_{ts}\right)^{4}&\to 0.
\end{align*}
Note that since
$E(\sum_{t}w_{it}\Zs_{t}'\gamma(\beta_{it}-\beta))^{4}\preceq
(\sum_{t}w_{it})^{4}\preceq 1$, it follows from \Cref{sitem:4nu,sitem:UZ-fourth},
and the $C_{r}$ inequality that the fourth moment of ${V}_{i}$ exists and is
bounded. Therefore, by arguments as in the proof of \Cref{theorem:wA-bound},
$\sum_{s}E_{W}[a_{s}^{4}]\preceq \sum_{s}n_{s}^{4}$, so that
\begin{equation}\label{eq:xs-as4}
  r_{\N}^{2}\sum_{s}E_{W}[\tilde{\Xs}_{s}^{4}a_{s}^{4}]=
  r_{\N}^{2}\sum_{s}E_{W}[E[\tilde{\Xs}_{s}^{4}\mid\mathcal{F}_{0}]a_{s}^{4}]
  \preceq r_{\N}^{2}\sum_{s}E_{W}[a_{s}^{4}]
  \preceq r_{\N}^{2}\sum_{s}n_{s}^{4}\leq \max_{s}n_{s}^{2}r_{\N}\to 0
\end{equation}
by \begin{NoHyper}\Cref{item:sector-size-normality}\end{NoHyper}. Second, since $\beta_{is}$ is bounded by
\Cref{sitem:beta-bounded}, we have $b_{ss}\preceq \sum_{i}w_{is}^{2}\leq n_{s}$, so that
\begin{equation*}
  r_{\N}^{1+\nu/4}\sum_{s}E_{W}[(\tilde{\Xs}_{s}^{2}-\sigma_{s}^{2})^{2+\nu/2}b_{ss}^{2+\nu/2}]
  \preceq r_{\N}^{1+\nu/4}\sum_{s}n_{s}^{2+\nu/2}\leq
  (r_{\N}\max_{s}n_{s}^{2})^{\nu/4}\to 0.
\end{equation*}
Third, by similar arguments
\begin{multline*}
  r_{\N}^{2}\sum_{s}E_{W}\left(\sum_{t=1}^{s-1}\tilde{\Xs}_{s}\tilde{\Xs}_{t}b_{st}\right)^{4}
  = r_{\N}^{2}\sum_{s}E_{W}
  E[\tilde{\Xs}_{s}^{4}\mid\mathcal{F}_{0}]E\left[\left(\sum_{t=1}^{s-1}\tilde{\Xs}_{t}b_{st}\right)^{4}
    \mid\mathcal{F}_{0}\right]\\
  \preceq
  r_{\N}^{2}\sum_{s}\left(\sum_{t=1}^{s-1}\sum_{i}w_{is}w_{it}\right)^{4} \leq
  r_{\N}^{2}\sum_{s}n_{s}^{4}\to 0.
\end{multline*}
The claim that $r_{\N}^{2}\sum_{s}E_{W}\left(\sum_{t=1}^{s-1}\tilde{\Xs}_{s}\tilde{\Xs}_{t}b_{ts}\right)^{4}\to
0$ follows by similar arguments.

It remains to verify that the conditional variance converges. Since
$\mathcal{V}_{\N}$ can be written as
\begin{multline*}
  \mathcal{V}_{\N}=\frac{1}{\sum_{s=1}^{S}n_{s}^{2}}\var\left(\sum_{i}(X_{i}-Z_{i}'\gamma)\epsilon_{i}\mid
    \mathcal{F}_{0} \right) =r_{\N}\sum_{s}E[\mathcal{Y}_{s}^{2}\mid\mathcal{F}_{0}]+o_{P}(1)\\
  =r_{\N}\sum_{s}\left[E\left[(\tilde{\Xs}_{s}a_{s}
      +(\tilde{\Xs}_{s}^{2}-\sigma_{s}^{2})b_{ss})^{2}\mid\mathcal{F}_{0}\right]
    +\sum_{t=1}^{s-1}\sigma^{2}_{s}\sigma^{2}_{t}(b_{st}+b_{ts})^{2}
  \right]+o_{p}(1),
\end{multline*}
we can decompose
\begin{equation*}
  r_{\N}
  \sum_{s}E[\mathcal{Y}_{s}^{2}\mid\mathcal{F}_{s-1}]-\mathcal{V}_{\N}
  =2D_{1}
  +D_{2}+2D_{3}+o_{p}(1),
\end{equation*}
where
\begin{align*}
  D_{1}&=r_{\N}\sum_{s}(\sigma^{2}_{s}a_{s}
  +E[\tilde{\Xs}_{s}^{3}\mid\mathcal{F}_{0}]
  b_{ss})\sum_{t=1}^{s-1}\tilde{\Xs}_{t}(b_{st}+b_{ts}), \\
  D_{2}&=r_{\N}\sum_{s}
  \sigma^{2}_{s}\sum_{t=1}^{s-1}(\tilde{\Xs}_{t}^{2}-\sigma^{2}_{t})
  (b_{st}+b_{ts})^{2}, \\
  D_{3}&=r_{\N}\sum_{s}
  \sigma^{2}_{s}\sum_{t=1}^{s-1}\sum_{u=1}^{t-1}\tilde{\Xs}_{t}\tilde{\Xs}_{u}
  (b_{st}+b_{ts})(b_{su}+b_{us}).
\end{align*}
It therefore suffices to show that $D_{j}=o_{p}(1)$ for $j=1,2,3$. Since
$E[D_{j}\mid\mathcal{F}_{0}]=0$, it suffices to show that
$\var(D_{j}\mid W)=E_{W}[\var(D_{j}\mid\mathcal{F}_{0})]$ converges to zero.
Since $b_{st}+b_{ts}\preceq \overbar{w}_{st}$, and since
$E_{W}[\abs{a_{s}a_{t}}] \preceq n_{s}n_{t}$, and
$\abs{b_{ss}}\preceq \overbar{w}_{ss}\leq n_{s}$, it follows that
\begin{multline*}
  \var(D_{1}\mid W)=r_{\N}^{2} \sum_{t}E_{W}\left[\sigma^{2}_{t}
    \left(\sum_{s=t+1}^{S}(b_{st}+b_{ts})(\sigma^{2}_{s}a_{s}
      +E[\tilde{\Xs}_{s}^{3}\mid\mathcal{F}_{0}]
      b_{ss})\right)^{2}\right]\\
  \preceq r_{\N}^{2} \sum_{t}
  \left(\sum_{s=t+1}^{S}\overbar{w}_{st}n_{s}\right)^{2} \leq
  r_{\N}^{2}\max_{s}n_{s}^{2} \sum_{t}
  \left(\sum_{s}\overbar{w}_{st}\right)^{2}= r_{\N}\max_{s}n_{s}^{2}\to 0,
\end{multline*}
where the convergence to zero follows by \begin{NoHyper}\Cref{item:sector-size-normality}\end{NoHyper}. By
similar arguments, since $\overbar{w}_{st}\leq n_{s}$
\begin{multline*}
  \var(D_{2}\mid W)=
  r_{\N}^{2}\sum_{t}E_{W}(\tilde{\Xs}_{t}^{2}-\sigma^{2}_{t})^{2}
  \left(\sum_{s=t+1}^{S}
    \sigma^{2}_{s}
    (b_{st}+b_{ts})^{2}\right)^{2}
  \preceq
  r_{\N}^{2}\sum_{t}
  \left(\sum_{s=t+1}^{S}\overbar{w}_{st}^{2}\right)^{2}\\
  \leq r_{\N}^{2}\sum_{t}
  \left(\sum_{s=1}^{S}n_{s}\overbar{w}_{st}\right)^{2}
  \leq r_{\N}\max_{s}n_{s}^{2}\to 0.
\end{multline*}
Finally,
\begin{multline*}
  \var(D_{3}\mid W)=r_{\N}^{2}\sum_{t}\sum_{u=t+1}^{S} E_{W} \sigma_{t}^{2}\sigma_{u}^{2}
  \left(\sum_{s=u+1}^{S}
    \sigma^{2}_{s}(b_{st}+b_{ts})(b_{su}+b_{us})\right)^{2}\\
  \preceq r_{\N}^{2}\sum_{t}\sum_{u=t+1}^{S} \left(\sum_{s=u+1}^{S}
    \overbar{w}_{st}\overbar{w}_{su}\right)^{2} \leq r_{\N}^{2} \sum_{s, t, u, v}
  \overbar{w}_{st}\overbar{w}_{su} \overbar{w}_{vt}\overbar{w}_{vu} \leq
  r_{\N}\max_{s}n_{s}^{2}\to 0,
\end{multline*}
where the last line follows from the fact that since
$\sum_{s}\overbar{w}_{st}=n_{t}$ and $\overbar{w}_{st}\leq n_{s}$,
\begin{multline}\label{eq:wstuv}
  \sum_{s, t, u, v} \overbar{w}_{st}\overbar{w}_{su}
  \overbar{w}_{vt}\overbar{w}_{vu} \leq \max_{s}n_{s}\sum_{s, t, u, v}
  \overbar{w}_{su} \overbar{w}_{vt}\overbar{w}_{vu}
  =\max_{s}n_{s}\sum_{u, v}n_{u}n_{v}\overbar{w}_{vu} \\
  \leq \max_{s}n_{s}^{2}\sum_{u, v}n_{v}\overbar{w}_{vu}
  =\max_{s}n_{s}^{2}/r_{\N}.
\end{multline}
Consequently, $D_{j}=o_{p}(1)$ for $j=1,2,3$, the conditional variance
converges, and the theorem follows.

\subsubsection{Proof of Proposition~5}\label{sec:proof-texorpdfstr-co}

We'll prove a more general result that doesn't assume constant treatment
effects. In particular, we will show that under the conditions of the
\namecref{theorem:se-consistency} when the condition $\beta_{is}=\beta$ is
dropped, the variance estimator
$\hat{\mathcal{V}}_{\N}=r_{\N}\sum_{s}\widehat{\Xs}_{s}\hat{R}_{s}^{2}$, where
$r_{\N}=1/\sum_{s=1}^{S}n_{s}^{2}$ satisfies
\begin{equation}\label{eq:hatv-consistency}
  \hat{\mathcal{V}}_{\N}=r_{\N}\sum_{s=1}^{S}E[\tilde{\Xs}_{s}^{2}R_{s}^{2}\mid\mathcal{F}_{0}]+o_{p}(1),
\end{equation}
where, using the definitions of $a_{s}$ and $b_{st}$ in \cref{eq:as-bs},
\begin{equation*}
  R_{s}=\sum_{i=1}^{\N}w_{is}\epsilon_{i}=a_{s}+\sum_{i=1}^{\N}w_{is}L_{i}+
\sum_{t=1}^{S}\tilde{\Xs}_{t}b_{st}.
\end{equation*}
Since under constant treatment effects,
$\mathcal{V}_{\N}=r_{\N}\sum_{s=1}^{S}E[\tilde{\Xs}_{s}^{2}R_{s}^{2}\mid\mathcal{F}_{0}]$,
the assertion of the \namecref{theorem:se-consistency} follows
from~\cref{eq:hatv-consistency}.

Throughout the proof, we write $E_{\mathcal{F}_{0}}[\cdot]$ and $E_{W}[\cdot]$
to denote expectations conditional on $\mathcal{F}_{0}$, and $W$, respectively.
Let $\tilde{\theta}=(\tilde{\beta}, \tilde{\delta}')'$, $\theta=(\beta, \delta)$,
$M_{i}=(X_{i}, Z_{i}')'$. We can decompose the variance estimator as
\begin{equation}\label{eq:hatV-decomposition}
  \hat{\mathcal{V}}_{\N}=
  r_{\N}\sum_{s}(\widehat{\Xs}_{s}^{2}-\tilde{\Xs}_{s}^{2})\hat{R}_{s}^{2}
  +r_{\N}\sum_{s}\tilde{\Xs}_{s}^{2}(\hat{R}_{s}^{2}-R_{s}^{2})
  +r_{\N}\sum_{s}(\tilde{\Xs}_{s}^{2}R_{s}^{2}-E_{\mathcal{F}_{0}}[\tilde{\Xs}_{s}^{2}R_{s}^{2}])+r_{\N}\sum_{s}
  E_{\mathcal{F}_{0}}[\tilde{\Xs}_{s}^{2}R_{s}^{2}].
\end{equation}
We need to show that the first three terms are $o_{p}(1)$. Since
$\tilde{\epsilon}_{i}=\epsilon_{i}+M_{i}'(\theta-\tilde{\theta})$, with
$\epsilon_{i}=V_{i}+L_{i}+\sum_{t}w_{it}\tilde{\Xs}_{t}(\beta_{it}-\beta)$, we
can decompose
\begin{equation}\label{eq:Rs-decomposition}
  \hat{R}_{s}^{2}=\sum_{i, j}w_{is}w_{js}\tilde{\epsilon}_{i}\tilde{\epsilon}_{j}
  =R_{s}^{2}+2\sum_{i, j}w_{js}w_{is}M_{i}'(\theta-\tilde{\theta})
  \epsilon_{j}
  +\sum_{i, j}w_{is}w_{js}M_{i}'(\theta-\tilde{\theta})M_{j}'(\theta-\tilde{\theta}).
\end{equation}
Therefore, the second term in \cref{eq:hatV-decomposition} satisfies
\begin{equation*}
  \begin{split}
    r_{\N}\sum_{s}\tilde{\Xs}_{s}^{2}(\hat{R}_{s}^{2}-R_{s}^{2}) &=
    2(\theta-\tilde{\theta})'\left[r_{\N}
      \sum_{s, i, j}w_{is}w_{js}\tilde{\Xs}_{s}^{2}M_{i}\epsilon_{j}\right]
    +(\theta-\tilde{\theta})'\left[r_{\N}\sum_{s, i, j}\tilde{\Xs}_{s}^{2}w_{is}w_{js}M_{i}M_{j}'\right](\theta-\tilde{\theta})
    \\
    &= (\theta-\tilde{\theta})'O_{p}(1)
    +(\theta-\tilde{\theta})'O_{p}(1)(\theta-\tilde{\theta})
    =o_{p}(1),
  \end{split}
\end{equation*}
where the second line follows by applying \Cref{theorem:Op1-variance} to the
terms in square brackets. Next, the third term in~\eqref{eq:hatV-decomposition}
can be decomposed as
\begin{multline}\label{eq:10-terms}
  r_{\N}\sum_{s}(\tilde{\Xs}_{s}^{2}R_{s}^{2}-E_{\mathcal{F}_{0}}[\tilde{\Xs}_{s}^{2}R_{s}^{2}])
  =\\
  +r_{\N}\sum_{s}b_{ss}^{2}(\tilde{\Xs}_{s}^{4}-E_{\mathcal{F}_{0}}[\Xs_{s}^{4}])
  +r_{\N}\sum_{s< t}(b_{st}^{2}+b_{ts}^{2})(\tilde{\Xs}_{s}^{2}\tilde{\Xs}_{t}^{2}-\sigma^{2}_{s}\sigma^{2}_{t})
  +2r_{\N}\sum_{s}\sum_{t< u}b_{st}b_{su}\tilde{\Xs}_{s}^{2}\tilde{\Xs}_{t}\tilde{\Xs}_{u}\\
+r_{\N}\sum_{s}(\tilde{\Xs}_{s}^{2}-\sigma^{2}_{s})a_{s}^{2}
  +r_{\N}\sum_{i, j, s}w_{js}w_{is}(\tilde{\Xs}_{s}^{2}L_{i}L_{j}-E_{\mathcal{F}_{0}}[\tilde{\Xs}_{s}^{2}L_{i}L_{j}])
  +2r_{\N}\sum_{i, s}w_{is}a_{s}(\tilde{\Xs}_{s}^{2}L_{i}-E_{\mathcal{F}_{0}}[\tilde{\Xs}_{s}^{2}L_{i}])\\
  +2r_{\N}\sum_{s<t}a_{s}b_{st}\tilde{\Xs}_{s}^{2}\tilde{\Xs}_{t}
  +2r_{\N}\sum_{s< t}a_{t}b_{ts}\tilde{\Xs}_{t}^{2}\tilde{\Xs}_{s}
  +2r_{\N}\sum_{s}a_{s}b_{ss}(\tilde{\Xs}_{s}^{3}-E_{\mathcal{F}_{0}}[\tilde{\Xs}_{s}^{3}])\\
  +r_{\N}\sum_{i, s, t}w_{is}b_{st}(\tilde{\Xs}_{s}^{2}\tilde{\Xs}_{t}L_{i}-E_{\mathcal{F}_{0}}[\tilde{\Xs}_{s}^{2}\tilde{\Xs}_{t}L_{i}]).
\end{multline}
We will show that all terms are of the order $o_{p}(1)$. By the inequality of
von Bahr and Esseen, since $b_{ss}$ is bounded by a constant times
$\overbar{w}_{ss}\leq n_{s}$,
\begin{equation*}
  E_{\mathcal{F}_{0}}\abs{r_{\N}\sum_{s}b_{ss}^{2}(\tilde{\Xs}_{s}^{4}-E_{\mathcal{F}_{0}}[\Xs_{s}^{4}])}^{1+\nu/4}
  \preceq
    r_{\N}^{1+\nu/4}\sum_{s}n_{s}^{2+\nu/2}
  E_{\mathcal{F}_{0}}\abs{(\tilde{\Xs}_{s}^{4}-E_{\mathcal{F}_{0}}[\Xs_{s}^{4}])}^{1+\nu/4}
  \leq (\max_{s}n_{s}^{2}r_{\N})^{\nu/4}\to 0
\end{equation*}
by \begin{NoHyper}\Cref{item:sector-size-normality}\end{NoHyper}, so that the
first term is $o_{p}(1)$. The second term can be written as
\begin{equation*}
  r_{\N}\sum_{s< t}(b_{st}^{2}+b_{ts}^{2})(\tilde{\Xs}_{s}^{2}-\sigma^{2}_{s})(\tilde{\Xs}_{t}^{2}-\sigma^{2}_{t})
  +r_{\N}\sum_{s\neq t}(b_{st}^{2}+b_{ts}^{2})(\tilde{\Xs}_{s}^{2}-\sigma^{2}_{s})\sigma^{2}_{t}
\end{equation*}
The conditional variance of both summands is bounded by a constant times
$r_{\N}^{2}\sum_{s}(\sum_{t}\overbar{w}_{st}^{2})^{2}
\leq r_{\N}^{2}\cdot\sum_{s}n_{s}^{4}\to 0$, so that the second term is also $o_{p}(1)$.
The third term admits the decomposition
\begin{multline*}
  2r_{\N}\sum_{s}\sum_{t<u}b_{st}b_{su}\tilde{\Xs}_{s}^{2}\tilde{\Xs}_{t}\tilde{\Xs}_{u}=
  2r_{\N}\sum_{s, t}\sum_{s\not\in\{t,
    u\}}b_{st}b_{su}\tilde{\Xs}_{s}^{2}\tilde{\Xs}_{t}\tilde{\Xs}_{u}
  + 2r_{\N}\sum_{t\neq u}b_{tt}b_{tu}E_{\mathcal{F}_{0}}[\tilde{\Xs}^{3}_{t}]\tilde{\Xs}_{u}\\
  2r_{\N}\sum_{u<t}b_{tt}b_{tu}(\tilde{\Xs}_{t}^{3}-E_{\mathcal{F}_{0}}[\tilde{\Xs}^{3}_{t}])\tilde{\Xs}_{u}
  +
  2r_{\N}\sum_{t<u}b_{tt}b_{tu}(\tilde{\Xs}_{t}^{3}-E_{\mathcal{F}_{0}}[\tilde{\Xs}^{3}_{t}])\tilde{\Xs}_{u}.
\end{multline*}
The conditional variance of the first summand is bounded by a constant times
$r_{\N}^{2}\sum_{t, u, s,
  v}\overbar{w}_{st}\overbar{w}_{su}\overbar{w}_{vt}\overbar{w}_{vu}$, which
converges to zero by the inequality in~\cref{eq:wstuv}. The conditional variance
of the second summand is bounded by a constant times
$r^{2}_{\N}\sum_{s, t,
  u}\overbar{w}_{tt}\overbar{w}_{tu}\overbar{w}_{ss}\overbar{w}_{su} \leq
r_{\N}^{2}\max_{s}n_{s}^{2}\sum_{s}n_{s}^{2}\to 0$. Since
$(\tilde{\Xs}_{t}^{3}-E_{\mathcal{F}_{0}}[\tilde{\Xs}^{3}_{t}])\sum_{u=1}^{t-1}b_{tt}b_{tu}\tilde{\Xs}_{u}$
and
$\tilde{\Xs}_{u}\sum_{t=1}^{u-1}b_{tt}b_{tu}(\tilde{\Xs}_{t}^{3}-E_{\mathcal{F}_{0}}[\tilde{\Xs}^{3}_{t}])$
are both martingale differences, by the inequality of von Bahr and Esseen, the
$4/3$-th absolute moment of the last two terms is bounded by a constant times
$r_{\N}^{4/3}\sum_{s, t}\overbar{w}_{tt}^{4/3}\overbar{w}_{ts}^{4/3} \leq
(\max_{s}n_{s}^{2}r_{\N})^{1/3} r_{\N}\sum_{t}n^{2}_{t}\to 0$. Thus, all
summands in the above display are of the order $o_{p}(1)$, and the third term in
\cref{eq:10-terms} is therefore also $o_{p}(1)$. The fourth term is $o_{p}(1)$
by arguments in \cref{eq:xs-as4}. By the triangle and Cauchy-Schwarz
inequalities, the conditional expectation of the absolute value of the fifth
term is bounded by
\begin{equation*}
2r_{\N}\sum_{i, j, s}w_{js}w_{is}E_{W}[\tilde{\Xs}_{s}^{4}]^{1/2}E_{W}[L_{i}^{4}]^{1/4}E_{W}[L_{j}^{4}]^{1/4}
\preceq \max_{i}E_{W}[L_{j}^{4}]^{1/2}\to 0.
\end{equation*}
Similarly, conditional expectation of the absolute
value of the sixth term is bounded by
\begin{equation*}
  4r_{\N}\sum_{i, j, s}w_{is}w_{js}E_{W}[V_{j}^{4}]^{1/4}E[\tilde{\Xs}_{s}^{4}]^{1/2}E_{W}[L_{i}^{4}]^{1/4}
  \preceq
  \max_{i}E_{W}[L_{j}^{4}]^{1/4}\to 0.
\end{equation*}
Thus, by the Markov inequality, the fifth and sixth terms are both of the order
$o_{p}(1)$. The conditional variance of the seventh and eighth terms is bounded
by a constant times
$r_{\N}^{2}\sum_{s, t, u}n_{s}n_{u}\overbar{w}_{st}\overbar{w}_{ut}\leq
r_{\N}\max_{s}n_{s}^{2}\to 0$, so that they are both $o_{p}(1)$ by Markov
inequality. By the inequality of von Bahr and Esseen, the $4/3$-th absolute
moment of the last ninth term is bounded by a constant times
$r_{\N}^{4/3}\sum_{s}E_{W}[\abs{a_{s}}^{4/3}]n_{s}^{4/3}\preceq
(\max_{s}n_{s}^{2}r_{\N})^{1/3}\to 0$, since by Jensen's inequality,
$E\abs{a_{s}}^{4/3}\leq (Ea_{s}^{2})^{2/3}$, which is bounded by a constant
times $n_{s}^{4/3}$. Finally, the expectation of the absolute value of the last
term in~\cref{eq:10-terms} is bounded by a constant times
\begin{equation*}
  r_{\N}\sum_{i, s, t}w_{is}\overbar{w}_{st}E_{W}[\tilde{\Xs}_{s}^{4}]^{1/2}E_{W}[\tilde{\Xs}^{4}_{t}]^{1/4}E_{W}[L_{i}^{4}]^{1/4}
  \preceq
  \max_{i}E_{W}[L_{i}^{4}]^{1/4} \to 0.
\end{equation*}

It remains to show that the first term in \cref{eq:hatV-decomposition} is
$o_{p}(1)$. It follows from~\cref{eq:dotx} and
\begin{NoHyper}\cref{eq:Z-proxy}\end{NoHyper}
that
\begin{equation*}
  \widehat{\Xs}=(W'W)^{-1}W'\ddot{X}=\tilde{\Xs}-
  (W'W)^{-1}W'U(\hat{\gamma}-\gamma)-\Zs(\hat{\gamma}-\gamma)-(W'W)^{-1}W'U\gamma,
\end{equation*}
where $\hat{\gamma}=(Z'Z)^{-1}Z'X$.
Let $\mathcal{U}=(W'W)^{-1}W'U$, and denote the $s$th row by $\mathcal{U}_{s}'$.
Since $\mathcal{U}_{sk}^{4}=(\sum_{i}((W'W)^{-1}W')_{si}U_{ik})^{4}$, it follows
by the Cauchy-Schwarz inequality that
\begin{equation*}
  E[\mathcal{U}_{sk}^{4}\mid W]\leq \max_{s}E[
(\sum_{i}((W'W)^{-1}W')_{si}U_{ik})^{4}\mid W]\preceq
\max_{s}(\sum_{i}\abs{((W'W)^{-1}W')_{si}})^{4},
\end{equation*}
which is bounded assumption of the \namecref{theorem:se-consistency}. Therefore,
the fourth moments of $\mathcal{U}_{s}$ are bounded uniformly over $s$. Observe
also that $E_{W}[\epsilon_{i}^{4}]$ is bounded uniformly over $s$ by assumptions
of the \namecref{theorem:se-consistency}. Therefore, by
applying~\Cref{theorem:Op1-variance} after using the expansion in
\cref{eq:Rs-decomposition}, we get
\begin{multline*}
  r_{\N}\sum_{s}(\widehat{\Xs}_{s}^{2}-\tilde{\Xs}_{s}^{2})\hat{R}_{s}^{2}=
  r_{\N}\sum_{s}\hat{R}_{s}^{2} (\mathcal{U}_{s}'\gamma)^{2}
  - 2r_{\N}\sum_{s}\hat{R}_{s}^{2}\tilde{\Xs}_{s}\mathcal{U}_{s}'\gamma\\
  +r_{\N}\sum_{s}\hat{R}_{s}^{2}\left[2\mathcal{U}_{s}'\gamma -2\tilde{\Xs}_{s}
    +(\Zs_{s}+\mathcal{U}_{s})'(\hat{\gamma}-\gamma)
  \right](\Zs_{s}+\mathcal{U}_{s})'(\hat{\gamma}-\gamma)\\
  = r_{\N}\sum_{s}{R}_{s}^{2} (\mathcal{U}_{s}'\gamma)^{2}-
   2r_{\N}\sum_{s}{R}_{s}^{2}\tilde{\Xs}_{s}\mathcal{U}_{s}'\gamma +O_{p}(1)(\hat{\gamma}-\gamma)+o_{p}(1).
\end{multline*}
By Cauchy-Schwarz inequality,
\begin{equation*}
  r_{\N}\sum_{s}E_{W}\abs{{R}_{s}^{2} (\mathcal{U}_{s}'\gamma)^{2}}
  \leq
  r_{\N}\sum_{s}(E_{W}[R_{s}^{4}])^{1/2}(E_{W}(\mathcal{U}_{s}'\gamma)^{4})^{1/2}
  \preceq
  \max_{s}(E_{W}(\mathcal{U}_{s}'\gamma)^{4})^{1/2} r_{\N}\sum_{s}n_{s}^{2}\to 0,
\end{equation*}
since
$\max_{s}E_{W}[(\mathcal{U}_{s}'\gamma)^{4}]\preceq
\max_{i}E_{W}(U_{i}'\gamma)^{4}\max_{s}(\sum_{i}\abs{((W'W)^{-1}W')_{si}})^{4}$,
which converges to zero by assumption of the \namecref{theorem:se-consistency}.
By similar arguments,
$2r_{\N}\sum_{s}E_{W}\abs{{R}_{s}^{2}\tilde{\Xs}_{s}\mathcal{U}_{s}'\gamma}\to 0
$ also, so that
\begin{equation*}
  r_{\N}\sum_{s}(\widehat{\Xs}_{s}^{2}-\tilde{\Xs}_{s}^{2})\hat{R}_{s}^{2}=
  o_{p}(1)+O_{p}(1)(\hat{\gamma}-\gamma)=o_{p}(1),
\end{equation*}
where the second equality follows from~\cref{eq:gamma-consistent}.

\subsubsection{Inference under heterogeneous effects}\label{sec:infer-under-heter}

For valid (but perhaps conservative) inference under heterogeneous effects, we
need to ensure that when $\beta_{is}\neq \beta$,
\begin{NoHyper}\cref{eq:se-consistency}\end{NoHyper}
holds with inequality, that is,
\begin{equation}\label{eq:se-consistency-conservative}
  \frac{\sum_{s=1}^{S}\widehat{\Xs}_{s}^{2}\hat{R}_{s}^{2}}{\sum_{s=1}^{S}n_{s}^{2}}\geq
  \mathcal{V}_{\N}+o_{p}(1).
\end{equation}
To discuss conditions under which this is the case, suppose, for simplicity,
that $L_{i}=0$ so that~\begin{NoHyper}\cref{eq:potential-outcomes}\end{NoHyper}
holds, and
$R_{s}=\sum_{s}w_{is}\epsilon_{i}$, where
$\epsilon_{i}=Y_{i}(0)-Z_{i}'\delta+\sum_{s}\Xs_{s}w_{is}(\beta_{is}-\beta)$ is
the regression residual. Then the ``middle sandwich'' in the asymptotic variance
sandwich formula, $\mathcal{V}_{\N}$, as defined in
\begin{NoHyper}\Cref{theorem:normality-Z}\end{NoHyper},
can be decomposed into three terms:
\begin{multline}\label{eq:sandwich-decomposition}
  \mathcal{V}_{\N}=\frac{\var\big(\sum_{s}\tilde{\Xs}_{s}R_{s}\mid\mathcal{F}_{0}\big)}{\sum_{s=1}^{S}n_{s}^{2}}
  =
  \frac{\sum_{s}E[\tilde{\Xs}^{2}_{s}R^{2}_{s}\mid\mathcal{F}_{0}]}{\sum_{s=1}^{S}n_{s}^{2}}
  -\frac{\sum_{s}E[\tilde{\Xs}_{s}R_{s}\mid\mathcal{F}_{0}]^{2}}{\sum_{s=1}^{S}n_{s}^{2}}
  + \frac{\sum_{s\neq t}\operatorname{cov}(\tilde{\Xs}_{s}R_{s},
    \tilde{\Xs}_{t}R_{t}\mid\mathcal{F}_{0})}{
    \sum_{s=1}^{S}n_{s}^{2}}\\
  =D_{1}+D_{2}+D_{3},
\end{multline}
where
\begin{align*}
  D_{1}&=\frac{\sum_{s}E[\tilde{\Xs}^{2}_{s}R^{2}_{s}\mid\mathcal{F}_{0}]}{\sum_{s=1}^{S}n_{s}^{2}},&
  D_{2}&=-\frac{\sum_{s}\left(\sum_{i}\sigma^{2}_{s}w_{is}^{2}(\beta_{is}-\beta)\right)^{2}}{\sum_{s=1}^{S}n_{s}^{2}},\\
 D_{3}&= \frac{\sum_{s\neq
      t}\sigma^{2}_{s}\sigma^{2}_{t}\sum_{i, j}w_{is}w_{it}(\beta_{it}-\beta)w_{jt}w_{js}(\beta_{js}-\beta)
  }{ \sum_{s=1}^{S}n_{s}^{2}}.
\end{align*}

As shown in the proof
of \begin{NoHyper}\Cref{theorem:se-consistency}\end{NoHyper} (see
\cref{eq:hatv-consistency}), the standard error estimator consistently estimates
$D_{1}$. Under homogeneous effects, $D_{2}=D_{3}=0$, and it follows that the standard error estimator is consistent. To ensure
valid inference under heterogeneous effects, one needs to ensure that
$D_{2}+D_{3}\leq o_{p}(1)$. This is the case under several sufficient conditions, and we give two
such conditions below.

The term $D_{2}$ reflects the variability of the treatment effect and it is
always negative. It therefore makes the variance estimate that we propose
conservative if $D_{3}=o_{p}(1)$. An analogous term, also reflecting
the variability of the treatment effect, is present in randomized, and
cluster-randomized trials, which is why the robust and cluster-robust standard
error estimators yield conservative inference in these settings \citep[see, for
example][Chapter 6]{imbens_causal_2015}. The term $D_{3}$ reflects correlation
between the treatment effects. It arises due to aggregating the sectoral shocks
$\Xs_{s}$ to a regional level to form the shifter $X_{i}$, and it has no analog
in cluster-randomized trials. Indeed, in the example with ``concentrated
sectors'', which is analogous to cluster-randomized trials if there are no
covariates, the term equals zero, since in that case $w_{is}w_{it}=0$ for
$s\neq t$. Our standard errors are thus valid, although conservative, in this
case.

More generally, a sufficient condition for validity of our standard error
estimator under treatment effect heterogeneity is that
$T_{\N}=\sum_{s\neq t}(\sum_{i}w_{is}w_{it})^{2}/\sum_{s}n_{s}^{2}\to 0$, since
$D_{3}=O_{p}(T_{\N})$. The condition $T_{\N}\to 0$ requires that the shares are
sufficiently concentrated so that not too many regions ``specialize'' in more
than one sector (in the sense that the sectoral share $w_{is}$ is bounded away
from zero as $S \to \infty$ for more than one sector). For example,
$T_{\N}\to 0$ if the share of the second-largest sector goes to zero as
$S\to\infty$, that is $\max_{i, s\neq s_{i}}w_{is}\to 0$, where $s_{i}$ denotes
the largest sector in region $i$. This follows from the inequalities
\begin{equation*}
  \begin{split}
    \sum_{i, j}\sum_{s\neq t}w_{is}w_{it}w_{js}w_{jt}&
    =\sum_{i, j, s, t}I(s=s_{i}, t\neq s_{i})w_{is}w_{it}w_{js}w_{jt}
    +\sum_{i, j}\sum_{s\neq t}I(s\neq s_{i})w_{is}w_{it}w_{js}w_{jt}\\
    & \leq\sum_{i, j, s, t}I(t\neq s_{i})w_{is}w_{it}w_{js}w_{jt}
    +\sum_{i, j, s, t}I(s\neq s_{i})w_{is}w_{it}w_{js}w_{jt}\\
    &\leq 2\max_{i, s\neq s_{i}}w_{is} \sum_{i, j, s, t}w_{it}w_{js}w_{jt}
    \leq2\max_{i, s\neq s_{i}}w_{is} \sum_{t}n_{t}^{2}=o(r_{\N}).
  \end{split}
\end{equation*}
For illustration, in the empirical application in
\begin{NoHyper}\Cref{sec:ADH}\end{NoHyper}, $T_{\N}=0.0014$.

A second sufficient condition for the asymptotic negligibility of $D_{3}$ is
that the conditional variance of the shifters $\Xs_{s}$,
$\sigma^{2}_{s}=E[(\Xs_{s}-\Zs_{s}'\gamma)^{2}\mid \mathcal{F}_{0}]$ and the
weighted treatment effects $\sigma^{2}_{s}\beta_{is}$ are mean-independent of
the shares $W$, provided some additional mild regularity conditions are
satisfied, as shown in the lemma below. Importantly, this condition still allows
the treatment effects to depend on the controls $Z$, or other aspects of the
model, such as $Y_{i}(0)$: the covariance assumptions in the lemma allow the
treatment effects $\beta_{is}$ to be correlated within a region and/or within a
sector. The assumption that
$\sum_{i}\sum_{s\neq t}w^{2}_{is}w^{2}_{it}/\sum_{s'}n_{s'}^{2}\to 0$ holds if
either a vanishing fraction of regions ``specialize'' in more than one sector
(in the sense that the sectoral share $w_{is}$ is bounded away from zero as
$S\to \infty$ for more than one sector). It also holds if
$S/\sum_{s}n_{s}\to 0$, that is, the number of regions grows faster than the
number of sectors.\footnote{This follows from the inequalities
  $\sum_{i, s, t}w^{2}_{is}w^{2}_{it}\leq \sum_{s}n_{s}$, and
  $\sum_{s}n_{s}^{2}\geq (\sum_{s}n_{s})^{2}/S$.} For illustration, the quantity
equals $0.00022$
in the empirical example in
\begin{NoHyper}\Cref{sec:ADH}\end{NoHyper}. The lemma uses the notation defined at the
beginning of
\begin{NoHyper}\Cref{sec:proof-texorpdfstr-co}\end{NoHyper}.
\begin{lemma}\label{lemma:Dop1}
  Suppose that the assumptions of
  \begin{NoHyper}\Cref{theorem:normality-Z}\end{NoHyper} hold. Suppose, in addition, that the conditional expectations
  $E[\sigma_{s}^{2}\beta_{is}\mid
  W]=E[(\Xs_{s}-\Zs_{s}'\gamma)^{2}\beta_{is}\mid W]$ and
  $E[\sigma^{2}_{s}\mid W]=E[(\Xs_{s}-\Zs_{s}'\gamma)^{2}\mid W]$ do not depend
  on $W$, $i$, or $s$. Suppose also that
  $\cov(\sigma_{s}^{2}\beta_{is}, \sigma_{t}^{2}\beta_{jt}\mid W)=0$ unless
  $i=j$ or $s=t$, that
  $\cov((\sigma_{s}^{2}\beta_{is}, \sigma^{2}_{s}), \sigma_{t}^{2}\mid W)=0$
  unless $s=t$, and that
  $\sum_{s\neq t}\sum_{i}w^{2}_{is}w^{2}_{it}/\sum_{s}n_{s}^{2}\to 0$. Then
  $D_{3}=o_{p}(1)$.
\end{lemma}

\begin{proof}
  By \Cref{sitem:beta-bounded,sitem:2nu},
   \begin{equation*}
    r_{\N}\sum_{s\neq t}\sum_{i}E_{W}\abs{\sigma^{2}_{s}\sigma^{2}_{t}w^{2}_{is}w^{2}_{it}(\beta_{it}-\beta)(\beta_{js}-\beta)}
    \preceq
    r_{\N}\sum_{s\neq t}\sum_{i}w_{is}^{2}w_{it}^{2},
  \end{equation*}
  and the right-hand side converges to zero by assumption of the
  \namecref{lemma:Dop1}. Therefore, by Markov inequality,
  $D_{3}=r_{\N}\sum_{s\neq t}\sum_{i\neq j}
  w_{is}w_{it}\sigma^{2}_{t}(\beta_{it}-\beta)w_{jt}w_{js}\sigma^{2}_{s}(\beta_{js}-\beta)+o_{p}(1)$.
 By
  \Cref{sitem:4nu,sitem:beta-bounded,,sitem:UZ-fourth},
  and assumptions of the \namecref{lemma:Dop1}, the variance of
  $\sum_{i, s}w_{is}^{2}\sigma_{s}^{2}\beta_{is}/\N$ and of
  $\sum_{i, s}w_{is}^{2}\sigma_{s}^{2}/\N$ conditional on $W$ is bounded by a
  constant times
  $\sum_{i, j, s}w_{is}^{2}w_{js}^{2}/\N^{2}+\sum_{i,
    s, t}w_{is}^{2}w_{it}^{2}/\N^{2}\leq 2\max_{s}n_{s}/N\to 0$. Therefore, by
  \Cref{sitem:detX}, $\beta=\mu/\sigma+o_{p}(1)$, where
  $\mu=E_{W}[(\Xs_{s}-\Zs_{s}'\gamma)^{2}\beta_{is}]$ and
  $\sigma=E_{W}[\sigma^{2}_{s}]$. It then follows that
  \begin{multline*}
    D_{3}=r_{\N}\sum_{s\neq t}\sum_{i\neq
      j}w_{is}w_{it}w_{jt}w_{js}(\sigma^{2}_{s}\beta_{js}-\mu)(\sigma^{2}_{t}\beta_{it}-\mu)
    -2r_{\N}\sum_{s\neq t}\sum_{i\neq j}w_{is}w_{it}w_{jt}w_{js}(\mu-\sigma^{2}_{t}\mu/\sigma)(\sigma^{2}_{s}\beta_{js}-\mu)\\
    +r_{\N}\sum_{s\neq t}\sum_{i\neq
      j}w_{is}w_{it}w_{jt}w_{js}(\mu-\sigma^{2}_{s}\mu/\sigma)(\mu-\sigma^{2}_{t}\mu/\sigma) +o_{p}(1).
  \end{multline*}
  Each term in the above display has mean zero, and variance bounded by a
  constant times
  \begin{multline*}
    r_{\N}^{2}\sum_{s\neq t}(\sum_{i\neq j}w_{is}w_{it}w_{jt}w_{js})^{2} +
    r_{\N}^{2}\sum_{i\neq j}(\sum_{s\neq
      t}w_{is}w_{it}w_{jt}w_{js})^{2}\\
    \leq r_{\N}^{2}\max_{s}n_{s}^{2} \sum_{i, j, s, t}w_{is}w_{it}w_{js}w_{jt} +
    r_{\N}^{2}\sum_{i, j, s, t}w_{it}w_{jt}w_{is}w_{js} \leq
    2r_{\N}\max_{s}n_{s}^{2}\to 0.
  \end{multline*}
  Therefore, $D_{3}=o_{p}(1)$ by Markov inequality and dominated convergence
  theorem.
\end{proof}

Although both the condition $T_{\N}\to 0$ and the conditions
in~\Cref{lemma:Dop1} may be restrictive in some applications, note that both of
these conditions are merely sufficient, but not necessary for $D_{3}+D_{2}\leq
o_{p}(1)$.

\subsection{Proofs and additional details for IV regression}\label{sec:proofs-iv}

We prove~\begin{NoHyper}\cref{eq:se-IV,eq:se-IV-psi}\end{NoHyper}, and show that the bias of the estimator
$\tilde{\alpha}$ is of the order
$\frac{1}{N}\sum_{i,s}w_{is}\check{w}_{is}/\check{n}_{s}$. We also discuss how
the case with estimated shifters relates to the literature on many instruments.

\subsubsection{Assumptions}

To compactly state the assumptions, let
$\mathcal{F}_{0}=(\Zs, U, Y_{1}(0), Y_{2}(0), B, W, \check{W})$, and put
$\check{W}=W$, and $\psi_{is}=0$ if the shifters $\Xs$ are observed.

We impose an instrumental variables version of the regularity conditions
\begin{NoHyper}\Cref{assumption:regularity,assumption:regularity-Z}\end{NoHyper}:
\begin{assumption}\label{assumption:regularity-IV}
  \aitem\label{item:consistency0-IV} For some $\nu>0$,
  $E[\Xs_{s}^{2+\nu}\mid \mathcal{F}_{0}]$ exists and is uniformly bounded. The
  support of $\beta_{is}$ is bounded. Conditional on $(W, \check{W})$, the second
  moments of ${Y}_{1i}(0), Y_{2i}(0), U_{i}$ and $\Zs_{s}$ exist, and are bounded
  uniformly over $i$ and $s$. $Z'Z/N$ converges in probability to a positive
  definite non-random limits; %
  \aitem\label{item:normality0-IV} For some $\nu>0$,
  $E[\abs{\Xs_{s}}^{4+\nu}\mid \mathcal{F}_{0}, \Psi]$ is uniformly bounded, and
  $\Xs_{s}$ are independent across $s$ conditional on $(\mathcal{F}_{0}, \Psi)$,
  with $E[\Xs_{s}\mid \mathcal{F}_{0}, \Psi]=E[\Xs_{s}\mid \Zs]$. Conditional on
  $(W, \check{W})$, the fourth moments of $Y_{1i}(0)$, $U_{i}$ and $\Zs_{s}$
  exist, and are bounded uniformly over $i$ and $s$.
  \Cref{sitem:delta} and \Cref{sitem:deltagamma}
  hold $\delta=E[Z'Z]^{-1}E[Z'Y_{1}(0)]$, $\check{\delta}=(Z'Z)^{-1}Z'Y_{1}(0)$,
  and $\epsilon_{i}=Y_{1i}-Y_{2i}\alpha-Z_{i}'\delta$.
\end{assumption}
\Cref{item:consistency0-IV} is needed for consistency, and
\Cref{item:normality0-IV} is needed for asymptotic normality. When the shifters
are observed, these assumptions are natural analogs of the regularity conditions
in the OLS case that are needed for consistency
(\Cref{sitem:beta-bounded,sitem:2nu} and
  \Cref{sitem:detZZ,sitem:UZ-moments}) and asymptotic normality
(\Cref{sitem:4nu} and
  \Cref{sitem:UZ-fourth,sitem:delta,sitem:deltagamma}). When the
shifters are not directly observed, \Cref{item:normality0-IV} strengthens
\begin{NoHyper}\Cref{item:random-assignment-IV}\end{NoHyper}
so that it holds conditionally on $\Psi$ also.

If $X_{i}$ is not observed, we need to impose additional
conditions on $\psi_{is}$ and the weights $\check{w}_{is}$:
\begin{assumption}\label{assumption:consistency-psi} Let $A_{-i}$ denote the
  vector $A$ with the $i$th element removed. Let
  $\mathcal{F}_{-i}=\sigma(\allowbreak Y_{1,-i}(0),\allowbreak Y_{2,-i}(0),
  \allowbreak U_{-i}, W,\check{W}, \Zs)$.
  \aitem\label{item:epsi} For all $s$ and $i$,
    $E[\check{w}_{is}\psi_{is}\mid \mathcal{F}_{-i}]=0$, and
    $E[\check{w}_{is}^{2}\psi_{is}^{2}\mid \mathcal{F}_{0}]$ is bounded by a
    universal constant times $\check{w}_{is}^{2}$;
  \aitem\label{item:covpsi} For all $s, t$, and all $i\neq j$,
    $E[\check{w}_{is}\check{w}_{jt}\psi_{is}\psi_{jt}\mid \mathcal{F}_{-i}]=0$;
  \aitem\label{item:two-regions}
    $\max_{i, s}\check{w}_{is}/\sum_{j=1}^{\N}\check{w}_{js}$ is bounded away
    from $1$;
  \aitem\label{item:L-bounded}
    $\max_{i}\sum_{s}\frac{n_{s}}{\check{n}_{s}}\check{w}_{is}$ is bounded;
  \aitem\label{item:split} There exist variables $\{C_{i}, \eta_{i}\}_{i=1}^{\N}$
    such that $(Y_{i1}(0), U_{i})=C_{i}+\eta_{i}$, and conditional on
    $(C, W,\Zs)$,
    $\{\check{w}_{i1}\psi_{i1}, \dotsc, \check{w}_{iS}\psi_{iS}, \eta_{i}\}$ are
    independent across $i$, with uniformly bounded second moments, and
    $E[(\check{w}_{is}\psi_{is}, \eta_{i})\mid C, \allowbreak W, \check{W},
    \Zs]=0$. Conditional on $(W, \check{W})$, the fourth moments of $\eta_{i}$
    and $C_{i}$ are uniformly bounded;
  \aitem\label{sec:psi-4} $E_{W, \check{W}}[\check{w}_{is}\psi_{js}]^{4}$ is bounded by a constant
    times $\check{w}_{is}^{4}$;
  \aitem\label{item:rns} $\N/(\sum_{s}n_{s}^{2})^{2} \to 0$.
\end{assumption}
\Cref{item:epsi} requires that the local shock $\psi_{is}$ in region $i$ is mean
zero, and unrelated to the regional variables $(Y_{1j}(0), Y_{2j}(0), U_{j})$ in
other regions. Importantly, it allows these local shocks to be correlated with
the regional variables in region $i$. In particular, in some applications, it
may be the case that $Y_{2i}=\sum_{s}w_{is}X_{is}+\eta_{i}$, with the additional
term $\eta_{i}$ potentially zero. In this case $\psi_{is}$ is always
mechanically correlated with $Y_{2i}$ (and hence also $Y_{1i}$ if there is
endogeneity). As we will show below, this correlation causes bias in the
estimator $\tilde{\alpha}$ that ignores the estimation error in the shifters.

\Cref{item:covpsi} requires that these local shocks are uncorrelated across
regions: this ensures consistency of the leave-one-out estimator. One could
relax this assumption and instead only require no correlation across clusters of
regions, in which case one would have to leave out region $i$'s cluster when
constructing an estimate of $X_{i}$. The local shocks are allowed to be
correlated across industries in the same region. The scaling by $\check{w}_{is}$
in the statement of the assumption allows for the possibility that $X_{is}$
gives an uninformative signal about $\Xs_{s}$ if $\check{w}_{is}=0$. %
\Cref{item:two-regions} imposes
two mild regularity conditions on the weights; it ensures that no single weight
$\check{w}_{is}$ is so large that it dominates a particular sector, which is
necessary for the leave-one-out estimator to be well-defined.

\Cref{item:L-bounded} ensures that the weights $\check{w}_{is}$ are balanced in
the sense that no single region $i$ is asymptotically non-negligible. The
condition holds under equal weighting, $\check{w}_{is}=1$, since in this case
$\sum_{s}n_{s}\check{w}_{is}/\check{n}_{s}=\sum_{s}n_{s}/\N\leq 1$. Oftentimes,
the weights $\check{w}_{is}$ take the form $\check{w}_{is}=L_{i}w_{is}$, where
$L_{i}$ is a measure of the size or region $i$. In this case,
$\sum_{s}n_{s}\check{w}_{is}/\check{n}_{s}=\sum_{s}\frac{L_{i}w_{is}}{\overbar{L}_{s}}$,
where $\overbar{L}_{s}=\check{n}_{s}/n_{s}=\sum_{i}L_{i}w_{is}/\sum_{j}w_{js}$
is the sector-weighted average size of a region. Thus, the condition requires
that the sector-weighted size of region $i$, $w_{is}L_{i}$, is non-negligible
relative to the national average for at most a fixed number of sectors. Since
$\sum_{s}n_{s}\check{w}_{is}/\check{n}_{s}\leq
\frac{\max_{i}L_{i}}{\min_{j}L_{j}}$, a sufficient condition is that the ratio
of the largest to the smallest region is bounded.

\Cref{item:rns,sec:psi-4,item:split} are only needed for asymptotic normality.
\Cref{item:split} effectively imposes that only the part of $(Y_{i1}(0), U_{i})$
that's independent of $\psi_{i}$ is allowed to be correlated across $i$; the
part that's related to $\psi_{i}$ must be independent across $i$.
\Cref{item:rns} imposes a very mild condition on the sector sizes, and holds,
for example, if $n_{s}\geq 1$.

\subsubsection{Asymptotic results}

When the shifters are observed, we obtain the following result, which implies
\begin{NoHyper}\cref{eq:se-IV}\end{NoHyper} in the main text:
\begin{proposition}\label{theorem:iv-asymptotics}
  Suppose that
  \begin{NoHyper}\Cref{item:sector-size,item:indepX}
  and \Cref{assumption:DGP-IV}\end{NoHyper} hold with
  $\mathcal{F}_{0}=(\Zs, U, Y_{1}(0), \allowbreak Y_{2}(0), \allowbreak B, W)$,
  and that \Cref{item:consistency0-IV} holds. Then the estimator $\hat{\alpha}$
  in
  \begin{NoHyper}\cref{eq:IVestimator}\end{NoHyper} is consistent. If, in addition,
  \begin{NoHyper}\Cref{item:sector-size-normality}\end{NoHyper} and \Cref{item:normality0-IV} hold, then
  $\hat{\alpha}$ satisfies
  \begin{NoHyper}\cref{eq:se-IV}\end{NoHyper},
  provided $\mathcal{V}_{\N}$ converges to a non-random limit.
\end{proposition}

The consistency result follows since by arguments analogous to those in the
proof of \begin{NoHyper}\Cref{theorem:consistency-Z}\end{NoHyper} (see, in
particular, \begin{NoHyper}\cref{eq:xy}\end{NoHyper}),
$N^{-1}\sum_{i}\ddot{X}_{i}Y_{1i}(0)=o_{p}(1)$, and
$N^{-1}\sum_{i}\ddot{X}_{i}Y_{2i}(0)=N^{-1}\sum_{i, s}\sigma^{2}_{s}w_{is}^{2}\beta_{is}+o_{p}(1)$.
Furthermore, since $N^{-1}\sum_{i, s}\sigma^{2}_{s}w_{is}^{2}\beta_{is} \neq 0$
by
\begin{NoHyper}\Cref{item:iv-relevance}\end{NoHyper}, it follows by Slutsky's
lemma that
\begin{equation*}
\hat{\alpha}-\alpha=\frac{N^{-1}\sum_{i}\ddot{X}_{i}Y_{1i}(0)}{N^{-1}\sum_{i}\ddot{X}_{i}Y_{2i}(0)}
=o_{p}(1).
\end{equation*}
The asymptotic normality result follows since
$r_{\N}^{1/2}\sum_{i}\ddot{X}_{i}Y_{1i}(0)=\mathcal{N}(0,\mathcal{V}_{\N})+o_{p}(1)$ by
arguments analogous to those in
proof of \begin{NoHyper}\Cref{theorem:normality-Z}\end{NoHyper} (see, in
particular, \begin{NoHyper}\cref{eq:OLS-middle}\end{NoHyper}).

\begin{proposition}\label{theorem:mismeasured-iv-consistency}
  Suppose that \begin{NoHyper}\Cref{item:sector-size,item:indepX}
  and \Cref{assumption:DGP-IV}\end{NoHyper} hold with
  $\mathcal{F}_{0}=(\Zs, U, Y_{1}(0), \allowbreak Y_{2}(0), \allowbreak B, W,
  \check{W})$, and that \Cref{item:consistency0-IV} and
  \Cref{item:L-bounded,item:two-regions,item:covpsi,item:epsi} hold. Then the
  estimator $\hat{\alpha}_{-}$ is consistent for $\alpha$. Furthermore, the
  estimator $\tilde{\alpha}$ satisfies
  $\tilde{\alpha}=\alpha+O_{p}\left(\frac{1}{\N}
    \sum_{i, s}\frac{w_{is}\check{w}_{is}}{\check{n}_{s}}\right)$, provided that
  $(\ddot{\hat{X}}'Y_{2}/\N)^{2}$ converges to a strictly positive probability
  limit.
\end{proposition}

The asymptotic bias $\tilde{\alpha}$ is analogous to the own observation bias of
the two-stage least squares (2SLS) estimator in settings with many instruments.
To see the connection, consider the special case in which
$Y_{2i}=\sum_{s}w_{is}X_{is}=\sum_{s}w_{is}\Xs_{s}+\sum_{s}w_{is}\psi_{is}$, and
each region specializes in a single sector, $w_{is}=\1{s(i)=s}$, with
$\check{w}_{is}=w_{is}$. Then we can write $Y_{2i}=\Xs_{s(i)}+\psi_{is(i)}$, and
$\hat{X}_{i}=\frac{1}{n_{s}}\sum_{i}\1{s(i)=s}Y_{2i}$. This setting is
isomorphic to a many instrument setting, where the instruments are group
indicators $\1{s(i)=s}$, individuals are assigned to groups, and the average
treatment intensity depends on group membership (for example, the endogenous
variable may be the length of a sentence, the groups are groups of individuals
assigned to the same judge, and judges differ in their average sentencing
severity $\Xs_{s}$). Then the first-stage predictor used by the 2SLS estimator
is $\hat{X}_{i}$. Since $\hat{X}_{i}$ puts weight $1/n_{s}$ on the first-stage
regression error $\psi_{is(i)}$, this generates a bias in the 2SLS estimate,
which persists in large samples unless the weight $1/n_{s}$ is negligible. In
our setting, \Cref{theorem:mismeasured-iv-consistency} shows that the bias is of
the order
$\frac{1}{\N} \sum_{i, s}\frac{w_{is}\check{w}_{is}}{\check{n}_{s}}\leq
\frac{1}{\N} \sum_{i, s}\frac{\check{w}_{is}}{\check{n}_{s}}= S/\N$. Thus, a
sufficient condition for consistency is that the number of sectors grows more
slowly than the number of regions. This is analogous to the requirement for 2SLS
consistency in the many instruments literature that the number of instruments
grows more slowly than the number of observations.

\begin{proposition}\label{theorem:normality-IV}
  Suppose that
  \begin{NoHyper}\Cref{assumption:inference,assumption:DGP-IV}\end{NoHyper} hold with
  $\mathcal{F}_{0}=(\Zs, U, Y_{1}(0), \allowbreak Y_{2}(0), \allowbreak B, W,
  \check{W})$, and that
  \Cref{assumption:regularity-IV,assumption:consistency-psi} hold. Suppose that
  $\mathcal{V}_{\N}$ and $\mathcal{W}_{\N}$, defined in
    \begin{NoHyper}\cref{eq:se-IV-psi}\end{NoHyper}, converge in probability to
    non-random limits. Then
  \begin{equation*}
    \frac{\N}{\sqrt{\sum_{s=1}^{S}n_{s}^{2}}}(\hat{\alpha}_{-}-\alpha)
    =\mathcal{N}\left(0,\frac{\mathcal{V}_{\N}+\mathcal{W}_{\N}}{
        \left(\frac{1}{\N}\sum_{i}\ddot{X}_{i}Y_{2i}\right)^{2}}\right)+o_{p}(1).
  \end{equation*}
\end{proposition}

The additional term $\mathcal{W}_{\N}$ in the expression for the asymptotic
variance of $\hat{\alpha}_{-}$, which is absent if $\Xs$ is observed, is of the
order
\begin{equation*}
  \frac{1}{\sum_{s}n_{s}^{2}}\sum_{j}
  \left(\sum_{s}\frac{n_{s}\check{w}_{js}}{\check{n}_{s}}\right)^{2}
  +\frac{1}{\sum_{s}n_{s}^{2}}\sum_{i, j,s, t}\frac{w_{is}\check{w}_{js}}{\check{n}_{s}}
  \frac{w_{jt}\check{w}_{it}}{\check{n}_{t}}\preceq
  \frac{\N+S}{\sum_{s}n_{s}^{2}}\preceq S/N+(S/N)^{2},
\end{equation*}
where the second inequality follows \Cref{item:L-bounded}, and the last
inequality follows by $\ell_{1}$-$\ell_{2}$ norm inequality
$\sqrt{S\sum_{s}n_{s}^{2}}\geq \sum_{s}n_{s}$, and we assume that
$\sum_{s}w_{is}$ is bounded away from zero so that $\sum_{s}n_{s}$ is of the
same order as $N$. Therefore, if the number of regions grows faster than the
number of sectors, the term will be asymptotically negligible. This is similar
to the result in the many IV literature that the usual standard error formula
for the jackknife IV estimator is valid if the number of instruments grows more
slowly than the sample size. The term $\mathcal{W}_{\N}$ also has a similar
structure to the many-instrument term in the standard error for jackknife IV
(see \citet{chao_asymptotic_2012}).

\subsubsection{Proof of
  \texorpdfstring{\Cref{theorem:mismeasured-iv-consistency}}{%
Proposition~\ref{theorem:mismeasured-iv-consistency}}}

By the arguments in the proof of
\begin{NoHyper}\Cref{theorem:consistency-Z}\end{NoHyper}, for the first
part of the \namecref{theorem:mismeasured-iv-consistency}, it suffices to show
that $(\ddot{\hat{X}}_{-}-\ddot{X})'Y_{1}/\N=o_{p}(1)$ and
$(\ddot{\hat{X}}_{-}-\ddot{X})'Y_{2}/\N=o_{p}(1)$, which in turn follows if we
can show that for $A_{i}\in\{Y_{1i}, Y_{2i}, Z_{i}\}$,
\begin{equation}
  \label{eq:xhat-a}
  \frac{1}{\N}\sum_{i}(\hat{X}_{i,-}-X_{i})A_{i}
  = \frac{1}{\N}\sum_{j, i,s}\1{j\neq i}
  \frac{w_{is}\check{w}_{js}}{\check{n}_{s, -i}}\psi_{js}A_{i}
  =o_{p}(1),
\end{equation}
where $\check{n}_{s, -i}=\sum_{j=1}^{\N}\check{w}_{js}-\check{w_{is}}$. By
\Cref{item:epsi}, conditional on $W$, this term has mean zero. Since by
\Cref{item:covpsi},
$\1{j\neq j'}\1{j\neq i}\1{j'\neq i'}E_{W, \check{W}}[w_{js}\psi_{js}A_{i}\cdot
w_{j't}\psi_{j't}A_{i'}]=0$ unless $j=i'$ and $j'=i$, the variance of this term
is given by
\begin{multline*}
  \frac{1}{\N^{2}}\sum_{j, i, i', s, t}\1{j\neq i, i'}w_{is}w_{i't}
  \frac{E_{W, \check{W}}[\check{w}_{js}
    \psi_{js}A_{i}\check{w}_{jt}\psi_{jt}A_{i'}]
  }{\check{n}_{s, -i}\check{n}_{t, -i'}}\\
  + \frac{1}{\N^{2}}\sum_{j, i, s, t}\1{j\neq i}w_{is}w_{jt} \frac{E_{W,
      \check{W}}[\check{w}_{js}\psi_{js}\check{w}_{it}\psi_{it}A_{i}A_{j}]}{
    \check{n}_{s, -i}\check{n}_{t, -j}}.
\end{multline*}
Now, by \Cref{item:consistency0-IV},
$E_{W, \check{W}}[\check{w}_{js}
\psi_{js}A_{i}\check{w}_{jt}\psi_{jt}A_{i'}]\preceq
\check{w}_{js}\check{w}_{jt}E_{W, \check{W}}[ A_{i}A_{i'}]$, which is bounded by
a constant times $\check{w}_{js}\check{w}_{jt}$ since the second moment of
$A_{i}$ is uniformly bounded by
\Cref{item:epsi}. Similarly,
$E_{W, \check{W}} [\check{w}_{js}\psi_{js}\check{w}_{it}\psi_{it}A_{i}A_{j}]$ is
bounded by a constant times $\check{w}_{js}\check{w}_{it}$. Therefore, the
expression in the preceding display is bounded by a constant times
\begin{multline*}
  \frac{1}{\N^{2}}\sum_{j, i, i', s, t}w_{is}w_{i't} \frac{\check{w}_{js}
    \check{w}_{jt}}{\check{n}_{s,-i}\check{n}_{t, -i'}} +
  \frac{1}{\N^{2}}\sum_{j, i, s, t}w_{is}w_{jt} \frac{
    \check{w}_{js}\check{w}_{it}}{
    \check{n}_{s, -i}\check{n}_{t, -j}}\\
  \leq \frac{1}{\N^{2}}
  \max_{is}\frac{\check{n}_{s}^{2}}{\check{n}_{s, -i}^{2}}\left[
    \sum_{j}\left(\sum_{s}n_{s}\frac{\check{w}_{js}}{\check{n}_{s}}\right)^{2}
    + \N\right]
  \preceq \frac{1}{N},
\end{multline*}
where the first inequality follows since
$\sum_{j, i, s, t}w_{is}w_{jt} \frac{\check{w}_{js}\check{w}_{it}}{
  \check{n}_{s}\check{n}_{t}} \leq \sum_{j, i, s, t}w_{is}w_{jt} \frac{
  \check{w}_{js}}{ \check{n}_{s}}\leq \sum_{j, s}n_{s} \frac{\check{w}_{js}}{
  \check{n}_{s}}=\N$, and the second inequality follows since
\Cref{item:two-regions} implies
$\max_{is}\check{n}_{s}/\check{n}_{s, -i}=
1/(1-\max_{is}\check{w}_{is}/\check{n}_{is})$ is bounded, and since
\Cref{item:L-bounded} implies that
$\sum_{j}\left(\sum_{s}n_{s}\frac{\check{w}_{js}
  }{\check{n}_{s}}\right)^{2}\preceq \sum_{j}1=\N$. Therefore, \cref{eq:xhat-a}
holds by Markov inequality and the dominated convergence theorem.

To show the second part of the proposition, decompose
\begin{equation*}
\frac{1}{\N}\sum_{i} A_{i}(\hat{X}_{i}-\hat{X}_{i,-})=
  \frac{1}{\N}\sum_{i, s}\frac{w_{is}\check{w}_{is}}{\check{n}_{s}}\psi_{is}A_{i}
  -  \frac{1}{\N}\sum_{i, j,s}\1{j\neq i}
  \frac{\check{w}_{is}}{\check{n}_{s}}\frac{w_{is}\check{w}_{js}}{\check{n}_{s, -i}} \psi_{js}A_{i}.
\end{equation*}
By arguments similar to those above, conditional on $(W, \check{W})$, the second
term has mean zero and variance that converges to zero. By
\Cref{item:epsi} and Jensen's inequality, the
mean of the first term is of the order
$\frac{1}{\N}\sum_{i, s}\frac{w_{is}\check{w}_{is}}{\check{n}_{s}}$.
Consequently, provided that $(\ddot{\hat{X}}'Y_{2}/\N)^{2}$ converges to a
strictly positive limit, we have
\begin{equation*}
  \tilde{\alpha}-\alpha=\frac{O_{p}(\frac{1}{\N}\sum_{i, s}\frac{w_{is}\check{w}_{is}}{\check{n}_{s}}
)}{\ddot{\hat{X}}'Y_{2}/\N}=O_{p}\left(\frac{1}{\N}
\sum_{i, s}\frac{w_{is}\check{w}_{is}}{\check{n}_{s}}\right),
\end{equation*}
as required.

\subsubsection{Proof of \texorpdfstring{\Cref{theorem:normality-IV}}{%
Proposition~\ref{theorem:normality-IV}}}

Since
$\N r_{\N}^{1/2}(\hat{\alpha}_{-}-\alpha)=r_{\N}^{1/2}\hat{\ddot{X}}_{-}'Y_{1}(0)/\hat{\ddot{X}}_{-}'Y_{2}/N
=r_{\N}^{1/2}\hat{\ddot{X}}_{-}'Y_{1}(0)\cdot
(\beta_{FS}\N^{-1}\sum_{i, s}w_{is}^{2}\sigma^{2}_{s})^{-1}(1+o_{P}(1))$, it
suffices to show that
\begin{equation*}
  r_{\N}^{1/2}\hat{\ddot{X}}_{-}'Y_{1}(0)=\mathcal{N}(0,\mathcal{V}_{\N}+\mathcal{W}_{\N})+o_{p}(1).
\end{equation*}
By arguments as in the proof of
\begin{NoHyper}\Cref{theorem:normality-Z}\end{NoHyper},
\begin{equation*}
  \begin{split}
    r_{\N}^{1/2}\hat{\ddot{X}}_{-}'Y_{1}(0) &=
    r_{\N}^{1/2}(W\tilde{\Xs}-U\gamma+(\hat{X}_{-}-X))'(Z(\delta-\check{\delta})+\epsilon_{\Delta})
    \\
    &= r_{\N}^{1/2}(W\tilde{\Xs})'\epsilon_{\Delta} +
    r_{\N}^{1/2}(\hat{X}_{-}-X)'(Z(\delta-\check{\delta})+\epsilon_{\Delta})
    +o_{p}(1)\\
    &=r_{\N}^{1/2}(W\tilde{\Xs}+(\hat{X}_{-}-X))'\epsilon_{\Delta} +o_{p}(1),
  \end{split}
\end{equation*}
where the last line follows since $(\hat{X}_{-}-X)'Z/\N=o_{p}(1)$ by
\cref{eq:xhat-a}. Let
$C_{\Delta, i}=C_{iY(0)}-C_{iU}'\delta-\sum_{s}w_{is}\Zs_{s}'\delta$ and
$\eta_{\Delta, i}=\eta_{iY(0)}-\eta_{iU}'\delta$, so that
$\epsilon_{\Delta, i}=Y_{i1}(0)-Z_{i}'\delta=\eta_{\Delta, i}+C_{\Delta, i}$. Then we can decompose
\begin{equation*}
  r_{\N}^{1/2} (W\tilde{\Xs}+(\hat{X}_{-}-X))'\epsilon_{\Delta}
  =r_{\N}^{1/2}\sum_{j=1}^{\N+S}\mathcal{Y}_{j},
\end{equation*}
where
\begin{equation*}
  \mathcal{Y}_{j}=
  \begin{cases}
    \sum_{i=1}^{\N}\sum_{s=1}^{S}w_{is}\check{w}_{js}\frac{\1{j\neq
        i}\psi_{js}C_{\Delta, i}}{\check{n}_{s, -i}}
    +\sum_{i=1}^{j-1}\sum_{s=1}^{S}\left[\frac{w_{is}\check{w}_{js}\psi_{js}
        \eta_{\Delta, i}}{\check{n}_{s, -i}}
      +\frac{\check{w}_{is}w_{js}\eta_{\Delta, j}\psi_{is}}{\check{n}_{s, -j}} \right],& j=1,\dotsc, \N,\\
    \tilde{\Xs}_{j-\N}\sum_{i}w_{i, j-\N}\epsilon_{\Delta, i}, &
    j=\N+1,\dotsc, \N+S.
  \end{cases}
\end{equation*}
Let $H$ denote the matrix with rows $\eta_{i}'$, and define the $\sigma$-fields
$\mathcal{G}_{i}=\sigma(W, \check{W}, \Zs, C,\eta_{1}, \dotsc, \allowbreak,
\eta_{i}, \allowbreak\psi_{1}, \allowbreak\dotsc, \psi_{i})$, $i=1,\dotsc, \N$,
$\mathcal{G}_{i}=\sigma(W, \check{W}, \Zs, C,H, \Psi, \Xs_{1}, \dotsc, \Xs_{j-\N})$,
$j=\N+1,\dotsc, \N+S$. Then, under
\Cref{item:split}, $\mathcal{Y}_{j}$ is a
martingale difference array with respect to the filtration $\mathcal{G}_{j}$.
Since by the arguments in the proof of
\begin{NoHyper}\Cref{theorem:normality-Z}\end{NoHyper},
$r_{\N}^{1+\nu/4}\sum_{j=\N+1}^{\N+S}E_{W, \check{W}}[\mathcal{Y}_{j}^{2+\nu/2}]\to 0$, and
$r_{\N}\sum_{j=\N+1}^{\N+S} E[\mathcal{Y}_{j}^{2}
\mid\mathcal{G}_{j-1}]-\mathcal{V}_{\N}=o_{p}(1)$, it suffices to show that
$r_{\N}^{2}\sum_{j=1}^{\N}E_{W, \check{W}}[\mathcal{Y}_{j}^{4}]\to 0$, and
$r_{\N}\sum_{j=1}^{\N} E[\mathcal{Y}_{j}^{2}
\mid\mathcal{G}_{j-1}]-\mathcal{W}_{\N}=o_{p}(1)$. The result then follows by a
martingale central limit theorem.

Since $\check{n}_{s}/\check{n}_{s, -i}$ is bounded, and $\sum_{s}w_{js}\leq 1$,
and since $\sum_{s=1}^{S}\frac{n_{s}\check{w}_{js}}{ \check{n}_{s}}$ is bounded
by \Cref{item:L-bounded}, we have the bound
\begin{equation*}
  r_{\N}^{2} \sum_{j=1}^{\N} E_{W, \check{W}}\left(
    \sum_{i=1}^{j-1}\sum_{s=1}^{S}w_{is}\check{w}_{js}\frac{\psi_{js}\eta_{\Delta, i}}{
      \check{n}_{s, -i}}\right)^{4}
  \preceq  r_{\N}^{2} \sum_{j}\left(
    \sum_{i=1}^{j-1}\sum_{s=1}^{S}\frac{w_{is}\check{w}_{js}}{
      \check{n}_{s}}\right)^{4}
  \leq r_{\N}^{2}\sum_{j=1}^{\N}  \left(
    \sum_{s=1}^{S}\frac{n_{s}\check{w}_{js}}{
      \check{n}_{s}}\right)^{4}
  \leq   r_{\N}^{2} \N,
\end{equation*}
which converges to zero by \Cref{item:rns}. By an
analogous argument, the conditional expectation of
$r_{\N}^{2}
\sum_{j=1}^{\N}\left(\sum_{i=1}^{\N}\sum_{s=1}^{S}w_{is}\check{w}_{js}\frac{\1{j\neq
      i}\psi_{js}C_{\Delta, i}}{\check{n}_{s, -i}}\right)^{4}$ and of
$r_{\N}^{2}
\sum_{j=1}^{\N}\left(\sum_{i=1}^{j-1}\sum_{s=1}^{S}\check{w}_{is}w_{js}\frac{\eta_{\Delta, j}\psi_{is}}{\check{n}_{s, -j}}
\right)^{4}$ is also bounded by $r_{\N}^{2} \N$, so that
$r_{\N}^{2}\sum_{j=1}^{\N}E_{W, \check{W}}[\mathcal{Y}_{j}^{4}]\to 0$ by
$C_{r}$-inequality.

It remains to show that the conditional variance
$r_{\N}\sum_{j=1}^{\N} E[\mathcal{Y}_{j}^{2} \mid\mathcal{G}_{j-1}]$ converges.
Expanding the expectation yields
\begin{multline*}
  r_{\N}\sum_{j=1}^{\N} E[\mathcal{Y}_{j}^{2} \mid\mathcal{G}_{j-1}]
  =2r_{\N}\sum_{i, j,s, t}\sum_{i'}^{j-1} \frac{\1{j\neq i}
    E_{\mathcal{G}_{0}}[\check{w}_{js}\check{w}_{jt}\psi_{js}\psi_{jt}]}{\check{n}_{s, -i}}
  \frac{w_{is}w_{i't} C_{\Delta, i}\eta_{\Delta, i'}}{\check{n}_{t, -i'}}\\
  +2r_{\N}\sum_{i, j,s, t}\sum_{i'=1}^{j-1}\frac{\1{j\neq
      i}E_{\mathcal{G}_{0}}[\check{w}_{js}w_{jt}\psi_{js}\eta_{\Delta, j}]}{\check{n}_{s, -i}}
  \frac{w_{is}\check{w}_{i't}C_{\Delta, i}\psi_{i't}}{\check{n}_{t, -j}}\\
  +r_{\N}\sum_{j, s, t} \sum_{i=1}^{j-1} \sum_{i'=1}^{j-1}\1{i\neq i'}
  \frac{E_{\mathcal{G}_{0}}[\check{w}_{js}\check{w}_{jt}\psi_{js}\psi_{jt}]}{
    \check{n}_{s, -i}} \frac{w_{i't}w_{is}\eta_{\Delta, i}\eta_{\Delta,
      i'}}{\check{n}_{t, -i'}}\\
  +2r_{\N}\sum_{j, s, t}\sum_{i=1}^{j-1}\sum_{i'=1}^{j-1}\1{i\neq i'}
  \frac{E_{\mathcal{G}_{0}}[\check{w}_{js}w_{jt}\psi_{js}\eta_{\Delta, j}]}{
    \check{n}_{s, -i}} \frac{w_{is}\check{w}_{i't} \eta_{\Delta,
      i}\psi_{i't}}{\check{n}_{t, -j}}  \\
  r_{\N}\sum_{j, s,t}\sum_{i=1}^{j-1} \sum_{i'=1}^{j-1}\1{i\neq
    i'}\frac{w_{jt}w_{js}E_{\mathcal{G}_{0}}[\eta_{\Delta, j}^{2}]}{\check{n}_{s, -j}}
  \frac{\check{w}_{i't}\check{w}_{is}\psi_{i't}\psi_{is}}{\check{n}_{t, -j}}\\ +
  r_{\N}\sum_{j, s,
    t}\sum_{i=1}^{j-1}\frac{w_{jt}w_{js}E_{\mathcal{G}_{0}}[\eta_{\Delta, j}\eta_{\Delta, j}]}{\check{n}_{s, -j}}
  \frac{\check{w}_{is}\check{w}_{it}\psi_{is}\psi_{it}}{\check{n}_{t, -j}}
  +2r_{\N}\sum_{j, s, t}\sum_{i=1}^{j-1}\frac{E_{\mathcal{G}_{0}}[\check{w}_{js}w_{jt}\psi_{js}\eta_{\Delta, j}]
    }{ \check{n}_{s, -i}}
  \frac{w_{is}\check{w}_{it}\eta_{\Delta, i}\psi_{it}}{\check{n}_{t, -j}}\\
  +r_{\N}\sum_{j, s, t} \sum_{i=1}^{j-1}\frac{E_{\mathcal{G}_{0}}[\check{w}_{js}\check{w}_{jt}\psi_{js}\psi_{jt}]
    }{ \check{n}_{s, -i}}
  \frac{w_{is}w_{it} \eta^{2}_{\Delta, i}}{\check{n}_{t, -i}}
+ r_{\N}\sum_{j=1}^{\N}E_{\mathcal{G}_{0}}\left(\sum_{i=1}^{\N}\sum_{s=1}^{S}\frac{\1{j\neq
        i}w_{is}\check{w}_{js}\psi_{js}C_{\Delta, i}}{\check{n}_{s, -i}}\right)^{2}.
\end{multline*}

Conditional on $(W, \check{W})$, the first five terms are mean zero. The variance
of the first term is bounded by a constant times
\begin{equation*}
  r^{2}_{\N}\sum_{i'}\left(\sum_{i, j,s, t}
    \frac{ w_{is}w_{i't}  \check{w}_{js}\check{w}_{jt}}{\check{n}_{s}\check{n}_{t}}\right)^{2}
  =  r^{2}_{\N}\sum_{i'}\left(\sum_{j, t}
    \frac{ w_{i't}  \check{w}_{jt}}{\check{n}_{t}}
    \sum_{s}\frac{n_{s}\check{w}_{js}}{\check{n}_{s}}
  \right)^{2}\preceq
  r^{2}_{\N}\N.
\end{equation*}
Similarly, the variance of the second, third, fourth, and fifth term can be
shown to be bounded by a constant times $r^{2}_{\N}\N$.
Next, the expectation conditional on $(W, \check{W})$ of the absolute value of
the sixth term is bounded by a constant times
\begin{multline*}
  r_{\N}\sum_{i, j}\left(\sum_{s}\frac{\check{w}_{is}w_{js}}{\check{n}_{s}}\right)
  \left(\sum_{t}\frac{w_{jt}\check{w}_{it}}{\check{n}_{t}}\right)
  \leq
  r_{\N}\sum_{i}\max_{i'}\sum_{j}\left(\sum_{s}\frac{\check{w}_{is}w_{js}}{\check{n}_{s}}\right)
  \left(\sum_{t}\frac{w_{jt}\check{w}_{i't}}{\check{n}_{t}}\right)\\
  =  r_{\N}\sum_{i}\max_{i'}\sum_{j}\left(\sum_{s}\frac{\check{w}_{is}w_{js}}{\check{n}_{s}}\right)
  \left(\sum_{t}\frac{w_{jt}\check{w}_{i't}}{\check{n}_{t}}\right)
\end{multline*}

Consequently, by Markov inequality,
\begin{multline}\label{eq:cond-var-2}
  r_{\N}\sum_{j=1}^{\N} E[\mathcal{Y}_{j}^{2} \mid\mathcal{G}_{j-1}]=\\
  r_{\N}\sum_{j, s,
    t}\sum_{i=1}^{j-1}\frac{w_{jt}w_{js}E_{\mathcal{G}_{0}}[\eta_{\Delta, j}\eta_{\Delta, j}]}{\check{n}_{s, -j}}
  \frac{\check{w}_{is}\check{w}_{it}\psi_{is}\psi_{it}}{\check{n}_{t, -j}}
  +2r_{\N}\sum_{j, s, t}\sum_{i=1}^{j-1}\frac{E_{\mathcal{G}_{0}}[\check{w}_{js}w_{jt}\psi_{js}\eta_{\Delta, j}]
    }{ \check{n}_{s, -i}}
  \frac{w_{is}\check{w}_{it}\eta_{\Delta, i}\psi_{it}}{\check{n}_{t, -j}}\\
  +r_{\N}\sum_{j, s, t} \sum_{i=1}^{j-1}\frac{E_{\mathcal{G}_{0}}[\check{w}_{js}\check{w}_{jt}\psi_{js}\psi_{jt}]
    }{ \check{n}_{s, -i}}
  \frac{w_{is}w_{it} \eta^{2}_{\Delta, i}}{\check{n}_{t, -i}}
+ r_{\N}\sum_{j=1}^{\N}E_{\mathcal{G}_{0}}\left(\sum_{i=1}^{\N}\sum_{s=1}^{S}\frac{\1{j\neq
        i}w_{is}\check{w}_{js}\psi_{js}C_{\Delta, i}}{\check{n}_{s, -i}}\right)^{2}+o_{p}(1).
\end{multline}
Similarly, expanding the expression for $\mathcal{W}_{\N}$ yields
\begin{multline*}
  \mathcal{W}_{\N}= \frac{1}{r_{\N}}\sum_{i, i', j, s, t}\1{j\neq i, i'}\1{i\neq
    i'} \frac{\check{w}_{js}\check{w}_{jt}\psi_{jt}\psi_{js}}{\check{n}_{s, -i}}
  \frac{w_{is}w_{i't}\eta_{\Delta, i}\eta_{\Delta, i'}}{\check{n}_{t, -i'}}\\
  + \frac{2}{r_{\N}}\sum_{i, i', j, s, t}\1{j\neq
    i, i'}\frac{\check{w}_{js}\check{w}_{jt}\psi_{jt}\psi_{js}}{\check{n}_{s, -i}}
  \frac{w_{is}w_{i't}C_{\Delta, i}\eta_{\Delta, i'}}{\check{n}_{t, -i'}}
  \\
  +\frac{1}{r_{\N}}\sum_{i, j,s, t}\1{i\neq
    j}\frac{w_{is}\check{w}_{js}\psi_{it}C_{\Delta, i}}{\check{n}_{s, -i}}
  \frac{w_{jt}\check{w}_{it}\psi_{js}\eta_{\Delta, j}}{\check{n}_{t, -j}}
  +\frac{1}{r_{\N}}\sum_{i, j,s, t}\1{i\neq
    j}\frac{w_{is}\check{w}_{js}\psi_{it}\eta_{\Delta, i}}{\check{n}_{s, -i}}
  \frac{w_{jt}\check{w}_{it}\psi_{js}C_{\Delta, j}}{\check{n}_{t, -j}}
  \\
  +\frac{1}{r_{\N}}\sum_{i, j,s, t}\1{i\neq
    j}\frac{w_{is}\check{w}_{js}\psi_{it}C_{\Delta, i}}{\check{n}_{s, -i}}
  \frac{w_{jt}\check{w}_{it}\psi_{js}C_{\Delta, j}}{\check{n}_{t, -j}}\\
  + \frac{1}{r_{\N}}\sum_{i, j, s, t}\1{j\neq i}
  \frac{\check{w}_{js}\psi_{js}\check{w}_{jt}\psi_{jt}}{\check{n}_{s, -i}}
  \frac{w_{is}w_{it}\eta^{2}_{\Delta, i}}{\check{n}_{t, -i}}
  +\frac{2}{r_{\N}}\sum_{i, j,s, t}\1{i<j}\frac{w_{is}\check{w}_{js}\psi_{it}\eta_{\Delta, i}}{\check{n}_{s, -i}}
  \frac{w_{jt}\check{w}_{it}\psi_{js}\eta_{\Delta, j}}{\check{n}_{t, -j}}\\
  +\frac{1}{r_{\N}}\sum_{j}\left(\sum_{i, s}\1{i\neq
      j}\frac{w_{is}\check{w}_{js}\psi_{js}C_{\Delta,
        i}}{\check{n}_{s, -i}}\right)^{2}.
\end{multline*}
Conditional on $(W, \check{W})$, the first five terms are mean zero. The variance
of the first term is bounded by a constant times
\begin{multline*}
  \frac{1}{r^{2}_{\N}}\sum_{i'}\left(\sum_{i, j, s, t} \frac{
      \check{w}_{js}\check{w}_{jt}}{\check{n}_{s}}
    \frac{w_{is}w_{i't}}{\check{n}_{t}}\right)^{2}+
  \frac{1}{r^{2}_{\N}}\sum_{i'}\left(\sum_{i, j, s, t} \frac{
      \check{w}_{js}\check{w}_{jt}}{\check{n}_{s}}
    \frac{w_{is}w_{i't}}{\check{n}_{t}}\right) \left(\sum_{i_{2}, j_{2}, s_{2},
      t_{2}} \frac{
      \check{w}_{j_{2}s_{2}}\check{w}_{j_{2}t_{2}}}{\check{n}_{s_{2}}}
    \frac{w_{i's_{2}}w_{i_{2}t_{2}}}{\check{n}_{t_{2}}}\right)\\
  + \frac{1}{r^{2}_{\N}}\sum_{i'}\left(\sum_{i, j, s, t} \frac{
      \check{w}_{js}\check{w}_{jt}}{\check{n}_{s}}
    \frac{w_{is}w_{i't}}{\check{n}_{t}}\right) \left(\sum_{i_{2}, j_{2}, s_{2},
      t_{2}} \frac{\check{w}_{i's_{2}}\check{w}_{i't_{2}}}{\check{n}_{s_{2}}}
    \frac{w_{is_{2}}w_{j_{2}t_{2}}}{\check{n}_{t_{2}}}\right)
  \preceq  \frac{\N}{r^{2}_{\N}}.
\end{multline*}
Similarly, the variance of the second, third, fourth and fifth term can also be
shown to be bounded by a constant times $Nr_{\N}^{2}$. Therefore by Markov
inequality, in view of \cref{eq:cond-var-2},
\begin{multline*}
  r_{\N}\sum_{j=1}^{\N} E[\mathcal{Y}_{j}^{2} \mid\mathcal{G}_{j-1}]-\mathcal{W}_{\N}=
  r_{\N}\sum_{j, s,
    t}\sum_{i=1}^{j-1}\frac{w_{jt}w_{js}E_{\mathcal{G}_{0}}([\eta_{\Delta, j}^{2}]-\eta_{\Delta, j}^{2})}{\check{n}_{s, -j}}
  \frac{\check{w}_{is}\check{w}_{it}\psi_{is}\psi_{it}}{\check{n}_{t, -j}}\\
  +2r_{\N}\sum_{j, s, t}\sum_{i=1}^{j-1}\frac{\check{w}_{js}w_{jt}(E_{\mathcal{G}_{0}}[\psi_{js}\eta_{\Delta, j}]-\psi_{js}\eta_{\Delta, j})
   }{ \check{n}_{s, -i}}
  \frac{w_{is}\check{w}_{it}\eta_{\Delta, i}\psi_{it}}{\check{n}_{t, -j}}\\
  +r_{\N}\sum_{j, s, t} \sum_{i=1}^{j-1}\frac{\check{w}_{js}\check{w}_{jt}(E_{\mathcal{G}_{0}}[\psi_{js}\psi_{jt}]-\psi_{js}\psi_{jt})
    }{ \check{n}_{s, -i}}
  \frac{w_{is}w_{it} \eta^{2}_{\Delta, i}}{\check{n}_{t, -i}}\\
  + r_{\N}\sum_{j=1}^{\N}\sum_{i, i', s,t}
  \1{j\neq i, i'}\frac{w_{is}C_{\Delta, i}w_{i't}C_{\Delta, i'}}{\check{n}_{s, -i}}
\frac{\check{w}_{jt}\check{w}_{js}(E_{\mathcal{G}_{0}}[\psi_{js}\psi_{jt}]-\psi_{js}\psi_{jt})}{\check{n}_{t, -i'}}
+o_{p}(1).
\end{multline*}
All terms in this expression have mean zero conditional on $W$, and the variance
of each term can be shown to be bounded by a constant times $r_{\N}N$, so that
$ r_{\N}\sum_{j=1}^{\N} E[\mathcal{Y}_{j}^{2}
\mid\mathcal{G}_{j-1}]-\mathcal{W}_{\N}=o_{p}(1)$ as required.

\section{Stylized economic model: baseline microfoundation}\label{app:general_model}

\Cref{sec:environment,sec:equilibrium} provide a microfoundation for the
stylized economic model presented in \begin{NoHyper}
  \Cref{sec:environment_main}\end{NoHyper}. In
\Cref{appsec:sector_region_employment}, we use this microfoundation to derive
expressions analogous to those
in \begin{NoHyper}\cref{eq:emp_change,eq:wage_change} in
  \Cref{sec:impacteconshocks}\end{NoHyper}. In
\Cref{appsec:sector_region_employment_small}, we exploit again our
microfoundation and outline a set of restrictions on the model fundamentals such
our main identification restriction,
 \begin{NoHyper}\Cref{item:random-assignment} in \Cref{sec:no-covariates}\end{NoHyper}, holds.

\subsection{Environment}\label{sec:environment}

We consider a model with multiple sectors $s=1, \dotsc, S$ and multiple regions
$i, j=1,\dotsc, \JN$. Regions are partitioned into countries indexed by
$c=1,\dotsc, C$, and we denote the set of regions located in a country $c$ by
$\JN_{c}$. Region $i$ has a population of $M_{i}$ individuals who cannot move
across regions. Each individual belongs to a different group, $g=1,\dotsc, G$.
The share of group $g$ in the population of region $i$ is $n_{ig}$.

\paragraph{Production.}
Each sector $s$ in region $i$ has a representative firm that produces a
differentiated good using only local labor. For simplicity, we assume that workers of different groups are perfect substitutes in production.
The quantity $Q_{is}$ produced by sector $s$ in region $i$ is produced using labor with productivity $A_{is}$; i.e.
\begin{equation}\label{eq:prod_function_main}
Q_{is} = A_{is}L_{is},
\end{equation}
where $L_{is}$ denotes the number of workers (irrespective of their group) employed by the representative firm
in this sector-region pair. Regions thus differ in terms of their
sector-specific productivity $A_{is}$.

\paragraph{Preferences for consumption goods.} Every individual has identical nested preferences over the sector- and region-specific differentiated goods. Specifically, we assume that individuals have Cobb-Douglas preferences over sectoral composite goods,
\begin{equation}\label{eq:upper_pref_main}
C_j = \prod_{s=1}^{S} \left(C_{js}\right)^{\gamma_s},
\end{equation}
where $C_{j}$ is the utility level of a worker located in region $j$ that obtains utility $C_{js}$ from consuming goods in sector $s$, and $C_{js}$ is a CES aggregator of the sector $s$ goods produced in different regions:
\begin{equation}\label{eq:sector_pref_main}
C_{js} = \left[\sum_{i=1}^{\JN}\left(c_{ijs}\right)^{\frac{\sigma_s-1}{\sigma_s}} \right]^{\frac{\sigma_s}{\sigma_s-1}}, \qquad\sigma_s\in(1,\infty),
\end{equation}
where $c_{ijs}$ denotes the consumption in region $j$ of the sector $s$ good
produced in region $i$. This preference structure has been previously
used in \citet{armington1969theory}, \citet{anderson1979theoretical} and
multiple papers since
\citep*[e.g.][]{andersonvanwincoop2003gravity,arkolakis2012new}.

\paragraph{Preferences for sectors and non-employment.} Individuals of every
group $g$ have the choice of being employed in one of the sectors $s=1, \dotsc, S$
of the economy or opting for non-employment, which we index as $s=0$.
Conditional on being employed, all workers of group $g$ have identical
homogeneous preferences over their sector of employment, but workers differ in
their preferences for non-employment. Specifically, conditional on obtaining
utility $C_{j}$ from the consumption of goods, the utility of a worker $\iota$
of group $g$ living in region $j$ is
\begin{equation}\label{eq:ucjiota}
  U(\iota \mid C_{j}) = \begin{cases}
    u(\iota)C_j & \text{if employed in any sector $s=1,\dotsc, S$,} \\
    C_j & \text{if not employed ($s=0$).}
\end{cases}
\end{equation}
We assume that each individual $\iota$ belonging to group $g$ and living in a
region located in country $c$ independently draws $u(\iota)$ from a Pareto
distribution with scale parameter $\nu_{cg}$ and shape parameter $\phi$, so that
the cumulative distribution function of $u(\iota)$ is given by
\begin{equation}\label{eq:dist_u}
  F_{ig}^u(u) = 1- \left(\frac{u}{\upsilon_{cg}}\right)^{-\phi}, \qquad
  u\geq\upsilon_{cg}, \qquad \phi > 1.
\end{equation}
If a worker living in region $j$ chooses to be employed, she will earn wage
$\omega_{j}$. In equilibrium, wages are equalized across sectors and groups because (i) firms are indifferent between workers of different groups, (ii) workers are indifferent about the sector of employment, and (iii) workers are freely mobile across sectors. If a worker chooses to not be employed, she receives a benefit $b_{j}$. We denote the total number of employed workers of group $g$ in region $j$ by $L_{jg}$, the total employment in region $j$ as $L_j = \sum_{g=1}^G L_{jg}$, and the employment rate in $j$ as $E_{j}\equiv L_{j}/M_{j}$.
\footnote{We assume that benefits are paid by a national
  government that imposes a flat tax $\chi_{c}$ on all income earned in country $c$. The budget constraint of the government is thus $\sum_{j\in \JN_{c}}\{\chi_{c}(\omega_j E_j + b_j(1-E_j))M_{j}\} = \sum_{j\in \JN_{c}}\{b_j(1-E_j) M_{j}\}$. Alternatively, we could think of the option
  $s=0$ as home production and assume that workers that opt for home production
  in region $j$ obtain $b_{j}$ units of the final good, which they consume. This
  alternative model is isomorphic to that in the main text.}

\paragraph{Market structure.} Goods and labor markets are perfectly competitive.

\paragraph{Trade costs.} We assume that there are no trade
costs, which implies that the equilibrium price of the good produced in a region
is the same in every other region; i.e.
$p_{ijs} = p_{is}$ for $j=1,\dotsc, \JN$. Thus, for every sector $s$
there is a composite sectoral good that has identical price $P_{s}$ in all
regions; i.e.
\begin{equation}\label{eq:Ps}
(P_{s})^{1-\sigma_{s}}=\sum_{s=1}^{S}(p_{is})^{1-\sigma_{s}},
\end{equation}
and the final good's price is $P=\prod_{s=1}^{S}(P_{s})^{\gamma_{s}}$.

\subsection{Equilibrium}\label{sec:equilibrium}

We now characterize the equilibrium wage $\omega_{j}$ and total employment
$L_{j}$ of all regions $j=1,\dots, \JN$.

\paragraph{Consumption.} We first solve the expenditure minimization problem of
an individual residing in region $j$. Given the sector-level utility in
\cref{eq:sector_pref_main} and the condition that $p_{ijs} = p_{is}$ for
$ j=1,\dots, \JN$, all regions $j$ have identical spending shares $x_{is}$ on goods
from region $i$, given by
\begin{equation}\label{eq:tradable_demand_main}
  x_{is} = \left(\frac{p_{is}}{P_s}\right)^{1-\sigma_s}.
\end{equation}

\paragraph{Labor supply.} Every worker maximizes the utility function in
\cref{eq:ucjiota} in order to decide whether to be employed. Consequently,
conditional on the wage $\omega_{i}$ and the non-employment benefit $b_i$, the
total employment of individuals of group $g$ in region $i$ is
$L_{ig} = n_{ig} M_i \Pr \left[u_i(\iota)\omega_i > b_i \right]$. It therefore follows from
\cref{eq:dist_u} that $L_{i} = \sum_{g=1}^G L_{jg}$ is
\begin{align}\label{eq:extensive_margin_main}
L_{i} = \omega_{i}^{\phi} v_i
\end{align}
such that
\begin{align}\label{eq:v_i_decomposition}
v_i = \nu_i \sum_{g=1}^G n_{ig} \nu_{cg}
\end{align}
with $\nu_i \equiv M_i b_i^{-\phi}$, and $\nu_{cg} \equiv \upsilon_{cg}^\phi$.

\paragraph{Producer's problem.} In perfect competition, firms must earn zero profits and, therefore,
\begin{equation}\label{eq:producers_problem}
p_{is} =\frac{\omega_i}{A_{is}}.
\end{equation}

\paragraph{Goods market clearing.} Given that labor is the only factor of production and firms earn no profits, the income of all individuals living in region $i$ is $W_i \equiv \sum_s \omega_i L_{is}$, and world income is $W \equiv \sum_i W_i$. We normalize world income to one, $W=1$. Given preferences in \cref{eq:upper_pref_main}, all individuals spend a share $\gamma_s$ of their income on sector $s$, so that world demand for the differentiated good $s$ produced in region $i$ is $x_{is} \gamma_s$. Goods market clearing requires world demand for good $s$ produced in region $i$ to equal total revenue of the representative firm operating in sector $s$ in region $i$, $\omega_i L_{is}$. Thus, using the expression in \cref{eq:tradable_demand_main}, we obtain
\begin{equation}\label{eq:labor_demand_tradable_main_sectoral}
L_{is} = (\omega_i)^{-\sigma_s} \left(A_{is} P_s\right)^{\sigma_s-1} \gamma_s.
\end{equation}
Note that this labor demand equation is analogous to that in \begin{NoHyper}\cref{labor_demand_maintext} of \Cref{sec:model}\end{NoHyper}, with the region- and sector-specific demand shifter $D_{is}$ defined as $D_{is}=\left(A_{is} P_s\right)^{\sigma_s-1} \gamma_s$.

If, without loss of generality, we split the region- and sector-specific productivity $A_{is}$ into a country and sector-specific component $A_{cs}$ and a residual $\tilde{A}_{is}$,
\begin{align}\label{eq: A_is_decomposition}
A_{is}=A_{cs}\tilde{A}_{is}.
\end{align}

\paragraph{Labor market clearing.} Given the sector- and region-specific labor
demand in \cref{eq:labor_demand_tradable_main_sectoral}, total labor demand in
region $i$ is
\begin{equation}\label{eq:labor_demand_tradable_main_total}
L_{i} = \sum_{s=1}^{S}(\omega_i)^{-\sigma_s}\left(A_{is} P_s\right)^{\sigma_s-1} \gamma_s.
\end{equation}
Labor market clearing requires labor supply in \cref{eq:extensive_margin_main} to equal labor demand in \cref{eq:labor_demand_tradable_main_total}:
\begin{equation}\label{eq:market_clearing_main}
v_i (\omega_i)^{\phi} = \sum_{s=1}^{S} (\omega_i)^{-\sigma_s} \left(A_{is} P_s\right)^{\sigma_s-1} \gamma_s.
\end{equation}

\paragraph{Equilibrium.} Given technology parameters
$\{A_{cs}\}_{c=1,s=1}^{C, S}$ and $\{\tilde{A}_{is}\}_{i=1,s=1}^{\JN, S}$,
preference parameters $\{(\sigma_{s}, \allowbreak \gamma_{s})\}_{s=1}^{S}$, labor supply
parameters $\phi$, $\{\nu_i\}_{i=1}^{\JN}$, $\{n_{ig}\}_{i=1,g=1}^{\JN, G}$ and
$\{\nu_{cg}\}_{c=1,g=1}^{C, G}$, and normalizing world income to equal 1, $W=1$,
we can use \cref{eq:Ps,eq:producers_problem,eq:v_i_decomposition,eq:
  A_is_decomposition,eq:market_clearing_main} to solve for the equilibrium wage
in every world region, $\{\omega_{i}\}_{i=1}^{\JN}$, the equilibrium price of
every sector-region specific good $\{p_{is}\}_{i=1,s=1}^{\JN, S}$, and the sectoral
price indices $\{P_{s}\}_{s=1}^{S}$. Given these equilibrium wages and sectoral
price indices, we can use \cref{eq:labor_demand_tradable_main_total} to solve
for the equilibrium level of employment in every region, $\{L_{i}\}_{i=1}^{\JN}$.

\subsection{Labor market impact of sectoral shocks: equilibrium relationships}\label{appsec:sector_region_employment}

We assume that, in every period, the model described in
\Cref{sec:environment,sec:equilibrium} characterizes the labor market
equilibrium in every region $i=1,\dots, \JN$. Across periods, we assume that the
parameters $\{\sigma_{s}\}_{s=1}^{S}$, $\{n_{ig}\}_{i=1,g=1}^{\JN, G}$, and
$\phi$ are fixed, and that all changes in the labor market outcomes
$\{\omega_{i}, L_{i}\}_{i=1}^{\JN}$ are generated by changes in technology
$\{A_{cs}\}_{c=1,s=1}^{C, S}$ and $\{\tilde{A}_{is}\}_{i=1,s=1}^{\JN, S}$, sectoral
preferences $\{\gamma_{s}\}_{s=1}^{S}$, and labor supply parameters
$\{\nu_i\}_{i=1}^{\JN}$ and $\{\nu_{cg}\}_{c=1,g=1}^{C, G}$.

We focus here on understanding how changes in these exogenous parameters affect the labor market equilibrium in all regions located in a given country $c$; i.e.\ all regions belonging to the set $\JN_{c}$.

In our model, the sectoral prices mediate the impact of all foreign technology
and labor supply shocks on the labor market equilibrium of every region in
country $c$; i.e.\ the changes in $\{(\omega_{i}, L_{i})\}_{i\in \JN_{c}}$ depend
on the changes in $\{\tilde{A}_{is}\}_{s=1,i\notin \JN_{c}}^{S}$,
$\{\nu_i\}_{i\notin \JN_{c}}$, and $\{\nu_{c'g}\}_{g=1,c'\neq c}^{G}$ only through
changes in $\{P_{s}\}_{s=1}^{S}$. Therefore, we can write the changes in wages
and employment in every region $i$ of the population of interest $\JN_{c}$ as a
function of the changes in the sectoral prices, and the changes in the
productivity and labor supply shocks in region $i$.

\paragraph{Isomorphism.} As in \begin{NoHyper}\Cref{sec:impacteconshocks}\end{NoHyper}, we use $\hat{z} = \log(z^{t}/z^{0})$ to denote log-changes in any given variable $z$ between
some initial period $t=0$ and any other period $t$. Up to a first-order approximation around the initial equilibrium, \cref{eq:labor_demand_tradable_main_total,eq:market_clearing_main} imply that
\begin{equation}\label{eq:emp_change_app0}
\hat{L}_i = \sum_{s=1}^{S} l_{is}^0 \left[\theta_{is} \hat{P}_s + \lambda_i((\sigma_{s}-1)\hat{A}_{cs}+\hat{\gamma}_{s}) + \lambda_i((\sigma_{s}-1)\hat{\tilde{A}}_{is})\right] + (1 - \lambda_i)(\sum_{g=1}^G \tilde{w}_{ig} \hat{\nu}_{cg}+\hat{\nu}_i),
\end{equation}
with $l_{is}^0\equiv L^{0}_{is}/L^{0}_{i}$,
$\tilde{w}_{ig} \equiv L_{ig}^0/L_i^0$, $\theta_{is}=(\sigma_{s}-1)\lambda_{i}$
and $\lambda_i \equiv \phi \left[\phi + \sum_s l_{is}^0 \sigma_s\right]^{-1}$.
Combining
\cref{eq:extensive_margin_main,eq:v_i_decomposition,eq:emp_change_app0}, we can
similarly obtain
\begin{equation}\label{eq:wage_change_app0}
\hat{\omega}_i = \sum_{s=1}^{S} l_{is}^0 \left[\theta_{is} \hat{P}_s + \lambda_i((\sigma_{s}-1)\hat{A}_{cs}+\hat{\gamma}_{s}) + \lambda_i((\sigma_{s}-1)\hat{\tilde{A}}_{is})\right] - \phi^{-1}\lambda_i(\sum_{g=1}^G \tilde{w}_{ig} \hat{\nu}_{cg}+\hat{\nu}_i).
\end{equation}
Given our emphasis on understanding the changes in labor market outcomes for regions located in the same country, all regions in the population of interest will share the same value of $A_{cs}$ for every sector $s$, and the same value of $\hat{\nu}_{cg}$ for every labor group $g$; thus, we can simplify the notation by writing $\hat{A}_{cs}=\hat{A}_{s}$ and $\hat{\nu}_{cg}=\hat{\nu}_{g}$ for all $s$ and $g$, respectively. Given this notational simplification and the following equivalences
\begin{align}
\chi_{s}&=P_{s},\label{eq:equival1}\\
\mu_{s}&=(A_{s})^{\sigma_s-1} \gamma_s,\label{eq:equival3}\\
\eta_{is}&=(\tilde{A}_{is})^{\sigma_s-1}\label{eq:equival4},
\end{align}
we can easily see that the expressions in
\cref{eq:emp_change_app0,eq:wage_change_app0} are identical to those
in \begin{NoHyper}\cref{eq:emp_change,eq:wage_change} in
  \Cref{sec:impacteconshocks}\end{NoHyper}, respectively. Consequently, the
environment described in \Cref{sec:environment,sec:equilibrium} does indeed
provide a microfoundation for the equilibrium relationships in
\begin{NoHyper}\cref{eq:emp_change,eq:wage_change}\end{NoHyper}.

\subsection{Identification of labor market impact of sectoral prices}\label{appsec:sector_region_employment_small}

As the mapping in \cref{eq:equival1} illustrates, we may think of the changes in
sectoral prices $\{\hat{P}_s\}_{s=1}^{S}$ as our sectoral shocks of interest.
Given data on changes in a labor market outcome (e.g.\ changes in the
employment rate $\hat{L}_{i}$) for all units of a population of interest formed
by all regions of a particular country $c$, and data on the changes in sectoral
prices
$\{\hat{P}_s\}_{s=1}^{S}$, \begin{NoHyper}\Cref{item:random-assignment}
  in \Cref{sec:no-covariates}\end{NoHyper} indicates that identifying the
coefficient in front of a shift-share term that aggregates these sectoral price
changes requires that these are as good as randomly allocated.

In the context of the equilibrium relationship in \cref{eq:emp_change_app0}, the sectoral price changes
$\{\hat{P}_s\}_{s=1}^{S}$ will
satisfy \begin{NoHyper}\Cref{item:random-assignment}\end{NoHyper} if they are mean independent of: country $c$-specific sectoral productivity changes
$\{\hat{A}_{cs}\}_{s=1}^{S}$; country $c$-specific labor-group supply
shocks $\{\hat{\nu}_{cg}\}_{g=1}^{G}$; region and sector-specific productivity
shocks, for all sectors and all regions in country $c$,
$\{\hat{\tilde{A}}_{is}\}_{s=1,i\in \JN_{c}}^{S}$; region-specific labor supply
shocks, for all regions in country $c$, $\{\hat{\nu}_i\}_{i\in \JN_{c}}$. This
mean independence restriction will hold if the following two conditions are
satisfied.

First, country $c$ is ``small''; i.e.\ all labor demand and labor supply shocks in country $c$ have no impact on the changes in sectoral prices $\{\hat{P}_s\}_{s=1}^{S}$.

Second, labor demand and labor supply shocks affecting any region $i$ in the country
or population of interest $c$ are mean independent of any labor demand and labor
supply shock affecting any other region of the world economy that is ``large''
(i.e.\ any other region whose labor demand and supply shocks have an impact on
the changes in sectoral prices).

In summary, if the vector of shifters of interest $\{\Xs_{s}\}_{s=1}^{S}$
corresponds to the sectoral price changes $\{\hat{P}_s\}_{s=1}^{S}$, the
researcher is interested on the impact of these shifters on a collection of
``small'' regions, and labor market shocks in these ``small'' regions are
independent of the corresponding shocks in any ``large'' region, then the
identification condition
in \begin{NoHyper}\Cref{item:random-assignment}\end{NoHyper} is satisfied.

\subsubsection{Impact of labor demand and supply shocks on sector-specific price indices}\label{appsec:princeindex}

In general equilibrium, the price change in every sector $s$, $\hat{P}_s$, depends on the shocks $A_{cs}$, $\hat{\tilde{A}}_{is}$, $\hat{\gamma}_s$, $\hat{\nu}_i$, and $\nu_{cg}$ of all sectors, labor groups, and regions in the world economy. Specifically, the change in the sector-specific price index is
\begin{equation}\label{eq:price_change_large}
\hat{P}_s = - \sum_{s'} \alpha_{ss'} \sum_{j=1}^{\JN} x_{js'}^0(\hat{A}_{js'} +\tilde{\lambda}_j \hat{v}_j - \tilde{\lambda}_j \sum_k l_{jk}^0 [\hat{\gamma}_k + (\sigma_k-1) \hat{A}_{jk}]),
\end{equation}
where $\tilde{\lambda}_j \equiv \left[\phi + \sum_s l_{is}^0 \sigma_s\right]^{-1}$, $\{\alpha_{ss'}\}_{s=1,s'=1}^{S, S}$ are positive constants, and $x^{0}_{js}$ is the share of the world production in sector $s$ that corresponds to region $j$ in the initial equilibrium; i.e.\ $x_{js}^0\equiv X^{0}_{js}/\sum_{i=1}^{\JN}X^{0}_{is}$. Imposing that all regions in a country $c$ verify that $x_{js}^0\approx 0$ for all $j\in \JN_{c}$ and for $s=1,\dots, S$, we can rewrite the change in the sector-specific price index as
\begin{equation}\label{eq:price_change_small}
\hat{P}_s = - \sum_{s'} \alpha_{ss'} \sum_{j\notin \JN_{c}} x_{js'}^0(\hat{A}_{js'} +\tilde{\lambda}_j \hat{v}_j - \tilde{\lambda}_j \sum_k l_{jk}^0 [\hat{\gamma}_k + (\sigma_k-1) \hat{A}_{jk}]).
\end{equation}
In this case, $\hat{P}_s$ does not depend on the labor supply shocks and
technology shocks in any region $j$ included in country $c$; i.e.\ $\hat{P}_s$
depends neither on $\{\hat{A}_{cs}\}_{s=1}^{S}$, nor $\{\hat{\nu}_{cg}\}_{g=1}^{G}$, nor $\{\hat{\tilde{A}}_{is}\}_{s=1,i\in \JN_{c}}^{S}$, nor $\{\hat{\nu}_i\}_{i\in \JN_{c}}$.

\paragraph{Proof of \cref{eq:price_change_large}.}
\Cref{eq:tradable_demand_main,eq:market_clearing_main} imply that
\begin{displaymath}
\hat{P}_s - \sum_k \tilde{\alpha}_{sk} \hat{P}_k = \sum_j x_{js}^0 (\tilde{\lambda}_j \sum_k l_{jk}^0 [\hat{\gamma}_k + (\sigma_k-1) \hat{A}_{jk}] - \tilde{\lambda}_j \hat{v}_j - \hat{A}_{js}),
\end{displaymath}
where $\tilde{\alpha}_{sk} \equiv \sum_j x_{js}^0 l_{jk}^0 \tilde{\lambda}_j (\sigma_k-1)$. Let us use bold variables to denote vectors, $\boldsymbol{y} \equiv [y_s]_s$, and bar bold variables to denote matrices, $\bar{\boldsymbol{a}} \equiv [a_{sk}]_{s, k}$. Thus, we can rewrite the equation above in matrix form as
\begin{displaymath}
\left(I - \boldsymbol{\bar{\alpha}}\right) \hat{\boldsymbol{P}} = \hat{\boldsymbol{\eta}},
\end{displaymath}
with $\hat{\eta}_s \equiv \sum_j x_{js}^0 \left(\tilde{\lambda}_j \sum_k l_{jk}^0 \left[\hat{\gamma}_k + (\sigma_k-1) \hat{A}_{jk} \right] - \tilde{\lambda}_j \hat{v}_j - \hat{A}_{js} \right)$. In order to obtain \cref{eq:price_change_large}, it is sufficient to show that $\left(I - \boldsymbol{\bar{\alpha}}\right)$ is a nonsingular m-matrix and, therefore, it has a positive inverse matrix. To establish this result, notice first that $\tilde{\alpha}_{sk} \in (0,1)$ for every $s$ and $k$; to show this, it is sufficient to show that, for every $j$, $k$, and $s$, it holds that $0<x_{js}^0<1$ and
\begin{displaymath}
0<l_{jk}^0 \tilde{\lambda}_j (\sigma_k-1) = \frac{l_{jk}^0 (\sigma_k -1)}{\phi + \sum_k l_{jk}^0 \sigma_k} < \frac{l_{jk}^0 \sigma_k}{\phi + \sum_k l_{jk}^0 \sigma_k} < 1,
\end{displaymath}
where the last two inequalities arise from $\sigma_k > 1$ and $\phi > 0$.

Finally, to show that $\left(I - \boldsymbol{\bar{\alpha}}\right)$ is nonsingular, it is sufficient to establish that it is diagonal dominant:
\begin{align*}
|1 - \tilde{\alpha}_{sk}| - \sum_{k \neq s} |\tilde{\alpha}_{sk}| &= 1 - \sum_j x_{js}^0 \frac{l_{js}^0 (\sigma_s -1)}{\phi + \sum_k l_{jk}^0 \sigma_k} - \sum_{k\neq s} \sum_j x_{js}^0 \frac{l_{jk}^0 (\sigma_k -1)}{\phi + \sum_k l_{jk}^0 \sigma_k}, \\
& = \sum_j x_{js}^0 \left(1 - \frac{\sum_{k} l_{jk}^0 (\sigma_k -1)}{\phi + \sum_k l_{jk}^0 \sigma_k} \right) \\
& = \sum_j x_{js}^0 \left(\frac{\phi + 1}{\phi + \sum_k l_{jk}^0 \sigma_k} \right) > 0. \ \blacksquare
\end{align*}

\section{Stylized economic model: Extensions}\label{sec:econ-model}

In \Cref{appsec:specific,appsec:specific_pref}, we provide alternative microfoundations for the equilibrium relationship in \begin{NoHyper}\cref{eq:emp_change}\end{NoHyper}. Finally, in \Cref{appsec:migration}, we incorporate migration into the baseline microfoundation described in \Cref{app:general_model}.

\subsection{Sector-specific factors of production}\label{appsec:specific}

We extend here the model described in \Cref{app:general_model} to incorporate other factors of production. In particular, we introduce a sector-specific factor, as in \citet*{jones1971specific} and, more recently, \citet{kovak2013regional}.

\subsubsection{Environment}\label{sec:environment_capital}

The only difference with respect to the setting described in \Cref{sec:environment} is that the production function in \cref{eq:prod_function_main} is substituted for a Cobb-Douglas production function that combines labor and capital inputs:
\begin{displaymath}
Q_{is} = A_{is} \left(L_{is}\right)^{1-\theta_{s}} \left(K_{is}\right)^{\theta_{s}}.
\end{displaymath}
We assume that capital is a sector-specific factor of production (sector-$s$ capital has no use in any other sector) and that, for every sector, each region has an endowment of sector-specific capital $\bar{K}_{is}$.

\subsubsection{Equilibrium}\label{sec:equilibrium_capital}

\paragraph{Consumption.} The consumer's problem is identical to that in \Cref{sec:equilibrium}.

\paragraph{Labor supply.} The labor supply decision is identical to that in \Cref{sec:equilibrium}.

\paragraph{Producer's problem.} Conditional on the region-$i$ equilibrium wage $\omega_{i}$ and rental rate of sector-$s$ capital $R_{is}$, the cost minimization problem of the sector-$s$ region-$i$ representative firm and the market clearing condition for sector-$s$ region-$i$ specific capital imply that
\begin{displaymath}
\frac{1-\alpha_{s}}{\alpha_{s}} \frac{\bar{K}_{is}}{L_{is}} = \frac{\omega_{i}}{R_{is}}.
\end{displaymath}
Conditional on the sector-$s$ region-$i$ final good price $p_{is}$, the firm's zero profit condition implies that
\begin{displaymath}
p_{is} A_{is} \tilde{\alpha}_{s} = \left(\omega_{i}\right)^{1-\theta_{is}}\left(R_{is}\right)^{\theta_{is}},
\end{displaymath}
where $\tilde{\alpha}_{s} \equiv \left(\alpha_{s}\right)^{\alpha_{s}}\left(1-\alpha_{s}\right)^{1-\alpha_{s}}$. The combination of these two conditions yields the demand for labor in sector $s$ and region $i$,
\begin{align}\label{eq:labordemand_specific}
L_{is} = \frac{1-\alpha_{s}}{\alpha_{s}} \bar{K}_{is} \left(\frac{p_{is} A_{is} \tilde{\alpha}_{s}}{\omega_{i}} \right)^{\frac{1}{\alpha_{s}}},
\end{align}
and the total sales of the sector-$s$ region-$i$ good as a function of the output price $p_{is}$,
\begin{align}\label{eq:goodsuppy_specific}
X_{is} = \frac{1}{1-\alpha_{s}} \omega_{i} L_{is} = \frac{\bar{K}_{is}}{\alpha_{s}} \left(p_{is} A_{is} \tilde{\alpha}_{s} \right)^{\frac{1}{\alpha_{s}}} \left(\omega_{i}\right)^{1 - \frac{1}{\alpha_{s}}}.
\end{align}

\paragraph{Goods market clearing.} Applying the same normalization as in
\Cref{sec:environment}, $W=1$, the total expenditure in the sector-$s$
region-$i$ good is equal to $x_{is}\gamma_{s}$, with $x_{is}$ defined in
\cref{eq:tradable_demand_main} as a function of the equilibrium prices $p_{is}$.
Equating $x_{is}\gamma_{s}$ and \cref{eq:goodsuppy_specific}, we can solve for
the equilibrium value of $p_{is}$ as a function of the sector-$s$ price index
$P_{s}$:
\begin{equation}\label{eq:producers_problem_specific}
  p_{is} = \left[\frac{\bar{K}_{is}}{\alpha_{s}} \left(A_{is} \tilde{\alpha}_{s} \right)^{\frac{1}{\alpha_{s}}} \left(\omega_{i}\right)^{1 - \frac{1}{\alpha_{s}}} \frac{(P_s)^{1-\sigma_s}}{\gamma_s} \right]^{-\theta_{is} \eta_{is}},
\end{equation}
where
$\delta_{s} \equiv \left(1+ \alpha_{s}(\sigma_s-1) \right)^{-1} \in (0,1)$.
Additionally, combining
\cref{eq:labordemand_specific,eq:producers_problem_specific}, we obtain an
expression for labor demand in sector-$s$ region-$i$ as a function of the
equilibrium wage $\omega_{i}$, the sector-$s$ price $P_{s}$ and other exogenous
determinants:
\begin{equation}\label{eq:labor_demand_capital}
L_{is} = \kappa_{is} \gamma_s^{\delta_{s}} \left(A_{is}P_{s} \right)^{(\sigma_s - 1)\delta_{s}} \left(\omega_{i} \right)^{-\sigma_s \delta_{s}},
\end{equation}
where $\kappa_{is} \equiv (1-\alpha_{s}) (\bar{K}_{is} \tilde{\alpha}_{s}^{\frac{1}{\alpha_{s}}} / \alpha_{s})^{1- \delta_{s}}$. Note that this labor demand equation is analogous to that in \begin{NoHyper}\cref{labor_demand_maintext}\end{NoHyper}, with the region- and sector-specific demand shifter $D_{is}$ defined as
\begin{align*}
D_{is}=\kappa_{is}(\gamma_s)^{\delta_{s}} \left(A_{is}P_{s} \right)^{(\sigma_s - 1)\delta_{s}},
\end{align*}
and with the labor demand elasticity now defined as $\sigma_s \delta_{s}$. Note that the labor demand elasticity in \begin{NoHyper}\cref{labor_demand_maintext}\end{NoHyper} is identical to that in \cref{eq:labor_demand_capital} in the specific case in which $\delta_{s}=1$, which will hold when $\alpha_{s}=0$. Without loss of generality, we split the region- and sector-specific productivity $A_{is}$ according to \cref{eq: A_is_decomposition}.

\paragraph{Labor market clearing.} Given the sector- and region-specific labor
demand in \cref{eq:labor_demand_capital}, total labor demand in region $i$ is
\begin{equation}\label{eq:labor_demand_tradable_capital_total}
L_{i} = \sum_{s=1}^{S}\kappa_{is} \gamma_s^{\delta_{s}} \left(A_{is}P_{s} \right)^{(\sigma_s - 1)\delta_{s}} \left(\omega_{i} \right)^{-\sigma_s \delta_{s}}.
\end{equation}
Labor market clearing requires labor supply in \begin{NoHyper}\cref{eq:extensive_margin_main}\end{NoHyper} to equal labor demand in \cref{eq:labor_demand_tradable_capital_total}:
\begin{align}\label{eq:market_clearing_capital}
v_i (\omega_i)^{\phi} = \sum_{s=1}^{S} \kappa_{is} \gamma_s^{\delta_{is}} \left(A_{is}P_{s} \right)^{(\sigma_s - 1)\delta_{is}}\left(\omega_{i} \right)^{-\sigma_s \delta_{is}}, \qquad j=1,\dots, \JN.
\end{align}

\paragraph{Equilibrium.} Given the technology parameters
$\{\alpha_{s}\}_{s=1}^{S}$, $\{A_{cs}\}_{c=1,s=1}^{C, S}$ and
$\{\tilde{A}_{is}\}_{i=1,s=1}^{\JN, S}$, sector- and region-specific capital inputs
$\{\bar{K}_{is}\}_{i=1,s=1}^{\JN, S}$, preference parameters
$\{(\sigma_{s}, \allowbreak \gamma_{s})\}_{s=1}^{S}$, labor supply parameters
$\phi$, $\{\nu_i\}_{i=1}^{\JN}$, $\{n_{ig}\}_{i=1,g=1}^{\JN, G}$ and
$\{\nu_{cg}\}_{c=1,g=1}^{C, G}$, and normalizing world income to equal 1, $W=1$,
we can use \cref{eq:Ps,eq:v_i_decomposition,eq:
  A_is_decomposition,eq:producers_problem_specific,eq:market_clearing_capital}
to solve for the equilibrium wage in every world region,
$\{\omega_{i}\}_{i=1}^{\JN}$, the equilibrium price of every sector-region
specific good $\{p_{is}\}_{i=1,s=1}^{\JN, S}$, and the sectoral price indices
$\{P_{s}\}_{s=1}^{S}$. Given these equilibrium wages and sectoral price indices,
we can use \cref{eq:labor_demand_tradable_capital_total} to solve for the
equilibrium level of employment in every region, $\{L_{i}\}_{i=1}^{\JN}$.

\subsubsection{Labor market impact of sectoral shocks}

We assume that, in every period, the model described in \Cref{sec:environment_capital,sec:equilibrium_capital} characterizes the labor market equilibrium in every region $i=1,\dots, \JN$. Across periods, we assume that
the parameters $\{(\sigma_{s}, \alpha_{s})\}_{s=1}^{S}$, $\{n_{ig}\}_{i=1,g=1}^{\JN, G}$, and $\phi$ are
fixed, and that all changes in the labor market outcomes $\{\omega_{i}, L_{i}\}_{i=1}^{\JN}$ are generated by changes in technology
$\{A_{cs}\}_{c=1,s=1}^{C, S}$ and $\{\tilde{A}_{is}\}_{i=1,s=1}^{\JN, S}$, sectoral preferences $\{\gamma_{s}\}_{s=1}^{S}$,
and labor supply parameters $\{\nu_i\}_{i=1}^{\JN}$ and $\{\nu_{cg}\}_{c=1,g=1}^{C, G}$. We focus here on understanding how changes in these exogenous parameters affect the labor market equilibrium in all regions located in a given country $c$; i.e.\ all regions belonging to the set $\JN_{c}$.

\paragraph{Isomorphism.} Following steps analogous to those in \Cref{appsec:sector_region_employment}, we can show that \cref{eq:labor_demand_tradable_capital_total,eq:market_clearing_capital} imply that
\begin{align}\label{eq:emp_change_capital}
\hat{L}_i& = \sum_{s=1}^{S} l_{is}^0 \left[\theta_{is} \hat{P}_s + \lambda_i((\sigma_{s}-1)\delta_{s}\hat{A}_{cs}+\delta_{s}\hat{\gamma}_{s}) + \lambda_i((\sigma_{s}-1)\delta_{s}\hat{\tilde{A}}_{is}+\hat{\kappa}_{is})\right]\nonumber\\
&+ \left(1 - \lambda_i\right)(\sum_{g=1}^G \tilde{w}_{ig} \hat{\nu}_{cg}+\hat{\nu}_i),
\end{align}
with $\theta_{is}=(\sigma_{s}-1)\delta_{s}\lambda_{i}$ and $\lambda_i \equiv \phi(\phi + \sum_s l_{is}^0 \sigma_s \delta_{s})^{-1}$. As in \Cref{appsec:sector_region_employment}, given our emphasis on understanding the changes in labor market outcomes for regions located in the same country, all regions in the population of interest will share the same value of $A_{cs}$ for every sector $s$, and the same value of $\hat{\nu}_{cg}$ for every labor group $g$; thus, we can simplify the notation by writing $\hat{A}_{cs}=\hat{A}_{s}$ and $\hat{\nu}_{cg}=\hat{\nu}_{g}$ for all $s$ and $g$, respectively. Given this notational simplification and the following equivalences
\begin{align}
\chi_{s}&=P_{s}, \label{eq:equival_capital1}\\
\mu_{s}&=(A_{s})^{(\sigma_s-1)\delta_{s}} (\gamma_s)^{\delta_{s}}, \label{eq:equival_capital3}\\
\eta_{is}&=\kappa_{is}(\tilde{A}_{is})^{(\sigma_s-1)\delta_{s}}.\label{eq:equival_capital4}
\end{align}
we can easily see that the expression in \cref{eq:emp_change_capital} is identical to that in \begin{NoHyper}\cref{eq:emp_change} in \Cref{sec:impacteconshocks}\end{NoHyper}. Consequently, the environment described in \Cref{sec:environment_capital,sec:equilibrium_capital} does indeed provide a microfoundation for the equilibrium relationship in \begin{NoHyper}\cref{eq:emp_change}\end{NoHyper}.

\subsection{Sector-specific preferences}\label{appsec:specific_pref}

We extend the model described in \Cref{app:general_model} to allow workers to have idiosyncratic preferences for being employed in the different $s=1,\dots, S$ sectors and for being non-employed $s=0$. In order to maintain the analysis simple, we assume here that there is a single worker group $G=1$.

\subsubsection{Environment}\label{sec:environment_roy}

The only difference with respect to the setting
described in \Cref{sec:environment} is that the utility function in
\cref{eq:ucjiota,eq:dist_u} is substituted by an alternative utility function
that features workers idiosyncratic preferences for being employed in the
different $s=1,\dots, S$ sectors and for being non-employed $s=0$. Specifically,
we assume here that, conditional on obtaining utility $C_{i}$ from the
consumption of goods, the utility of a worker $\iota$ living in region $i$ is
\begin{equation}\label{eq:ucjiota_specific_pref}
U_{is} = u_{s}(\iota) C_i,
\end{equation}
and, to simplify the analysis, we assume that $u_{s}(\iota)$ is i.i.d.\ across
individuals $\iota$ and sectors $s$ with a Fréchet cumulative distribution
function; i.e.\ for every region $i=1,\dots, \JN$ and sector $s=0,\dots, S$,
\begin{equation}\label{eq:dist_u_specific_pref}
F_u(u) = e^{- v_{is} u^{-\phi}},\qquad \phi>1.
\end{equation}
This modeling of workers' sorting patterns across sectors is similar to that in
\citet*{galle2017slicing} and \citet*{burstein2018inequality}. See
\citet*{adao2016} for a framework that relaxes the distributional assumption in
\cref{eq:dist_u_specific_pref}. Given that individuals have heterogeneous
preferences for employment in different sectors, workers are no longer
indifferent across sectors and, thus, equilibrium wages
$\{\omega_{is}\}_{s=1}^{S}$ may vary across sectors within a region $i$. As in
the main text, we assume that workers that choose the non-employment sector
$s=0$ in region $i$ receive non-employment benefits $b_{i}$.

\subsubsection{Equilibrium}\label{sec:equilibrium_roy}

\paragraph{Consumption.} The consumer's problem is identical to that in \Cref{sec:equilibrium}.

\paragraph{Labor supply.} Conditional on the equilibrium wages
$\{\omega_{is}\}_{s=1}^{S}$, the labor supply in sector $s=1,\dots, S$ of region
$i$ is
\begin{align}\label{eq:extensive_margin_roy}
L_{is} = M_i\frac{v_{is} (\omega_{is})^\phi}{\Phi_i} \quad \textrm{with} \quad \Phi_i \equiv v_{i0} b_{i}^\phi +\sum_{s = 1}^S v_{is} (\omega_{is})^\phi,
\end{align}
and the labor supply in the non-employment sector $s=0$ is
\begin{align}\label{eq:extensive_margin_roy2}
L_{i0}=M_i\frac{v_{i0} (b_{i})^\phi}{\Phi_i}.
\end{align}

\paragraph{Producer's problem.} In perfect competition, firms must earn zero profits and, therefore,
\begin{equation}\label{eq:producers_problem_roy}
p_{is} =\frac{\omega_{is}}{A_{is}}.
\end{equation}

\paragraph{Goods market clearing.} The conditions determining the equilibrium in the good's market and, consequently, the region- and sector-specific labor demand equations are identical to those in \Cref{sec:equilibrium}.

\paragraph{Labor market clearing.} Combining the region- and sector-specific
labor supply in \cref{eq:extensive_margin_roy} with the region- and
sector-specific labor demand in \cref{eq:labor_demand_tradable_main_sectoral},
and imposing the normalization $W=1$, the labor market clearing condition in
every sector $s=1,\dots, S$ and region $i=1,\dots, \JN$ is
\begin{align}\label{eq:market_clearing_roy}
M_i\frac{v_{is} (\omega_{is})^\phi}{\Phi_i} = (\omega_{is})^{-\sigma_s}\left(A_{is} P_s\right)^{\sigma_s-1} \gamma_s.
\end{align}

\paragraph{Equilibrium.} Given productivity parameters
$\{A_{cs}\}_{c=1,s=1}^{C, S}$ and $\{\tilde{A}_{is}\}_{i=1,s=1}^{\JN, S}$, preference parameters
$\{\sigma_{s}, \gamma_{s}\}_{s=1}^{S}$, labor supply
parameters $\phi$ and $\{v_{is}\}_{i=1,s=0}^{\JN, S}$, and normalizing world income to equal 1,
$W=1$, we can use \cref{eq:Ps,eq: A_is_decomposition,eq:producers_problem_roy,eq:market_clearing_roy} to
solve for the equilibrium wage in every sector and region, $\{\omega_{is}\}_{i=1,s=1}^{\JN, S}$, the equilibrium price of every sector- and region-specific good $\{p_{is}\}_{i=1,s=1}^{\JN, S}$, and the sectoral price indices $\{P_{s}\}_{s=1}^{S}$. Given these equilibrium wages and sectoral price indices,
we can use \cref{eq:extensive_margin_roy,eq:extensive_margin_roy2} to solve for the
equilibrium level of employment in every sector and region, $\{L_{is}\}_{i=1,s=0}^{\JN, S}$.

\subsubsection{Labor market impact of sectoral shocks}

We assume that, in every period, the model described in \Cref{sec:environment_roy,sec:equilibrium_roy} characterizes the labor market equilibrium in every region $i=1,\dots, \JN$. Across periods, we assume that
the parameters $\{\sigma_{s}\}_{s=1}^{S}$, and $\phi$ are
fixed, and that all changes in the labor market outcomes $\{\omega_{i}, L_{i}\}_{i=1}^{\JN}$ are generated by changes in technology
$\{A_{cs}\}_{c=1,s=1}^{C, S}$ and $\{\tilde{A}_{is}\}_{i=1,s=1}^{\JN, S}$, sectoral preferences $\{\gamma_{s}\}_{s=1}^{S}$,
and labor supply parameters $\{v_{is}\}_{i=1,s=1}^{\JN, S}$. We focus here on understanding how changes in these exogenous parameters affect the labor market equilibrium in all regions located in a given country $c$; i.e.\ all regions belonging to the set $\JN_{c}$.

\paragraph{Isomorphism.} Given that the total population of a region, $M_{i}$, is fixed across time periods, it holds that, to a first-order approximation, $l^{0}_{i0}\hat{L}_{i0}+(1-l^{0}_{i0})\hat{L}_{i}=0$, where $\hat{L}_{i}$ denotes the log-change in total population in region $i$. Therefore, the change in total employment in region $i$ may be written as
\begin{align}
\hat{L}_{i}&= -\frac{l_{i0}^0}{1-l_{i0}^0} \hat{L}_{i0}\nonumber\\
&=\frac{l_{i0}^0}{1-l_{i0}^0} (\hat{\Phi}_i - \phi \hat{b}_i - \hat{v}_{i0})\nonumber\\
&=\frac{l_{i0}^0}{1-l_{i0}^0}(\sum_{s = 0}^S l_{is}^0 \hat{v}_{is} + \phi l_{i0}^0\hat{b}_{i} + \phi\sum_{s = 1}^S l_{is}^0\hat{\omega}_{is}- \phi \hat{b}_i - \hat{v}_{i0}).
\label{eq:emp_change_specific_pref}
\end{align}
From \cref{eq:market_clearing_roy}, we can express the changes in wages in every sector and every region of country $c$ as
\begin{align}\label{eq:wage_change_specific_pref}
\hat{\omega}_{is}  = (\phi + \sigma_s)^{-1} \left(\hat{\Phi}_i +\hat{\gamma}_s + (\sigma_s - 1) (\hat{A}_{is} + \hat{P}_s)  - \hat{v}_{is}\right).
\end{align}
Combining \cref{eq:emp_change_specific_pref,eq:wage_change_specific_pref}, we can re-express the change in total employment in region $i$ as
\begin{align}\label{eq:emp_change_specific_pref2}
\hat{L}_{i}&=\sum_{s = 1}^S  l_{is}^0[\theta_{is}\hat{P}_s + \lambda_{i}(\phi + \sigma_s)^{-1}((\sigma_s - 1)\hat{A}_{cs}+\hat{\gamma}_s) + \lambda_{i}(\phi + \sigma_s)^{-1}(\sigma_s - 1)\hat{\tilde{A}}_{is}]+\hat{\nu}_i,
\end{align}
where $\hat{\nu}_i= l_{i0}^0(1-l_{i0}^0)^{-1}(\hat{v}_{i} - \phi \hat{b}_i-\hat{v}_{i0})$, $\hat{v}_{i}=(1 - \phi \sum_{s = 1}^S  l_{is}^0(\phi + \sigma_s)^{-1})^{-1}(\phi l_{i0}^0\hat{b}_{i}+l_{i0}^0\hat{v}_{i0}+\sum_{s = 1}^S  l_{is}^0 \sigma_s (\phi + \sigma_s)^{-1}\hat{v}_{is})$, $\beta_{is}=(\sigma_s - 1)(\phi + \sigma_s)^{-1}\lambda_{i}$, and $\lambda_i=\phi l_{i0}^0(1-l_{i0}^0)^{-1}(1 - \phi \sum_{s = 1}^S  l_{is}^0(\phi + \sigma_s)^{-1})^{-1}$.

As in \Cref{appsec:sector_region_employment}, given our emphasis on understanding the changes in labor market outcomes for regions located in the same country, all regions in the population of interest will share the same value of $A_{cs}$ for every sector $s$; thus, we can simplify the notation by writing $\hat{A}_{cs}=\hat{A}_{s}$ for all $s$. Given this notational simplification, the following equivalences
\begin{align*}
\chi_{s}&=P_{s},\\
\mu_{s}&=(A_{s})^{(\sigma_s-1)(\phi + \sigma_s)^{-1}} (\gamma_s)^{(\phi + \sigma_s)^{-1}},\\
\eta_{is}&=(\tilde{A}_{is})^{(\sigma_s-1)(\phi + \sigma_s)^{-1}},
\end{align*}
and the adjustment of the expression for $\lambda_{i}$ and $\hat{\nu}_{i}$, the expression in \cref{eq:emp_change_specific_pref2} is identical to that in \begin{NoHyper}\cref{eq:emp_change} in \Cref{sec:impacteconshocks}\end{NoHyper}. Consequently, the environment described in \Cref{sec:environment_roy,sec:equilibrium_roy} does indeed provide a microfoundation for the equilibrium relationship in \begin{NoHyper}\cref{eq:emp_change}\end{NoHyper}.

\subsection{Allowing for regional migration}\label{appsec:migration}

We extend here the baseline environment described in
\Cref{sec:environment}
to allow for mobility of individuals across regions within a single country $c$. As in \Cref{appsec:specific_pref}, to maintain the analysis simple, we focus on the special case with a single worker group, $G=1$.

\subsubsection{Environment}\label{sec:environment_migration}

We still assume that the number of individuals living in each country $c$ is
fixed and equal to $M_{c}$. The only difference with respect to the setting described in
\Cref{sec:environment} is that the mass of individuals living in a region $i$,
$M_{i}$, is no longer fixed. We assume that, before the realization of
the shock $u(\iota)$ in \cref{eq:ucjiota}, individuals must decide their
preferred region of residence taking into account their idiosyncratic
preferences for local amenities in each region. Specifically, we assume that the
utility to individual $\iota$ of residing in region $i$ is
\begin{equation}\label{eq:pref_location}
U(\iota) =  \tilde{u}_i(\iota)\left(\bar{U}_i (\omega_i/P, b_i/P) - 1\right)
\end{equation}
where $\bar{U}_i (\omega_i/P, b_i/P)$ is the expected utility of residing in
region $i$, as determined by \cref{eq:ucjiota,eq:dist_u}, and
$\tilde{u}_i(\iota)$ is the idiosyncratic amenity level of region $i$ for
individual $\iota$. For simplicity, we assume that individuals draw their
idiosyncratic amenity level independently (across individuals and regions) from
a Type I extreme value distribution:
\begin{equation}\label{eq:frechet_dist}
\tilde{u}_i(\iota) \sim F_{\tilde{u}}(\tilde{u}) = e^{-\tilde{u}^{-\tilde{\phi}}}, \qquad\tilde{\phi}>0.
\end{equation}
A similar modeling of labor mobility has been previously imposed, among others, in \citet*{allenarkolakis2016topography}, \citet*{redding2016}, \citet*{allen2018universal}, and \citet{fajgelbaum2018statetaxes}, among others. See \citet*{reddingrossi2017} for additional references.

\subsubsection{Equilibrium}\label{sec:equilibrium_migration}

\paragraph{Consumption.} The consumer's problem is identical to that in \Cref{sec:equilibrium}.

\paragraph{Labor supply.} To characterize the labor supply in region $i$, we first compute $\bar{U}_i (w_i/P, b_i/P)$:
\begin{align*}
\bar{U}_i (\omega_i/P, b_i/P)&= \frac{\omega_i}{P} \int_{b_i/\omega_i}^\infty u d F_u(u) + \frac{b_i}{P} \int_{\nu_i}^{b_i/\omega_i} d F_u(u),\\
&= \phi \frac{\omega_i}{P} \int_{b_i/\omega_i}^\infty  \Big(\frac{u}{\nu_i}\Big)^{-\phi} du + \frac{b_i}{P} \int_{\nu_i}^{b_i/\omega_i} \frac{\phi}{\nu_i} \Big(\frac{u}{\nu_i}\Big)^{-\phi - 1} du,\\
&= \frac{\phi}{\phi-1} \frac{\omega_i}{P} \nu_i^{\phi} \Big(\frac{\omega_i}{b_i}\Big)^{\phi-1}  + \frac{b_i}{P}  \Big(1 - \nu_i^\phi \Big(\frac{\omega_i}{b_i}\Big)^{\phi}  \Big),\\
&= \frac{b_i}{P} \Big(1 + \frac{1}{\phi-1} \nu_i^\phi  \Big(\frac{\omega_i}{b_i}\Big)^{\phi} \Big).
\end{align*}
To simplify the analysis, we assume that the unemployment benefit is identical
in all regions and equal to the price index $P$; i.e.\ $b_i = P$ for all
$i\in \JN$. Defining $v_{i}\equiv(\nu_{i}/b_{i})^{\phi}$ as in
\cref{eq:extensive_margin_main}, the assumption that $b_i = P$ for all $i\in \JN$
implies that $v_{i}\equiv\nu_{i}/P$ and, thus,
\begin{align*}
\bar{U}_i (\omega_i/P, b_i/P)&=1 + \frac{1}{\phi-1} v_i  \left(\frac{\omega_i}{P}\right)^{\phi},
\end{align*}
and the share of national population in region $i$ is
\begin{align*}
M_i&= \Pr\left[\tilde{u}_i(\iota)\left(\bar{U}_i (\omega_i/P, b_i/P) - 1\right) > \tilde{u}_j(\iota)\left(\bar{U}_j (\omega_j/P, b_j/P) - 1\right), \ \  \forall j \in \JN_c\right]\\
&= \Pr\left[\tilde{u}_i(\iota) v_i (\omega_i)^\phi > \tilde{u}_j(\iota) v_j (\omega_j)^\phi, \ \  \forall j \in \JN_c\right].
\end{align*}
Given the distributional assumption in \cref{eq:frechet_dist}, it holds that
\begin{equation}\label{eq:labor_supply_region}
M_i = \frac{v_{i}(\omega_i)^{\phi_m}}{\Phi_c}M_{c} \ \ \textrm{such that} \ \ \Phi_c = \sum_{j\in \JN_c} v_{j}(\omega_j)^{\phi_m} \ \ \textrm{and} \ \ \phi_m\equiv\tilde{\phi}\phi.
\end{equation}
Given the value of $M_{i}$, total employment in region $i$ is determined as in \cref{eq:extensive_margin_main}. Therefore, the total labor supply in region $i$ is
\begin{equation}\label{eq:extensive_margin_migration}
L_{i}=\frac{v_{i}(\omega_i)^{\phi_m}}{\sum_{j\in \JN_c}v_{j} (\omega_j)^{\phi_m}}M_{c} v_i (\omega_i)^{\phi}.
\end{equation}

\paragraph{Producer's problem.} The producer's problem is identical to that in \Cref{sec:equilibrium}.

\paragraph{Goods market clearing.} The conditions determining the equilibrium in the good's market and, consequently, the region- and sector-specific labor demand equations are identical to those in \Cref{sec:equilibrium}.

\paragraph{Labor market clearing.} Combining the region- and sector-specific
labor supply in \cref{eq:extensive_margin_migration} with the aggregate labor
demand in \cref{eq:labor_demand_tradable_main_total}, and imposing the
normalization $W=1$, the labor market clearing condition in every region
$i\in \JN_{c}$ is
\begin{equation}\label{eq:market_clearing_main_migration}
\frac{v_{i}(\omega_i)^{\phi_m}}{\sum_{j\in \JN_c}v_{j} (\omega_j)^{\phi_m}}M_{c} v_i (\omega_i)^{\phi}  = \sum_s  (\omega_i)^{-\sigma_s}  \left(A_{is} P_s\right)^{\sigma_s-1} \gamma_s,
\end{equation}
or, equivalently,
\begin{equation}\label{eq:market_clearing_main_migration2}
(\Phi_c)^{-1}M_{c} v_i (\omega_i)^{\phi+\phi_{m}}  = \sum_s  (\omega_i)^{-\sigma_s}  \left(A_{is} P_s\right)^{\sigma_s-1} \gamma_s,
\end{equation}
for every region $i$ in every country $c$.

\paragraph{Equilibrium.} Given productivity parameters
$\{A_{cs}\}_{c=1,s=1}^{C, S}$ and $\{\tilde{A}_{is}\}_{i=1,s=1}^{\JN, S}$, preference parameters
$\{\sigma_{s}, \gamma_{s}\}_{s=1}^{S}$, labor supply
parameters, $\phi$, $\phi_{m}$, and $\{v_{i}\}_{i=1}^{\JN}$, and normalizing world income to equal 1,
$W=1$, we can use \cref{eq:Ps,eq: A_is_decomposition,eq:producers_problem,eq:market_clearing_main_migration2} to
solve for the equilibrium wage in every region, $\{\omega_{i}\}_{j=1}^{\JN}$, the equilibrium price of every sector- and region-specific good $\{p_{is}\}_{i=1,s=1}^{\JN,S}$, and the sectoral price indices $\{P_{s}\}_{s=1}^{S}$. Given these equilibrium wages and sectoral price indices,
we can use \cref{eq:extensive_margin_migration} to solve for the
equilibrium level of employment in every region, $\{L_{i}\}_{i=1}^{\JN}$.

\subsubsection{Labor market impact of sectoral shocks}

We assume that, in every period, the model described in \Cref{sec:environment_migration,sec:equilibrium_migration} characterizes the labor market equilibrium in every region $i=1,\dots, \JN$. Across periods, we assume that the parameters $\{\sigma_{s}\}_{s=1}^{S}$, $\phi$ and $\phi_{m}$ are fixed, and that all changes in the labor market outcomes $\{\omega_{i}, L_{i}\}_{i=1}^{\JN}$ are generated by changes in technology
$\{A_{cs}\}_{c=1,s=1}^{C, S}$ and $\{\tilde{A}_{is}\}_{i=1,s=1}^{\JN, S}$, sectoral preferences $\{\gamma_{s}\}_{s=1}^{S}$,
and labor supply parameters $\{v_{i}\}_{i=1}^{\JN}$. We focus here on understanding how changes in these exogenous parameters affect the labor market equilibrium in all regions located in a given country $c$; i.e.\ all regions belonging to the set $\JN_{c}$.

\paragraph{Isomorphism.} According to \cref{eq:extensive_margin_migration}, the change in employment in any region $i$ in country $c$ is
\begin{align}\label{eq:employment_change_migration}
\hat{L}_{i} = 2\hat{v}_{i}+(\phi+\phi_{m})\hat{\omega}_i-\hat{\Phi}_c.
\end{align}
Assuming that $\{M_{c}\}_{c=1}^{C}$, $\{\sigma_{s}\}_{s=1}^{S}$, and $(\phi, \phi_{m})$ are fixed and totally differentiating \cref{eq:market_clearing_main_migration} with respect to the remaining determinants of $\hat{\omega}_i$, we can express the changes in wages in every region $i$ of country $c$ as
\begin{align}\label{eq:wage_change_migration}
\hat{\omega}_i  = \lambda_i \hat{\Phi}_c + \lambda_i \sum_{s=1}^{S} l_{is}^0 \left[\hat{\gamma}_s + (\sigma_s - 1)(\hat{A}_{is} + \hat{P}_s) \right] - \lambda_i \hat{v}_i,
\end{align}
where $\lambda_i \equiv (\phi + \phi_m + \sum_s l_{is}^0 \sigma_s)^{-1}$. Using the expression in \cref{eq:labor_supply_region}, we can also express
\begin{align}\label{eq:Phichanges}
\hat{\Phi}_c &=  \sum_{i\in \JN_c} m_i^0 \left(\phi_m \hat{\omega}_i + \hat{v}_i\right),
\end{align}
where $m_{i}^{0}$ is the share of individuals living in country $c$ that had residence in region $i$ at the initial period $0$; i.e.\ $m^{0}_{i}\equiv M^{0}_{i}/M^{0}_{c}$, with $M^{0}_{c}\equiv\sum_{i\in \JN_{c}}M^{0}_{i}$.

Combining \cref{eq:employment_change_migration,eq:wage_change_migration}, we can express the change in total employment in region $i$ as
\begin{align}\label{eq:emp_change_migration}
\hat{L}_{i}&= [(\phi+\phi_{m})\lambda_i-1]\hat{\Phi}_c + \sum_{s=1}^{S} l_{is}^0[\theta_{is}\hat{P}_s+ \lambda_i(\phi+\phi_{m})((\sigma_s - 1)\hat{A}_{cs} + \hat{\gamma}_s)+ \lambda_i(\phi+\phi_{m})(\sigma_s - 1)\hat{\tilde{A}}_{is}]\nonumber\\
&+[2-  (\phi+\phi_{m})\lambda_i]\hat{v}_i
\end{align}
where $\theta_{is}=(\sigma_{s}-1)(\phi+\phi_{m})\lambda_i$. As in \Cref{appsec:sector_region_employment}, given our emphasis on understanding the changes in labor market outcomes for regions located in the same country, all regions in the population of interest will share the same value of $A_{cs}$ for every sector $s$; thus, we can simplify the notation used in \cref{eq:emp_change_migration} by writing $\hat{A}_{cs}=\hat{A}_{s}$ for all $s$. Given this notational simplification, if it were to be the case that $\hat{\Phi}_c=0$, the expression in \cref{eq:emp_change_migration} would be analogous to that in \begin{NoHyper}\cref{eq:emp_change}\end{NoHyper} under the following equivalences
\begin{align*}
\chi_{s}&=P_{s},\\
\mu_{s}&=(A_{s})^{(\sigma_s-1)(\phi+\phi_{m})} (\gamma_s)^{(\phi+\phi_{m})},\\
\eta_{is}&=(\tilde{A}_{is})^{(\sigma_s-1)(\phi+\phi_{m})},
\end{align*}
and the necessary adjustment of the expression for $\lambda_{i}$ and $\hat{\nu}_{i}$. However, the term $\hat{\Phi}_c$ will generally not be zero and, as indicated in \cref{eq:Phichanges}, it will generally capture the effect of shocks to all regions in the same country $c$ as the region of interest $i$. In the specific case in which $\sigma_{s}=\sigma$ for all sectors $s$, it will be the case that $\lambda_{i}=\lambda$ for all regions $i$, and, consequently, the term $[(\phi+\phi_{m})\lambda_i-1]\hat{\Phi}_c$ will be common to all regions $i$ belonging to the same country $c$. In this special case, the parameter $\theta_{is}$ will no longer capture the total effect of the price shifters $\hat{P}_s$ but the differential effect of this price shifter on region $i$ relative to all other regions in the same country $c$.

\section{Additional placebo exercises}\label{appsec:placebo}

This section presents additional placebo exercises that complement the results
in \begin{NoHyper}\Cref{sec:overr-usual-stand,sec:placebo-exercise}\end{NoHyper}.
\cref{sec:empirical_distribution_placebo} reports the empirical distribution of
the estimated coefficients and standard errors of the baseline placebo exercise
in \begin{NoHyper}\Cref{sec:overr-usual-stand,sec:placebo-exercise}\end{NoHyper}.
\Cref{sec:control_residual_sector} investigates the importance of controlling
for the size of the residual sector in shift-share specifications. In
\Cref{sec:placebo_confounding}, we present results illustrating the impact of
confounding sector-level shocks on different estimators of the coefficient on
the shift-share covariate of interest. \Cref{app:panel_placebo} investigates the
consequences of serial correlation in panel data applications of shift-share
specifications. \Cref{app:approximation_error} analyzes the consequences of
misspecification of our baseline linearly additive potential outcome framework.
\Cref{app: placebo_other_share} reports results investigating the performance of
inference procedures in the presence of unobserved shift-share components whose
shares differ from those of the shift-share variable of interest.
\Cref{sec:heter-treatm-effects} studies the consequences of treatment
heterogeneity. In \Cref{app:placebo_additional_results}, we provide additional
results for the placebo exercises described
in \begin{NoHyper}\Cref{sec:overr-usual-stand,sec:placebo-exercise}\end{NoHyper}.
<
\subsection{Placebo exercise: empirical distributions}\label{sec:empirical_distribution_placebo}

\begin{figure}[!tp]
\centering
\includegraphics[scale=0.8, trim = 1in 3.2in 0 3.2in]{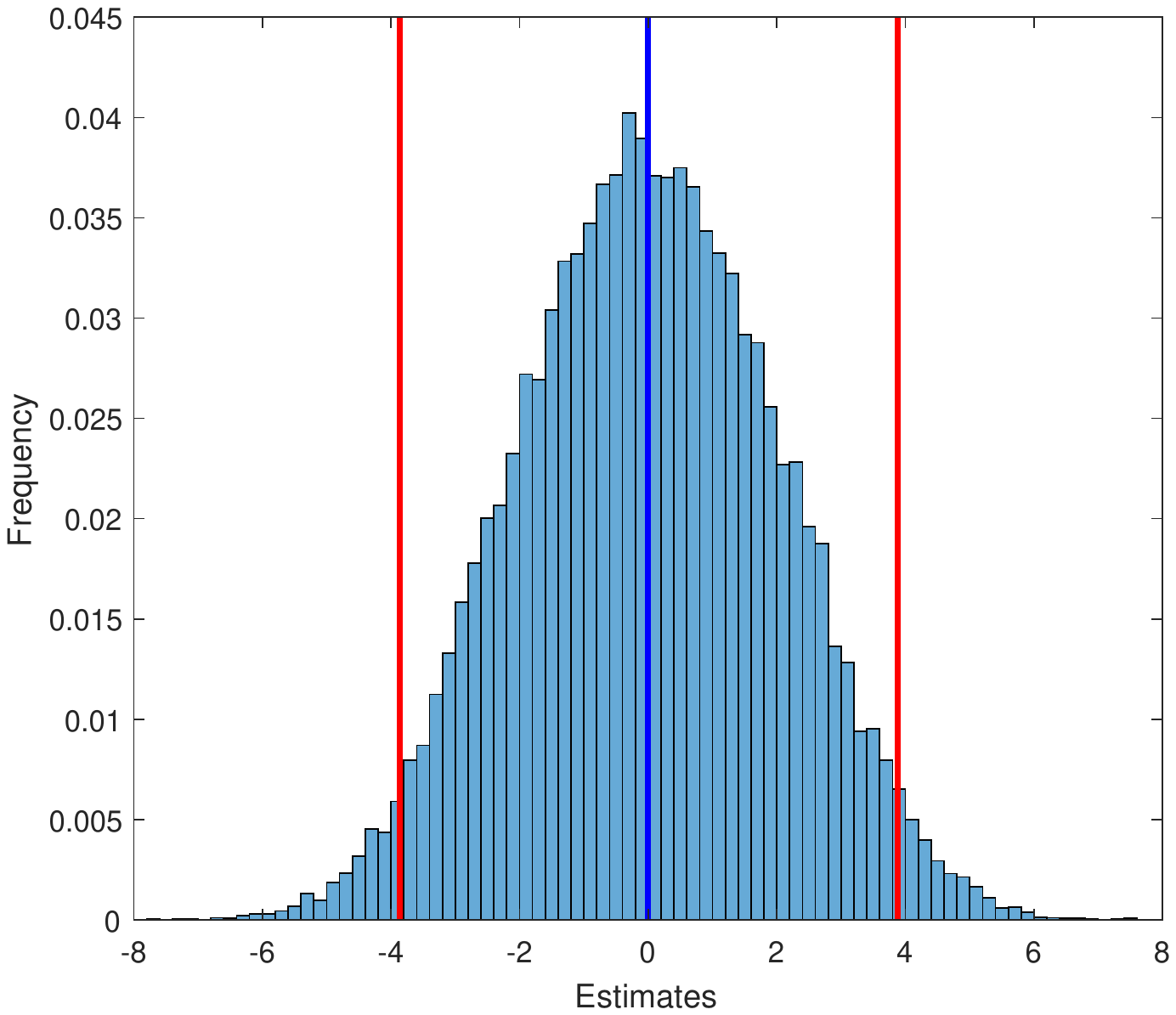}
\caption{Empirical distribution of estimated coefficients in the placebo exercise.}\label{fig:EmpDist}
\captionsetup{width=0.9\linewidth}
\caption*{\scriptsize Notes: The blue line indicates the average estimated coefficient; the red lines indicate the 2.5\% and 97.5\% percentiles of distribution of $\hat{\beta}^{m}$ across the $m=1,\dots,30,000$ simulations. The dependent variable is the 2000--2007 change in the employment rate.}
\end{figure}

\Cref{fig:EmpDist} reports the empirical distribution of the estimated
coefficients when: (a) the dependent variable is the 2000--2007 change in each
CZ's employment rate; in each simulation draw $m$, we draw a random vector
$(\Xs_{1}^m, \dots, \Xs_{S-1}^m)$ of i.i.d.\ normal random variables with zero
mean and variance $\var(\Xs_{s}^{m})=5$, and set $\Xs^{m}_{S}=0$; and (c) the
vector of controls $Z_{i}$ only includes a constant. The empirical distribution
of the estimated coefficients resembles a normal distribution centered around
$\beta=0$. For more details in the placebo exercise that generates this
distribution of estimated coefficients, see \begin{NoHyper}\Cref{sec:overr-usual-stand}\end{NoHyper}.

\FloatBarrier%

\subsection{Controlling for size of the residual sector}\label{sec:control_residual_sector}

In the placebo simulations described in
\begin{NoHyper}\Cref{tab:simple_placebo,tab:simple_placebo_akm,tab:departures_placebo}\end{NoHyper}, we have drawn the
shifters from a mean-zero distribution. In
\Cref{tab:residual_sector_control_placebo}, we depart from the mean-zero
assumption.

\begin{table}[!t]
\centering
\caption{Controlling for the size of the residual sector in each CZ}\label{tab:residual_sector_control_placebo}
\tabcolsep=3pt
\begin{tabular}{@{}l rc cccc cccr @{}}
  \toprule
  &\multicolumn{2}{c}{Estimate} & \multicolumn{4}{c}{Median eff.\ s.e.} &
\multicolumn{4}{c}{Rejection rate}\\ [1pt]
\cmidrule(rl){2-3} \cmidrule(rl){4-7}\cmidrule(rl){8-11}
  &Mean & Std.\ dev  & Robust & Cluster & AKM  & AKM0 & Robust & Cluster & AKM  & AKM0 \\
  & \multicolumn{1}{c}{(1)} & \multicolumn{1}{c}{(2)} & \multicolumn{1}{c}{(3)} & \multicolumn{1}{c}{(4)}
& \multicolumn{1}{c}{(5)} & \multicolumn{1}{c}{(6)} & \multicolumn{1}{c}{(7)} & \multicolumn{1}{c}{(8)} & \multicolumn{1}{c}{(9)}
& \multicolumn{1}{c}{(10)} \\
  \midrule

\multicolumn{9}{@{}l}{\textbf{Panel A\@: Shifters with mean equal to zero}} \\ [2pt]

No controls     &       $0.01$  &       $1.99$  &       $0.74$  &       $0.92$  &       $1.91$  &       $2.23$  &       $48.0$\%        &       $37.7$\%        &       $7.6$\% &       $4.5$\% \\
Control: $1-\sum_{s=1}^{S} w_{is}$     &       $-0.02$ &       $1.43$  &       $0.74$  &       $0.84$  &       $1.31$  &       $1.52$  &       $33.6$\%        &       $28.4$\%        &       $11.2$\%        &       $4.7$\% \\ [6pt]

\multicolumn{9}{@{}l}{\textbf{Panel B\@: Shifters with mean different from zero}} \\ [2pt]

  No controls     &       $-4.67$ &       $1.28$  &       $0.71$  &       $0.94$  &       $1.48$  &       $1.66$  &       $99.1$\%        &       $97.8$\%        &       $85.4$\%        &       $87.6$\%        \\
  Control: $1-\sum_{s=1}^{S} w_{is}$     &       $0.00$  &       $1.43$  &       $0.74$  &       $0.84$  &       $1.31$  &       $1.52$  &       $33.3$\%        &       $27.8$\%        &       $11.1$\%        &       $4.6$\% \\ [2pt]
  \bottomrule
  \multicolumn{11}{p{6.7in}}{\scriptsize{Notes: All estimates in this table use
  the change in the share of the working-age population employed in each CZ as
  the outcome variable $Y_{i}$ in~\begin{NoHyper}\cref{eq:bartikreg}\end{NoHyper}. This table indicates the   median and standard deviation of the OLS estimates of $\beta$
  in~\begin{NoHyper}\cref{eq:bartikreg} across the placebo samples (columns (1) and (2)), the
  median effective standard error estimates (columns (3) to (6)), and the
  percentage of placebo samples for which we reject the null hypothesis
  $H_{0}\colon\beta=0$ using a 5\% significance level test (columns (7) to
  (10)). \textit{Robust} is the Eicker-Huber-White standard error;
  \textit{Cluster} is the standard error that clusters CZs in the same state;
  \textit{AKM} is the standard error in \Cref{remark:akm-construction};
  \textit{AKM0} is the confidence interval in \Cref{remark:akm0-construction}.
  For each inference procedure, the median effective standard error is equal to
  the median length of the corresponding 95\% confidence interval divided by
  $2\times1.96$. Results are based on 30,000 placebo samples. In Panel A,
  $(\Xs_{1}^m, \dotsc, \Xs_{S}^m)$ is drawn i.i.d.\ from a normal distribution
  with zero mean and variance equal to five in each placebo sample. In Panel B,
  $(\Xs_{1}^m, \dotsc, \Xs_{S}^m)$ is drawn i.i.d.\ from a normal distribution with mean equal to one and variance equal to five in each placebo sample. For each of the two panels, the first row presents results in which no control is accounted for in the estimating equation; the second row presents results in which we control for the size of the residual sector.\end{NoHyper}}}
\end{tabular}
\end{table}

As discussed in \begin{NoHyper}\Cref{sec:adding-covariates}\end{NoHyper},
controlling for the region-specific sum of shares, $\sum_{s=1}^{S}w_{is}$, is
important if the shifters have non-zero mean. In our placebo setting, this is
equivalent to controlling for the CZ-specific share of employment in the
non-manufacturing sector in 1990, $1-\sum_{s=1}^{S}w_{is}$; we refer to this
control here as the ``residual sector control''. Panel A in
\Cref{tab:residual_sector_control_placebo} shows that, when the shifters are
mean zero, the mean of $\hat{\beta}$ is not affected by whether we include the
residual sector control. However, including the residual sector control
attenuates the overrejection problem of traditional inference methods.
Intuitively, this control soaks part of the correlation in residuals that
traditional inference methods do not take into account. Panel B in
\Cref{tab:residual_sector_control_placebo} shows that, if the shifter mean is
non-zero, the OLS estimate of $\beta$
in~\begin{NoHyper}\cref{eq:bartikreg}\end{NoHyper} suffers from
substantial bias when the residual sector control is not included in the
regression; this bias disappears once it is included. Specifically, in Panel B,
  $\Xs_{s}^m \sim \mathcal{N}(1,5)$, and the estimator in the first row of this
  panel suffers from negative bias because the positive mean of the shifters
  creates a positive correlation between the shift-share regressor of interest
  and the control $\sum_{s=1}^{S}w_{is}$, which captures the larger secular
  decline in the employment rate in regions initially specialized in
  manufacturing production.

\subsection{Confounding sector-level shocks: omitted variable bias and solutions}\label{sec:placebo_confounding}

In this appendix, we illustrate the consequences of violations of the assumption that the shifters $(\Xs_{1}, \dots, \Xs_{S})$ are independent of other shocks affecting the outcome variable of interest. Specifically, we show the impact that the presence of latent sector-specific shocks correlated with the shifters $(\Xs_{1}, \dots, \Xs_{S})$ has on the properties of the OLS estimator of the coefficient on the shift-share regressor of interest $X_{i}\equiv\sum_{s=1}^{S}\Xs_{s}$. We also illustrate the properties of two solutions to this problem: (i) the inclusion of regional controls as a proxy for sector-level unobserved shocks (see \begin{NoHyper}\Cref{sec:adding-covariates}\end{NoHyper}), and (ii) the use of a shift-share instrumental variable constructed as a weighted average of exogenous sector-level shocks (see \begin{NoHyper}\Cref{sec:ivreg}\end{NoHyper}).

To generate the shifters of interest, the confounding sectoral shocks, and the exogenous sector-specific shocks that will enter the instrumental variable, we extend the baseline placebo exercise and, for each sector $s$ and simulation $m$, we take a draw of a three-dimensional vector
\begin{displaymath}
(\Xs_s^{a, m}, \Xs_s^{b, m}, \Xs_s^{c, m}) \sim N(0; \tilde{\Sigma}),
\end{displaymath}
where $\Xs_{s}^{a}$ is the shifter of interest, $\Xs_{s}^{b}$ is the unobserved confounding shock, $\Xs_{s}^{c}$ is an exogenous shifter. Specifically, the matrix $\tilde{\Sigma}$ is such that $var(\Xs_s^{a})=var(\Xs_s^{b})=var(\Xs_s^{c})=\tilde{\sigma}$, $cov(\Xs_s^{a}, \Xs_s^{b})=cov(\Xs_s^{a}, \Xs_s^{c})=\tilde{\rho}\tilde{\sigma}$, and $cov(\Xs_s^{b}, \Xs_s^{c})=0$. Thus, we impose that $\Xs_s^{a}$ has a correlation of $\tilde{\rho}$ with both $\Xs_s^{b}$ and $\Xs_s^{c}$, but $\Xs_s^{b}$ and $\Xs_s^{c}$ are independent. In our simulations, we set $\tilde{\rho} = 0.7$ and $\tilde{\sigma} = 12$.

To assign the role of a confounding effect to $\Xs_s^{b}$, we generate an outcome variable as
\begin{displaymath}
Y_i^m = Y^{obs}_i + \delta \sum_{s=1}^{S}w_{is}\Xs_s^{b, m},
\end{displaymath}
where $Y^{obs}_i$ is the observed 2000--2007 change in the employment rate in CZ $i$, and $\delta$ is a parameter controlling the impact of the unobserved sectoral shocks ($\Xs_{1}^{b}, \dots, \Xs_{S}^{b}$) on the simulated outcome $Y_i^m$. Thus, the parameter $\delta$ captures the magnitude of the impact that the unobserved shocks $(\Xs_{1}^{b}, \dots, \Xs_{S}^{b})$ have on the outcome variable. We simulate data both with $\delta = 0$ and with $\delta = 6$.

In addition, we assume that we observe a regional variable that is a noisy measure of CZ $i$'s exposure to the unobserved sectoral shocks ($\Xs_{1}^{b}, \dots, \Xs_{S}^{b}$),
\begin{displaymath}
X_i^{b, m} = u_i^m + \sum_s w_{is}  \Xs_s^{b,m} \quad \textrm{where} \quad u_i^m \sim N(0,\sigma_u).
\end{displaymath}
The parameter $\sigma_{u}$ thus modulates the measurement error in $X_i^{b}$ as
a proxy for the impact of the unobserved shocks
($\Xs_{1}^{b},\dots,\Xs_{S}^{b}$) on CZ $i$. We simulate data both with
$\sigma_{u} = 0$ and with $\sigma_{u} = 6$.\footnote{Using the notation
  in \begin{NoHyper}\Cref{sec:adding-covariates}\end{NoHyper}, the simulated
  variable $\Xs_s^{a}$ corresponds to $\Xs_{s}$, the simulated variable
  $\Xs_{s}^{b}$ is an element of $\Zs_{s}$, $u_{i}$ corresponds to $U_{i}$, and
  $X^{b}_{i}$ to $Z_{i}$. The value of the parameter $\gamma$
  in \begin{NoHyper}\cref{eq:linear-controls}\end{NoHyper} is thus equal to
  $\tilde{\rho}$.}

For each set of parameters $(\delta,\sigma_{u})$ and for each simulation draw,
we compute three estimators of the impact of
$X_{i}^{a}\equiv\sum_{s=1}^{S}w_{is}\Xs_s^{a}$ on $Y_{i}$. First, we ignore the
possible endogeneity problem and compute the OLS estimator without controls;
i.e.\ the estimator in \begin{NoHyper}\cref{eq:OLS-no-covariates}\end{NoHyper}.
Second, we consider the OLS estimator of the coefficient on $X_{i}^{a}$ in a
regression that includes $X_i^{b}$ as a proxy for the vector of unobserved
confounding sectoral shocks; i.e.\ the estimator
in \begin{NoHyper}\cref{eq:OLS-adding-covariates}\end{NoHyper}. Third, we
consider the IV estimator that uses $X^{c}_i \equiv \sum_i w_{is} \Xs_s^{c}$ as
the instrumental variable; i.e.\ the estimator
in \begin{NoHyper}\cref{eq:IVestimator}\end{NoHyper}. For each of these three
estimators, we implement four inference procedures: \textit{Robust},
\textit{Cluster}, \textit{AKM} and \textit{AKM0}. All results are reported in
\Cref{tab: IV_placebo}.

When there is no confounding sectoral shock ($\delta = 0$), Panel A shows that all three estimators yield an average coefficient close to zero. Panels B and C report results in the presence of confounding sectoral shocks ($\delta>0$); in this case, the OLS estimator of the coefficient on $X^{a}_{i}$ in a simple regression of $Y_{i}$ on $X_{i}^{a}$ without additional covariates is positively biased ($\hat{\beta}=4.2$). The introduction of the regional control only yields unbiased estimates when it is a good proxy for the latent confounding sectoral shock (i.e.\ if $\sigma_u = 0$ as in Panel B). In contrast, the IV estimate always yields an average estimated coefficient close to zero.

As illustrated in \Cref{tab: IV_placebo}, traditional inference methods always under-predict the dispersion in the estimated coefficient. As discussed in \begin{NoHyper}\cref{eq:Vakm_Vcl}\end{NoHyper}, this is driven by the correlation between the unobservable residuals of regions with similar sector employment compositions. The \textit{AKM} and \textit{AKM0} inference procedures impose no assumption on the cross-regional pattern of correlation in the regression residuals and yield, on average, estimates of the median length of the 95\% confidence interval that are equal or higher to the standard deviation of the empirical distribution of estimates. As a result, as \Cref{tab: IV_placebo} reports, while traditional methods overreject the null $H_{0}:\beta=0$ in the context of both OLS and IV estimation procedures, our methods yield the correct test size for both estimators.

\begin{table}[!tp]
\centering
\caption{Magnitude of standard errors and rejection rates---Confounding effects}\label{tab: IV_placebo}
\tabcolsep=3pt
\begin{tabular}{@{}l rc cccc ccrr}
  \toprule
  &\multicolumn{2}{c}{Estimate} & \multicolumn{4}{c}{Median eff.\ s.e.} &
\multicolumn{4}{c}{Reject. $H_{0}\colon \beta = 0$ at 5\%}\\ [1pt]
\cmidrule(rl){2-3} \cmidrule(rl){4-7}\cmidrule(rl){8-11}
  &Mean & Std.\ dev  & Robust & Cluster & AKM  & AKM0 & Robust & Cluster & AKM  & \multicolumn{1}{c}{AKM0} \\
  & \multicolumn{1}{c}{(1)} & \multicolumn{1}{c}{(2)} & \multicolumn{1}{c}{(3)} & \multicolumn{1}{c}{(4)}
& \multicolumn{1}{c}{(5)} & \multicolumn{1}{c}{(6)} & \multicolumn{1}{c}{(7)} & \multicolumn{1}{c}{(8)} & \multicolumn{1}{c}{(9)}
& \multicolumn{1}{c}{(10)} \\
  \midrule

\multicolumn{11}{@{}l}{\textbf{Panel A\@: No confounding effect ($\delta = 0$)}} \\ [2pt]

OLS no controls &       $0.00$  &       $1.28$  &       $0.47$  &       $0.59$  &       $1.23$  &       $1.43$  &       $48.2$\%        &       $37.6$\%        &       $7.7$\% &       $4.5$\% \\
OLS with controls       &       $0.00$  &       $1.80$  &       $0.67$  &       $0.83$  &       $1.72$  &       $1.97$  &       $47.6$\%        &       $37.9$\%        &       $7.9$\% &       $4.7$\% \\
2SLS    &       $0.00$  &       $1.84$  &       $0.69$  &       $0.85$  &       $1.76$  &       $2.02$  &       $47.7$\%        &       $37.7$\%        &       $7.7$\% &       $4.6$\% \\ [6pt]

\multicolumn{11}{@{}l}{\textbf{Panel B\@: Confounding effect ($\delta = 6$) and perfect regional control ($\sigma_u = 0$)}} \\

OLS no controls &       $4.19$  &       $1.47$  &       $0.58$  &       $0.70$  &       $1.38$  &       $1.60$  &       $97.9$\%        &       $96.8$\%        &       $80.9$\%        &       $72.2$\%        \\
OLS with controls       &       $-0.01$ &       $1.81$  &       $0.67$  &       $0.83$  &       $1.72$  &       $1.97$  &       $48.2$\%        &       $38.3$\%        &       $8.1$\% &       $4.6$\% \\
2SLS    &       $-0.01$ &       $1.85$  &       $0.69$  &       $0.85$  &       $1.75$  &       $2.02$  &       $48.1$\%        &       $38.3$\%        &       $8.0$\% &       $4.7$\% \\ [6pt]

\multicolumn{11}{@{}l}{\textbf{Panel C\@: Confounding effect ($\delta = 6$) and imperfect regional control ($\sigma_u = 2$)}} \\

OLS no controls &       $4.20$  &       $1.47$  &       $0.58$  &       $0.70$  &       $1.37$  &       $1.60$  &       $97.9$\%        &       $96.8$\%        &       $81.4$\%        &       $72.6$\%        \\
OLS with controls       &       $4.10$  &       $1.46$  &       $0.58$  &       $0.70$  &       $1.39$  &       $1.61$  &       $97.7$\%        &       $96.3$\%        &       $79.4$\%        &       $71.3$\%        \\
2SLS    &       $-0.22$ &       $2.46$  &       $0.93$  &       $1.10$  &       $2.12$  &       $2.66$  &       $41.7$\%        &       $34.0$\%        &       $8.1$\% &       $4.6$\% \\ [6pt]

  \bottomrule
  \multicolumn{11}{p{6.5in}}{\scriptsize{Notes: All estimates in this table use
  the change in the share of the working-age population employed in each CZ as
  the outcome variable $Y_{i}$
  in \begin{NoHyper}\cref{eq:bartikreg}\end{NoHyper}. This table indicates the
  median and standard deviation of the OLS estimates of $\beta$ in
  \begin{NoHyper}\cref{eq:bartikreg}\end{NoHyper}
  across the placebo samples (columns (1) and (2)), the median effective
  standard error estimates (columns (3) to (6)), and the percentage of placebo
  samples for which we reject the null hypothesis $H_{0}\colon\beta=0$ using a
  5\% significance level test (columns (7) to (10)). The median
  effective standard error refers to the median length of the 95\% confidence
  interval across the simulated datasets divided by $2\times 1.96$
  \textit{Robust} is the Eicker-Huber-White standard error; \textit{Cluster} is
  the standard error that clusters CZs in the same state; \textit{AKM} is the
  standard error in \begin{NoHyper}\Cref{remark:akm-construction}\end{NoHyper};
  \textit{AKM0} is the confidence interval in
  \begin{NoHyper}\Cref{remark:akm0-construction}\end{NoHyper}. All results are based on 30,000 simulation draws.}}
\end{tabular}
\end{table}

\FloatBarrier%

\subsection{Panel data: serial correlation in residuals and shifters}\label{app:panel_placebo}

In this appendix, we focus on panel data applications and perform several placebo exercises that illustrate the consequences of serial correlation in either the shifters $(\Xs_{1},\dots,\Xs_{S})$ or the regression residuals on the properties of several standard error estimates. For each of our placebo exercises, we generate $30,000$ placebo samples indexed by $m$. Each of them contains $722$ regions, $397$ sectors, and $2$ periods: the first period corresponds to 1990--2000 changes, and the second period corresponds to 2000--2007 changes. As in the baseline placebo, each region corresponds to a U.S.\ Commuting Zone (CZ), and each sector corresponds to a 4-digit SIC manufacturing industry. We index each region by $j$ and each sector by $k$. When implementing the \textit{AKM} and \textit{AKM0} in this context, we follow the approach in \begin{NoHyper}\Cref{sec:panel}\end{NoHyper} by defining ``generalized regions'' as $i=(j,t)$, ``generalized sectors'' as $s=(k,t)$, and shares $w_{is}$ as in \begin{NoHyper}\cref{eq:panel_share}\end{NoHyper}.

As in our baseline placebo, each simulated sample $m$ has identical values of
the shares $\{w_{is}\}_{i=1,s=1}^{N,S}$. Specifically, the shares in periods 1
and 2 correspond to employment shares in 1990 and 2000, respectively. Depending
on the placebo exercise, the placebo samples may differ across simulated samples
in terms of the outcomes $\{Y_{i}\}_{i=1}^{N}$. Finally, all placebo samples
always differ in the shifters $(\Xs_{1}, \dotsc, \Xs_{S})$.

For each simulated sample $m$, we draw the random vector of shifters $(\Xs_{1}^m, \dots, \Xs_{S}^m)$ from the joint distribution
\begin{align}\label{eq:Sigma2}
  (\Xs_{1}^m, \dotsc, \Xs_{S}^m) \sim \mathcal{N}\left(0,\Sigma^2\right),
\end{align}
where $\Sigma^2$ is a $S\times S$ covariance matrix with
$\Sigma^2_{sk}=(1-\rho^2)\sigma^2\1{s=k}+\rho^2\sigma^2\1{c(s)=c(k)}$ and, for every $s$, $c(s)$ indicates the ``cluster'' that the generalized sector $s$ belongs to. We incorporate serial correlation in the sector-level shocks by defining clusters of generalized sectors associated with the same underlying sector in different periods. We follow the baseline placebo by setting $\sigma^2 = 5$. The value of $\rho^{2}$ controls the degree of correlation across shifters of different generalized sectors that correspond to the same underlying industrial sector at different points in time.

For each simulated sample $m$, we generate the outcome of region $i$ in the placebo sample $m$ as
\begin{align}\label{eq:distYplacebo}
Y_{i}^m = Y_i + \eta_i^m,
\end{align}
where $Y_i$ denotes the change in the employment rate in the generalized region $i$. By changing the distribution from which the term $\eta_i^m$ is drawn, we change the distribution of the regression residuals. We implement different placebo exercises in which $\{\eta_i^m\}_{i=1}^{N}$ is drawn from different distributions.

In some placebo exercises, we allow for serial correlation in $\eta_i$ for every region $i$ but impose that $\eta_{i}$ is independent of $\eta_{j}$ for any two different regions $i$ and $j$; specifically,
\begin{align}\label{eq:disteta1}
(\eta_1^m,\dotsc,\eta_{1444}^m) \sim \mathcal{N}(0,\Sigma^1),
\end{align}
where $\Sigma^1$ is a $1444\times 1444$ covariance matrix with $\Sigma^1_{ii'}=(1-\rho^1)\sigma^1\1{i=i'}+\rho^1\sigma^1\1{j(i)=j(i')}$ and $j(i)$ is the region associated with the generalized observation $i$. We set $\sigma^1 = Var(Y_i)/2$ and generate different placebo samples for different values of $\rho^{1}$. The value of $\rho^{1}$ controls the degree of correlation across regression residuals of different generalized regions that correspond to the same geographic region at different points in time.

In some other placebo exercises, we assume that $\eta_i^m$ has a shift-share structure with shares identical to those entering the shift-share component of interest. Specifically, we assume that
\begin{align}\label{eq:disteta2}
\eta_i^m = \sum_{s=1}^{S} w_{is} \mu_s^m \quad \text{such that} \quad  (\mu_{1}^m, \dots, \mu_{S}^m) \sim \mathcal{N}\left(0,\Sigma^2\right),
\end{align}
where $\Sigma^2$ is identical to the variance matrix of the shifters $(\Xs_{1}^m, \dots, \Xs_{S}^m)$ introduced in \cref{eq:Sigma2}.

We start by evaluating the robustness of our results to the existence of serial
correlation in regional outcomes or regression residuals. In Panel A of
\Cref{tab: panel_placebo}, we implement a placebo exercise in which the shifters
are drawn according to \cref{eq:Sigma2} with $\rho^2 = 0$ (i.e.\ no serial
correlation in sectoral shifters) and the outcome variables are drawn according
to \cref{eq:distYplacebo,eq:disteta1} with three different values of $\rho^{1}$
(i.e.\ different degree of serial correlation in the regression residuals). The
rejection rates of all six inference procedures we consider (\textit{Robust},
\textit{Cluster}, \textit{AKM} and \textit{AKM0}, the last two both in a version
that assumes that the shifters are independent, and in a version that allows
them to be serially correlated) are robust to different degrees of serial
correlation in the regression residuals. The reason is that, as illustrated in
column (4) of \Cref{tab: panel_placebo}, the standard deviation of the estimator
$\hat{\beta}$ is invariant to these patterns of serial correlation in the
regression residuals.

In Panel B, we implement a placebo exercise in which the shifters are drawn according to \cref{eq:Sigma2} with $\rho^2$ equal to either 0, 0.5 or 1 (i.e.\ different degrees of serial correlation in sectoral shifters) and the distribution of the simulated outcome variables is identical to their empirical distribution (i.e.\ $\eta_{i}^{m}=0$ for every region $i$ and placebo sample $m$). The results indicate that the larger the serial correlation in the sector-level shifters, the larger the rejection rates implied by the \textit{Robust} and \textit{Cluster} standard errors, as well as those implied by an implementation of the \textit{AKM} and \textit{AKM0} inference procedures that wrongly assumes that the shifters are independent across generalized sectors. Conversely, as illustrated in columns (15) and (16) in Panel B of \Cref{tab: panel_placebo}, the \textit{AKM} and \textit{AKM0} become very close to the nominal rejection rate of 5\% once we cluster across generalized sectors that correspond to the same underlying sector at different points in time.

In Panel C, we depart from the setting described in Panel B in that we draw values of $\eta_{i}^{m}$ according to the distribution described in \cref{eq:disteta2}. The sector-level shifters entering the shift-share covariate of interest $X^{m}_{i}$ and the term $\eta_i^m$ are thus drawn from the same distribution. The results are very similar to those in Panel B.

Finally, in Panel D, we draw shifters $(\Xs_{1}^m, \dots, \Xs_{S}^m)$ that are
not only serially correlated but also correlated across 4-digit industries
belonging to the same 3-digit sector. Columns (11) to (14) show that, when
ignored by the corresponding inference procedure, such correlation patterns in
the shifters of interest lead to an overrejection problem, the severity of which
depends on the correlation in the shifters. Columns (15) and (16) show that this
overrejection problem disappears when we implement the \textit{AKM} and
\textit{AKM0} inference procedures clustering across all generalized shifters
that correspond to pairs of a 4-digit sectors and time period such that the
4-digit sector is associated to the same 3-digit industry.

\afterpage{%
\begin{landscape}
\begin{table}[!tp]
\centering
\caption{Magnitude of standard errors and rejection rates: panel data with serially correlated shifters}\label{tab: panel_placebo}
\tabcolsep=3pt
\begin{tabular}{cc rc cccccc cccccc }
  \toprule
 \multicolumn{2}{c}{Serial Correl.} &\multicolumn{2}{c}{Estimate} & \multicolumn{6}{c}{Median eff.\ s.e.} &
\multicolumn{6}{c}{Rejection rate of $H_{0}\colon \beta = 0$ at 5\%}\\ [1pt]
\cmidrule(rl){1-2} \cmidrule(rl){3-4} \cmidrule(rl){5-10}\cmidrule(rl){11-16}
 $\rho^1$& $\rho^2$ & Mean & Std.\ dev  & Robust & Cluster & AKM  & AKM0 & AKM  & AKM0 & Robust & Cluster & AKM  & AKM0 & AKM  & AKM0 \\
 & &  &   &  &  &   &  & (cluster)  & (cluster) &  &  &   &  & (cluster)  & (cluster) \\
 \multicolumn{1}{c}{(1)} & \multicolumn{1}{c}{(2)} & \multicolumn{1}{c}{(3)} & \multicolumn{1}{c}{(4)}
& \multicolumn{1}{c}{(5)} & \multicolumn{1}{c}{(6)} & \multicolumn{1}{c}{(7)} & \multicolumn{1}{c}{(8)} & \multicolumn{1}{c}{(9)}
& \multicolumn{1}{c}{(10)} & \multicolumn{1}{c}{(11)} & \multicolumn{1}{c}{(12)} & \multicolumn{1}{c}{(13)} & \multicolumn{1}{c}{(14)}  & \multicolumn{1}{c}{(15)} & \multicolumn{1}{c}{(16)} \\
  \midrule

\multicolumn{16}{@{}l}{\textbf{Panel A\@: Correlation over time in a residual region-level component}} \\ [2pt]
$0$     &       $0$     &       $-0.01$ &       $1.12$  &       $0.55$  &       $0.67$  &       $1.06$  &       $1.15$  &       $1.05$  &       $1.17$  &       $34.3$\%        &       $24.0$\%        &       $6.7$\% &       $5.0$\% &       $7.0$\% &       $4.8$\% \\
$0.5$   &       $0$     &       $-0.01$ &       $1.10$  &       $0.55$  &       $0.66$  &       $1.07$  &       $0.00$  &       $1.06$  &       $1.18$  &       $32.7$\%        &       $24.6$\%        &       $6.6$\% &       $5.0$\% &       $6.7$\% &       $4.8$\% \\
$1$     &       $0$     &       $-0.01$ &       $1.09$  &       $0.55$  &       $0.66$  &       $1.05$  &       $0.00$  &       $1.05$  &       $1.16$  &       $32.7$\%        &       $24.2$\%        &       $5.9$\% &       $4.7$\% &       $6.5$\% &       $4.3$\% \\ [6pt]

\multicolumn{16}{@{}l}{\textbf{Panel B\@: Correlation over time in shifter of interest}} \\ [2pt]
$0$     &       $0$     &       $-0.01$ &       $1.05$  &       $0.46$  &       $0.60$  &       $1.02$  &       $1.1$   &       $1.01$  &       $1.12$  &       $39.4$\%        &       $27.0$\%        &       $6.4$\% &       $4.6$\% &       $6.7$\% &       $4.5$\% \\
$0$     &       $0.5$   &       $0.00$  &       $1.13$  &       $0.47$  &       $0.59$  &       $1.04$  &       $1.12$  &       $1.09$  &       $1.22$  &       $42.9$\%        &       $31.6$\%        &       $8.1$\% &       $6.1$\% &       $6.9$\% &       $4.7$\% \\
$0$     &       $1$     &       $0.00$  &       $1.22$  &       $0.47$  &       $0.57$  &       $1.07$  &       $1.14$  &       $1.18$  &       $1.37$  &       $46.2$\%        &       $36.5$\%        &       $10.1$\%        &       $7.9$\% &       $7.6$\% &       $4.5$\% \\ [6pt]

\multicolumn{16}{@{}l}{\textbf{Panel C\@: Correlation over time in shifter of interest and in a residual shift-share component}} \\ [2pt]
$0$     &       $0$     &       $0.00$  &       $1.06$  &       $0.47$  &       $0.60$  &       $1.03$  &       $1.11$  &       $1.02$  &       $1.14$  &       $39.7$\%        &       $27.5$\%        &       $6.7$\% &       $4.8$\% &       $6.9$\% &       $4.7$\% \\
$0$     &       $0.5$   &       $0.00$  &       $1.14$  &       $0.47$  &       $0.59$  &       $1.05$  &       $1.12$  &       $1.1$   &       $1.23$  &       $42.6$\%        &       $31.5$\%        &       $8.0$\% &       $6.2$\% &       $6.8$\% &       $4.5$\% \\
$0$     &       $1$     &       $0.01$  &       $1.22$  &       $0.48$  &       $0.58$  &       $1.07$  &       $1.14$  &       $1.19$  &       $1.38$  &       $45.8$\%        &       $36.1$\%        &       $9.6$\% &       $7.5$\% &       $7.2$\% &       $4.2$\% \\ [6pt]

\multicolumn{16}{@{}l}{\textbf{Panel D\@: Correlation over time and within 3-digit sectors in shifter of interest and in a residual shift-share component}} \\ [2pt]
$0$     &       $0$     &       $0.00$  &       $1.06$  &       $0.46$  &       $0.60$  &       $1.03$  &       $1.11$  &       $1.01$  &       $1.16$  &       $39.6$\%        &       $27.1$\%        &       $6.6$\% &       $4.9$\% &       $7.2$\% &       $4.7$\% \\
$0$     &       $0.5$   &       $0.02$  &       $1.24$  &       $0.47$  &       $0.61$  &       $1.03$  &       $1.1$   &       $1.19$  &       $1.39$  &       $47.3$\%        &       $34.6$\%        &       $11.8$\%        &       $9.8$\% &       $7.4$\% &       $4.5$\% \\
$0$     &       $1$     &       $0.00$  &       $1.40$  &       $0.47$  &       $0.61$  &       $1.02$  &       $1.09$  &       $1.35$  &       $1.68$  &       $53.5$\%        &       $41.7$\%        &       $16.9$\%        &       $14.5$\%        &       $8.1$\% &       $4.2$\% \\ [2pt]
\bottomrule
\multicolumn{16}{p{8.7in}}{\scriptsize{Notes: All estimates in this table use
  the change in the share of the working-age population employed in each CZ as
  the outcome variable $Y_{i}$
  in \begin{NoHyper}\cref{eq:bartikreg}\end{NoHyper}. This table indicates the
  median and standard deviation of the OLS estimates of $\beta$ in \begin{NoHyper}\cref{eq:bartikreg}\end{NoHyper} across the placebo samples (columns (1) and (2)), the median effective standard error estimates (columns (3) to (8)), and the percentage of placebo samples for which we reject the null hypothesis $H_{0}\colon\beta=0$ using a 5\% significance level test (columns (8) to (15)). The median effective standard error refers to the median length of the 95\% confidence interval across the simulated datasets divided by $2\times 1.96$ \textit{Robust} is the Eicker-Huber-White standard error; \textit{Cluster} is the standard error that clusters CZs in the same state; \textit{AKM} is the standard error in \begin{NoHyper}\Cref{remark:akm-construction}\end{NoHyper}; \textit{AKM0} is the confidence interval in
  \begin{NoHyper}\Cref{remark:akm0-construction}\end{NoHyper}. In Panels A, B, and C, \textit{AKM (cluster)} and \textit{AKM0 (cluster)} assume that the shifter corresponding to each 4-digit SIC shifter is distributed independently of those corresponding to other 4-digit shifter, but allow for correlation over time in these 4-digit SIC shifters. In Panel D, \textit{AKM (cluster)} and \textit{AKM0 (cluster)} additionally allow for correlation across 4-digit SIC shifters that belong to the same 3-digit SIC sector. All results are based on 30,000 simulation draws.}}
\end{tabular}
\end{table}
\end{landscape}}

\subsection{Misspecification in linearly additive potential outcome framework}\label{app:approximation_error}

In this appendix section, we study the consequences of potential misspecification in the linearly additive potential outcome framework introduced in \begin{NoHyper}\cref{eq:potential-outcomes} in \Cref{sec:transition_econometrics}\end{NoHyper}. The extent to which this linearly additive framework is misspecified obviously depends on what the true potential outcome framework is. Inspired by the economic model described in \begin{NoHyper}\Cref{sec:model}\end{NoHyper}, we outline a nonlinear potential outcome framework in \Cref{app:nonlinearframework}. In \Cref{app:linearframework}, we determine theoretically the asymptotic properties of the OLS estimator of the coefficient on the shift-share component in the linearly additive potential outcome framework; specifically, we compare the treatment effects implied by the linear framework to those implied by the nonlinear one. In \Cref{app:simul_approx_error}, we present simulation results that quantify the bias in the estimation of treatment effects that arise from assuming a linearly additive potential outcome framework when the true one corresponds to the nonlinear framework described in \Cref{app:nonlinearframework}.

\subsubsection{Nonlinear potential outcome framework}\label{app:nonlinearframework}

Consider the special case of the model of \begin{NoHyper}\Cref{sec:model}\end{NoHyper} in which the labor demand elasticity is identical in all sectors, i.e.\ $\sigma_s = \sigma$ for all $s$. We also set $\rho_s \equiv 1$ for all $s$. In this case, region $i$'s labor demand in sector $s$ is
\begin{displaymath}
L_{is}=\left(\omega_{i}\right)^{-\sigma}\left(\chi_{s}\mu_{s}\eta_{is}\right),
\end{displaymath}
which implies that the total labor demand in region $i$ is
\begin{displaymath}
L_{i}=\left(\omega_{i}\right)^{-\sigma}\sum_{s=1}^S\left(\chi_{s}\mu_{s}\eta_{is}\right).
\end{displaymath}
By equalizing this expression with the expression for region $i$'s labor supply in \begin{NoHyper}\cref{labor_supply_maintext} in \Cref{sec:model}\end{NoHyper}, we obtain the following relationship between equilibrium wages in region $i$ and both labor supply and labor demand shocks in $i$:
\begin{equation}\label{eq_wage_homog}
\log\omega_{i}= \check{\beta} \log\left(\sum_{s=1}^S\left(\chi_{s}\mu_{s}\eta_{is}\right)\right)- \check{\beta}\log\nu_{i}
\end{equation}
where $\check{\beta}\equiv (\phi + \sigma)^{-1}$.

We focus here on determining the impact on log-changes in regional wages $\omega_i$ of log-changes in the sectoral demand shifters $\{\chi_{s}\}_{s=1}^{S}$; i.e.\ using the notation introduced in \begin{NoHyper}\Cref{sec:impacteconshocks}\end{NoHyper}, we focus on characterizing the impact of $\{\hat{\chi}_{s}\}_{s=1}^{S}$ on $\hat{\omega}_i$. Because of the nonlinear nature of the relationship between labor demand shocks and wages in \cref{eq_wage_homog}, the impact of $\{\hat{\chi}_{s}\}_{s=1}^{S}$ on $\hat{\omega}_i$ depends on the changes in all other labor demand and supply shocks. For simplicity, we focus on the case in which all these other labor demand and supply shocks remain constant at their initial level. From \cref{eq_wage_homog}, the wages in the new and old equilibria are given by
\begin{align*}
\log\omega_{i}& =\check{\beta}\log\left(\sum_{s=1}^S \chi_{s}^0\mu_{s}^0\eta_{is}^0 e^{\hat{\chi}_{s}}\right)- \check{\beta} \log\nu_{i}^0,\\
\log\omega_{i}^0& =\check{\beta} \log\left(\sum_{s=1}^S \chi_{s}^0\mu_{s}^0\eta_{is}^0 \right)- \check{\beta} \log\nu_{i}^0,
\end{align*}
where we use a superscript zero to denote the value of the variables in the initial equilibrium and the absence of superscript denotes the value of the corresponding variable in the new equilibrium. By taking the difference between these two expressions,
\begin{equation}\label{eq:eqchawnonlin}
\hat{\omega}_{i}  =\check{\beta} \log\left(\sum_{s=1}^S \frac{\chi_{s}^0\mu_{s}^0\eta_{is}^0}{\sum_{k=1}^S \chi_{k}^0\mu_{k}^0\eta_{ik}^0} e^{\hat{\chi}_{s}} \right) = \check{\beta} \log\left(\sum_{s=1}^S \frac{L_{is}^{0}\left(\omega_{i}^{0}\right)^{\sigma}}{\sum_{k=1}^S L_{ik}^{0}\left(\omega_{i}^{0}\right)^{\sigma}} e^{\hat{\chi}_{s}} \right) = \check{\beta}\log\left(\sum_{s=1}^S \frac{L_{is}^{0}}{L_i^{0}} e^{\hat{\chi}_{s}} \right)
\end{equation}
where the second equality follows from rearranging the terms in the labor demand
expression in \begin{NoHyper}\cref{labor_demand_maintext} in
  \Cref{sec:model}\end{NoHyper} to obtain the equality
$\chi_{s}^0\mu_{s}^0\eta_{is}^0=L_{is}^{0}\left(\omega_{i}^{0}\right)^{\sigma}$
for every region and sector, and the third equality follows from the fact that
labor market clearing yields $L_i^{0} = \sum_{s=1}^{S} L_{is}^{0}$.

Note that,
by using data on the labor allocation across sectors for every region in some
initial equilibrium (i.e.\ $L^{0}_{is}/L^{0}_{i}$, for every $i$ and $s$), the
expression in \cref{eq:eqchawnonlin} allows to compute the effect of changes in
the sector-specific labor demand shifters $\{\chi_{s}\}_{s=1}^{S}$ while
calibrating the value of the overall labor demand shifter
$(\chi_{s}^0)^{\rho_s}\mu_{s}^0\eta_{is}^0$ at the initial equilibrium.
Furthermore, the last expression in \cref{eq:eqchawnonlin} has the advantage
that, conditional on values of $\{\hat{\chi}_{s}\}_{s=1}^{S}$ that are of
interest, it depends exclusively on the parameter $\check{\beta}$; specifically,
it does not depend on the labor demand parameter $\sigma$.

We can map the expression in \cref{eq:eqchawnonlin} to a nonlinear potential outcome framework by setting $\Xs_{s} = \hat{\chi}_s$, $Y_i = \hat{\omega}_i$, and $w_{is} = L_{is}^{0}/L_i^{0}$ for every region and sector; i.e.
\begin{align}\label{eq:potential_outcome_nonlinear}
Y_i(\Xs_1,\dotsc, \Xs_{S}) = \check{\beta}\log\left(\sum_{s=1}^{S} w_{is} e^{\Xs_{s}}\right).
\end{align}
According to the model in \begin{NoHyper}\Cref{sec:model}\end{NoHyper}, this nonlinear potential outcome function yields the exact expression for the change in wages implied by a change in the labor demand shifters $\{\hat{\chi}_{s}\}_{s=1}^{S}$. Using \cref{eq:potential_outcome_nonlinear} we can also compute the treatment effect on region $i$ of changing the shifters from $\{\Xs_{s}\}_{s=1}^{S}$ to $\{\Xs'_{s}\}_{s=1}^{S}$,
\begin{align}\label{eq:indtreateffect_nonlinear}
Y_i(\Xs_1,\dotsc,\Xs_{S})-Y_i(\Xs'_1,\dotsc,\Xs'_{S}) = \check{\beta}\Big[\log\Big(\sum_{s=1}^{S} w_{is} e^{\Xs_{s}}\Big)-\log\Big(\sum_{s=1}^{S} w_{is} e^{\Xs_{s}}\Big)\Big].
\end{align}
and the average treatment effect
\begin{align}\label{eq:avgtreateffect_nonlinear}
\bar{Y}(\Xs_1,\dotsc,\Xs_{S})-\bar{Y}(\Xs'_1,\dotsc,\Xs'_{S}) = \check{\beta}\frac{1}{N}\sum_{i=1}^{N}\Big[\log\Big(\sum_{s=1}^{S} w_{is} e^{\Xs_{s}}\Big)-\log\Big(\sum_{s=1}^{S} w_{is} e^{\Xs_{s}}\Big)\Big].
\end{align}

The linearly additive function in \begin{NoHyper}\cref{eq:potential-outcomes} in \Cref{sec:transition_econometrics}\end{NoHyper} provides a first-order approximation to the nonlinear function in \cref{eq:indtreateffect_nonlinear}. In the next two subsections, we study the extent to which the linear expression in \begin{NoHyper}\cref{eq:potential-outcomes}\end{NoHyper} provides an accurate approximation to the nonlinear one in \cref{eq:potential_outcome_nonlinear}. Specifically, we explore the extent to which the treatment effects in \cref{eq:indtreateffect_nonlinear,eq:avgtreateffect_nonlinear} are well approximated by those computed on the basis of the linear potential outcome framework introduced in \begin{NoHyper}\Cref{sec:transition_econometrics}\end{NoHyper}.

The extent to which the linear approximation is accurate will depend on the
distribution of $\{\Xs_{s}\}_{s=1}^{S}$. Throughout this section, we assume that
$\{\Xs_{s}\}_{s=1}^{S}$ are independently drawn from a normal distribution,
\begin{equation}\label{eq:DGP_nonlinear}
\Xs_{s}\sim \mathcal{N}(0,\gamma^2),
\end{equation}
so that $e^{\Xs_{s}}$ is log-normally distributed with $E[(e^{\Xs_{s}})^k] = e^{k^2 \gamma^2 / 2}$.

\subsubsection{Asymptotic properties of the shift-share linear specification}\label{app:linearframework}

We consider here the asymptotic properties of the OLS estimator of $\beta$ in the linear shift-share regression,
\begin{equation}\label{reg_nonlinear}
Y_i = \alpha + \beta \sum_{s=1}^{S}w_{is}\Xs_{s} + \epsilon_i,
\end{equation}
when the distribution of $\Xs_{s}$ for every sector $s$ is given by \cref{eq:DGP_nonlinear}, and the distribution of $Y_i$ for every region $i$ is given by the potential outcome framework in \cref{eq:potential_outcome_nonlinear}. Since $\Xs_{s}$ has mean zero, the constant does not affect the regression estimand, which is given by
\begin{equation}\label{eq:betapp}
  \beta =  \frac{\sum_{i=1}^{N}E[{X}_{i}Y_{i}]}{\sum_{i=1}^{N}E[{X}_{i}^{2}]},
\end{equation}
where, under \cref{eq:potential_outcome_nonlinear,eq:DGP_nonlinear},
\begin{equation}\label{eq:betapp_den}
  \sum_{i=1}^{N} E[X_{i}^{2}]=\gamma^{2}\sum_{i=1}^{N}\sum_{s=1}^{S}w^{2}_{is},
\end{equation}
and
\begin{equation}\label{eq:betapp_num}
\sum_{i=1}^{N}E[{X}_{i}Y_{i}]=
\check{\beta} E\sum_{i=1}^{N}\sum_{s=1}^{S}w_{is}\Xs_{s}\log\left(\sum_{k=1}^{S}w_{ik}e^{\Xs_{k}}\right).
\end{equation}
Using \cref{eq:betapp,eq:betapp_den,eq:betapp_num}, we can obtain an expression for $\beta$, the OLS estimand in a regression of $Y_{i}$ on $\sum_{s=1}^{S}w_{is}\Xs_{s}$,
\begin{align}\label{eq: betaanalytic}
\beta = \check{\beta}\frac{\sum_{i=1}^{N}\sum_{s=1}^{S}w_{is}E[\gamma Z_{s}\log(\sum_{k=1}^{S}w_{ik}e^{\gamma Z_{k}})]}{\gamma^{2}\sum_{i=1}^{N}\sum_{s=1}^{S}w_{is}^{2}},
\end{align}
as well as for the difference between this value of $\beta$ and the parameter from the nonlinear model in \cref{eq_wage_homog}:
\begin{equation}\label{eq:analytic-bias}
\beta-\check{\beta}=\check{\beta}\frac{\sum_{i=1}^{N}\sum_{s=1}^{S}w_{is}E[\gamma Z_{s}\log(\sum_{k=1}^{S}w_{ik}e^{\gamma Z_{k}})]}{\gamma^{2}\sum_{i=1}^{N}\sum_{s=1}^{S}w_{is}^{2}}-\check{\beta},
\end{equation}
where $\{Z_{s}\}_{s=1}^{S}$ are $i.i.d$ standard normal. As it is clear from this expression, the difference between $\beta$ and $\check{\beta}$ depends on the shares $\{w_{is}\}_{i=1,s=1,}^{N, S}$, the value of the $\gamma$ (i.e.\ the standard deviation of $\Xs_{s}$ for every $s$, according to \cref{eq:DGP_nonlinear}), and the value of $\check{\beta}$ itself.

The expression analogous to that in \cref{eq:indtreateffect_nonlinear} when the linear potential outcome framework in \begin{NoHyper}\cref{eq:potential-outcomes} in \Cref{sec:transition_econometrics}\end{NoHyper} is assumed is the following,
\begin{align}\label{eq:indtreateffect_linear}
Y_i(\Xs_1,\dotsc,\Xs_{S})-Y_i(\Xs'_1,\dotsc,\Xs'_{S}) = \beta\Big(\sum_{s=1}^{S} w_{is} (\Xs_{s}-\Xs'_{s})\Big).
\end{align}
and the expression analogous to that in \cref{eq:indtreateffect_nonlinear} is
\begin{align}\label{eq:avgtreateffect_linear}
\bar{Y}(\Xs_1,\dotsc, \Xs_{S})-\bar{Y}(\Xs'_1,\dotsc,\Xs'_{S}) = \beta\frac{1}{N}\sum_{i=1}^{N}\Big(\sum_{s=1}^{S} w_{is} (\Xs_{s}-\Xs'_{s})\Big).
\end{align}

\subsubsection{Simulation}\label{app:simul_approx_error}

In this section, we construct a simulation exercise to quantify: (a) the difference between $\beta$ and $\check{\beta}$, using \cref{eq:analytic-bias} to compute such difference; (b) the correlation coefficient between the $i$-specific treatment effects in \cref{eq:indtreateffect_nonlinear} and those in \cref{eq:indtreateffect_linear}; and, (c) the difference between the average treatment effect in \cref{eq:avgtreateffect_nonlinear} and that in \cref{eq:avgtreateffect_linear}.

In all  simulations, we calibrate the labor supply elasticity to equal 2, $\sigma = 2$, and the inverse labor supply elasticity to equal 0.5, $\phi =0.5$, implying that $\check{\beta} = 0.4$.  To remain close to our baseline placebo exercise, we calibrate the shares $\{w_{is}\}_{i=1,s=1}^{N,S}$ using 1990 data on sector-region employment shares for 722 US CZs and 396 4-digit manufacturing sectors. Concerning the value of the variance of the sectoral shifters, we present results for five different values of $var(\Xs_{s}) = \gamma^{2}$ varying between $\gamma^{2}=0.5$ and $\gamma^{2}=10$. For each value of $\gamma$, we then generate $30,000$ samples indexed by $m$ such that $\{\Xs_s^m\}_{s=1}^{396}$ are independently drawn according to \cref{eq:DGP_nonlinear} and $\{Y_i^m\}_{i=1}^{722}$ are constructed according to \cref{eq:potential_outcome_nonlinear}.

For each placebo sample $m$, we compute the OLS estimator $\hat{\beta}$ of the parameter $\beta$ defined in \cref{eq:betapp}, confidence intervals for $\beta$ according to the \textit{Robust}, \textit{Cluster}, \textit{AKM} and \textit{AKM0} inference procedures, the true linear approximation to the $i$-specific treatment effect and to the average treatment effect (i.e.\ the expressions in \cref{eq:indtreateffect_linear,eq:avgtreateffect_linear} with $\check{\beta}$ instead of $\beta$), the estimated linear approximation to the $i$-specific treatment effect and to the average treatment effect (i.e.\ the expressions in \cref{eq:indtreateffect_linear,eq:avgtreateffect_linear} with $\hat{\beta}$ instead of $\beta$), and the true $i$-specific treatment effects and their average (i.e.\ the expressions in \cref{eq:indtreateffect_nonlinear,eq:avgtreateffect_nonlinear} with $\check{\beta}=0.4$).

A comparison of columns (2) and (3) in \Cref{tab: nonlinear_estimate_placebo}
illustrates that the average across the placebo samples generated under the same
value of $\gamma$ of the OLS estimates of $\beta$,
$\overline{\hat{\beta}}\equiv (30,000)^{-1}\sum_{m=1}^{30,000}\hat{\beta}^{m}$
(reported in column (3)) is very close to the true value of the parameter
$\beta$ (reported in column (2)). We compute this true value of $\beta$ using
the expression in \cref{eq: betaanalytic} and Monte Carlo integration based on
50,000 draws of $(Z_{1}, \dotsc, Z_{S})$ from the distribution in
\cref{eq:DGP_nonlinear}. Thus, as expected, the average value of
$\hat{\beta}^{m}$ is very close to its theoretical value.

Columns (4)--(7) of \Cref{tab: nonlinear_estimate_placebo} report different
measures of the average treatment effect across simulated samples. Specifically,
we compute in these three columns, in this order, the average across the 30,000
placebo samples of: (a) the true linear approximation to the average treatment
effect (i.e.\ the expression in \cref{eq:avgtreateffect_linear} with the value
$\beta$ set to the expression in \cref{eq: betaanalytic}); the estimated linear
approximation to the average treatment effect (i.e.\ the expression in
\cref{eq:avgtreateffect_linear} with $\hat{\beta}^{m}$ instead of $\beta$); and
the true average treatment effect (i.e.\ the expression in
\cref{eq:avgtreateffect_nonlinear}). When the variance of sector-level shocks is
low ($\gamma^2 = 0.1)$, the first row in \Cref{tab: nonlinear_estimate_placebo}
shows that all these three averages are very close to each other. As the
variance of sector-level shocks grows, the remaining rows in \Cref{tab:
  nonlinear_estimate_placebo} show that the bias in the linear approximations to
the average treatment effect grows. Columns (6) and (7) of \Cref{tab:
  nonlinear_estimate_placebo} illustrate that not only the linear approximation
to the average treatment effects worsen as $\gamma^{2}$ increases, but the
average (across the 30,000 placebo samples) correlation coefficient between the
$i$-specific linear treatment effects in \cref{eq:indtreateffect_linear}
(computed with $\check{\beta}$ instead of $\beta$) and the nonlinear ones in
\cref{eq:indtreateffect_nonlinear} becomes much lower.

In summary, \Cref{tab: nonlinear_estimate_placebo} shows that, when the value of the variance of the sector-level shocks is small, the difference between $\beta$ and $\check{\beta}$ reported in \cref{eq:analytic-bias} is small, and the linear approximations to the treatment effects in \cref{eq:indtreateffect_linear,eq:avgtreateffect_linear} remain very close to their non-linear counterparts in \cref{eq:indtreateffect_nonlinear,eq:avgtreateffect_nonlinear}. Conversely, these approximations become much worse as the variance of the sector-level shocks increases.

In \Cref{tab: nonlinear_placebo}, we study the performance of different
inference methods in their capacity to provide information about the value of
$\beta$ in \cref{eq: betaanalytic} or about the parameter $\check{\beta}$.
Columns (2)--(6) report the standard deviation of the OLS estimated coefficients
$\hat{\beta}^{m}$ and the average estimated standard errors obtained with
different inference procedures. Columns (7)--(10) report the rejection rate of
the null hypothesis that $\beta = \check{\beta}$ and columns (11)--(14) report
the rejection rate of the null hypothesis that $\beta$ coincides with the
expression in \cref{eq: betaanalytic}. Results are similar for all levels of
$\gamma^2$: robust and state-clustered standard errors significantly
underestimate the standard deviation of the OLS estimator, while the
\textit{AKM} and \textit{AKM0} are much closer to this standard deviation. In
line with these results, when testing the null that $\beta$ coincides with the
expression in \cref{eq: betaanalytic} at the 5\% significance level, columns
(13)-(14) show that the rejection rates are close to 5\% for \textit{AKM} and
\textit{AKM0} inference procedures, but columns (11)--(12) show that the
analogous rejection rates are around 50\% for the \textit{Robust} and
\textit{Cluster} inference procedures. Given the difference (reported in
\Cref{tab: nonlinear_estimate_placebo}) between the value of $\beta$ in
\cref{eq: betaanalytic} and the value of $\check{\beta}$, it is not surprising
that, as illustrated in columns (7)--(10) of \Cref{tab: nonlinear_placebo},
rejection rates for the null that $\beta$ equals $\check{\beta}$ are larger than
for the null that $\beta$ equals the expression in \cref{eq: betaanalytic}, no
matter what inference procedure we use. However, it is remarkable that, when the
\textit{AKM0} inference procedure is used, these rejection rates remain quite
close to 5\% and always below 10\%.

In summary, \Cref{tab: nonlinear_placebo} shows that, no matter what the value of the variance of the sector-level shocks is, the relative performance of the four different inference procedures that we consider in all our placebo simulations is consistent with what we have documented in \begin{NoHyper}\Cref{sec:overr-usual-stand,sec:placebo-baseline}\end{NoHyper}. \textit{Robust} and \textit{Cluster} lead to overrejection of the estimand of the OLS estimator, while \textit{AKM} and \textit{AKM0} maintain their good coverage properties for this estimand. Interestingly, even when the OLS estimated does not coincide with the structural parameter $\check{\beta}$, the \textit{AKM0} inference procedure maintains good coverage for this structural parameter; the reason is that, as the variance of the sector-level shocks increases and the OLS estimand becomes more different from $\check{\beta}$, the length of the \textit{AKM0} confidence interval also increases, and it does so at a rate such that it contains $\check{\beta}$ in a fraction of placebo samples that is always between 5\% and 10\%.

\begin{table}[!tp]
\centering
\caption{First-order approximation error: bias in $\hat{\beta}$ and in estimated average treatment effect}\label{tab: nonlinear_estimate_placebo}
\begin{tabular}{@{}ccccccc@{}}
\toprule
$var(\Xs_{s})$  & $\cref{eq: betaanalytic}$  & $\overline{\hat{\beta}}$                         &  \multicolumn{3}{c}{Avg. Treatment Effect}                                                    & Correlation between                                   \\
\cmidrule(rl){1-1} \cmidrule(rl){2-2} \cmidrule(rl){3-3} \cmidrule(rl){4-6}
                        &                                       &                                               &  \multicolumn{2}{c}{Linear}           & Non-linear                                    & linear \& non-linear                                                  \\
 \cmidrule(rl){4-5} \cmidrule(rl){6-6}
                        &                                               &                                                               & Estimated             & True          & True                                                  & avg.\ treatment effect                                 \\
\cmidrule(rl){4-4} \cmidrule(rl){5-5} \cmidrule(rl){6-6} \cmidrule(rl){7-7}
 (1)                                    & (2)                                                   & (3)                   & (4)           & (5)                                                   & (6)                           &  (7)  \\
\midrule
0.1     &       $0.41$  &       $0.41$  &       $0.00$  &       $0.00$  &       $0.00$  &       $0.96$  \\
1       &       $0.48$  &       $0.48$  &       $0.00$  &       $0.00$  &       $0.05$  &       $0.76$  \\
2       &       $0.54$  &       $0.53$  &       $0.00$  &       $0.00$  &       $0.11$  &       $0.64$  \\
5       &       $0.63$  &       $0.62$  &       $0.01$  &       $0.00$  &       $0.34$  &       $0.47$  \\
10      &       $0.65$  &       $0.62$  &       $0.01$  &       $0.00$  &       $0.81$  &       $0.36$  \\ [2pt]

\bottomrule
\multicolumn{7}{p{6in}}{\scriptsize{Notes: The sectoral shifters $\Xs_{s}$ are $i.i.d$, drawn from a normal distribution with mean zero and variance $var(\Xs_{s})$. Column (1) indicates the different values of $var(\Xs_{s})$ that we consider in our simulation exercise; for each value of $var(\Xs_{s})$ listed in column (1), we generate 30,000 simulated samples. Given a set of draws of the shifters $(\Xs^{m}_{1}, \dots, \Xs^{m}_{s}, \dots, \Xs^{m}_{S})$ for a simulated sample indexed by $m$, their true impact on the outcome of a region $i$ is $\check{\beta}\log(\sum_{s}w_{is}\exp(\Xs^{m}_{s}))$ and the first-order approximation to this expression is $\beta\sum_{s}w_{is}\Xs^{m}_{s}$. We set $\check{\beta}=0.4$ for all our simulation exercises. Given this value of $\check{\beta}$ and the value of $var(\Xs_{s})$ in column (1), we report in column (2) the value of $\beta$, the estimand of the OLS estimator in a regression of $Y_{i}$ on $X_{i}$ computed according to the expression in \cref{eq: betaanalytic}. We report in column (3) the average (across the simulated samples) value of this OLS estimator $\hat{\beta}^{m}$. Column (4) and (5) reports the average (across the simulated samples) value of the linearly approximated average treatment effect in \cref{eq:avgtreateffect_linear}, with the only difference being whether the value of $\beta$ in this expression is set to the value in \cref{eq: betaanalytic} or to the average of the OLS estimator $\hat{\beta}^{m}$. Column (6) reports the average (across the simulated samples) value of the true average treatment effect in \cref{eq:avgtreateffect_nonlinear}. Column (7) reports the median (across the simulated samples) value of the correlation coefficient between the true treatment effect in \cref{eq:avgtreateffect_nonlinear} and that arising from the first-order approximation in \cref{eq:avgtreateffect_linear}. See the description in \Cref{app:simul_approx_error} for additional details.}}
\end{tabular}
\end{table}

\afterpage{\begin{landscape}
\begin{table}[!tp]
\centering
\caption{First-order approximation error: impact on standard errors and rejection rates.}\label{tab: nonlinear_placebo}
\tabcolsep=5pt
\begin{tabular}{@{}cccccccccrcccrc@{}}
  \toprule
$var(\Xs_{s})$  &\multicolumn{2}{c}{Estimate} & \multicolumn{4}{c}{Median eff.\ s.e.} &
\multicolumn{4}{c}{Rejection rate of $H_{0}\colon\beta=\check{\beta}$} & \multicolumn{4}{c}{Rejection rate of $H_{0}\colon\beta=\cref{eq: betaanalytic}$} \\ [1pt]
\cmidrule(rl){2-3} \cmidrule(rl){4-7}\cmidrule(rl){8-11} \cmidrule(rl){12-15}
  &Mean & Std.\ dev  & Robust & Cluster & AKM  & AKM0 & Robust & Cluster & \multicolumn{1}{c}{AKM}  & AKM0 & Robust & Cluster & \multicolumn{1}{c}{AKM}  & AKM0 \\
  & \multicolumn{1}{c}{(1)} & \multicolumn{1}{c}{(2)} & \multicolumn{1}{c}{(3)} & \multicolumn{1}{c}{(4)}
& \multicolumn{1}{c}{(5)} & \multicolumn{1}{c}{(6)} & \multicolumn{1}{c}{(7)} & \multicolumn{1}{c}{(8)} & \multicolumn{1}{c}{(9)}
& \multicolumn{1}{c}{(10)} & \multicolumn{1}{c}{(11)} & \multicolumn{1}{c}{(12)} & \multicolumn{1}{c}{(13)}
& \multicolumn{1}{c}{(14)} \\
\midrule

  0.1     &       $0.41$  &       $0.07$  &       $0.03$  &       $0.03$  &       $0.07$  &       $0.08$  &       $45.3$\%        &       $38.7$\%        &       $9.6$\% &       $3.9$\% &       $45.1$\%        &       $38.4$\%        &       $9.5$\% &       $4.0$\% \\
  1       &       $0.48$  &       $0.10$  &       $0.03$  &       $0.04$  &       $0.09$  &       $0.10$  &       $60.7$\%        &       $56.3$\%        &       $15.3$\%        &       $5.9$\% &       $52.8$\%        &       $47.9$\%        &       $10.4$\%        &       $2.8$\% \\
  2       &       $0.53$  &       $0.14$  &       $0.04$  &       $0.05$  &       $0.12$  &       $0.14$  &       $66.7$\%        &       $62.6$\%        &       $17.3$\%        &       $7.7$\% &       $54.1$\%        &       $49.1$\%        &       $9.1$\% &       $2.5$\% \\
  5       &       $0.62$  &       $0.19$  &       $0.06$  &       $0.07$  &       $0.18$  &       $0.21$  &       $72.5$\%        &       $68.4$\%        &       $18.2$\%        &       $9.1$\% &       $53.9$\%        &       $48.6$\%        &       $7.7$\% &       $2.4$\% \\
  10      &       $0.62$  &       $0.22$  &       $0.07$  &       $0.08$  &       $0.22$  &       $0.25$  &       $67.8$\%        &       $63.0$\%        &       $14.2$\%        &       $6.8$\% &       $53.5$\%        &       $47.4$\%        &       $7.3$\% &       $3.0$\% \\
  \bottomrule
  \multicolumn{15}{p{8.8in}}{\scriptsize{Notes: The sectoral shifters $\Xs_{s}$
  are $i.i.d$ drawn from a normal distribution with mean zero and variance
  $var(\Xs_{s})$. Column (1) indicates the different values of $var(\Xs_{s})$
  that we consider in our simulation exercise; for each value of $var(\Xs_{s})$
  listed in column (1), we generate 30,000 simulated samples. This table
  indicates the median and standard deviation of the OLS estimates of $\beta$
  in \begin{NoHyper}\cref{eq:bartikreg}\end{NoHyper} across the placebo samples
  (columns (1) and (2)), the median effective standard error estimates (columns
  (3) to (6)), the percentage of placebo samples for which we reject the null
  hypothesis $H_{0}\colon\beta=\check{\beta}$ using a 5\% significance level
  test (columns (7) to (10)), and the percentage of placebo samples for which we
  reject the null hypothesis that $\beta$ coincides with the expression in
  \cref{eq: betaanalytic} using a 5\% significance level test (columns (11) to
  (14)). \textit{Robust} is the Eicker-Huber-White standard error;
  \textit{Cluster} is the standard error that clusters CZs in the same state;
  \textit{AKM} is the standard error
  in \begin{NoHyper}\Cref{remark:akm-construction}\end{NoHyper}; \textit{AKM0}
  is the confidence interval
  in \begin{NoHyper}\Cref{remark:akm0-construction}\end{NoHyper}.
  For each inference procedure, the median effective standard error is equal
  to the median length of the corresponding 95\% confidence interval divided by
  $2\times 1.96$.}}
\end{tabular}
\end{table}
\end{landscape}}

\subsection{Unobserved shift-share components with different shares}\label{app: placebo_other_share}

\begin{NoHyper}\Cref{eq:Vakm_Vcl} in
  \Cref{sec:no-covariates}\end{NoHyper} %
characterizes the source of the overrejection problem affecting traditional
inference methods in shift-share specifications, showing that
heteroskedasticity-robust and cluster-robust standard errors overreject whenever
the correlation between residuals is positive. This positive correlation arises
when the residual has a shift-share structure
in \begin{NoHyper}\cref{eq:Eepseps}\end{NoHyper}, the unobserved shifters may
vary at the same level as the shift-share covariate of interest (e.g.\ sectors)
or a different one (e.g.\ countries of origin of immigrants). In this section,
we conduct a placebo simulation to illustrate the bias in both robust and state
clustered standard errors that arises when the regression residual has a
shift-share component.

We generate $30,000$ placebo samples indexed by $m$ with $722$ US CZs and $396$ 4-digit SIC manufacturing industries. As in the baseline placebo exercise discussed in \begin{NoHyper}\Cref{sec:overr-usual-stand,sec:placebo-baseline}\end{NoHyper}, we compute the shift-share covariate of interest using the sectoral employment shares of US CZs in 1990 and sectoral shifters that are drawn independently from a normal distribution with mean equal zero and variance equal to five; i.e.
\begin{displaymath}
X_i^m = \sum_{s=1}^{396} w_{is} \Xs_s^m \quad \text{such that} \quad  \Xs_s^m \sim N(0,5).
\end{displaymath}
The difference between the simulation exercise we consider here and the baseline placebo simulation in \begin{NoHyper}\Cref{sec:overr-usual-stand,sec:placebo-baseline}\end{NoHyper} is that the outcome variable is no longer taken from the observed data. Instead, this outcome variable varies across placebo samples and it is drawn randomly for each simulated sample $m$ as
\begin{displaymath}
Y^{m}_{i}=\sum_{s=1}^{396}\tilde{w}_{is} \mathcal{A}^{m}_{s}
 \qquad \text{such that} \quad  \mathcal{A}^{m}_{s}\sim N(0,5),
\end{displaymath}
where $\tilde{w}_{is}$ are shares that may be different from (but possibly correlated with) the baseline sectoral employment shares in each CZ\@; i.e.\ $\tilde{w}_{is}$ may be different from $w_{is}$. Specifically, for all placebo samples, we generate a single set of alternative shares as
\begin{align}\label{eq:tildew}
\tilde{w}_{is}= \frac{\exp \left(u_{is}+\ln(w_{is}+v_{is}) \right)}{\sum_{k=1}^{396} \exp \left(u_{ik}+\ln(w_{ik}+v_{ik}) \right)} \left(\sum_{k=1}^{396} w_{ik} \right)
\end{align}
where $u_{is}$ and $v_{is}$ drawn randomly such that $u_{is}\sim N(0,\sigma^{2}_{u})$ and $v_{is}\sim U[0,\sigma_{v}]$.

Given a pair of values $(\sigma_{u}, \sigma_{v})$, for each placebo sample we
compute: (a) the OLS estimator of the regression of $Y_i^m$ on $X_i^m$ and a
constant; (b) effective standard errors according to the robust,
state-clustered, \textit{AKM} and \textit{AKM0} inference procedures; (c) for
each of these inference procedures, the outcome of a 5\% significance level test
of hypothesis of the null hypothesis $H_{0}\colon\beta=0$. Each row of
\Cref{tab:simulation_twoshiftshares} reports several summary statistics of the
distribution of these quantities across the 30,000 placebo samples. Each row
does so for placebo samples generated by different values of $\sigma_u$ and
$\sigma_v$.

The first row of \Cref{tab:simulation_twoshiftshares} considers the case in
which $\sigma_v = \sigma_u = 0$. In this case, $w_{is} = \tilde{w}_{is}$ for
every $i$ and $s$ and, thus, the correlation coefficient between the shares
entering the covariate of interest and those entering the regression residual
equal 1 (see column (3) in \Cref{tab:simulation_twoshiftshares}). In this case,
as in our baseline placebo, robust and state-cluster standard errors have
rejection rates for a 5\% significance level test that are around 30\%--35\%. In
contrast, the \textit{AKM} and \textit{AKM0} inference procedures exhibit
rejection rates that are 10\% and 4\%, respectively. The
remaining rows of \Cref{tab:simulation_twoshiftshares} show that, as we increase
the value of $\sigma_v$ and $\sigma_u$, the correlation between $w_{is}$ and
$\tilde{w}_{is}$ declines, which attenuates the overrejection problem affecting
testing procedures that rely on robust and state-clustered standard errors.
However, the rejection rates of these two inference methods are still above 10\%
even when the correlation between $w_{is}$ and $\tilde{w}_{is}$ is as low as
0.18. For all cases, the rejection rates of the \textit{AKM} and \textit{AKM0}
testing procedures remain stable and close to 5\%.

\begin{table}[!tp]
\centering
\caption{Bias in standard errors when regression residual is a shift-share term
  with shares correlated with those entering the shift-share covariate of
  interest}\label{tab:simulation_twoshiftshares}
\tabcolsep=3.5pt
\begin{tabular}{@{}ccc rc cccc rrrc @{}}
  \toprule
  & & &\multicolumn{2}{c}{Estimate} & \multicolumn{4}{c}{Median eff.\ s.e.} &
\multicolumn{4}{c}{Rejection rate of $H_{0}\colon\beta=0$}\\ [1pt]
\cmidrule(rl){4-5} \cmidrule(rl){6-9}\cmidrule(rl){10-13}
   $\sigma_{u}^2$ & $\sigma_{v}$ & $\rho_{w_{is}, \tilde{w}_{is}}$ &Mean & Std.\ dev  & Robust & Cluster & AKM  & AKM0 & \multicolumn{1}{c}{Robust} & \multicolumn{1}{c}{Cluster} & \multicolumn{1}{c}{AKM}  & \multicolumn{1}{c}{AKM0} \\
 \multicolumn{1}{c}{(1)} & \multicolumn{1}{c}{(2)} & \multicolumn{1}{c}{(3)} & \multicolumn{1}{c}{(4)}
& \multicolumn{1}{c}{(5)} & \multicolumn{1}{c}{(6)} & \multicolumn{1}{c}{(7)} & \multicolumn{1}{c}{(8)} & \multicolumn{1}{c}{(9)}
& \multicolumn{1}{c}{(10)} & \multicolumn{1}{c}{(11)} & \multicolumn{1}{c}{(12)} & \multicolumn{1}{c}{(13)}    \\
  \midrule

$0$     &       $0$     &       $1.00$  &       $0.00$  &       $0.17$  &       $0.08$  &       $0.08$  &       $0.14$  &       $0.16$  &       $34.8$\%        &       $31.0$\%        &       $10.2$\%        &       $3.7$\% \\
$1$     &       $0$     &       $0.77$  &       $0.00$  &       $0.16$  &       $0.08$  &       $0.09$  &       $0.14$  &       $0.16$  &       $31.5$\%        &       $27.3$\%        &       $9.9$\% &       $4.0$\% \\
$3$     &       $0$     &       $0.55$  &       $0.00$  &       $0.15$  &       $0.09$  &       $0.09$  &       $0.13$  &       $0.15$  &       $23.8$\%        &       $22.7$\%        &       $9.8$\% &       $4.1$\% \\
$5$     &       $0$     &       $0.44$  &       $0.00$  &       $0.14$  &       $0.10$  &       $0.10$  &       $0.13$  &       $0.15$  &       $18.0$\%        &       $17.4$\%        &       $9.6$\% &       $4.3$\% \\
$0$     &       $0.001$ &       $1.00$  &       $0.00$  &       $0.11$  &       $0.06$  &       $0.06$  &       $0.10$  &       $0.11$  &       $31.6$\%        &       $28.7$\%        &       $10.0$\%        &       $3.7$\% \\
$1$     &       $0.001$ &       $0.70$  &       $0.00$  &       $0.10$  &       $0.05$  &       $0.06$  &       $0.09$  &       $0.10$  &       $28.1$\%        &       $26.5$\%        &       $9.8$\% &       $4.2$\% \\
$3$     &       $0.001$ &       $0.41$  &       $0.00$  &       $0.09$  &       $0.06$  &       $0.06$  &       $0.08$  &       $0.09$  &       $19.0$\%        &       $19.3$\%        &       $8.8$\% &       $4.1$\% \\
$5$     &       $0.001$ &       $0.28$  &       $0.00$  &       $0.09$  &       $0.06$  &       $0.06$  &       $0.08$  &       $0.09$  &       $13.4$\%        &       $14.5$\%        &       $8.1$\% &       $4.4$\% \\
$0$     &       $0.01$  &       $1.00$  &       $0.00$  &       $0.04$  &       $0.02$  &       $0.02$  &       $0.04$  &       $0.04$  &       $25.3$\%        &       $23.4$\%        &       $9.4$\% &       $3.6$\% \\
$1$     &       $0.01$  &       $0.38$  &       $0.00$  &       $0.05$  &       $0.03$  &       $0.04$  &       $0.04$  &       $0.05$  &       $14.3$\%        &       $14.2$\%        &       $7.6$\% &       $3.8$\% \\
$3$     &       $0.01$  &       $0.18$  &       $0.00$  &       $0.04$  &       $0.03$  &       $0.03$  &       $0.04$  &       $0.05$  &       $11.6$\%        &       $12.3$\%        &       $7.7$\% &       $4.4$\% \\
$5$     &       $0.01$  &       $0.10$  &       $0.00$  &       $0.05$  &       $0.05$  &       $0.04$  &       $0.05$  &       $0.06$  &       $7.9$\% &       $9.0$\% &       $7.5$\% &       $4.2$\% \\ [2pt]

\bottomrule
\multicolumn{13}{p{6.6in}}{\scriptsize{Notes: We impose that, for every
  simulated sample $m=1,\dots,30000$, the outcome variable is
  $Y^{m}_{i}=\sum_{s}\tilde{w}_{is}\mathcal{A}^{m}_{s}$, with
  $\mathcal{A}^{m}_{s}$ drawn from a normal distribution with mean zero and
  variance equal to five. The shares $\{\tilde{w}_{is}\}_{i, s}$ vary across the
  cases described in each of the rows in the table above but, for each of these
  rows, are fixed across the 30,000 simulated samples. Specifically, given
  shares $\{w_{is}\}_{i, s}$ that capture the employment share in CZ $i$ employed
  in sector $s$ in 1990, we generate each $\tilde{w}_{is}$ according to the
  expression in \cref{eq:tildew}, with $u_{is}$ and $v_{is}$ drawn randomly
  according to the distributions $u_{is}\sim\mathcal{N}(0,\sigma^{2}_{u})$ and
  $U[0,\sigma_{v}]$. The first two columns in the table above indicate the
  values of $\sigma_{u}$ and $\sigma_{v}$ used to generate
  $\{\tilde{w}_{is}\}_{i, s}$ in each case. As illustrated in the third column,
  the larger the value of either $\sigma_{u}$ or $\sigma_{v}$, the lower the
  correlation coefficient $\rho_{w_{is}, \tilde{w}_{is}}$ between $w_{is}$ and
  $\tilde{w}_{is}$ across regions and sectors. Given the generated outcome
  variables $\{Y^{m}_{i}\}_{i}$ for each simulated sample $m$, we compute the
  OLS estimate of $\beta$ in the regression $Y^{m}_{i}=\beta
  X^{m}_{i}+\epsilon^{m}_{i}$, with
  $X^{m}_{i}=\sum_{s}w_{is}\mathcal{X}^{m}_{s}$ and each $\mathcal{X}^{m}_{s}$
  drawn randomly from a normal distribution with mean zero and variance equal to
  5. We indicate the mean and standard deviation of the OLS estimates of $\beta$
  across the simulated samples (columns (4) and (5)), the median effective
  standard error estimates (columns (6) to (9)), and the percentage of placebo
  samples for which we reject the null hypothesis $H_{0}\colon\beta=0$ using a
  5\% significance level test (columns (10) to (13)). \textit{Robust} is the
  Eicker-Huber-White standard error; \textit{Cluster} is the standard error that
  clusters CZs in the same state; \textit{AKM} is the standard error
  in \begin{NoHyper}\Cref{remark:akm-construction}\end{NoHyper}; \textit{AKM0}
  is the confidence interval in \begin{NoHyper}\Cref{remark:akm0-construction}\end{NoHyper}.}}
\end{tabular}
\end{table}

\subsection{Heterogeneous treatment effects}\label{sec:heter-treatm-effects}

We now present a placebo exercise to evaluate the performance of our inference procedures in the presence of heterogeneous treatment effects. For each placebo sample $m$, we construct the dependent variable as
\begin{displaymath}
Y_{i}^m=Y_{i}+\sum_{s}w_{is}\Xs_{s}^m\beta_{is} \quad \text{such that} \quad \beta_{is}=\lambda w_{is}.
\end{displaymath}
In all placebo samples, $Y_{i}$ is the change in the share of working-age population employed in CZ $i$ and $w_{is}$ is the share of sector $s$ in total employment of CZ $i$. As before, in each placebo sample, we take independent draws of the sector-level shifters from a normal distribution with a mean of zero and a variance of 5.

The parameter $\lambda$ controls the degree of  heterogeneity in the treatment effect of
the sector-level shifters. When $\lambda = 0$, this placebo exercise is
identical to our baseline placebo exercise
in \begin{NoHyper}\Cref{sec:overr-usual-stand}\end{NoHyper}. We are interested in
inference on the OLS estimand. By
\begin{NoHyper}\Cref{theorem:consistency-noZ}\end{NoHyper}, it is
given by
\begin{equation}\label{eq:beta0-het-te}
\beta_0=\sum_{i, s}w_{is}^{2}\beta_{is}/\sum_{i, s}w_{is}^{2}=\lambda
\sum_{i, s}w_{is}^{3}/\sum_{i, s}w_{is}^{2},
\end{equation}
which is linear in $\lambda$.

\Cref{tab:heterog_te} presents the results of the placebo exercise for different values of $\lambda$. For all values of $\lambda$, the average OLS estimate in column (3) is similar to $\beta_0$. Results indicate that both the standard deviation of the OLS estimator and the performance of the inference procedures are not sensitive to the value of $\lambda$.

\begin{table}[!t]
\centering
\caption{Heterogeneous treatment effects}\label{tab:heterog_te}
\tabcolsep=3pt
\begin{tabular}{@{}c cc cccc ccccc @{}}
  \toprule
  &&\multicolumn{2}{c}{Estimate} & \multicolumn{4}{c}{Median eff.\ s.e.} &
\multicolumn{4}{c}{Rejection rate of $H_0: \beta =\beta_0$}\\ [1pt]
\cmidrule(rl){3-4} \cmidrule(rl){5-8}\cmidrule(rl){9-12}
$\lambda$& $\beta_{0}$  &Mean & Std.\ dev  & Robust & Cluster & AKM  & AKM0 & Robust & Cluster & AKM  & AKM0 \\
 \multicolumn{1}{c}{(1)} & \multicolumn{1}{c}{(2)} & \multicolumn{1}{c}{(3)} & \multicolumn{1}{c}{(4)}
& \multicolumn{1}{c}{(5)} & \multicolumn{1}{c}{(6)} & \multicolumn{1}{c}{(7)} & \multicolumn{1}{c}{(8)} & \multicolumn{1}{c}{(9)}
& \multicolumn{1}{c}{(10)} & \multicolumn{1}{c}{(11)} & \multicolumn{1}{c}{(12)}\\
  \midrule
 0&       0.00&    0.00&    1.98&    0.73&    0.92&    1.91&    2.22&    0.48&    0.38&    0.07&    0.04\\
 1&       0.14&    0.15&    1.98&    0.73&    0.92&    1.91&    2.22&    0.48&    0.38&    0.07&    0.04\\
 3&       0.43&    0.45&    1.98&    0.74&    0.92&    1.91&    2.22&    0.48&    0.38&    0.07&    0.04\\
 5&       0.72&    0.74&    1.98&    0.74&    0.93&    1.91&    2.23&    0.48&    0.37&    0.08&    0.04\\
  \bottomrule
  \multicolumn{12}{p{5.7in}}{\scriptsize{Notes: This table indicates the median
  and standard deviation of the OLS estimates of  $\beta$ in \begin{NoHyper}\cref{eq:bartikreg}\end{NoHyper} across the placebo  samples (columns (3) and (4)), the median effective standard error estimates (columns (5) to (8)), and the percentage of placebo samples for which we
  reject the null hypothesis $H_{0}\colon\beta=\beta_0$ using a 5\% significance level
  test (columns (9) to (12)) where the true value of $\beta_0$  shown in column
  (2) is given in~\cref{eq:beta0-het-te}. \textit{Robust} is the Eicker-Huber-White standard   error; \textit{Cluster} is the standard error that clusters CZs in the same
  state; \textit{AKM} is the standard error in   \begin{NoHyper}\Cref{remark:akm-construction}\end{NoHyper}; \textit{AKM0} is   the confidence interval   in \begin{NoHyper}\Cref{remark:akm0-construction}\end{NoHyper}. For each inference procedure, the median effective standard error is equal to the median length of the corresponding 95\% confidence interval divided by
  $2\times1.96$. Results are based on 30,000 placebo samples.}}
\end{tabular}
\end{table}

\subsection{Other extensions}\label{app:placebo_additional_results}

In \Cref{tab:residual_sector_control_placebo_other_outcomes}, we report results
analogous to those
in \begin{NoHyper}\Cref{tab:residual_sector_control_placebo}\end{NoHyper} for
outcome variables $Y_{i}$ other than the employment rate in CZ $i$. The
rejection rates that we obtain are very similar to those reported
in \begin{NoHyper}\Cref{tab:residual_sector_control_placebo}\end{NoHyper} and
discussed in \begin{NoHyper}\Cref{sec:altenative_placebo}\end{NoHyper}.

In \Cref{TabDepVarRejection_county}, we investigate the sensitivity of our
results to an alternative definition of ``region''. We report results for a
placebo exercise that is analogous to the baseline placebo exercise discussed
in \begin{NoHyper}\Cref{sec:overr-usual-stand,sec:placebo-baseline}\end{NoHyper}
except for the use of counties instead of CZs as regions. We use the County
Business Patterns data to construct employment by county and sector using the
imputation procedure in \citet*{autordornhanson2013}. Since this procedure does
not yield wage bill information at the county level, we only implement the
placebo exercise for the outcome variables used in Panel A
of \begin{NoHyper}\Cref{tab:simple_placebo,tab:simple_placebo_akm}\end{NoHyper}:
employment rate; employment rate in manufacturing; and, employment rate in
non-manufacturing. The results show that the rejection rates of all four
inference procedures we consider are very similar to those obtained in the
baseline placebo exercise, which are reported precisely in Panel A
of \begin{NoHyper}\Cref{tab:simple_placebo,tab:simple_placebo_akm}\end{NoHyper}.

In \Cref{TabDepVarRejection_occ}, we investigate the sensitivity of our results
to an alternative definition of ``sector''. We report results for a placebo
exercise that is analogous to the baseline placebo exercise discussed
in \begin{NoHyper}\Cref{sec:overr-usual-stand,sec:placebo-baseline}\end{NoHyper}
except for the use of 331 occupations instead of 396 sectors as the unit of
observation at which the shifters vary. The results in
\Cref{TabDepVarRejection_occ} show that the overrejection problem affecting
tradition inference procedures is even more severe when the shift-share
covariate aggregates occupation-specific shifters than when it aggregates
sectoral shifters. Actually, only the \textit{AKM0} inference procedure yields
rejection rates for the null hypothesis $H_{0}\colon\beta=0$ that are below the
5\% significance level of the test.

\begin{table}[!tp]
\centering
\caption{Controlling for the size of the residual sector in each CZ}\label{tab:residual_sector_control_placebo_other_outcomes}
\tabcolsep=3pt
\begin{tabular}{@{}l rc cccc rrrr @{}}
  \toprule
  &\multicolumn{2}{c}{Estimate} & \multicolumn{4}{c}{Median eff.\ s.e.} &
\multicolumn{4}{c}{Rejection rate of $H_{0}\colon\beta=0$}\\ [1pt]
\cmidrule(rl){2-3} \cmidrule(rl){4-7}\cmidrule(rl){8-11}
  &Mean & Std.\ dev  & Robust & Cluster & AKM  & AKM0 & \multicolumn{1}{c}{Robust} & \multicolumn{1}{c}{Cluster} & \multicolumn{1}{c}{AKM}  & \multicolumn{1}{c}{AKM0} \\
  & \multicolumn{1}{c}{(1)} & \multicolumn{1}{c}{(2)} & \multicolumn{1}{c}{(3)} & \multicolumn{1}{c}{(4)}
& \multicolumn{1}{c}{(5)} & \multicolumn{1}{c}{(6)} & \multicolumn{1}{c}{(7)} & \multicolumn{1}{c}{(8)} & \multicolumn{1}{c}{(9)}
& \multicolumn{1}{c}{(10)} \\
  \midrule

\multicolumn{11}{@{}l}{\textbf{Panel A\@: Shifters with zero mean}} \\ [2pt]

\multicolumn{11}{@{}l}{\textit{Outcome variable: change in the share of working-age population in manufacturing}} \\ [2pt]

No controls     &       $-0.02$ &       $1.87$  &       $0.60$  &       $0.76$  &       $1.78$  &       $2.06$  &       $55.5$\%        &       $44.2$\%        &       $8.1$\% &       $4.2$\% \\
Control: $1-\sum_{s}w_{is}$     &       $0.00$  &       $1.03$  &       $0.56$  &       $0.63$  &       $0.97$  &       $1.12$  &       $30.1$\%        &       $25.8$\%        &       $10.0$\%        &       $4.4$\% \\ [3pt]

\multicolumn{11}{@{}l}{\textit{Change in the share of working-age population in non-manufacturing}} \\ [2pt]

No controls     &       $0.00$  &       $0.94$  &       $0.58$  &       $0.67$  &       $0.89$  &       $1.04$  &       $23.0$\%        &       $17.5$\%        &       $8.1$\% &       $4.5$\% \\
Control: $1-\sum_{s}w_{is}$     &       $0.00$  &       $1.05$  &       $0.60$  &       $0.68$  &       $0.97$  &       $1.12$  &       $27.5$\%        &       $22.6$\%        &       $9.8$\% &       $5.4$\% \\ [3pt]

\multicolumn{11}{@{}l}{\textit{Outcome variable: change in average log-weekly wage of all employees}} \\ [2pt]

No controls     &       $0.05$  &       $2.67$  &       $1.02$  &       $1.34$  &       $2.58$  &       $3.00$  &       $47.0$\%        &       $33.9$\%        &       $7.8$\% &       $4.4$\% \\
Control: $1-\sum_{s}w_{is}$     &       $0.00$  &       $1.21$  &       $0.95$  &       $1.07$  &       $1.15$  &       $1.33$  &       $12.9$\%        &       $8.9$\% &       $7.9$\% &       $4.8$\% \\ [3pt]

\multicolumn{11}{@{}l}{\textit{Outcome variable: change in average log-weekly wage of all employees in manufacturing}} \\ [2pt]

No controls     &       $0.02$  &       $2.94$  &       $1.69$  &       $2.11$  &       $2.75$  &       $3.19$  &       $27.0$\%        &       $17.3$\%        &       $9.3$\% &       $4.5$\% \\
Control: $1-\sum_{s}w_{is}$     &       $0.01$  &       $2.13$  &       $1.66$  &       $1.92$  &       $1.98$  &       $2.28$  &       $12.5$\%        &       $8.0$\% &       $7.7$\% &       $4.5$\% \\ [3pt]

\multicolumn{11}{@{}l}{\textit{Outcome variable: change in average log-weekly wage of all employees in non-manufacturing}} \\ [2pt]

No controls     &       $0.00$  &       $2.62$  &       $1.05$  &       $1.33$  &       $2.56$  &       $2.98$  &       $44.5$\%        &       $32.8$\%        &       $7.6$\% &       $4.4$\% \\
Control: $1-\sum_{s}w_{is}$     &       $0.00$  &       $1.24$  &       $0.98$  &       $1.08$  &       $1.17$  &       $1.35$  &       $12.8$\%        &       $9.5$\% &       $8.5$\% &       $4.7$\% \\ [8pt]

\multicolumn{11}{@{}l}{\textbf{Panel B\@: Shifters with non-zero mean}} \\ [2pt]

\multicolumn{11}{@{}l}{\textit{Outcome variable: change in the share of working-age population in manufacturing}} \\ [2pt]

No controls     &       $-3.92$ &       $1.12$  &       $0.57$  &       $0.81$  &       $1.34$  &       $1.51$  &       $98.7$\%        &       $97.6$\%        &       $80.6$\%        &       $78.7$\%        \\
Control: $1-\sum_{s}w_{is}$     &       $0.00$  &       $1.05$  &       $0.56$  &       $0.63$  &       $0.97$  &       $1.12$  &       $31.1$\%        &       $26.4$\%        &       $10.3$\%        &       $4.6$\% \\ [3pt]

\multicolumn{11}{@{}l}{\textit{Outcome variable: change in the share of working-age population in non-manufacturing}} \\ [2pt]

No controls     &       $-0.75$ &       $0.71$  &       $0.48$  &       $0.64$  &       $0.76$  &       $0.86$  &       $37.2$\%        &       $22.2$\%        &       $14.2$\%        &       $13.9$\%        \\
Control: $1-\sum_{s}w_{is}$     &       $0.01$  &       $1.05$  &       $0.60$  &       $0.68$  &       $0.97$  &       $1.13$  &       $27.6$\%        &       $22.5$\%        &       $9.7$\% &       $5.2$\% \\ [3pt]

\multicolumn{11}{@{}l}{\textit{Outcome variable: change in average log-weekly wage of all employees}} \\ [2pt]

No controls     &       $-6.52$ &       $1.55$  &       $0.97$  &       $1.58$  &       $1.91$  &       $2.15$  &       $99.6$\%        &       $98.3$\%        &       $90.9$\%        &       $90.5$\%        \\
Control: $1-\sum_{s}w_{is}$     &       $0.01$  &       $1.22$  &       $0.95$  &       $1.08$  &       $1.15$  &       $1.33$  &       $13.4$\%        &       $9.1$\% &       $8.1$\% &       $4.9$\% \\ [3pt]

\multicolumn{11}{@{}l}{\textit{Outcome variable: change in average log-weekly wage of all employees in manufacturing}} \\ [2pt]

No controls     &       $-5.38$ &       $1.88$  &       $1.54$  &       $2.29$  &       $1.94$  &       $2.17$  &       $89.3$\%        &       $69.8$\%        &       $75.1$\%        &       $71.0$\%        \\
Control: $1-\sum_{s}w_{is}$     &       $-0.02$ &       $2.13$  &       $1.66$  &       $1.91$  &       $1.98$  &       $2.28$  &       $12.5$\%        &       $8.1$\% &       $7.8$\% &       $4.7$\% \\ [3pt]

\multicolumn{11}{@{}l}{\textit{Outcome variable: change in average log-weekly wage of all employees in non-manufacturing}} \\ [2pt]

No controls     &       $-6.31$ &       $1.54$  &       $0.99$  &       $1.58$  &       $1.90$  &       $2.15$  &       $99.4$\%        &       $97.8$\%        &       $89.0$\%        &       $88.6$\%        \\
Control: $1-\sum_{s}w_{is}$     &       $0.01$  &       $1.24$  &       $0.98$  &       $1.08$  &       $1.17$  &       $1.35$  &       $12.6$\%        &       $9.4$\% &       $8.3$\% &       $4.7$\% \\ [2pt]
\bottomrule
\multicolumn{11}{p{6.6in}}{\scriptsize{Notes: This table indicates the median
  and standard deviation of the OLS estimates of
  $\beta$ in \begin{NoHyper}\cref{eq:bartikreg}\end{NoHyper} across the placebo
  samples (columns (1) and (2)), the median effective standard error estimates
  (columns (3) to (6)), and the percentage of placebo samples for which we
  reject the null hypothesis $H_{0}\colon\beta=0$ using a 5\% significance level
  test (columns (7) to (10)). \textit{Robust} is the Eicker-Huber-White standard
  error; \textit{Cluster} is the standard error that clusters CZs in the same
  state; \textit{AKM} is the standard error in
  \begin{NoHyper}\Cref{remark:akm-construction}\end{NoHyper}; \textit{AKM0} is
  the confidence interval
  in \begin{NoHyper}\Cref{remark:akm0-construction}\end{NoHyper}. For each
  inference procedure, the median effective standard error is equal to the
  median length of the corresponding 95\% confidence interval divided by
  $2\times1.96$. Results are based on 30,000 placebo samples. In Panel A,
  $(\Xs_{1}^m, \dots, \Xs_{S-1}^m)$ is drawn i.i.d.\ from a normal distribution
  with zero mean and variance equal to 5 in each placebo sample. In Panel B,
  $(\Xs_{1}^m, \dots, \Xs_{S-1}^m)$ is drawn i.i.d.\ from a normal distribution
  with mean equal to one and variance equal to 5 in each placebo sample. For
  each of the two panels, the first row presents results in which no control is
  accounted for in the estimating equation; the second row presents results in
  which we control for the size of the residual sector, $1-\sum_{s}w_{is}$.}}
\end{tabular}
\end{table}

\begin{table}[!tp]
\centering
\caption{Magnitude of standard errors and rejection rates: county-level analysis}\label{TabDepVarRejection_county}
\tabcolsep=2.5pt
\begin{tabular}{@{}lrccccccccc@{}}
  \toprule
  &\multicolumn{2}{c}{Estimate} & \multicolumn{4}{c}{Median eff.\ s.e.} &
\multicolumn{4}{c}{Rejection rate of $H_{0}\colon\beta=0$}\\ [1pt]
\cmidrule(rl){2-3} \cmidrule(rl){4-7}\cmidrule(rl){8-11}
  &Mean & Std.\ dev  & Robust & Cluster & AKM  & AKM0 & Robust & Cluster & AKM  & AKM0 \\
  & \multicolumn{1}{c}{(1)} & \multicolumn{1}{c}{(2)} & \multicolumn{1}{c}{(3)} & \multicolumn{1}{c}{(4)}
& \multicolumn{1}{c}{(5)} & \multicolumn{1}{c}{(6)} & \multicolumn{1}{c}{(7)} & \multicolumn{1}{c}{(8)} & \multicolumn{1}{c}{(9)}
& \multicolumn{1}{c}{(10)} \\
  \midrule

\multicolumn{11}{@{}l}{\textbf{Panel A\@: Change in the share of working-age population}}  \\ [2pt]

  employed (all)  &       $0.00$  &       $0.65$  &       $0.24$  &       $0.30$  &       $0.61$  &       $0.67$  &       $47.3$\%        &       $36.3$\%        &       $8.0$\% &       $4.8$\% \\
  employed (manuf.)       &       $0.00$  &       $0.77$  &       $0.18$  &       $0.27$  &       $0.71$  &       $0.78$  &       $65.5$\%        &       $51.4$\%        &       $8.1$\% &       $4.6$\% \\
  employed (non-manuf.)   &       $0.00$  &       $0.37$  &       $0.21$  &       $0.22$  &       $0.35$  &       $0.39$  &       $27.9$\%        &       $25.3$\%        &       $8.8$\% &       $4.6$\% \\ [2pt]
  \bottomrule
  \multicolumn{11}{p{6.7in}}{\scriptsize{Notes: This table indicates the median
  and standard deviation of the OLS estimates of $\beta$ in
  \begin{NoHyper}\cref{eq:bartikreg}\end{NoHyper} across the placebo samples
  (columns (1) and (2)), the median effective standard error estimates (columns
  (3) to (6)), and the percentage of placebo samples for which we reject the
  null hypothesis $H_{0}\colon\beta=0$ using a 5\% significance level test
  (columns (7) to (10)). \textit{Robust} is the Eicker-Huber-White standard
  error; \textit{Cluster} is the standard error that clusters CZs in the same
  state; \textit{AKM} is the standard error in
  \begin{NoHyper}\Cref{remark:akm-construction}\end{NoHyper}; \textit{AKM0}
  is the confidence interval in
  \begin{NoHyper}\Cref{remark:akm0-construction}\end{NoHyper}. For each
  inference procedure, the median effective standard error is equal to the
  median length of the corresponding 95\% confidence interval divided by
  $2\times1.96$. Results are based on 30,000 placebo samples.}}
\end{tabular}
\end{table}

\begin{table}[!tp]
\centering
\caption{Magnitude of standard errors and rejection rates: occupation-specific shifters}\label{TabDepVarRejection_occ}
\tabcolsep=2pt
\begin{tabular}{@{}lr rrrr rrcc r@{}}
  \toprule
  &\multicolumn{2}{c}{Estimate} & \multicolumn{4}{c}{Median eff.\ s.e.} &
\multicolumn{4}{c}{Rejection rate of $H_{0}\colon\beta=0$}\\ [1pt]
\cmidrule(rl){2-3} \cmidrule(rl){4-7}\cmidrule(rl){8-11}
  &Mean & Std.\ dev  & Robust & Cluster & AKM  & AKM0 & Robust & Cluster & AKM  & AKM0 \\
  & \multicolumn{1}{c}{(1)} & \multicolumn{1}{c}{(2)} & \multicolumn{1}{c}{(3)} & \multicolumn{1}{c}{(4)}
& \multicolumn{1}{c}{(5)} & \multicolumn{1}{c}{(6)} & \multicolumn{1}{c}{(7)} & \multicolumn{1}{c}{(8)} & \multicolumn{1}{c}{(9)}
& \multicolumn{1}{c}{(10)} \\
\midrule

\multicolumn{11}{@{}l}{\textbf{Panel A\@: Change in the share of working-age population}}  \\ [2pt]

employed (all)  &       $0.01$  &       $8.59$  &       $1.13$  &       $2.45$  &       $7.46$  &       $27.82$ &       $83.5$\%        &       $62.4$\%        &       $24.9$\%        &       $4.0$\% \\
employed (manuf.)       &       $0.02$  &       $8.13$  &       $0.80$  &       $1.82$  &       $6.55$  &       $25.50$ &       $89.7$\%        &       $75.3$\%        &       $32.9$\%        &       $3.2$\% \\
employed (non-manuf.)   &       $-0.01$ &       $4.03$  &       $0.96$  &       $1.76$  &       $3.06$  &       $9.86$  &       $65.1$\%        &       $38.4$\%        &       $17.9$\%        &       $3.8$\% \\ [6pt]

\multicolumn{11}{@{}l}{\textbf{Panel B\@: Change in average log weekly wage}}  \\ [2pt]

  employed (all)  &       $0.00$  &       $12.58$ &       $1.74$  &       $4.23$  &       $10.1$  &       $38.10$ &       $84.8$\%        &       $62.6$\%        &       $30.6$\%        &       $3.4$\% \\
  employed (manuf.)       &       $-0.07$ &       $11.11$ &       $3.24$  &       $6.18$  &       $9.41$  &       $31.96$ &       $56.2$\%        &       $27.2$\%        &       $11.7$\%        &       $4.9$\% \\
  employed (non-manuf.)   &       $0.01$  &       $12.60$ &       $1.77$  &       $4.19$  &       $9.96$  &       $37.96$ &       $84.9$\%        &       $64.1$\%        &       $31.8$\%        &       $3.2$\% \\ [2pt]
  \bottomrule
  \multicolumn{11}{p{6.7in}}{\scriptsize{Notes: This table indicates the median and standard deviation of the OLS estimates of $\beta$ in
  \begin{NoHyper}\cref{eq:bartikreg}\end{NoHyper} across the placebo samples
  (columns (1) and (2)), the median effective standard error estimates (columns
  (3) to (6)), and the percentage of placebo samples for which we reject the
  null hypothesis $H_{0}\colon\beta=0$ using a 5\% significance level test
  (columns (7) to (10)). \textit{Robust} is the Eicker-Huber-White standard
  error; \textit{Cluster} is the standard error that clusters CZs in the same
  state; \textit{AKM} is the standard error
  in \begin{NoHyper}\Cref{remark:akm-construction}\end{NoHyper}; \textit{AKM0}
  is the confidence interval
  in \begin{NoHyper}\Cref{remark:akm0-construction}\end{NoHyper}. For each
  inference procedure, the median effective standard error is equal to the
  median length of the corresponding 95\% confidence interval divided by
  $2\times1.96$. Results are based on 30,000 placebo samples.}}
\end{tabular}
\end{table}

\FloatBarrier%

\section{Empirical applications: additional results}\label{appsec:application}

\subsection{Effect of Chinese exports on U.S. labor market outcomes}\label{appsec:applicationADH}

This section presents additional results that complement the estimates
in \begin{NoHyper}\Cref{sec:ADH}\end{NoHyper} of the effect of Chinese import
competition on US local labor markets following the approach in
\citet*[ADH hereafter]{autordornhanson2013}.

\subsubsection{Placebo exercise: alternative distributions of shifters}\label{sec:plac-exerc-altern}

The reduced-form and the first-stage specifications have a panel data structure
discussed \begin{NoHyper}\Cref{sec:panel}\end{NoHyper}. Since the outcome data
and the share matrix $W$ is the same as in the placebo exercise in
\Cref{app:panel_placebo}, the results of that
placebo exercise are informative about the finite-sample properties of the four
inference procedures that we consider (robust standard errors, state-clustered
standard errors, and the \textit{AKM} and \textit{AKM0} procedures) in the ADH
empirical application. In this section, we investigate the robustness of the
results in \Cref{app:panel_placebo} to alternative distributions of the sectoral
shifters. In particular, instead of assuming that the shifters are i.i.d.\
according to a normal distribution, we consider distributions that are arguably
closer to the distribution of the actual shifters employed in ADH (the growth in
sectoral Chinese exports to high-income countries other than the US).

First, we consider a placebo exercise that differs from that in
\Cref{app:panel_placebo} only in that the sectoral shifters are drawn
independently from the empirical distribution of the shifters used in ADH\@. The
results are presented in Panel A of \Cref{tab:ADHshifters_placebo}. As in the
analysis in \Cref{app:panel_placebo}, although the data generating process for
our placebo exercise implies that $\beta=0$, the rejection rates of a 5\%
significance level test of the null hypothesis $H_{0}\colon\beta=0$ are
substantially above 5\% when robust and state-clustered standard errors are
used. The rejection rates implied by the \textit{AKM} and \textit{AKM0}
procedures are much closer to 5\%, with rejection rates are close to 10\%.

Second, to get closer to the specification in ADH, we incorporate into our placebo
specification the baseline set of controls that ADH use (see, e.g., column (6)
of Table 3 in ADH). In particular, we draw the sectoral shifters from the
empirical distribution of shifters used in ADH \textit{after partialling out the
  baseline set of controls used in ADH}.\footnote{To partial out a set of
  controls (which vary by region) from the shifters (which vary by sector), we
  implement the following two-step procedure. First, we obtain the residual of a
  regression of the shift-share instrumental variable $X_{i}$ used in ADH on the
  set of controls listed in column (6) of Table 3 in ADH\@; let $\ddot{X}_{i}$
  denote this residual. We then draw the shifters from the empirical
  distribution of the residualized sectoral shifters $\Xs^{res}$, which
  correspond to the regression coefficients from regressing $\ddot{X}_{i}$ onto
  the vector of shares $(w_{i1}, \dotsc, w_{iS})$, i.e.\
  $\Xs^{res} = (W'W)^{-1} W'\ddot{X}$.} Panel B of
\Cref{tab:ADHshifters_placebo} reports the results. For the \textit{Robust},
\textit{Cluster} and \textit{AKM} testing procedures, the rejection rates in
Panel B are very similar to those in Panel A\@, while the \textit{AKM0}
rejection rate is much closer to the nominal level.

\begin{table}[!tp]
\centering
\caption{Alternative distributions of sectoral shifters: placebo}\label{tab:ADHshifters_placebo}
\tabcolsep=3pt
\begin{tabular}{@{}l rc cccc cccc @{}}
  \toprule
  &\multicolumn{2}{c}{Estimate} & \multicolumn{4}{c}{Median eff.\ s.e.} &
\multicolumn{4}{c}{Rejection rate of $H_{0}\colon\beta=0$}\\ [1pt]
\cmidrule(rl){2-3} \cmidrule(rl){4-7}\cmidrule(rl){8-11}
  &Mean & Std.\ dev  & Robust & Cluster & AKM  & AKM0 & Robust & Cluster & AKM  & AKM0 \\
  & \multicolumn{1}{c}{(1)} & \multicolumn{1}{c}{(2)} & \multicolumn{1}{c}{(3)} & \multicolumn{1}{c}{(4)}
& \multicolumn{1}{c}{(5)} & \multicolumn{1}{c}{(6)} & \multicolumn{1}{c}{(7)} & \multicolumn{1}{c}{(8)} & \multicolumn{1}{c}{(9)}
& \multicolumn{1}{c}{(10)} \\
  \midrule

\multicolumn{11}{@{}l}{\textbf{Panel A\@: Empirical distribution of ADH (2013) shocks}} \\ [2pt]

period: 1990--2000       &       $0.11$  &       $0.49$  &       $0.16$  &       $0.19$  &       $0.38$  &       $0.85$  &       $48.5$\%        &       $39.9$\%        &       $10.7$\%        &       $9.3$\% \\
period: 2000--2007       &       $0.04$  &       $0.16$  &       $0.05$  &       $0.06$  &       $0.13$  &       $0.31$  &       $47.9$\%        &       $39.4$\%        &       $11.0$\%        &       $9.5$\% \\ [6pt]

\multicolumn{11}{@{}l}{\textbf{Panel B\@: Empirical distribution of residualized ADH (2013) shocks}} \\ [2pt]

period: 1990--2000       &       $0.00$  &       $0.14$  &       $0.05$  &       $0.06$  &       $0.12$  &       $0.21$  &       $46.5$\%        &       $37.9$\%        &       $10.4$\%        &       $3.7$\% \\
period: 2000--2007       &       $0.00$  &       $0.07$  &       $0.03$  &       $0.03$  &       $0.06$  &       $0.11$  &       $46.4$\%        &       $37.7$\%        &       $10.9$\%        &       $3.7$\% \\ [6pt]
\bottomrule
\multicolumn{11}{p{6.6in}}{\scriptsize{Notes: This table indicates the median
  and standard deviation of the OLS estimates of $\beta$ in
  \begin{NoHyper}\cref{eq:bartikreg}\end{NoHyper}
  across the placebo samples (columns (1) and (2)), the median effective
  standard error estimates (columns (3) to (6)), and the percentage of placebo
  samples for which we reject the null hypothesis $H_{0}\colon\beta=0$ using a
  5\% significance level test (columns (7) to (10)). \textit{Robust} is the
  Eicker-Huber-White standard error; \textit{Cluster} is the standard error that
  clusters CZs in the same state; \textit{AKM} is the standard error in
  \begin{NoHyper}\Cref{remark:akm-construction}\end{NoHyper}; \textit{AKM0} is
  the confidence interval in
  \begin{NoHyper}\Cref{remark:akm0-construction}\end{NoHyper}. For each
  inference procedure, the median effective standard error is equal to the
  median length of the corresponding 95\% confidence interval divided by $2\times1.96$. Results are based on 30,000
  placebo samples. In Panel A, each $\Xs_{s}^m$ is drawn from the empirical
  distribution of shifters $\Xs_{s}$ observed in the data; i.e.\ from the
  empirical distribution of changes in sectoral exports from China to
  high-income countries other than the US\@. In Panel B, each $\Xs_{s}^m$ is drawn
  from the empirical distribution of residualized shifters $\Xs_{s}$ observed in
  the data; i.e.\ from the empirical distribution of the residuals of projecting
  the changes in sectoral exports from China to high-income countries other than
  the US on the full vector of baseline controls in ADH\@; i.e.\ those in column 6
  of Table 3 in \citet*{autordornhanson2013}.}}
\end{tabular}
\end{table}

Next, we consider relaxing the assumption that the sectoral shifters are
independent, or independent across clusters. This specification is motivated by
the concern that the 1990--2000 and 2000--2007 sector-specific growth rates in
Chinese exports to high-income countries other than the US were determined at
least partly by a common factor that had possibly heterogeneous effects across
sectors. We formalize this by modeling year-$t$ imports from China of goods in
sector $s$ by high-income countries other than the US, $IMP_{st}$, as
\begin{equation}\label{China_shock1}
IMP_{st} =  X_{st}^{Ch} + \epsilon_{st},
\end{equation}
where $X_{st}^{Ch}$ is a sectoral component of Chinese exports common to all
destinations (i.e.\ it accounts for export supply factors), and
$ \epsilon_{st}$ is sector- and destination-specific component (i.e.\ it
accounts for export demand factors). We impose the following factor structure on
$X_{st}^{Ch}$:
\begin{equation}\label{China_shock2}
X_{st}^{Ch} = \eta_{s}\bar{X}_{t}^{Ch} +e_{st}.
\end{equation}
The term $\bar{X}_{t}^{Ch}$ captures unobserved factors that may potentially
impact Chinese exports across all sectors (e.g.\ growth in Chinese labor
productivity). The row-vector of sector-specific loadings $\eta_{s}$ indicates
how Chinese exports in each sector $s$ react to changes in the common unobserved
factors captured by $\bar{X}_{t}^{Ch}$ (e.g.\ how sensitive each sector $s$ is
to growth in Chinese labor productivity). Finally, $e_{st}$ is a sector- and
year-specific idiosyncratic component of Chinese exports. Note that, as long as
the distribution of $\bar{X}_{t}^{Ch}$ is not degenerate, the shifter $IMP_{st}$
will be correlated across any two sectors $s$ and $s'$ unless the loadings
$\eta_{s}$ and $\eta_{s'}$ are orthogonal. This correlation in shifters violates
the independence assumption imposed by \begin{NoHyper}\Cref{item:indepX} in
  \Cref{sec:no-covariates}\end{NoHyper} in a way that is not accounted for
by the clustering extension considered
in \begin{NoHyper}\Cref{sec:clustersectors}\end{NoHyper}. In the placebo
simulations that follow, we explore the consequences of the violation of this
assumption, as well as modifications of the AKM and AKM0 procedures that account
for the potential factor structure in the shifters.

Combining \cref{China_shock1,China_shock2} yields
\begin{align}\label{eq:impChinadecomp}
IMP_{st} = \eta_{s}\bar{X}_{t}^{Ch} + \varepsilon_{st},\qquad\text{with}\qquad \varepsilon_{st}=\epsilon_{st}+e_{st}.
\end{align}
To remain as close as possible to the empirical application in ADH, we use
annual data on sector-specific exports from China to other high-income countries
between 1991 and 2007 (which corresponds to the variable $IMP_{st}$ above) to
estimate the common factor $\bar{X}_{t}^{Ch}$, the factor loadings
$\{\eta_{s}\}_{s=1}^{S}$, and the residuals $\{\varepsilon_{st}\}_{s=1}^{S}$ for
every year $t$ and 4-digit SIC manufacturing sectors $s$ using the
interactive fixed effects estimator in \citet{bai2009}, as implemented by
\citet{gomez2015}.

\begin{figure}[!tp]
\centering
\caption{Histogram of estimates of $\{\eta_{s}\}_{s=1}^{S}$}\label{fig:histo_eta}
\includegraphics[width=9.5cm,height=5.4cm,trim={4cm 8.7cm 4cm 9.03cm},clip]{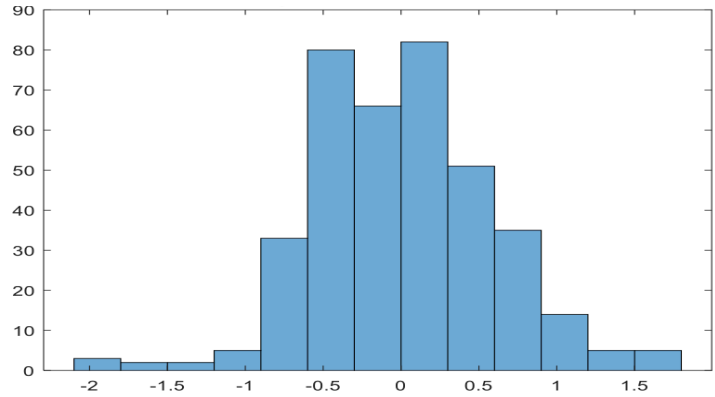}
\end{figure}

\begin{figure}[!tp]
\centering
\caption{Histogram of estimates of $\{u_{s,2007}-u_{s,1991}\}_{s=1}^{S}$}\label{fig:histo_u}
\includegraphics[width=9.5cm,height=5.4cm,trim={4cm 8.7cm 4cm 8.7cm},clip]{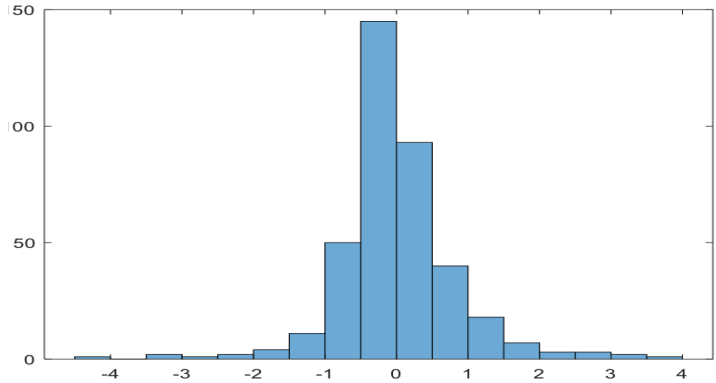}
\end{figure}

\begin{figure}[!tp]
\centering
\caption{Scatterplot of $\{IMP_{s,2007}-IMP_{s,1991}\}_{s=1}^{S}$ against $\{\eta_{s}(\bar{X}_{2007}^{Ch}-\bar{X}_{1991}^{Ch})\}_{s=1}^{S}$}\label{fig:scatter_china}
\includegraphics[width=10cm,height=6.3cm,trim={4cm 9.22cm 4cm 8.7cm},clip]{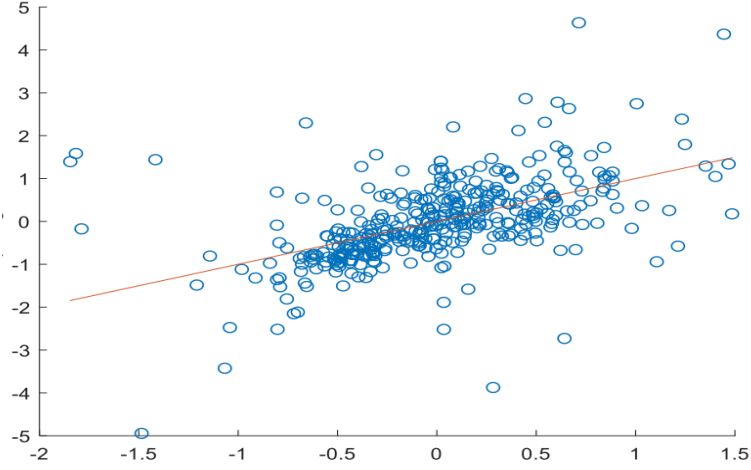}
\begin{minipage}{11cm}
  \vspace{1ex}\scriptsize Notes: Observed data on sector-specific export flows
  from China to high-income countries other than the US (i.e.\
  $IMP_{s,2007}-IMP_{s,1991}$) appear in the vertical axis; estimates of
  $\eta_{s}(\bar{X}_{2007}^{Ch}-\bar{X}_{1991}^{Ch})$ appear in the horizontal
  axis. The $R^{2}$ of this regression is 0.273.
\end{minipage}
\end{figure}

\Cref{fig:histo_eta} reports the histogram of the estimates of
$\{\eta_s\}_{s=1}^{S}$. There is considerable dispersion in the factor loadings
across sectors. The estimates also reveal substantial variation across sectors
and years in the idiosyncratic component of Chinese export growth
$\varepsilon_{st}$; this can be seen in \Cref{fig:histo_u}, which presents a
histogram of the sector-specific changes in $\varepsilon_{st}$ between 1991 and
2007. To provide a graphical illustration of the relative importance of the two
terms entering the right-hand side of \cref{eq:impChinadecomp},
\Cref{fig:scatter_china} provides a scatterplot of the variables
$\{IMP_{s,2007}-IMP_{s,1991}\}_{s=1}^{S}$ against the estimates of the terms
$\{\eta_{s}(\bar{X}_{2007}^{Ch}-\bar{X}_{1991}^{Ch})\}_{s=1}^{S}$; these terms
explain only 27\% of the cross-sectoral variation in export growth from China to
high-income countries other than the US between 1991 and 2007.


\Cref{tab:simulation_china_factor} reports the results of a placebo exercise
illustrating the effects of the correlation in sectoral shifters implied by the
estimated version of the model in \cref{eq:impChinadecomp} on the finite-sample
properties of the \textit{AKM} and \textit{AKM0} procedures. Specifically, we
modify the baseline placebo exercise described
in \begin{NoHyper}\Cref{sec:placebo-baseline}\end{NoHyper} by instead generating
the simulated sectoral shifters as
\begin{equation}\label{eq: shiftstrucapp}
\Xs_s^m = \kappa\eta^{m}_s\Delta\hat{\bar{X}}_{Ch} + u_s^m,\qquad\text{with}\qquad\Delta\hat{\bar{X}}^{Ch}=\hat{\bar{X}}_{2007}^{Ch} - \hat{\bar{X}}_{1991}^{Ch}
\end{equation}
where $\hat{\bar{X}}^{Ch}_{t}$ denotes the estimate of $\bar{X}^{Ch}_{t}$ for
$t=1991$ and $t=2007$. The parameter $\kappa$ controls the relative importance
of the factor component in the simulated shifters. For each simulated sample
$m$, the residuals $u_{s}^{m}$ are drawn independently from a distribution that
we vary across specifications. The term $\eta^{m}_{s}$ is either fixed across
the placebo samples $m$ and set to equal to the estimate $\hat{\eta}_{s}$, or
else drawn independently from the empirical distribution of $\hat{\eta}_{s}$.
Whether the factor loadings $\eta_{s}$ are fixed across the placebo samples or
random (and independent across $s$) is important for the properties of the
\textit{AKM} and \textit{AKM0} inference procedures. If the loadings are random
and independent, the shifters $\Xs_{s}$ will also be independent across $s$, so
that \begin{NoHyper}\Cref{item:indepX} in
  \Cref{sec:no-covariates}\end{NoHyper} holds, and we expect the
\textit{AKM} and \textit{AKM0} inference procedures to have good asymptotic
properties even if conditionally on the loadings, the interactive fixed effects
structure in \cref{eq:impChinadecomp} applies. On the other hand, if the
loadings are fixed across simulation samples, the shifters will be correlated,
so that the asymptotic results
in \begin{NoHyper}\Cref{sec:asympt-prop-bart}\end{NoHyper} do not apply.

In Panels A and B in \Cref{tab:simulation_china_factor}, we fix
$\eta^{m}_s=\hat{\eta}_s$ for every sector $s$ and placebo sample $m$, with
$u_{s}^{m}$ drawn i.i.d.\ from mean-zero normal distribution with variance $5$
in Panel A, and from the empirical the distribution of
$\hat{\varepsilon}_{s,2007}-\hat{\varepsilon}_{s,1991}$ in Panel B, where
$\hat{\varepsilon}_{st}$ is the interactive fixed effects estimate of the term
$\varepsilon_{st}$ in \cref{eq:impChinadecomp}. In the first three rows of each
panel, when no controls are included, larger values of $\kappa$ (which imply a
larger weight on the interactive fixed effects component
$\eta^{m}_s\Delta\hat{\bar{X}}_{Ch}$ in \cref{eq: shiftstrucapp}) imply larger
rejection rates of the null $H_{0}\colon\beta=0$ when we use either the
\textit{AKM} or the \textit{AKM0} inference procedures. For $\kappa=1$, which
corresponds to the specification in ADH, the rejection rates for \textit{AKM0}
are close to the nominal rates, and \textit{AKM} suffers from moderate
overrejection. Importantly, this overrejection problem can be fixed by
controlling for the term $\eta^{m}_s\Delta\hat{\bar{X}}_{Ch}$ as an additional
covariate in our regression specification (see rows 4 to 6 in Panels A and B in
\Cref{tab:simulation_china_factor}). This is in line with our theory, since
conditioning on this control restores the independence assumption on the
shifters. The takeaway form the results in Panels A and B in
\Cref{tab:simulation_china_factor} is thus that, if one thinks that the true
data generating process for the sectoral shifters $\{\Xs_{s}\}_{s=1}^{S}$
corresponds to the model in \cref{eq: shiftstrucapp}, then one should obtain a
consistent estimate of $\eta_s\Delta\hat{\bar{X}}_{Ch}$ and control for it in
the regression specification in order to ensure that the shifters are
independent conditional on the controls, so that
\begin{NoHyper}\Cref{item:indepX}\end{NoHyper} holds once we condition on the
control vector $Z_{i}$.

\begin{table}[!tp]
\centering
\caption{Simulation for common China shock with heterogeneous sectoral exposure}\label{tab:simulation_china_factor}
\tabcolsep=3pt
\begin{tabular}{@{}cc rc cccc cccc @{}}
  \toprule
  &  &\multicolumn{2}{c}{Estimate} & \multicolumn{4}{c}{Median eff.\ s.e.} &
\multicolumn{4}{c}{Rejection rate of $H_{0}\colon\beta=0$}\\ [1pt]
\cmidrule(rl){3-4} \cmidrule(rl){5-8}\cmidrule(rl){9-12}
   $\kappa$ & Control for &Mean & Std.\ dev  & Robust & Cluster & AKM  & AKM0 & Robust & Cluster & AKM  & AKM0 \\
  & $\eta^{m}_s\Delta\hat{\bar{X}}_{Ch}$  & \multicolumn{1}{c}{(1)} & \multicolumn{1}{c}{(2)} & \multicolumn{1}{c}{(3)} & \multicolumn{1}{c}{(4)}
& \multicolumn{1}{c}{(5)} & \multicolumn{1}{c}{(6)} & \multicolumn{1}{c}{(7)} & \multicolumn{1}{c}{(8)} & \multicolumn{1}{c}{(9)}
& \multicolumn{1}{c}{(10)} \\
  \midrule

\multicolumn{12}{@{}l}{\textbf{Panel A\@: $\eta^{m}_{s}=\hat{\eta}_{s}$ for all $m$ and $s$; $u^{m}_{s}\sim \mathcal{N}(0,5)$}} \\ [2pt]

0       &       No      &       $0.00$  &       $0.17$  &       $0.08$  &       $0.09$  &       $0.14$  &       $0.17$  &       $35.4$\%        &       $31.3$\%        &       $10.3$\%        &       $3.9$\% \\
1       &       No      &       $0.00$  &       $0.15$  &       $0.07$  &       $0.07$  &       $0.12$  &       $0.14$  &       $38.2$\%        &       $33.6$\%        &       $12.4$\%        &       $5.1$\% \\
3       &       No      &       $0.00$  &       $0.09$  &       $0.04$  &       $0.04$  &       $0.06$  &       $0.07$  &       $42.2$\%        &       $35.9$\%        &       $17.5$\%        &       $8.4$\% \\
0       &       Yes     &       $0.00$  &       $0.16$  &       $0.08$  &       $0.08$  &       $0.14$  &       $0.16$  &       $34.9$\%        &       $31.6$\%        &       $10.4$\%        &       $4.2$\% \\
1       &       Yes     &       $0.00$  &       $0.16$  &       $0.08$  &       $0.08$  &       $0.14$  &       $0.16$  &       $35.1$\%        &       $31.8$\%        &       $10.5$\%        &       $4.3$\% \\
3       &       Yes     &       $0.00$  &       $0.16$  &       $0.08$  &       $0.08$  &       $0.14$  &       $0.16$  &       $34.9$\%        &       $31.9$\%        &       $10.5$\%        &       $4.3$\% \\ [6pt]

\multicolumn{12}{@{}l}{\textbf{Panel B\@: $\eta^{m}_{s}=\hat{\eta}_{s}$ for all $m$ and $s$; $u^{m}_{s}\sim F_{emp}$}} \\ [2pt]

0       &       No      &       $0.00$  &       $0.43$  &       $0.20$  &       $0.21$  &       $0.35$  &       $0.49$  &       $36.5$\%        &       $33.2$\%        &       $12.1$\%        &       $3.5$\% \\
1       &       No      &       $0.00$  &       $0.26$  &       $0.11$  &       $0.12$  &       $0.18$  &       $0.21$  &       $42.3$\%        &       $36.3$\%        &       $17.3$\%        &       $8.2$\% \\
3       &       No      &       $0.00$  &       $0.10$  &       $0.04$  &       $0.05$  &       $0.07$  &       $0.08$  &       $43.9$\%        &       $37.3$\%        &       $18.8$\%        &       $9.4$\% \\
0       &       Yes     &       $0.00$  &       $0.43$  &       $0.19$  &       $0.21$  &       $0.34$  &       $0.46$  &       $36.7$\%        &       $33.7$\%        &       $12.7$\%        &       $3.9$\% \\
1       &       Yes     &       $0.00$  &       $0.43$  &       $0.19$  &       $0.21$  &       $0.34$  &       $0.46$  &       $36.0$\%        &       $33.1$\%        &       $12.3$\%        &       $3.7$\% \\
3       &       Yes     &       $0.00$  &       $0.43$  &       $0.19$  &       $0.21$  &       $0.34$  &       $0.46$  &       $36.3$\%        &       $33.4$\%        &       $12.3$\%        &       $3.6$\% \\ [6pt]

\multicolumn{12}{@{}l}{\textbf{Panel C\@: $(\eta^{m}_{s},u^{m}_{s})\sim F_{emp}$}} \\ [2pt]

0       &       No      &       $0.00$  &       $0.43$  &       $0.20$  &       $0.21$  &       $0.35$  &       $0.49$  &       $36.7$\%        &       $33.1$\%        &       $12.0$\%        &       $3.5$\% \\
1       &       No      &       $0.00$  &       $0.26$  &       $0.12$  &       $0.13$  &       $0.22$  &       $0.26$  &       $36.0$\%        &       $32.1$\%        &       $10.5$\%        &       $3.8$\% \\
3       &       No      &       $0.00$  &       $0.10$  &       $0.05$  &       $0.05$  &       $0.09$  &       $0.11$  &       $35.3$\%        &       $31.4$\%        &       $10.3$\%        &       $3.7$\% \\
0       &       Yes     &       $0.00$  &       $0.43$  &       $0.19$  &       $0.21$  &       $0.34$  &       $0.46$  &       $36.2$\%        &       $33.1$\%        &       $12.1$\%        &       $3.5$\% \\
1       &       Yes     &       $0.00$  &       $0.43$  &       $0.19$  &       $0.21$  &       $0.34$  &       $0.46$  &       $37.1$\%        &       $33.5$\%        &       $12.4$\%        &       $3.9$\% \\
3       &       Yes     &       $0.00$  &       $0.42$  &       $0.18$  &       $0.20$  &       $0.32$  &       $0.42$  &       $37.8$\%        &       $34.4$\%        &       $13.5$\%        &       $5.2$\% \\ [2pt]

\bottomrule
\multicolumn{12}{p{6.6in}}{\scriptsize{Notes: We impose the assumption that the year-specific sectoral shifters $IMP_{st}$ are generated from the model in \cref{eq:impChinadecomp}. We compute the estimates of the parameters in this model using \citet{gomez2015}, which implements the estimation approach in \citet{bai2009}. To compute these estimates, we use annual data on exports from China to high-income countries other than the US, $IMP_{st}$, between 1991 and 2007 (i.e.\ the same sectoral exports used to construct the instrumental variable in \citet*{autordornhanson2013}) for all sectors used in our baseline placebo exercise. We use these estimates to construct a treatment variable $X^{m}_{i}\equiv\sum_{s}w_{is}\Xs^{m}_{s}$, with each $\Xs^{m}_{s}$ defined as in \cref{eq: shiftstrucapp}, for every simulated sample $m=1,\dots,30,000$. The different panels impose different assumptions on the distribution of $(\eta^{m}_{s}, u^{m}_{s})$ across sectors and simulated samples. In Panels A and B in \Cref{tab:simulation_china_factor}, we fix $\eta^{m}_s=\hat{\eta}_s$ for every sector $s$ and placebo sample $m$. The placebo simulations whose results we present in these two panels differ in the distribution from which $u_{s}^{m}$ is drawn. In Panel A, we draw $u_s^m$ independently across sectors and placebo samples either from a normal distribution with mean zero and variance equal to five. In Panel B, we draw $u_s^m$ independently  from the distribution of $\hat{\varepsilon}_{s,2007}-\hat{\varepsilon}_{s,1991}$ across sectors, where, for $t=2007$ and $t=1991$, $\hat{\varepsilon}_{st}$ is the estimate of the term $\varepsilon_{st}$ in \cref{eq:impChinadecomp} (in Panel B). The placebo exercises in Panel C of \Cref{tab:simulation_china_factor} differs from that in Panel B in that, in the former, each $\eta_{s}^{m}$ is independently drawn across sectors $s$ and placebo samples $m$ from the distribution of $\hat{\eta}_{s}$ across sectors, where $\hat{\eta}_{s}$ is our estimate of the term $\eta_{s}$ in \cref{eq:impChinadecomp}. In all three panels, we compute the outcome variable as $Y^{m}_{i}=\sum_{s}w_{is}\mu^{m}_{s}$, with $\mu^{m}_{s}$ drawn randomly from a normal distribution with mean zero and variance equal to 5. Given the variables $Y^{m}_{i}$ and $X^{m}_{i}$ for each simulated sample $m$, we compute an estimate of $\beta$ in the regression $Y_{i}=\beta X^{m}_{i}+\epsilon_{i}$ (whenever there is a `No' in the second column) or in the regression $Y_{i}=\beta X^{m}_{i}+\gamma \sum_{s}\eta^{m}_s\Delta\hat{\bar{X}}_{Ch}+\epsilon_{i}$ (whenever there is a `Yes' in the second column). We indicate the median and standard deviation of the OLS estimates of $\beta$ across the simulated samples (columns (1) and (2)), the median effective standard error estimates (columns (3) to (6)), and the percentage of placebo samples for which we reject the null hypothesis $H_{0}\colon\beta=0$ using a 5\% significance level test (columns (7) to (10)). \textit{Robust} is the Eicker-Huber-White standard error; \textit{Cluster} is the standard error that clusters CZs in the same state; \textit{AKM} is the standard error in \begin{NoHyper}\Cref{remark:akm-construction}\end{NoHyper}; \textit{AKM0} is the confidence interval in \begin{NoHyper}\Cref{remark:akm0-construction}\end{NoHyper}.}}
\end{tabular}
\end{table}

In Panel C of \Cref{tab:simulation_china_factor}, instead of holding the loadings
fixed, we draw both $\eta_{s}^{m}$ and $\nu_{s}^{m}$ in each placebo sample $m$
from the empirical distribution of the interactive fixed effects estimates,
independently across $s$. This makes the shifters independent across $s$, so
that, as discussed above, \begin{NoHyper}\Cref{item:indepX} in
  \Cref{sec:no-covariates}\end{NoHyper} holds even without conditioning on
$\eta_s\Delta\hat{\bar{X}}_{Ch}$. As a result, the rejection rates for the
\textit{AKM} and \textit{AKM0} inference procedures reported in Panel C are
similar to those reported in \begin{NoHyper}\Cref{tab:simple_placebo_akm} in
  \Cref{sec:placebo-baseline}\end{NoHyper} and unaffected by the value of the
parameter $\kappa$ in \cref{eq: shiftstrucapp}. In particular, the \textit{AKM0}
inference procedure yields always rejection rates that are very close to 5\%.

\subsubsection{Placebo exercise: accounting for controls in the first-stage regression}

The placebo exercise described
in \begin{NoHyper}\Cref{sec:placeboresults,sec:placebo-exercise}\end{NoHyper}
use the outcome variables $Y_{i}$ and the shares $w_{is}$ used in
\citet*{autordornhanson2013} for the period 2000--2007. The placebo exercise
discussed in \begin{NoHyper}\Cref{app:panel_placebo}\end{NoHyper} gets closer to
the reduced-form empirical specification in \citet*{autordornhanson2013} by
incorporating information on outcome variables and shares both for the period
1990--2000 and for the period 2000--2007. However, these two placebo exercises
implement a specification that differs from that in
\citet*{autordornhanson2013} in that it includes no controls. As argued
in \begin{NoHyper}\Cref{sec:transition_econometrics}\end{NoHyper}, the
overrejection problem affecting robust and state-clustered standard errors that
is documented in the simulations is caused by cross-regional correlation in
residuals across observations with similar shares. The inclusion of controls may
improve the performance these methods, since the controls may soak up some (or
even most) of the cross-regional correlation in the residuals.

\begin{table}[!tp]
\centering
\caption{Placebo exercise for the first-stage regression in \citet*{autordornhanson2013}}\label{tab:ADH_FS_placebo}
\tabcolsep=7pt
\begin{tabular}{l cc cccc cccc}
  \toprule
  &\multicolumn{2}{c}{Estimate} & \multicolumn{4}{c}{Median eff.\ s.e.} &
\multicolumn{4}{c}{Rejection rate $H_{0}\colon \beta=0$}\\ [1pt]
\cmidrule(rl){2-3} \cmidrule(rl){4-7}\cmidrule(rl){8-11}
  &Mean & Std.\ dev  & Robust & Cluster & AKM  & AKM0 & Robust & Cluster & AKM  & AKM0 \\
  & \multicolumn{1}{c}{(1)} & \multicolumn{1}{c}{(2)} & \multicolumn{1}{c}{(3)} & \multicolumn{1}{c}{(4)}
& \multicolumn{1}{c}{(5)} & \multicolumn{1}{c}{(6)} & \multicolumn{1}{c}{(7)} & \multicolumn{1}{c}{(8)} & \multicolumn{1}{c}{(9)}
& \multicolumn{1}{c}{(10)} \\
  \midrule

\multicolumn{11}{l}{\textbf{Panel A\@: No controls}} \\ [2pt]
&       $0.01$  &       $1.73$  &       $0.72$  &       $0.81$  &       $1.63$  &       $1.88$  &       $41.5$\%        &       $36.7$\%        &       $6.5$\% &       $4.0$\% \\ [6pt]

\multicolumn{11}{l}{\textbf{Panel B\@: Controls: ADH IV}} \\ [2pt]
&       $0.01$  &       $1.01$  &       $0.63$  &       $0.63$  &       $0.93$  &       $1.06$  &       $20.6$\%        &       $21.3$\%        &       $7.8$\% &       $4.3$\% \\ [6pt]

\multicolumn{11}{l}{\textbf{Panel C\@: Controls: ADH IV and all controls included in Table 3, col. 6 of  in \citet{autordornhanson2013}}} \\ [2pt]
&       $0.00$  &       $0.68$  &       $0.51$  &       $0.51$  &       $0.64$  &       $0.72$  &       $14.4$\%        &       $14.1$\%        &       $5.6$\% &       $3.8$\% \\ [2pt]

  \bottomrule
  \multicolumn{11}{p{6.5in}}{\scriptsize{Notes: This table indicates the median
  and standard deviation of the OLS estimates of $\beta$ in \begin{NoHyper}\cref{eq:bartikreg}\end{NoHyper} across the placebo samples (columns (1) and (2)), the median effective standard error estimates (columns (3) to (6)), and the percentage of placebo samples for which we reject the null hypothesis $H_{0}\colon\beta=0$ using a 5\% significance level test (columns (7) to (10)). \textit{Robust} is the Eicker-Huber-White standard error; \textit{Cluster} is the standard error that clusters CZs in the same state; \textit{AKM} is the standard error in \begin{NoHyper}\Cref{remark:akm-construction}\end{NoHyper}; \textit{AKM0} is the confidence interval in \begin{NoHyper}\Cref{remark:akm0-construction}\end{NoHyper}. For each inference procedure, the median effective standard error is equal to the median length of the corresponding 95\% confidence interval divided by $2\times1.96$. Results are based on 30,000 placebo samples. In all three panels, each $\Xs_{s}^m$ is $i.i.d$ drawn from a normal distribution with mean zero and variance equal to 5. In Panel A, we introduce no controls in the regression equation. In Panel B, we control for the instrumental variable used in \citet*{autordornhanson2013}; i.e.\ the shift-share aggregator of changes in sectoral exports from China to high-income countries other than the US\@. In Panel C, we control for the instrumental variable used in \citet*{autordornhanson2013} and for the broadest set of controls used in that paper; i.e.\ the set of controls used in column 6 of Table 3 of \citet*{autordornhanson2013}.}}
\end{tabular}
\end{table}

In \Cref{tab:ADH_FS_placebo}, we introduce a placebo sample for the first-stage
regression in \citet*{autordornhanson2013}. In Panel A, when we do not include
any controls, both robust and state-clustered standard errors over-reject the
null hypothesis $H_{0}\colon\beta_{1}=0$. In Panel B, we include as a control
the shift-share instrumental variable used in \citet*{autordornhanson2013}, and
the rejection rate for these procedures decreases to about 20\%. Finally, in
Panel C, we additionally include all controls used in the baseline specification
in \citet*{autordornhanson2013}, and the \textit{Robust} and \textit{Cluster}
rejection rates get closer to 14\%. It can also be seen from
\Cref{tab:ADH_FS_placebo} that the rejection rates for the \textit{AKM} and
\textit{AKM0} procedures are always very close to the 5\% nominal level.

\subsubsection{Additional empirical results}

In \Cref{TabADH_RF,TabADH_2SLS} we extend the results presented
in \begin{NoHyper}\Cref{TabADH} in \Cref{sec:ADH}\end{NoHyper}. Specifically,
\Cref{TabADH_RF,TabADH_2SLS} present results not only for all workers (in Panel
A), but also two subsets of workers: college graduates (in Panel B) and
non-college graduates (in Panel C). Additionally, while the \textit{AKM} and
\textit{AKM0} confidence intervals presented
in \begin{NoHyper}\Cref{TabADH}\end{NoHyper} cluster observations belonging to
the same 3-digit sector in different periods (which we denote in
\Cref{TabADH_RF,TabADH_2SLS} as \textit{AKM (3d cluster)} and \textit{AKM0 (3d
  cluster)}), \Cref{TabADH_RF,TabADH_2SLS} also present \textit{AKM} and
\textit{AKM0} confidence intervals that only cluster on time (denoted as
\textit{AKM (4d cluster)} and \textit{AKM0 (4d cluster)}), and \textit{AKM} and
\textit{AKM0} that treat shifters as independent both across 4-digit sectors and
across time periods (denoted as \textit{AKM (indep.)} and \textit{AKM0
  (indep.)})

There are several takeaways from the results in \Cref{TabADH_RF,TabADH_2SLS}.
First, accounting for the possible correlation in the shifters has only a
minimal impact on the \textit{AKM} confidence intervals (i.e.\ the \textit{AKM
  (indep.)}, \textit{AKM (4d cluster)}, and \textit{AKM (3d cluster)} confidence
intervals are always very similar); the impact on the \textit{AKM0} confidence
intervals is a bit larger but also quite small. Second, while the \textit{AKM}
and \textit{AKM0} confidence intervals are quite similar to the \textit{Robust}
and \textit{Cluster} ones in the case of college graduates (Panel B), they are
much larger for non-college graduates (Panel C). Finally, similarly to what we observed
in \begin{NoHyper}\Cref{TabADH} in \Cref{sec:ADH}\end{NoHyper}, the
\textit{AKM0} confidence interval is not centered around the point estimate: it
includes more values of the parameter to the left of the point estimate than it
does to the right.

\begin{table}[!tp]
\centering
\caption{Effect of Chinese on U.S. Commuting Zones in \citet*{autordornhanson2013}: Reduced-Form Regression}\label{TabADH_RF}
\tabcolsep=0.15cm
\begin{tabular}{@{}lcccccc@{}}
\toprule
&
\multicolumn{3}{c}{Change in the employment share}
&\multicolumn{3}{c}{Change in avg.\ log weekly wage}\\
 \cmidrule(rl){2-4} \cmidrule(rl){5-7}
& All & Manuf. & Non-Manuf. & All & Manuf. & Non-Manuf. \\
& (1) & (2) & (3) & (4) & (5) & (6) \\
\midrule

\multicolumn{7}{l}{\textbf{Panel A\@: All Workers}} \\ [2pt]

$\hat{\beta}$   &       -0.49   &       -0.38   &       -0.11   &       -0.48   &       0.10    &       -0.48   \\
Robust  &       [-0.71,-0.27]   &       [-0.48,-0.28]   &       [-0.31,0.08]    &       [-0.80,-0.16]   &       [-0.50,0.69]    &       [-0.83,-0.13]   \\
Cluster &       [-0.64,-0.34]   &       [-0.45,-0.30]   &       [-0.27,0.05]    &       [-0.78,-0.18]   &       [-0.51,0.70]    &       [-0.81,-0.15]   \\
AKM (indep.)    &       [-0.79,-0.18]   &       [-0.52,-0.24]   &       [-0.33,0.10]    &       [-0.84,-0.12]   &       [-0.47,0.66]    &       [-0.88,-0.08]   \\
AKM0  (indep.)  &       [-1.08,-0.25]   &       [-0.63,-0.26]   &       [-0.51,0.07]    &       [-1.08,-0.15]   &       [-0.91,0.58]    &       [-1.22,-0.15]   \\
AKM (4d cluster)        &       [-0.79,-0.19]   &       [-0.52,-0.23]   &       [-0.33,0.10]    &       [-0.87,-0.09]   &       [-0.49,0.68]    &       [-0.90,-0.07]   \\
AKM0 (4d cluster)       &       [-1.10,-0.26]   &       [-0.66,-0.25]   &       [-0.52,0.07]    &       [-1.16,-0.13]   &       [-0.99,0.59]    &       [-1.28,-0.14]   \\
AKM (3d cluster)        &       [-0.81,-0.17]   &       [-0.52,-0.23]   &       [-0.35,0.12]    &       [-0.88,-0.07]   &       [-0.50,0.69]    &       [-0.93,-0.03]   \\
AKM0 (3d cluster)       &       [-1.24,-0.24]   &       [-0.67,-0.25]   &       [-0.64,0.08]    &       [-1.27,-0.10]   &       [-1.16,0.61]    &       [-1.47,-0.11]   \\ [6pt]

\multicolumn{7}{l}{\textbf{Panel B\@: College Graduates}} \\ [2pt]

$\hat{\beta}$   &       -0.27   &       -0.37   &       0.11    &       -0.48   &       0.29    &       -0.47   \\
Robust  &       [-0.42,-0.12]   &       [-0.48,-0.26]   &       [-0.04,0.25]    &       [-0.82,-0.13]   &       [-0.10,0.68]    &       [-0.83,-0.11]   \\
Cluster &       [-0.39,-0.14]   &       [-0.48,-0.27]   &       [-0.04,0.26]    &       [-0.83,-0.13]   &       [-0.14,0.72]    &       [-0.81,-0.12]   \\
AKM (indep.)    &       [-0.45,-0.09]   &       [-0.50,-0.25]   &       [-0.03,0.24]    &       [-0.82,-0.13]   &       [-0.11,0.69]    &       [-0.83,-0.11]   \\
AKM0  (indep.)  &       [-0.57,-0.11]   &       [-0.56,-0.24]   &       [-0.11,0.24]    &       [-1.00,-0.13]   &       [-0.35,0.68]    &       [-1.07,-0.14]   \\
AKM (4d cluster)        &       [-0.45,-0.09]   &       [-0.51,-0.23]   &       [-0.04,0.25]    &       [-0.85,-0.10]   &       [-0.14,0.72]    &       [-0.85,-0.09]   \\
AKM0 (4d cluster)       &       [-0.58,-0.11]   &       [-0.59,-0.23]   &       [-0.11,0.25]    &       [-1.08,-0.11]   &       [-0.41,0.70]    &       [-1.14,-0.13]   \\
AKM (3d cluster)        &       [-0.45,-0.08]   &       [-0.52,-0.23]   &       [-0.04,0.25]    &       [-0.88,-0.08]   &       [-0.14,0.72]    &       [-0.89,-0.05]   \\
AKM0 (3d cluster)       &       [-0.62,-0.09]   &       [-0.59,-0.20]   &       [-0.17,0.25]    &       [-1.20,-0.08]   &       [-0.46,0.73]    &       [-1.32,-0.09]   \\ [6pt]

\multicolumn{7}{l}{\textbf{Panel C\@: Non-College Graduates}} \\ [2pt]

$\hat{\beta}$   &       -0.70   &       -0.37   &       -0.34   &       -0.51   &       -0.06   &       -0.52   \\
Robust  &       [-1.02,-0.38]   &       [-0.48,-0.25]   &       [-0.60,-0.07]   &       [-0.90,-0.13]   &       [-0.69,0.56]    &       [-0.94,-0.10]   \\
Cluster &       [-0.92,-0.48]   &       [-0.47,-0.26]   &       [-0.55,-0.12]   &       [-0.84,-0.19]   &       [-0.53,0.40]    &       [-0.87,-0.17]   \\
AKM (indep.)    &       [-1.18,-0.22]   &       [-0.55,-0.19]   &       [-0.68,0.01]    &       [-1.08,0.05]    &       [-0.70,0.57]    &       [-1.15,0.11]    \\
AKM0  (indep.)  &       [-1.68,-0.34]   &       [-0.72,-0.23]   &       [-1.01,-0.06]   &       [-1.59,-0.06]   &       [-1.26,0.45]    &       [-1.78,-0.04]   \\
AKM (4d cluster)        &       [-1.17,-0.23]   &       [-0.55,-0.18]   &       [-0.67,0.00]    &       [-1.09,0.06]    &       [-0.70,0.57]    &       [-1.14,0.11]    \\
AKM0 (4d cluster)       &       [-1.69,-0.35]   &       [-0.74,-0.23]   &       [-1.01,-0.07]   &       [-1.64,-0.05]   &       [-1.30,0.45]    &       [-1.80,-0.04]   \\
AKM (3d cluster)        &       [-1.22,-0.18]   &       [-0.55,-0.18]   &       [-0.71,0.04]    &       [-1.10,0.07]    &       [-0.72,0.60]    &       [-1.16,0.13]    \\
AKM0 (3d cluster)       &       [-1.95,-0.32]   &       [-0.79,-0.23]   &       [-1.21,-0.04]   &       [-1.80,-0.04]   &       [-1.55,0.46]    &       [-2.02,-0.02]   \\ [2pt]

  \bottomrule
  \multicolumn{7}{p{6.6in}}{\scriptsize{Notes:  $N = 1,444$ (722 CZs $\times$ two
  time periods). Models are weighted by start of period CZ share of national
  population. All regressions include the full vector of baseline controls in
  ADH\@; i.e.\ those in column 6 of Table 3 in \citet*{autordornhanson2013}. 95\% confidence intervals in square brackets.
  \textit{Robust} is the Eicker-Huber-White standard error; \textit{Cluster} is
  the standard error that clusters of CZs in the same state; \textit{AKM (indep.)} is the
  standard error in \begin{NoHyper}\Cref{remark:akm-construction}\end{NoHyper};
  \textit{AKM (4d cluster)} is the standard error in
  \begin{NoHyper}\cref{eq:se-AKM-cluster}\end{NoHyper}
  with 4-digit SIC clusters; \textit{AKM (3d cluster)} is the standard error in
  \begin{NoHyper}\cref{eq:se-AKM-cluster}\end{NoHyper}
  with 3-digit SIC clusters; \textit{AKM0 (indep.)} is
  the confidence interval in
  \begin{NoHyper}\Cref{remark:akm0-construction}\end{NoHyper}; \textit{AKM0 (4d
  cluster)} is the confidence interval with 4-digit SIC
  clusters described in the last sentence of~\begin{NoHyper}\Cref{sec:clustersectors}\end{NoHyper}; and
  \textit{AKM0 (3d cluster)} is the confidence interval with 3-digit SIC
  clusters described in the last sentence of~\begin{NoHyper}\Cref{sec:clustersectors}\end{NoHyper}.} }
\end{tabular}
\end{table}

\begin{table}[!tp]
\centering
\caption{Effect of Chinese on U.S. Commuting Zones in \citet*{autordornhanson2013}: 2SLS Regression}\label{TabADH_2SLS}
\tabcolsep=0.15cm
\begin{tabular}{@{}lcccccc@{}}

\toprule
&
\multicolumn{3}{c}{Change in the employment share}
&\multicolumn{3}{c}{Change in avg.\ log weekly wage}\\
 \cmidrule(rl){2-4} \cmidrule(rl){5-7}
& All & Manuf. & Non-Manuf. & All & Manuf. & Non-Manuf. \\
& (1) & (2) & (3) & (4) & (5) & (6) \\
\midrule

\multicolumn{7}{l}{\textbf{Panel A\@: All Workers}} \\ [2pt]

$\hat{\beta}$   &       -0.77   &       -0.60   &       -0.18   &       -0.76   &       0.15    &       -0.76   \\
Robust  &       [-1.10,-0.45]   &       [-0.78,-0.41]   &       [-0.47,0.12]    &       [-1.23,-0.29]   &       [-0.81,1.11]    &       [-1.27,-0.25]   \\
Cluster &       [-1.12,-0.42]   &       [-0.79,-0.40]   &       [-0.45,0.10]    &       [-1.26,-0.26]   &       [-0.81,1.11]    &       [-1.28,-0.24]   \\
AKM (indep.)    &       [-1.19,-0.36]   &       [-0.81,-0.38]   &       [-0.50,0.15]    &       [-1.30,-0.22]   &       [-0.76,1.06]    &       [-1.32,-0.20]   \\
AKM0  (indep.)  &       [-1.40,-0.42]   &       [-0.89,-0.39]   &       [-0.65,0.11]    &       [-1.48,-0.23]   &       [-1.14,0.99]    &       [-1.58,-0.25]   \\
AKM (4d cluster)        &       [-1.19,-0.36]   &       [-0.84,-0.36]   &       [-0.50,0.15]    &       [-1.35,-0.17]   &       [-0.80,1.10]    &       [-1.36,-0.17]   \\
AKM0 (4d cluster)       &       [-1.46,-0.43]   &       [-0.96,-0.38]   &       [-0.66,0.12]    &       [-1.61,-0.21]   &       [-1.24,1.03]    &       [-1.69,-0.24]   \\
AKM (3d cluster)        &       [-1.25,-0.30]   &       [-0.84,-0.35]   &       [-0.54,0.18]    &       [-1.37,-0.15]   &       [-0.81,1.11]    &       [-1.42,-0.10]   \\
AKM0 (3d cluster)       &       [-1.69,-0.39]   &       [-1.01,-0.36]   &       [-0.84,0.14]    &       [-1.77,-0.17]   &       [-1.49,1.05]    &       [-1.97,-0.19]   \\ [6pt]

\multicolumn{7}{l}{\textbf{Panel B\@: College Graduates}} \\ [2pt]

$\hat{\beta}$   &       -0.42   &       -0.59   &       0.17    &       -0.76   &       0.46    &       -0.74   \\
Robust  &       [-0.64,-0.20]   &       [-0.81,-0.37]   &       [-0.08,0.41]    &       [-1.29,-0.22]   &       [-0.19,1.11]    &       [-1.29,-0.20]   \\
Cluster &       [-0.67,-0.18]   &       [-0.84,-0.34]   &       [-0.07,0.41]    &       [-1.37,-0.14]   &       [-0.22,1.14]    &       [-1.34,-0.15]   \\
AKM (indep.)    &       [-0.69,-0.16]   &       [-0.83,-0.36]   &       [-0.07,0.40]    &       [-1.30,-0.22]   &       [-0.22,1.14]    &       [-1.28,-0.20]   \\
AKM0  (indep.)  &       [-0.78,-0.16]   &       [-0.87,-0.33]   &       [-0.14,0.40]    &       [-1.44,-0.19]   &       [-0.45,1.13]    &       [-1.47,-0.21]   \\
AKM (4d cluster)        &       [-0.70,-0.15]   &       [-0.85,-0.33]   &       [-0.07,0.41]    &       [-1.34,-0.17]   &       [-0.27,1.18]    &       [-1.31,-0.18]   \\
AKM0 (4d cluster)       &       [-0.82,-0.17]   &       [-0.93,-0.32]   &       [-0.15,0.42]    &       [-1.56,-0.17]   &       [-0.53,1.18]    &       [-1.57,-0.21]   \\
AKM (3d cluster)        &       [-0.71,-0.13]   &       [-0.86,-0.32]   &       [-0.08,0.42]    &       [-1.37,-0.14]   &       [-0.25,1.17]    &       [-1.37,-0.11]   \\
AKM0 (3d cluster)       &       [-0.90,-0.14]   &       [-0.96,-0.27]   &       [-0.23,0.42]    &       [-1.71,-0.13]   &       [-0.61,1.21]    &       [-1.82,-0.15]   \\ [6pt]

\multicolumn{7}{l}{\textbf{Panel C\@: Non-College Graduates}} \\ [2pt]

$\hat{\beta}$   &       -1.11   &       -0.58   &       -0.53   &       -0.81   &       -0.10   &       -0.82   \\
Robust  &       [-1.58,-0.64]   &       [-0.76,-0.40]   &       [-0.93,-0.13]   &       [-1.35,-0.28]   &       [-1.07,0.87]    &       [-1.41,-0.23]   \\
Cluster &       [-1.61,-0.61]   &       [-0.77,-0.39]   &       [-0.94,-0.13]   &       [-1.28,-0.34]   &       [-0.84,0.63]    &       [-1.31,-0.33]   \\
AKM (indep.)    &       [-1.76,-0.47]   &       [-0.83,-0.33]   &       [-1.02,-0.04]   &       [-1.62,0.00]    &       [-1.09,0.89]    &       [-1.71,0.07]    \\
AKM0  (indep.)  &       [-2.12,-0.58]   &       [-0.95,-0.37]   &       [-1.27,-0.11]   &       [-2.01,-0.10]   &       [-1.55,0.79]    &       [-2.20,-0.07]   \\
AKM (4d cluster)        &       [-1.75,-0.47]   &       [-0.85,-0.32]   &       [-1.01,-0.05]   &       [-1.64,0.02]    &       [-1.10,0.90]    &       [-1.72,0.07]    \\
AKM0 (4d cluster)       &       [-2.19,-0.59]   &       [-1.02,-0.36]   &       [-1.29,-0.12]   &       [-2.12,-0.09]   &       [-1.63,0.79]    &       [-2.28,-0.07]   \\
AKM (3d cluster)        &       [-1.86,-0.36]   &       [-0.86,-0.30]   &       [-1.09,0.03]    &       [-1.68,0.06]    &       [-1.14,0.93]    &       [-1.78,0.14]    \\
AKM0 (3d cluster)       &       [-2.62,-0.52]   &       [-1.13,-0.35]   &       [-1.59,-0.07]   &       [-2.43,-0.07]   &       [-2.00,0.79]    &       [-2.69,-0.04]   \\ [2pt]

\bottomrule

  \multicolumn{7}{p{6.6in}}{\scriptsize{Notes: $N = 1,444$ (722 CZs $\times$ two
  time periods). Models are weighted by start of period CZ share of national
  population. All regressions include the full vector of baseline controls in
  ADH\@; i.e.\ those in column 6 of Table 3 in \citet*{autordornhanson2013}. 95\% confidence intervals in square brackets.
  \textit{Robust} is the Eicker-Huber-White standard error; \textit{Cluster} is
  the standard error that clusters of CZs in the same state; \textit{AKM (indep.)} is the
  standard error in \begin{NoHyper}\cref{eq:se-AKM-IV}\end{NoHyper}; \textit{AKM
  (4d cluster)} is the standard error in \begin{NoHyper}\cref{eq:se-AKM-IV}\end{NoHyper} with an adjustment
  analogous to that in \begin{NoHyper}\cref{eq:se-AKM-cluster}\end{NoHyper}
  with 4-digit SIC clusters;
  \textit{AKM (3d cluster)} is the standard error in
  \begin{NoHyper}\cref{eq:se-AKM-IV}\end{NoHyper}
  with
  an adjustment analogous to that in
  \begin{NoHyper}\cref{eq:se-AKM-cluster}\end{NoHyper}
  with d-digit SIC
  clusters; \textit{AKM0 (indep.)} is the confidence interval built using the
  standard error in \begin{NoHyper}\cref{eq:se-AKM-IV}\end{NoHyper} with the residual
  $(I-Z'(Z'Z)^{-1}Z')(Y_{1}-Y_{2}\alpha_{0})$ instead of the estimate
  $\hat{\epsilon}_{\Delta}=(I-Z'(Z'Z)^{-1}Z')(Y_{1}-Y_{2}\hat{\alpha})$;
  \textit{AKM0 (4d cluster)} and \textit{AKM0 (3d cluster)} impose the same
  adjustment to the procedure in \textit{AKM (4d cluster)} and \textit{AKM (3d
  cluster)}, respectively.} }
\end{tabular}
\end{table}

\FloatBarrier%

\subsection{Estimation of inverse labor supply elasticity}\label{app:estlaborsupply}

Shift-share IV regressions have been used extensively to
estimate inverse local labor supply elasticities. Using the notation in \begin{NoHyper}\Cref{sec:model}\end{NoHyper}, we can write the inverse
labor supply in each region $i$ as
\begin{equation}\label{eq:labor_sup_app}
\log \omega_{i} = \tilde{\phi}\log L_{i} - \tilde{\phi}\log v_{i},\qquad\text{with}\qquad\tilde{\phi}\equiv\phi^{-1},
\end{equation}
and, consequently, we can relate log changes in wages and log changes employment rates (or number of employees) for each region $i$ between any two time periods as
\begin{align}\label{eq:labor_sup_app_ch}
\hat{\omega}_{i} = \tilde{\phi}\hat{L}_{i} - \tilde{\phi}(\sum_{g=1}^G \tilde{w}_{ig} \hat{\nu}_g + \hat{\nu}_i).
\end{align}

\subsubsection{Bias in OLS estimate of inverse labor supply elasticity}\label{app:biasOLSestlabsup}

Using data on log changes in wages and employment rates for a set of regions, $\{(\hat{\omega}_{i}, \hat{L}_{i})\}_{i}$, one may consider using OLS to compute an estimate of $\tilde{\phi}$. However, such estimator will be inconsistent. To show this formally, note that, up to a first-order approximation around the initial equilibrium, we can write the change in employment in any given region $i$ as
\begin{equation}\label{eq:emp_change_app}
\hat{L}_i =   \sum_{s=1}^{S} l_{is}^0 \left[ \theta_{is} \hat{\chi}_s + \lambda_i \hat{\mu}_s + \lambda_i \hat{\eta}_{is} \right] + \left(1 - \lambda_i\right)(\sum_{g=1}^G \tilde{w}_{ig} \hat{\nu}_g + \hat{\nu}_i),
\end{equation}
and the change in wages as
\begin{align}\label{eq:wage_change_app}
\hat{\omega}_i=  \tilde{\phi} \sum_{s=1}^{S} l_{is}^0 (\theta_{is} \hat{\chi}_s + \lambda_i \hat{\mu}_s + \lambda_i \hat{\eta}_{is})-\tilde{\phi}\lambda_i(\sum_{g=1}^G \tilde{w}_{ig} \hat{\nu}_g + \hat{\nu}_i).
\end{align}
Using \cref{eq:labor_sup_app_ch}, the probability limit of the OLS estimator of $\tilde{\phi}$,
$\hat{\tilde{\phi}}_{OLS}$, can be written as
\begin{align}\label{eq: plimOLSesttilphi}
plim(\hat{\tilde{\phi}}_{OLS})&=\frac{cov(\hat{\omega}_{i},\hat{L}_{i})}{var(\hat{L}_i)}=\tilde{\phi}+\frac{cov(-\tilde{\phi}(\sum_{g=1}^G \tilde{w}_{ig} \hat{\nu}_g + \hat{\nu}_i), \hat{L}_{i})}{var(\hat{L}_i)},
\end{align}
where $cov(-\tilde{\phi}(\sum_{g=1}^G \tilde{w}_{ig} \hat{\nu}_g + \hat{\nu}_i),\hat{L}_{i})/var(\hat{L}_i)$ captures the asymptotic
bias in $\hat{\tilde{\phi}}_{OLS}$ as an estimator of $\tilde{\phi}$. To
characterize this term, we assume here that the of labor supply shocks
$\{\hat{\nu}_g\}_{g}$ and $\{\hat{\nu}_{i}\}_{i}$ are independent of the vector of all labor demand shocks
$(\{\hat{\chi}_s\}_{s}$, $\{\hat{\mu}_s\}_{s}$, $\{\hat{\eta}_{is}\}_{i, s})$
conditional on the matrix of weights $W\equiv\{l^{0}_{is}\}_{i, s}$ and the
matrix of parameters $B\equiv(\{\beta_{is}\}_{i, s}, \{\lambda_{i}\}_{i})$
\begin{equation}\label{eq:indepshocks}
(\{\hat{\chi}_s\}_{s}, \{\hat{\mu}_s\}_{s}, \{\hat{\eta}_{is}\}_{i, s})\indep (\{\hat{\nu}_g\}_{g}, \{\hat{\nu}_{i}\}_{i})
 \mid (W, B).
\end{equation}
Given this assumption and \cref{eq:emp_change_app}, we can rewrite
$plim(\hat{\tilde{\phi}}_{OLS})$ in \cref{eq: plimOLSesttilphi} as
\begin{align}\label{eq: plimOLSesttilphi_3}
plim(\hat{\tilde{\phi}}_{OLS})&=\tilde{\phi}+\frac{cov(-\tilde{\phi}(\sum_{g=1}^G \tilde{w}_{ig} \hat{\nu}_g + \hat{\nu}_i),(1-\lambda_{i})(\sum_{g=1}^G \tilde{w}_{ig} \hat{\nu}_g + \hat{\nu}_i))}{var(\hat{L}_i)}\nonumber\\
&=\tilde{\phi}-\tilde{\phi}(1-\lambda)\frac{var(\sum_{g=1}^G \tilde{w}_{ig} \hat{\nu}_g + \hat{\nu}_i)}{var(\hat{L}_i)},
\end{align}
where the second equality follows if we additionally assume that elasticity of
labor demand in \begin{NoHyper}\cref{labor_demand_maintext}\end{NoHyper} does
not vary across sectors, $\sigma_{s}=\sigma$ for all $s$, so that
$\lambda_{i}=\lambda$ for all $i$. As indicated
in \begin{NoHyper}\Cref{sec:impacteconshocks}\end{NoHyper}, in this case,
$\lambda\equiv\phi[\phi +\sigma\sum_{s=1}^{S} l_{is}^0]^{-1}$. Thus, if $\sigma>0$ (which guarantees that
$\lambda<1$) and $\tilde{\phi}>0$, then the OLS will underestimate the inverse labor supply elasticity
in the sense that $plim(\hat{\tilde{\phi}}_{OLS})<\tilde{\phi}$.

\subsubsection{Consistency of IV estimate of inverse labor supply elasticity}\label{app:consIVestlabsup}

Using data for a set of regions and sectors on log changes in wages and employment rates $\{(\hat{\omega}_{i}, \hat{L}_{i})\}_{i}$, initial employment shares $\{l^{0}_{is}\}_{i, s}$, and sectoral labor demand shifters $\{\hat{\chi}_{s}\}_{s}$, we can write the probability limit of the IV estimator of $\tilde{\phi}$ that uses $X_{i}\equiv\sum_{s=1}^{S}l^{0}_{is}\hat{\chi}_{s}$ as IV, $\hat{\tilde{\phi}}_{IV}$, as
\begin{align*}
plim(\hat{\tilde{\phi}}_{IV})&=\frac{cov(\hat{\omega}_{i}, X_{i})}{cov(\hat{L}_{i},X_{i})}.
\end{align*}
Given the expressions for $\hat{L}_{i}$ and $\hat{\omega}_{i}$ in \cref{eq:emp_change_app,eq:wage_change_app}, respectively, and the independence assumption in \cref{eq:indepshocks}, we can rewrite
\begin{align*}
plim(\hat{\tilde{\phi}}_{IV})&=\frac{cov(\tilde{\phi}\sum_{s=1}^{S} l_{is}^0 \left[ \theta_{is} \hat{\chi}_s + \lambda_i \hat{\mu}_s + \lambda_i \hat{\eta}_{is} \right] -\tilde{\phi}\lambda_i(\sum_{g=1}^G \tilde{w}_{ig} \hat{\nu}_g + \hat{\nu}_i),X_{i})}{cov(\sum_{s=1}^{S} l_{is}^0 \left[ \theta_{is} \hat{\chi}_s + \lambda_i \hat{\mu}_s + \lambda_i \hat{\eta}_{is} \right] + \left(1 - \lambda_i\right)(\sum_{g=1}^G \tilde{w}_{ig} \hat{\nu}_g + \hat{\nu}_i),X_{i})}\\\
&=\tilde{\phi}\frac{cov(\sum_{s=1}^{S} l_{is}^0 \left[ \theta_{is} \hat{\chi}_s + \lambda_i \hat{\mu}_s +\lambda_i \hat{\eta}_{is} \right],X_{i})}{cov(\sum_{s=1}^{S} l_{is}^0 \left[ \theta_{is} \hat{\chi}_s + \lambda_i \hat{\mu}_s + \lambda_i \hat{\eta}_{is} \right],X_{i})}\\
&=\tilde{\phi}.
\end{align*}
Therefore, under the distributional assumptions in \cref{eq:indepshocks}, the IV
estimator that uses a shift-share instrument that aggregates sector-specific
labor demand shifters is a consistent estimator of the inverse labor supply
elasticity. Notice that the heterogeneity in $\theta_{is}$ does not affect the consistency of $\hat{\tilde{\phi}}$. However, the consistency of $\hat{\tilde{\phi}}$ will depend on the specific labor demand shock being employed by the researcher to construct its shift-share IV being independent of the specific labor supply shocks that have been prevalent in the set of regions belonging the population of interest.

\subsubsection{Evaluation of leave-one-out IV through the lens of the model in Section 3}\label{app:missIVandModel}

We describe in this section how one may use the model
in \begin{NoHyper}\Cref{sec:model}\end{NoHyper} to frame the approach to the
estimation of the inverse labor supply elasticity described
in \begin{NoHyper}\Cref{sec:estlaborsupply}\end{NoHyper}. This approach is
described in general terms
in \begin{NoHyper}\Cref{sec:mismeasuredIV}\end{NoHyper}.

In \begin{NoHyper}\Cref{sec:estlaborsupply}\end{NoHyper}, we focus on the estimation of the inverse labor supply elasticity $\tilde{\phi}$ and we base the estimation of this parameter on the estimating equation
\begin{align}\label{eq:labor_sup_empirical_app}
\hat{\omega}_{i}=\tilde{\phi}\hat{L}_{i}+\delta Z_{i}+\epsilon_{i},\qquad\text{with}\qquad\tilde{\phi}=\phi^{-1}.
\end{align}
For simplicity, we assume here that we use no controls (i.e.\ $\delta=0$) and that, thus, we can rewrite the estimating equation above as
\begin{align}\label{eq:labor_sup_empirical_app_2}
\hat{\omega}_{i}=\tilde{\phi}\hat{L}_{i}+\epsilon_{i},\qquad\text{with}\qquad\tilde{\phi}=\phi^{-1}.
\end{align}
The advantage of focusing on the version without controls is that, in this case, the model in \begin{NoHyper}\Cref{sec:model}\end{NoHyper} clarifies that $\epsilon_{i}=- \tilde{\phi}(\sum_{g=1}^G \tilde{w}_{ig} \hat{\nu}_g + \hat{\nu}_i)$, where $\{\hat{\nu}_{g}\}_{g}$ and $\{\hat{\nu}_{i}\}_{i}$ are labor supply shocks. Thus, in the version without controls, there is a clear mapping between the regression residual of the structural equation, $\epsilon_{i}$, and the labor supply shocks in our economic model.

As discussed in \Cref{app:biasOLSestlabsup}, the OLS estimator of $\tilde{\phi}$
will be biased. However, as discussed in \Cref{app:consIVestlabsup}, one may
obtain a consistent estimate of $\tilde{\phi}$ by computing an IV estimator that
instruments for the log change in employment in region $i$, $\hat{L}_{i}$, using
as an instrument a shift-share aggregator of labor demand shocks
$\{\Xs_{s}\}_{s}$. In terms of the model
in \begin{NoHyper}\Cref{sec:model}\end{NoHyper}, $\Xs_{s}$ is any (possibly
sector $s$-specific) function of the sector $s$-specific labor demand shocks
$\chi_{s}$ and $\mu_{s}$
(see \begin{NoHyper}\cref{labor_demand_maintext,technology_maintext}\end{NoHyper}).
These sector-specific labor demand shocks are in many cases unobserved to the
researcher. In these cases, following \citet{bartik1991benefits} and the
subsequent literature on the estimation of inverse local labor supply
elasticities, it has become typical to estimate $\tilde{\phi}$ using as instruments one of two
different IVs: either a shift-share aggregator of the growth
in national employment in every sector $s$,
\begin{equation}\label{eq:measIVapp}
  X_{i}=\sum_{s=1}^{S}l^{0}_{is}\hat{L}_{s},\qquad\text{with}\qquad\hat{L}_{s}=\sum_{j=1}^{N}\frac{L^{0}_{js}}{\sum_{j'=1}^{N}L^{0}_{j's}}
  \frac{L_{js}^t-L_{js}^0}{L_{js}^0},
\end{equation}
or a shift-share aggregator of the leave-one-out measure of the growth  in national employment in sector $s$,
\begin{align}\label{eq:measIVapp_i}
X_{i,-}=\sum_{s=1}^{S}l^{0}_{is}\hat{L}_{s,-i},\qquad\text{with}\qquad
\hat{L}_{s,-i}=\sum_{j=1, j\neq i}^{N}\frac{L^{0}_{js}}{\sum_{j'=1, j'\neq i}^{N}L^{0}_{j's}}\frac{L_{js}^t-L_{js}^0}{L_{js}^0}.
\end{align}

We focus here on outlining the restrictions that one should impose on the sector-specific labor demand shifters $\{(\hat{\chi}_{s}, \hat{\mu}_{s})\}_{s}$, region- and sector-specific labor demand shifters $\{\hat{\eta}_{is}\}_{i, s}$, group-specific labor supply shifters, $\{\hat{\nu}_{g}\}_{g}$, and region-specific labor supply shifters $\{\hat{\nu}_{i}\}_{i}$ (all of them introduced in the model in \begin{NoHyper}\Cref{sec:model}\end{NoHyper}) so that the IV estimator that uses $X_{i,-}$ as an instrument yields a consistent estimate of $\tilde{\phi}$.

The variable $X_{i,-}$
in \cref{eq:measIVapp_i} is a valid instrument as long as we can write
\begin{align}\label{eq:xisapp}
\hat{L}_{is}=\Xs_{s}+\psi_{is},
\end{align}
and the following restrictions hold
\begin{align}
E[\Xs_{s}|\hat{\omega}(0), \hat{L}(0), L^{0}]&=E[\Xs_{s}], &\qquad&\text{for all $s$,}\label{eq:rest1}\\
E[l^{0}_{is}\psi_{is}|\hat{\omega}_{-i}(0), \hat{L}_{-i}(0), L^{0}]&=0,&\qquad&\text{for all $i$ and $s$,}\label{eq:rest2}\\
E[l^{0}_{is}\psi_{is}l^{0}_{js}\psi_{js}|\hat{\omega}_{-i}(0), \hat{L}_{-i}(0), L^{0}]&=0,&\qquad&\text{for all $i\neq j$ and $s$,}\label{eq:rest3}
\end{align}
where $\hat{\omega}_{-i}(0)$ ($\hat{L}_{-i}(0)$) denotes the change in wages (employment shares) in every region other than $i$ when the sectoral shock of interest equals 0 for all sectors (i.e.\ $\Xs_{s}=0$ for all $s$), and $L^{0}$ is the vector all region- and sector-specific shares in the initial equilibrium (i.e.\ $L^{0}=\{l^{0}_{is}\}_{i,s}$).

According to the model in \begin{NoHyper}\Cref{sec:model}\end{NoHyper}, we can express the changes in employment in sector $s$ in a region $i$ as
\begin{equation*}
\hat{L}_{is}=-\sigma_{s}\hat{\omega}_{i}+\rho_{s}\hat{\chi}_{s}+\hat{\mu}_{s}+\hat{\eta}_{is}.
\end{equation*}
Combining this expression with the expression for $\hat{\omega}_{i}$ in \cref{eq:wage_change_app} in \Cref{app:biasOLSestlabsup}, we can rewrite the change in employment in sector $s$ and region $i$ approximately as
\begin{equation}\label{eq:appLis}
\hat{L}_{is}=-\sigma_{s}\tilde{\phi}\sum_{s'=1}^{S} l_{is'}^0 \left[\theta_{is'} \hat{\chi}_{s'} + \lambda_i \hat{\mu}_{s'}+\lambda_i \hat{\eta}_{is'} \right] +\sigma_{s}\tilde{\phi}\lambda_i(\sum_{g=1}^G \tilde{w}_{ig} \hat{\nu}_g + \hat{\nu}_i)+\rho_{s}\hat{\chi}_{s}+\hat{\mu}_{s}+\hat{\eta}_{is},
\end{equation}
with $\lambda_i \equiv \phi \left[\phi + \sum_{s=1}^{S} l_{is}^0 \sigma_s\right]^{-1}$, $\theta_{is} \equiv \rho_s \lambda_i$, and $\tilde{\phi}=\phi^{-1}$.

Without imposing any restrictions on the values of the labor demand and supply elasticities, the expression for $\hat{L}_{is}$ in \cref{eq:appLis} will not satisfy the restrictions in \cref{eq:xisapp} to \cref{eq:rest3}. To illustrate this point, we can map the different terms in \cref{eq:appLis} into those in \cref{eq:xisapp} as
\begin{align}
\Xs_{s}&=  \rho_{s}\hat{\chi}_{s}+\hat{\mu}_{s},\label{eq: Xss_1}\\
\psi_{is}&= -\sigma_{s}\tilde{\phi}\sum_{s'=1}^{S} l_{is'}^0 \left[\theta_{is'} \hat{\chi}_{s'} + \lambda_i \hat{\mu}_{s'}+\lambda_i \hat{\eta}_{is'} \right] +\sigma_{s}\tilde{\phi}\lambda_i(\sum_{g=1}^G \tilde{w}_{ig} \hat{\nu}_g + \hat{\nu}_i)+\hat{\eta}_{is}.\label{eq: psiis_1}
\end{align}
Under this definition of the labor demand shock $\Xs_{s}$, the potential outcomes $\hat{\omega}_{i}(0)$ and $\hat{L}_{i}(0)$ are
\begin{align}
\hat{\omega}_{i}(0)&=\tilde{\phi}\sum_{s=1}^{S} l_{is}^0\lambda_i \hat{\eta}_{is}  -\tilde{\phi}\lambda_i(\sum_{g=1}^G \tilde{w}_{ig} \hat{\nu}_g + \hat{\nu}_i),\label{eq: omega0i_1}\\
\hat{L}_{i}(0)&=\sum_{s=1}^{S} l_{is}^0\lambda_i \hat{\eta}_{is} + \left(1 - \lambda_i\right)(\sum_{g=1}^G \tilde{w}_{ig} \hat{\nu}_g + \hat{\nu}_i).\label{eq: L0i_1}
\end{align}
Given the expressions in \cref{eq: psiis_1,eq: omega0i_1,eq: L0i_1}, the restriction on $\psi_{is}$ in \cref{eq:rest3} will not be satisfied: for any two regions $i$ and $i'$, $\psi_{is}$ and $\psi_{i's}$ are a function of the same set of sectoral demand shocks $\{\hat{\chi}_s\}_{s}$ and $\{\hat{\mu}_s\}_{s}$ and, thus, $\psi_{is}$ and $\psi_{i's}$ will generally be correlated with each other. Thus, unless additional restrictions are imposed, the IV estimator that uses the variable described in \cref{eq:measIVapp_i} as instrument for $\hat{L}_{i}$ in \cref{eq:labor_sup_empirical_app_2} will not be a consistent estimator of $\tilde{\phi}$.

However, under the restriction that $\sigma_{s}=0$ for every sector $s$,
the expression for $\hat{L}_{is}$ in \cref{eq:appLis} will satisfy the
restrictions in \cref{eq:xisapp} to \cref{eq:rest3}. In this case,
\begin{align}
\Xs_{s}&=  \rho_{s}\hat{\chi}_{s}+\hat{\mu}_{s},\label{eq: Xss_2}\\
\psi_{is}&= \hat{\eta}_{is},\label{eq: psiis_2}
\end{align}
and $\hat{\omega}_{i}(0)$ and $\hat{L}_{i}(0)$ correspond to the expressions in \cref{eq: omega0i_1} and \cref{eq: L0i_1}. Thus, if the sector-specific labor demand shocks $\{(\hat{\chi}_{s},\hat{\mu}_{s})\}_{s}$ are mean independent of the region-specific labor supply shocks $\{\hat{\nu}_{g}\}_{g}$ and $\{\hat{\nu}_{i}\}_{i}$ as well as of the region- and sector-specific labor demand shocks $\{\hat{\eta}_{is}\}_{i, s}$, the restriction in \cref{eq:rest1} will hold. Additionally, under the additional assumption that $\eta_{is}$ is mean zero and uncorrelated with $\eta_{js}$ for every $i\neq j$ and $s$, the restrictions in \cref{eq:rest2,eq:rest3} will hold. Thus, if these additional restrictions on the model in \begin{NoHyper}\Cref{sec:model}\end{NoHyper} hold, the IV estimator that uses the variable described in \cref{eq:measIVapp_i} as instrument to estimate $\tilde{\phi}$ in \cref{eq:labor_sup_empirical_app_2} will be consistent.

There are two alternative instrumental variables that do not use data on
any specific labor demand shock and that lead to consistent estimates
of the inverse labor supply elasticity $\tilde{\phi}$ under weaker restrictions
than those needed for the instrument in \cref{eq:measIVapp_i} to be valid.

First, conditional on a calibrated value of $\sigma_{s}$ for every sector $s$,
one may estimate $\tilde{\phi}$ using as an instrument for $\hat{L}_{i}$ the
following leave-one-out estimator:
\begin{equation}\label{eq:measIVapp_i_2}
  \tilde{X}_{i,-}=\sum_{s=1}^{S}l^{0}_{is}\hat{\tilde{L}}_{s, -i},\qquad\text{with}
  \qquad\hat{\tilde{L}}_{s, -i}=\sum_{j=1, j\neq i}^{N}\frac{L^{0}_{js}}{\sum_{j'=1, j'\neq i}^{N}L^{0}_{j's}}\hat{\tilde{L}}_{js},
  \quad\text{and}\quad\hat{\tilde{L}}_{is}=\hat{L}_{is}-\sigma_{s}\hat{\omega}_{i}.
\end{equation}
Combining the expression for $\hat{L}_{is}$ in \cref{eq:appLis} and the
expression for $\hat{\omega}_i$ in \cref{eq:wage_change_app}, we can write
\begin{equation}\label{eq:hattildeLis}
  \hat{\tilde{L}}_{is}=\rho_{s}\hat{\chi}_{s}+\hat{\mu}_{s}+\hat{\eta}_{is}.
\end{equation}
Thus, we can define $\Xs_{s}$ and $\psi_{is}$ as in \cref{eq: Xss_2} and
\cref{eq: psiis_2}. Consequently, as discussed above,
\cref{eq:rest1,eq:rest2,eq:rest3} will hold if: (a) the sector-specific labor
demand shocks $\{(\hat{\chi}_{s}, \hat{\mu}_{s})\}_{s}$ are mean independent of
the region-specific labor supply shocks $\{\hat{\nu}_{g}\}_{g}$ and $\{\hat{\nu}_{i}\}_{i}$ as well as of the region-
and sector-specific labor demand shocks $\{\hat{\eta}_{is}\}_{i, s}$; and (b)
$\eta_{is}$ is mean zero and uncorrelated with $\eta_{js}$ for every $i\neq j$
and $s$. Thus, under these two sets of assumptions, the IV estimator that
uses the variable described in \cref{eq:measIVapp_i_2} as instrument for
$\hat{L}_{i}$ in \cref{eq:labor_sup_empirical_app_2} will be a consistent
estimator of $\tilde{\phi}$ no matter what the value of the labor demand elasticities $\{\sigma_{s}\}_{s}$ is.

Second, under the assumption that the labor demand elasticity is constant across
sectors (i.e.\ $\sigma_{s}=\sigma$ for every $s$), the residual from projecting
$\hat{L}_{is}$, as defined in \cref{eq:appLis}, on a set of region-specific fixed
effects is equivalent to $\hat{\tilde{L}}_{is}$, as defined in
\cref{eq:hattildeLis}. Therefore, once we define $\Xs_{s}$ and $\psi_{is}$ as in
\cref{eq: Xss_2} and \cref{eq: psiis_2}, the IV estimator that uses the
variable described in \cref{eq:measIVapp_i_2} as an instrument for $\hat{L}_{i}$ in
\cref{eq:labor_sup_empirical_app_2} will be a consistent estimator of
$\tilde{\phi}$ if two assumptions hold: (a) the sector-specific labor demand
shocks $\{(\hat{\chi}_{s}, \hat{\mu}_{s})\}_{s}$ are mean independent of the
region-specific labor supply shocks $\{\hat{\nu}_{g}\}_{g}$ and $\{\hat{\nu}_{i}\}_{i}$ as well as of the region- and
sector-specific labor demand shocks $\{\hat{\eta}_{is}\}_{i, s}$; and (b)
$\eta_{is}$ is mean zero and uncorrelated with $\eta_{js}$ for every $i$, $j$,
and $s$.

\subsubsection{Placebo exercise}\label{app:placebo_leave_one_out}

In this section, we implement a placebo exercise to evaluate the finite-sample properties of our suggested inference procedures when
using the shift-share IVs introduced in \begin{NoHyper}\Cref{sec:mismeasuredIV}\end{NoHyper}. For each placebo sample $m=1,\dots,30,000$, we
construct sector- and region-specific shocks $X_{is}^m = \Xs_s^m + \psi_{is}^m$, where
$\Xs_s^m$ and $\psi_{is}^m$ are independently drawn from normal distributions
with variances equal to 5 and 10, respectively. We then use data on employment
shares of U.S. CZs by 4-digit manufacturing sectors, $\{w_{is}\}_{i=1,s=1}^{N, S}$
to compute
\begin{align*}
Y_{i2} &= \sum_{s=1}^{S}w_{is}X_{is}^m,&
Y_{i1} &= \rho\sum_{s=1}^{S}w_{is}\psi_{is}^m+\sum_{s=1}^{S}w_{is}A_{s}^m,
\end{align*}
where $A_{s}^m$ is independently drawn from a normal distribution with variance equal to 20.

Our goal is to estimate the effect  $\alpha$ of $Y_{i2}^{m}$ on $Y_{i1}^{m}$,
\begin{equation}\label{eq:leave_one_out_reg}
Y_{i1}^m = Y_{i2}^m\alpha + \epsilon_i^m.
\end{equation}
Note that, by the above construction, $\alpha = 0$. Therefore, the residual is
$\epsilon_i^m = Y^{m}_{i1}= \rho\sum_{s=1}^{S} w_{is}\psi_{is}^m+\sum_{s=1}^{S}w_{is}A_{s}^m$,
which indicates that there is a potential endogeneity problem stemming from the
fact that $\psi^{m}_{is}$ affects both $Y^{m}_{i1}$ and $Y^{m}_{i2}$ whenever $\rho \neq 0$.

We consider three different shift-share IVs. First, we consider the IV constructed directly with the shock $\Xs_s^m$:
\begin{equation*}
X_{i}^m =\sum_{s=1}^S w_{is}\Xs_{s}^m.
\end{equation*}
Second, we consider an IV constructed with the aggregate growth in $X_{is}$:
\begin{equation*}
\hat{X}_{i}^m =\sum_{s=1}^S w_{is} \hat{\Xs}_{s}^m \quad \text{such that} \quad \hat{\Xs}_{s}^m \equiv \sum_{i=1}^N\left(\frac{\check{w}_{is}}{\sum_{j=1}^N\check{w}_{js}}\right)X_{is}^m
\end{equation*}
where $\check{w}_{is} = L_{is}^0/\sum_{j=1}^{N} L_{js}^0$ is the share of CZ $i$ in the national employment of sector $s$ in 1990. Third, we consider an IV constructed with leave-one-out aggregate growth in $X_{is}$:
\begin{equation*}
\hat{X}_{i,-}^m =\sum_{s=1}^S w_{is} \hat{\Xs}_{s, -i}^m \quad \text{such that} \quad \hat{\Xs}_{s, -i}^m \equiv \sum_{j=1, j\neq i}^N\left(\frac{\check{w}_{js}}{\sum_{o=1,j\neq i}^N\check{w}_{os}}\right)X_{js}^m.
\end{equation*}

The instruments $X_{i}^m$ and $\hat{X}_{i,-}^m$ are always valid in our setting. However, whenever $\rho \neq 0$, the instrument $\hat{X}_{i}^m$ is invalid since $\{\psi^{m}_{is}\}_{s=1}^{S}$ affect
$\hat{X}_{i}^m$ and $\epsilon_i$.

\afterpage{\begin{landscape}
\begin{table}[!tp]
\centering
\caption{Mismeasured shifter: impact on standard errors and rejection rates.}\label{tab: leave_one_out_placebo}
\tabcolsep=5pt
\begin{tabular}{cccccccccrcccrc}
  \toprule
 $\rho$  &\multicolumn{2}{c}{Estimate} & \multicolumn{6}{c}{Median eff.\ s.e.} &
\multicolumn{6}{c}{Rejection rate of $H_{0}\colon\beta=0$}  \\ [1pt]
\cmidrule(rl){1-1} \cmidrule(rl){2-3} \cmidrule(rl){4-9}\cmidrule(rl){10-15} \cmidrule(rl){12-15}
 & Median & eff.\ s.e. & Robust & Cluster & AKM  & AKM0 & AKM & AKM0 & Robust & Cluster & AKM  & AKM0 & AKM & AKM0 \\
  & &   &  &  &   &  & \multicolumn{2}{c}{Leave-one-out} &  &  &  &  & \multicolumn{2}{c}{Leave-one-out} \\
 \cmidrule(rl){8-9}  \cmidrule(rl){14-15}
  & \multicolumn{1}{c}{(1)} & \multicolumn{1}{c}{(2)} & \multicolumn{1}{c}{(3)} & \multicolumn{1}{c}{(4)}
& \multicolumn{1}{c}{(5)} & \multicolumn{1}{c}{(6)} & \multicolumn{1}{c}{(7)} & \multicolumn{1}{c}{(8)} & \multicolumn{1}{c}{(9)}
& \multicolumn{1}{c}{(10)} & \multicolumn{1}{c}{(11)} & \multicolumn{1}{c}{(12)} & \multicolumn{1}{c}{(13)} & \multicolumn{1}{c}{(14)} \\
\midrule

\multicolumn{15}{@{}l}{\textbf{Panel A\@: Unfeasible shift-share IV}} \\ [2pt]
$0$     &       $0.00$  &       $0.35$  &       $0.16$  &       $0.18$  &       $0.29$  &       $0.37$  &       $-$     &       $-$     &       $0.33$  &       $0.29$  &       $0.09$  &       $0.04$  &       $-$     &       $-$     \\
$5$     &       $0.01$  &       $0.97$  &       $0.77$  &       $0.76$  &       $0.80$  &       $1.05$  &       $-$     &       $-$     &       $0.08$  &       $0.09$  &       $0.08$  &       $0.04$  &       $-$     &       $-$     \\
$10$    &       $-0.02$ &       $1.86$  &       $1.53$  &       $1.50$  &       $1.52$  &       $2.01$  &       $-$     &       $-$     &       $0.07$  &       $0.08$  &       $0.08$  &       $0.04$  &       $-$     &       $-$     \\ [6pt]

\multicolumn{15}{@{}l}{\textbf{Panel B\@: Shift-share IV with aggregate sector-level growth}} \\ [2pt]
$0$     &       $0.00$  &       $0.23$  &       $0.11$  &       $0.12$  &       $0.20$  &       $0.22$  &       $-$     &       $-$     &       $0.33$  &       $0.29$  &       $0.09$  &       $0.04$  &       $-$     &       $-$     \\
$5$     &       $1.66$  &       $0.50$  &       $0.39$  &       $0.40$  &       $0.45$  &       $0.51$  &       $-$     &       $-$     &       $0.89$  &       $0.89$  &       $0.85$  &       $0.79$  &       $-$     &       $-$     \\
$10$    &       $3.30$  &       $0.93$  &       $0.77$  &       $0.77$  &       $0.84$  &       $0.95$  &       $-$     &       $-$     &       $0.91$  &       $0.90$  &       $0.88$  &       $0.83$  &       $-$     &       $-$     \\ [6pt]

\multicolumn{15}{@{}l}{\textbf{Panel C\@: Shift-share IV with aggregate sector-level growth (leave-one-out)}} \\ [2pt]
$0$     &       $0.00$  &       $0.36$  &       $0.17$  &       $0.18$  &       $0.30$  &       $0.40$  &       $0.31$  &       $0.32$  &       $0.32$  &       $0.29$  &       $0.09$  &       $0.04$  &       $0.08$  &       $0.03$  \\
$5$     &       $-0.07$ &       $1.06$  &       $0.82$  &       $0.80$  &       $0.87$  &       $1.24$  &       $0.93$  &       $0.95$  &       $0.07$  &       $0.08$  &       $0.06$  &       $0.05$  &       $0.05$  &       $0.03$  \\
$10$    &       $-0.16$ &       $2.03$  &       $1.62$  &       $1.59$  &       $1.66$  &       $2.37$  &       $1.78$  &       $1.80$  &       $0.06$  &       $0.07$  &       $0.06$  &       $0.05$  &       $0.05$  &       $0.03$  \\ [6pt]

\bottomrule
\multicolumn{15}{p{8.2in}}{\scriptsize{Notes: This table reports the median
  and effective standard error estimates of the IV estimates of $\alpha$ in
  \cref{eq:leave_one_out_reg} across the placebo samples (columns (1) and (2)),
  the median effective standard error estimates (columns (3) to (8)), and the
  percentage of placebo samples for which we reject the null hypothesis
  $H_{0}\colon\beta=0$ using a 5\% significance level test (columns (9) to
  (14)). For each value of $\rho$, we generate 30,000 simulated samples with
  sector-region shocks $X_{is}^m = \Xs_s^m + \psi_{is}^m$, where $\Xs_s^m \sim
  \mathcal{N}(0,5)$  and $\psi_{is}^m \sim \mathcal{N}(0,10)$. We then construct
  $Y_{i2} = \sum_{s=1}^{S} w_{is}X_{is}^m$ and $Y_{i1} = \rho\sum_s w_{is}\psi_{is}^m+\sum_s w_{is}A_{s}^m$, where $A_{s}^m \sim \mathcal{N}(0,20)$.
\textit{Robust} is the Eicker-Huber-White standard error; \textit{Cluster} is the standard error that clusters CZs in the same state; \textit{AKM} is the standard error in \begin{NoHyper}\Cref{remark:akm-construction}\end{NoHyper}; \textit{AKM0} is the confidence interval in \begin{NoHyper}\Cref{remark:akm0-construction}\end{NoHyper}. \textit{AKM (leave-one-out)} and \textit{AKM0 (leave-one-out)} are the versions of \textit{AKM} and \textit{AKM0} in \begin{NoHyper}\Cref{sec:mismeasuredIV}\end{NoHyper}. For each inference procedure, the median effective standard error is equal to the median length of the corresponding 95\% confidence interval divided by $2\times1.96$.}}
\end{tabular}
\end{table}
\end{landscape}}

\Cref{tab: leave_one_out_placebo} reports the results of this placebo exercise
for different values of $\rho$. In Panel
A, we report results using $X_i^m$ as an instrument; we denote this instrument as the ``infeasible'' IV, as its construction requires observing the shifters $\{\Xs_{s}\}_{s}$. As
expected, for all values of $\rho$, the median $\hat{\alpha}^{m}$ across placebo samples is zero.
Because of the shift-share structure of $\epsilon_i$, robust and state-clustered
standard error estimators underestimate the variability of the estimates, while
\textit{AKM} and \textit{AKM0} inference procedures yield good coverage. Panel B presents the results based on the
feasible shift-share IV $\hat{X}_{i}^m$. When
using this IV, higher levels of $\rho$ yield higher average estimates of
$\alpha$. This follows from the endogeneity problem created by the fact that
$\{\psi^{m}_{is}\}_{s}$ are part of both the dependent variable, $Y^{m}_{i1}$, and the instrument, $\hat{X}_{i}^m$. Finally, Panel C presents results based on the leave-one-out IV $\hat{X}_{i,-}^m$. This instrument is not affected by an endogeneity problem, as it does not use information on region $i$-specific shocks $\{\psi^{m}_{is}\}_{s}$ when constructing the region $i$-specific variable $\hat{X}_{i,-}^m$. Thus, the average of the IV estimates of $\alpha$ that use $\hat{X}_{i,-}^m$ as an instrument is also very close to zero for all values of $\rho$. The results in Panel C also show that the leave-one-out versions of the \textit{AKM} and \textit{AKM0} inference procedures (see \begin{NoHyper}\Cref{sec:mismeasuredIV}\end{NoHyper}) yield slightly larger median effective standard errors than the baseline versions of the \textit{AKM} and \textit{AKM0} procedures (see \begin{NoHyper}\Cref{sec:ivreg}\end{NoHyper}).  In this particular application, the magnitude of the adjustment is modest: the implied rejection rates for the null hypothesis $H_{0}\colon\alpha=0$ differ by less than 2 percentage points.

\subsubsection{Additional results}\label{sec:additional-results-elasticity}

\Cref{tab:laborsupply_controls} reports estimates of the inverse labor supply
elasticity with alternative sets of controls. Column (2) replicates the
estimates of Panels A and B in column (3)
of \begin{NoHyper}\Cref{tab:laborsupply}\end{NoHyper}.
\Cref{tab:laborsupply_controls} shows that these results are robust to
controlling (a) only for period dummies (column (1)); (b) for period dummies and the proxies for region-specific labor supply shocks included in
\citet{amiormanning2018} (column (3)); (c) for period dummies, the controls included in \citet*{autordornhanson2013} and the proxies for region-specific labor supply shocks in
\citet{amiormanning2018} (column (4)).

\begin{table}[!tp]
\centering
\caption{Estimation of inverse labor supply elasticity: robustness with different control sets}\label{tab:laborsupply_controls}
\begin{tabular}{@{}lcccc@{}}
\toprule
 & (1) & (2) & (3) & (4) \\

\midrule

\multicolumn{4}{@{}l}{\textbf{Panel A\@: Bartik IV, Not leave-one-out estimator}} \\ [2pt]

 $\hat{\beta}$  &       $0.75$  &       $0.8$   &       $0.83$  &       $0.8$   \\
Robust  &       $[0.48,1.03]$   &       $[0.64,0.97]$   &       $[0.56,1.1]$    &       $[0.64,0.96]$   \\
Cluster &       $[0.44,1.07]$   &       $[0.60,1.01]$   &       $[0.55,1.11]$   &       $[0.59,1.02]$   \\
AKM     &       $[0.59,0.92]$   &       $[0.62,0.98]$   &       $[0.60,1.06]$   &       $[0.62,0.98]$   \\
AKM0            &       $[0.56,0.95]$   &       $[0.59,1.02]$   &       $[0.61,1.21]$   &       $[0.59,1.01]$   \\ [6pt]

\multicolumn{4}{@{}l}{\textbf{Panel B\@: Bartik IV, Leave-one-out estimator}} \\ [2pt]

 $\hat{\beta}$  &       $0.76$  &       $0.82$  &       $0.83$  &       $0.81$  \\
Robust  &       $[0.48,1.03]$   &       $[0.65,0.98]$   &       $[0.56,1.10]$   &       $[0.65,0.98]$   \\
Cluster &       $[0.43,1.08]$   &       $[0.60,1.03]$   &       $[0.55,1.11]$   &       $[0.59,1.04]$   \\
AKM     &       $[0.59,0.92]$   &       $[0.62,1.01]$   &       $[0.58,1.08]$   &       $[0.63,1.00]$   \\
AKM0            &       $[0.57,0.96]$   &       $[0.60,1.07]$   &       $[0.59,1.28]$   &       $[0.60,1.04]$   \\
AKM (\textit{leave-one-out})    &       $[0.59,0.92]$   &       $[0.61,1.02]$   &       $[0.58,1.08]$   &       $[0.62,1.01]$   \\
AKM0  (\textit{leave-one-out})  &       $[0.56,0.97]$   &       $[0.59,1.09]$   &       $[0.59,1.29]$   &       $[0.59,1.06]$   \\ [6pt]

\textbf{Controls:} \\   [2pt]
$\quad$ Period dummies  &       Yes     &       Yes     &       Yes     &       Yes     \\
$\quad$ Controls in \citet{autordornhanson2013}        &       No      &       Yes     &       No      &       Yes     \\
$\quad$ Controls in \citet{amiormanning2018}    &       No      &       No      &       Yes     &       Yes     \\ [6pt]
\bottomrule
  \multicolumn{5}{p{6.3in}}{\scriptsize{Notes:  $N = 1,444$ (722 CZs $\times$ 2
  time periods). The dependent variable is the log-change in mean weekly earnings in CZ $i$, and the regressor is the log-change in the employment rate in CZ $i$. Observations are weighted by the 1980 CZ share of national population. 95\% confidence intervals in square brackets. \textit{Robust} is the Eicker-Huber-White standard error; \textit{Cluster} is the standard error that clusters of CZs in the same state; \textit{AKM} is the standard error in \begin{NoHyper}\cref{eq:se-AKM-cluster}\end{NoHyper}
  with 3-digit SIC clusters; \textit{AKM0} is the confidence interval with 3-digit
  SIC clusters described in \begin{NoHyper}\Cref{sec:clustersectors}\end{NoHyper}; \textit{AKM (leave-one-out)} is the standard error in
  \begin{NoHyper}\cref{sec:mismeasuredIV}\end{NoHyper} with 3-digit SIC clusters; \textit{AKM0 (leave-one-out)} is the confidence interval with 3-digit SIC clusters described in
  \begin{NoHyper}\cref{sec:mismeasuredIV}\end{NoHyper}.
  Baseline controls in \citet{autordornhanson2013} are the controls in column 6
  of Table 3 in ADH\@. Amenity controls in \citet{amiormanning2018}: binary
  indicator for presence of coastline, three temperature indicators, log
  population density in 1900, log distance to the closest CZ\@.}}
\end{tabular}
\end{table}

\FloatBarrier%

\section{Effect of immigration on U.S. local labor markets}\label{appsec:applicationBP}

To complement the empirical applications discussed
in \begin{NoHyper}\Cref{sec:empapp}\end{NoHyper}, we present here the results of
estimating of the impact of immigration on labor market outcomes in the US\@. To
this end, we estimate the model
\begin{equation}\label{imm_empirical}
  Y_{it} = \beta \Delta ImmShare_{it} + Z_{it}'\delta + \epsilon_{it},
\end{equation}
where, for observation or cell $i$, $Y_{it}$ is the change in a labor market
outcome for native workers between years $t$ and $t-10$, $\Delta ImmShare_{it}$
is the change in the share of immigrants in total employment between years
$t$ and $t-10$, and $Z_{it}$ is a control vector that includes fixed effects.

Following \citet*{dustmann2016different}, one may classify different approaches to the estimation
of $\beta$ in \cref{imm_empirical} on the basis of the definition of the cell $i$: in
the \textit{skill-cell approach}, $i$ corresponds to an education-experience
cell defined at the national level \citep*[e.g.][]{borjas2003}; in the
\textit{spatial approach}, $i$ corresponds to a region
\citep*[e.g.][]{altonji1991effects}; in the \textit{mixed approach}, $i$
corresponds to the intersection of a region and an occupation, or a region and
an education group \citep*[e.g.][]{card2001immigrant}.

In the \textit{spatial}
and \textit{mixed} approaches, since \citet*{altonji1991effects} and
\citet*{card2001immigrant}, it has become common to instrument for the change
in the immigrant share $\Delta ImmShare_{it}$ using a shift-share IV\@:
\begin{equation}\label{ImmIV}
  X_{it} = \sum_{g=1}^{G} ImmShare_{igt_{0}}\frac{\Delta Imm_{gt}}{Imm_{gt_{0}}},
\end{equation}
where $g$ indexes countries (or groups of countries) of origin of immigrants,
and $t_{0}$ is some pre-sample or beginning-of-the-sample time period. The
variable $ImmShare_{igt_{0}}$ plays the role of the share $w_{is}$ in
\begin{NoHyper}\cref{eq:bartikreg}\end{NoHyper} and denotes the share of immigrants from origin $g$ in total
immigrant employment in cell $i$ in year $t_{0}$; the ratio
$\Delta Imm_{gt}/Imm_{gt_{0}}$ plays the role of the shifter $\Xs_{s}$ in
\begin{NoHyper}\cref{eq:bartikreg}\end{NoHyper}, with $\Delta Imm_{gt}$ denoting the change in the total
number of immigrants coming from origin $g$ between years $t$ and $t-10$, and
$Imm_{gt_{0}}$ denoting the total number of immigrants from region $g$ at the
national level in year $t_{0}$.

When estimating the parameter of interest $\beta$ in \cref{imm_empirical}, the researcher must make a choice on the sample period or time frame of the analysis, and on the $G$ countries (or areas) of origin used to construct the shift-share IV\@. In \Cref{appsec:listcountries}, we discuss two different sample periods previously used in the literature, and present a list of areas of origin of immigrants for which information is available in each of the two sample periods. In \Cref{appsec:placeboimm}, we present placebo evidence that illustrates the finite-sample properties of the different inference procedures when applied to the two sample periods discussed in \Cref{appsec:listcountries} and when using different sets of countries of origin of immigrants to construct the shift-share IV in \cref{ImmIV}. The main conclusion that arises from these placebo simulations is that restricting the set of countries of origin used in the construction of the shift-share IV to those with a relatively small value of $ImmShare_{igt_{0}}$ generally improves the finite-sample coverage of all different inference procedures. Consequently, in \Cref{appsec:resrescountries}, we present estimates of $\beta$ in \cref{imm_empirical} that use information on a restricted set of countries when building the shift-share IV in \cref{ImmIV}. For the sake of comparison, in \Cref{appsec:resallcountries}, we present estimates that use information on all countries for which information on immigration flows into the US is available for the relevant sample period.

In \Cref{appsec:resrescountries,appsec:resallcountries}, we present estimates of specifications that follow either the \textit{spatial approach} or the \textit{mixed approach}. In all specifications, information on all variables entering \cref{imm_empirical,ImmIV} comes from the Census Integrated Public Use Micro Samples for 1980--2000 and the American Community Survey for 2008--2012. In all regressions, the vector of controls $Z_{it}$ includes period dummies and, when implementing the \textit{mixed approach}, we also add occupation- or education-group-specific dummies to the vector $Z_{i}$. All tables referenced in this section are included at the end.

\subsection{Sample periods and list of countries of origin of immigrants}\label{appsec:listcountries}

The results we present use one of two time frames. The first one uses information on
immigrant shares (i.e.\ the variable $ImmShare_{ig t_{0}}$
in \cref{ImmIV}) measured in
1980, and information on the outcome variables, endogenous treatment, and
shifters of interest (i.e.\ the variables $Y_{it}$, $\Delta ImmShare_{it}$ and
$\Delta Imm_{gt}$ in \cref{imm_empirical,ImmIV}) for the periods 1980--1990, 1990--2000, and
2000--2010. \Cref{tab:origin_country_1980} lists all countries or areas of origin that we consider for
which information on the number of immigrants in the U.S. is available for all
periods in this time frame (i.e. 1980, 1990, 2000, and 2010).

The second time frames uses information on immigrant shares measured in 1960,
and information on the outcome variables, endogenous treatment, and
shifters of interest for the period 1970--1980. \Cref{tab:origin_country_1960}
lists all countries or areas of origin that we consider for which information on the number of immigrants
in the U.S. is available for all periods in this time frame (i.e. 1960,
1970 and 1980).

In both \Cref{tab:origin_country_1980,tab:origin_country_1960}, we have marked
in italics those countries or areas of origin that account for a relatively large share
(larger than 3\%) of the overall immigrant U.S. population in the corresponding
base year (this base year is 1980 for \Cref{tab:origin_country_1980} and 1960
for \Cref{tab:origin_country_1960}).

\subsection{Placebo simulations}\label{appsec:placeboimm}

In \Cref{tab:imm_3percent_placebo,tab:imm_all_placebo}, we present the results
of placebo exercises that illustrate the properties of different inference
procedures for the parameter on the shift-share covariate
in \cref{ImmIV} in regressions
of labor market outcomes for native workers on this
shift-share covariate. The only difference
between the analysis in \Cref{tab:imm_3percent_placebo} and the analysis in
\Cref{tab:imm_all_placebo} is in the set of areas of origin of immigrants used to
construct the shift-share covariate in \cref{ImmIV}.
While the former uses information only on those countries of origin whose total
share of immigrants in the corresponding baseline year $t_{0}$ (either 1960 or
1980, depending on the specification) is below 3\% (i.e.\ it uses information only on those countries of
origin $g$ that satisfy
$\sum_{i=1}^{N}ImmShare_{igt_{0}}/\sum_{i=1}^{N}\sum_{g'=1}^{G}ImmShare_{ig't_{0}}\leq
0.03$),
the latter uses information on all areas of origin of immigrants listed in the tables described in \Cref{appsec:listcountries}.

We present results for four outcome variables: the change in employment
($\Delta\log E_{i}$) and average wages ($\Delta\log w_{i}$) across all native
workers, and the change in average wages for high-skill and low-skill workers.
For each of the four outcome variables, we consider several
regressions in which we vary both the definition of a cell or unit of observation, and the sample
period. The first four rows of each panel in
\Cref{tab:imm_3percent_placebo,tab:imm_all_placebo} implement a purely
\textit{spatial approach}, defining each unit of observation as a commuting zone
(CZ) or as a metropolitan statistical area (MSA). The last four rows follow a
\textit{mixed approach}, defining each unit as the intersection of a CZ and either one of the fifty occupations defined in
\citet{bursteinhansontianvogel2018} (CZ-50 Occ.), one of seven aggregate
occupations defined similarly to \citet*{card2001immigrant} (CZ-7 Occ.), or
one of two education groups (CZ-Educ.). In terms of sample periods, we explore
two alternatives. We either define the weights in 1980 and measure the outcome
variable as the 1980--1990, 1990--2000, and 2000--2010 changes in log employment or log wages or, alternatively, we measure the weights in 1960 and measure
the outcome variable as the 1970--1980 change in the variable of interest.

\Cref{tab:imm_3percent_placebo,tab:imm_all_placebo} yield three key takeaways.
First, robust standard errors are generally biased downward,
leading frequently to an overrejection problem.

Second, when we construct the shift-share covariate in \cref{ImmIV} relying only on countries of
origin with relatively small shares of U.S. immigrant population in the baseline year,
state-clustered standard errors yield adequate rejection rates when the unit of
observation is defined as the intersection of a CZ and fifty detailed
occupation groups, shares are measured in 1980, and the outcome is defined as
the subsequent three decadal changes. In all other cases, inference procedures
based on state-clustered standard errors tend to overreject.

Third, the
\textit{AKM} and \textit{AKM0} inference procedures perform much better when the
shift-share covariate in \cref{ImmIV} is constructed using only countries of origin
with relatively small shares of U.S. immigrant population in the baseline year, so
that \begin{NoHyper}\Cref{item:sector-size,item:sector-size-normality}\end{NoHyper}
more plausibly hold. Furthermore, these inference procedures also tend to perform better in
specifications that apply a \textit{mixed approach} than in those that apply a
purely \textit{spatial approach}. One possible explanation for this pattern is that our asymptotics require that
the number of observations $N\to\infty$; thus, the behavior of the \textit{AKM} and \textit{AKM0}  inference
procedures is generally better in samples with a larger number of observations,
and the \textit{mixed approach}, which intersects each region with several
occupations or education groups, yields larger sample sizes. Importantly, while the \textit{AKM} inference procedure may still lead to confidence intervals that are too short in several specifications, the \textit{AKM0} inference procedure generally yields
accurate rejection rates. However, confidence intervals based on the \textit{AKM0} inference
procedure may be very conservative for certain specifications.

\subsection{Results with a restricted set of origin countries}\label{appsec:resrescountries}

All results presented in this section exploit information only on those
countries of origin whose total share of immigrants in the corresponding
baseline year $t_{0}$ (either 1960 or 1980, depending on the specification) is
below 3\%. More precisely, these results presented here are computed using an IV
such as that in \cref{ImmIV} constructed excluding those countries of origin $g$
for which
$\sum_{i=1}^{N}ImmShare_{igt_{0}}/\sum_{i=1}^{N}\sum_{g'=1}^{G}ImmShare_{ig't_{0}}>
0.03$. We exclude large origin countries so
that \begin{NoHyper}\Cref{item:sector-size,item:sector-size-normality}\end{NoHyper}
more plausibly hold. The simulations in \Cref{appsec:placeboimm} also suggest
that excluding large origin countries should lead to better finite-sample
performance of the inference procedures that we
propose.\footnote{\label{fn:3pc}Our theory currently does not provide guidance
  on the particular threshold that one should choose. While we find that the 3\%
  threshold works well in the placebo exercises in this particular application,
  we leave the question of what threshold one should in general pick to ensure
  that \begin{NoHyper}\Cref{item:sector-size,item:sector-size-normality}\end{NoHyper}
  plausibly hold to future research.}

\Cref{tab:immigration_3percent} presents results for three different
implementations of the \textit{mixed approach}. In all three
cases, the data comes from a three-period panel with $t=\{1990, 2000, 2010\}$ and
$t_{0}=1980$. The implementations differ in the
definition of a cell. In columns (1) to (4) of
\Cref{tab:immigration_3percent}, a cell corresponds to the intersection of a CZ
and one of the 50 occupations defined in Appendix F of
\citet{bursteinhansontianvogel2018}. In columns (5) and (6), we define a cell as
the intersection of a CZ and one of two education groups: high school-equivalent
or college-equivalent educated workers \citep*[see][]{card2009immigrant}. In
columns (7) to (10), a cell corresponds to the intersection of a CZ and one of
seven aggregate occupations \citep*[see][]{card2001immigrant}.\footnote{We group
  the 50 disaggregated occupations used in \citet{bursteinhansontianvogel2018}
  into seven aggregate occupations: laborers, farm workers and low-skilled
  service workers; operatives and craft workers; clerical workers; sales
  workers; managers; professional and technical workers; and others.\label{fn:busteinocc}}

Although \Cref{tab:immigration_3percent} adopts occupational definitions that build on
  those in \citet{bursteinhansontianvogel2018} and \citet{card2001immigrant},
  our specifications do not exactly match their definition of shares and
  shifters. Thus, our estimates should not be viewed as a test of the robustness
  of the results presented in these studies. Furthermore, no matter which definition of cell we use, when interpreting our
  estimates, one should bear in mind that, as discussed in
  \citet*{jaeger2018shift}, these may conflate the short- and the long-run
  responses to immigration shocks.

The magnitude and statistical significance of the estimates of $\beta$ in \cref{imm_empirical} is generally consistent across the
specifications studied in \Cref{tab:immigration_3percent}. In terms of the impact of immigration on native employment, we find that a one percentage point increase in the share of immigrants in
total employment reduces the number of native workers employed by 1.19--1.49\%,
with all estimates of $\beta$ being statistically different from zero at the 5\%
level for all four inference procedures that we consider. In terms of the impact of immigration on natives' average weekly wages, we find that the estimated impact of an
increase in the immigrant share is not statistically different from zero at the
5\% significance level according to the \textit{AKM} and \textit{AKM0} CIs; this is true for all three cell definitions and no matter whether we compute average wages for all workers, only for high-skill workers or only for low-skill workers. \textit{Robust} and \textit{Cluster} CIs also indicate that the effect of immigration on natives' average weekly wages is not statistically different from zero at the 5\% significance level when each cell corresponds to the intersection of a CZ and an education group, but these standard inference procedures sometimes predict that immigration has a positive effect on the wages of high-skill workers when occupations are used to define the unit of analysis (see columns (3) and (9) in \Cref{tab:immigration_3percent}).

While all inference procedures broadly agree in the statistical significance (at the 5\% significance level) of the impact of immigration on natives' labor market outcomes, there is considerable heterogeneity across specifications in the length of the \textit{AKM} and \textit{AKM0} confidence intervals
relative to those based on \textit{Robust} and \textit{Cluster} standard errors.
In columns (1) to (4), which use detailed occupations to define cells, $AKM$ and
$AKM0$ CIs tend to be very similar (in some cases, even
slightly smaller) to those based on state-clustered standard errors, although
they are generally much larger than those based on robust standard errors. In
contrast, for the other two cell definitions, the IV $AKM$ and $AKM0$ CIs are
on average, 200\% and 356\% wider than those based on state-clustered standard
errors, and the reduced-form $AKM$ and $AKM0$ CIs are on average 228\% and 358\%
wider than those based on state-clustered standard errors. Similarly, the CIs
for the first-stage coefficient, reported in Panel C, $AKM$ and $AKM0$ CIs are more than twice as
wide as \textit{Robust} and \textit{Cluster} CIs.\footnote{The results in \Cref{tab:immigration_3percent} are consistent with the placebo simulation results
in \Cref{tab:imm_3percent_placebo}, which show that state-clustered
standard errors lead to rejection rates that are very close to the nominal level
when a cell is defined as the intersection of CZs and 50 occupations,
but lead to overrejection for the other two cell definitions.}

To understand why standard inference procedures may lead to overrejection of the null hypothesis of no effect in certain cases, recall from the discussion in
\begin{NoHyper}\Cref{sec:no-covariates}\end{NoHyper} that robust and
state-clustered standard errors may be biased downward even if there is no shock
in the structural residual that varies exactly at the same
level as the shifters of interest; a downward bias will arise so long as there
is a shift-share component in the residual with shares that have a correlation
structure similar to that of the shares used to construct the shift-share
instrument. We present simulations that illustrate this point in \Cref{app: placebo_other_share}.

\Cref{tab:immigration_3percent_addres_CZ,tab:immigration_3percent_addres_MSA}
present results for different versions of the \textit{spatial approach}, using
CZs and MSAs as unit of observation, respectively. We present both estimates
that measure the immigrant shares $ImmShare_{igt_{0}}$ in 1980 and use data on
shifters and outcomes for the periods 1980--1990, 1990--2000, and 2000--2010,
and estimates that measure the immigrant shares in 1960 and use data on outcomes
only for the period 1970--1980. While the first sample definition mimics
that in \Cref{tab:immigration_3percent}, the second one is suggested in
\citet*{jaegerruiststuhler2018} as being more robust to potential bias in the
estimates of $\beta$ that arise from the combination of serial correlation in
the shifters $\Delta Imm_{gt}$ and the potentially slow adjustment of labor
market outcomes to these immigration shocks.

The placebo simulation results for the different specifications
considered in
\Cref{tab:immigration_3percent_addres_CZ,tab:immigration_3percent_addres_MSA}
(see \Cref{tab:imm_3percent_placebo}) reveal that, due to the relatively
small number of observations (i.e.\ small number of MSAs and CZs; small value of
$N$) and, in the case of the specification that relies on immigrants shares
measured in 1960, the relatively small number of countries of origin of
immigrants (i.e.\ small value of $G$), only the \textit{AKM0} inference
procedure consistently yields rejection rates that are close to the nominal level
of 5\%. However, the \textit{AKM0} procedure yields CIs with an
implied median effective standard error that is much larger than the true
standard deviation of the estimator. It is thus conservative. Thus, the
placebo results suggest that, for most of the specifications considered in
\Cref{tab:immigration_3percent_addres_CZ,tab:immigration_3percent_addres_MSA},
the \textit{AKM0} CIs may be conservative and the \textit{Robust}, \textit{Cluster} and \textit{AKM} CIs may be
too small. It is thus not surprising that, for the
different specifications considered in
\Cref{tab:immigration_3percent_addres_CZ,tab:immigration_3percent_addres_MSA},
the \textit{AKM0} CIs are much larger than than those implied by the other three
inference procedures.\footnote{This is particularly noticeable for the IV results in
Panel A\@; however, to interpret these CIs, one should bear in mind that, as the
first-stage results in Panel C show, the shift-share IV is weak in these
specifications. In the presence of weak IVs, only the \textit{AKM0} confidence
interval remains valid in general (see discussion
in \begin{NoHyper}\Cref{sec:ivreg}\end{NoHyper}).}

Finally, \Cref{tab:immigration_3percent_pvalue,tab:immigration_3percent_pvalues_addres_CZ,tab:immigration_3percent_pvalues_addres_MSA}
report p-values for the null hypothesis of no effect for all specifications considered
in \Cref{tab:immigration_3percent,tab:immigration_3percent_addres_CZ,tab:immigration_3percent_addres_MSA}, respectively.

\subsection{Results with all origin countries}\label{appsec:resallcountries}

\Cref{tab:immigration_all,tab:immigration_all_addres_CZ,tab:immigration_all_addres_MSA,tab:immigration_all_pvalue,tab:immigration_all_pvalues_addres_CZ,tab:immigration_all_pvalues_addres_MSA}
present results analogous to those
in \Cref{tab:immigration_3percent,tab:immigration_3percent_addres_CZ,tab:immigration_3percent_addres_MSA,tab:immigration_3percent_pvalue,tab:immigration_3percent_pvalues_addres_CZ,tab:immigration_3percent_pvalues_addres_MSA},
respectively. While the latter set of tables, as described in
\Cref{appsec:resrescountries}, use a shift-share instrument that
excludes countries of origin that account for more than 3\% of the overall
immigrant population in the baseline year, the former uses all areas of
origin of immigrants listed in \Cref{tab:origin_country_1980,tab:origin_country_1960}.

As the results of placebo simulations presented in \Cref{tab:imm_all_placebo}
show, using all countries of origin to construct the shift-share instrumental
variable of interest results in the \textit{Robust}, \textit{Cluster} and
\textit{AKM} standard errors underestimating the sampling variability of the
estimator of interest. \Cref{tab:imm_all_placebo} also shows that not excluding
any country of origin from the construction of the instrument
in \cref{ImmIV} results in the \textit{AKM0}
inference procedure being too conservative: the 95\% confidence interval often
has an infinite length and the rejection rates are generally much smaller than the 5\% nominal rate.

Given the relatively poor performance of all inference procedures in the placebo
simulations, one should use caution when extracting conclusions from the
estimates presented in
\Cref{tab:immigration_all,tab:immigration_all_addres_CZ,tab:immigration_all_addres_MSA,tab:immigration_all_pvalue,tab:immigration_all_pvalues_addres_CZ,tab:immigration_all_pvalues_addres_MSA}.

\begin{table}[t!]
        \centering
        \caption{Origin countries (1980 weights)}\label{tab:origin_country_1980}
        \begin{tabular}{llll}
          \toprule
          Afghanistan     &       France  &       Liechtenstein and Lux.  &       Scandinavia     \\
          Africa  &       Greece  &       Malaysia        &       Scotland        \\
          Albania         &       Gulf States     &       Maldives        &       Singapore       \\
          Andorra and Gibraltar   &       India   &       Malta   &       South America   \\
          Austria         &       Indonesia       &       \textit{Mexico}  &       Spain   \\
          Belgium         &       Iran    &       Nepal   &       Switzerland     \\
          Brunei  &       Iraq    &       Netherlands     &       Syria   \\
          Cambodia        &       Ireland         &       Oceania         &       Thailand        \\
          \textit{Canada}  &       Israel/Palestine        &       \textit{Other}   &       Turkey  \\
          Central America         &       Italy   &       Other Europe    &       Vietnam         \\
          China   &       Japan   &       Other USSR and Russia   &       Wales   \\
          \textit{Cuba and West Indies}    &       Jordan  &       Philippines     &       Yemen   \\
          Cyprus  &       Korea   &       Portugal        &               \\
          \textit{Eastern Europe}  &       Laos    &       Rest of Asia    &               \\
          \textit{England}         &       Lebanon         &       Saudi Arabia    &               \\
          \bottomrule
          \multicolumn{4}{p{5.8in}}{\scriptsize{Notes: In italics, countries
          that are dropped from the sample when considering only countries whose
          share is below 3\%; i.e.\ those countries in italics are countries of
          origin $g$ such that $(\sum_{i=1}^{N}ImmShare_{ig,1980}/\sum_{i=1}^{N}\sum_{g'=1}^{G}ImmShare_{ig',1980})>0.03$.}}
\end{tabular}
\end{table}%

\bigskip

\begin{table}[!tp]
        \centering
        \caption{Origin countries (1960 weights)}\label{tab:origin_country_1960}
        \begin{tabular}{llll}
        \toprule
     &       France  &       Liechtenstein and Lux.  &       Scandinavia     \\
Africa  &       Greece  &               &       Scotland        \\
Albania         &           &               &              \\
   &       India   &         &       South America   \\
Austria         &              &       \textit{Mexico}  &       Spain   \\
Belgium         &           &         &       Switzerland     \\
  &          &       Netherlands     &       Syria   \\
        &       Ireland         &       Oceania         &              \\
\textit{Canada}  &       Israel/Palestine        &       \textit{Other}   &       Turkey  \\
Central America         &       \textit{Italy}   &       Other Europe    &                \\
China   &       Japan   &       \textit{Other USSR and Russia}   &       Wales   \\
Cuba and West Indies    &        &       Philippines     &          \\
  &       Korea   &       Portugal        &               \\
\textit{Eastern Europe}  &           &       Rest of Asia    &               \\
\textit{England}         &       Lebanon         &         &               \\
\bottomrule
\multicolumn{4}{p{5.8in}}{\scriptsize{Notes: In italics, countries that are dropped from the sample when considering only countries whose share is below 3\%; i.e.\ those countries in italics are countries of origin $g$ such that $(\sum_{i=1}^{N}ImmShare_{ig1960}/\sum_{i=1}^{N}\sum_{g'=1}^{G}ImmShare_{ig'1960})>0.03$.}}

\end{tabular}
\end{table}%

\begin{table}[!tp]
\centering
\caption{Reduced-form placebo with origin countries below 3\% of total immigrant share}\label{tab:imm_3percent_placebo}
\footnotesize
\begin{tabular}{lcccccrrrr}
\toprule
                && \multicolumn{4}{c}{Median eff.\ s.e.} & \multicolumn{4}{c}{Rejection rate for $H_{0}\colon \beta=0$ at 5\%}                  \\
 \cmidrule(rl){3-6} \cmidrule(rl){7-10}
 && \multicolumn{1}{c}{{$\Delta \log E_i$}} & \multicolumn{3}{c}{{$\Delta \log w_i$}} & \multicolumn{1}{c}{{$\Delta \log E_i$}} & \multicolumn{3}{c}{{$\Delta \log w_i$}}         \\
 \cmidrule(rl){3-3} \cmidrule(rl){4-6} \cmidrule(rl){7-7} \cmidrule(rl){8-10}
 && All & All  & High-Skill & Low-Skill & \multicolumn{1}{c}{All} & \multicolumn{1}{c}{All}  & \multicolumn{1}{c}{High-Skill} & \multicolumn{1}{c}{Low-Skill} \\
Unit obs. & Weights & (1) & (2) & (3) & (4) & \multicolumn{1}{c}{(5)} & \multicolumn{1}{c}{(6)} & \multicolumn{1}{c}{(7)} & \multicolumn{1}{c}{(8)} \\
 \midrule

\multicolumn{9}{l}{\textbf{Panel A\@: Standard deviation of placebo estimate}} \\ [1pt]

CZ      &       1980    &       $1.14$  &       $0.46$  &       $0.30$  &       $1.04$  &               &               &               &               \\
CZ      &       1960    &       $0.96$  &       $0.81$  &       $0.80$  &       $1.51$  &               &               &               &               \\
MSA     &       1980    &       $1.94$  &       $0.49$  &       $0.74$  &       $0.83$  &               &               &               &               \\
MSA     &       1960    &       $1.99$  &       $0.25$  &       $0.33$  &       $0.71$  &               &               &               &               \\
CZ-50 Occ.      &       1980    &       $0.17$  &       $0.06$  &       $0.07$  &       $0.06$  &               &               &               &               \\
CZ-7 Occ.       &       1980    &       $1.59$  &       $0.40$  &       $0.47$  &       $0.42$  &               &               &               &               \\
CZ-Educ.        &       1980    &       $2.88$  &       $0.54$  &       ---       &       ---       &               &               &               &               \\ [2pt]

\multicolumn{9}{l}{\textbf{Panel B\@: Robust standard error}} \\ [1pt]

CZ      &       1980    &       $0.49$  &       $0.21$  &       $0.20$  &       $0.25$  &       $37.82$\%       &       $43.17$\%       &       $12.45$\%       &       $73.21$\%       \\
CZ      &       1960    &       $0.77$  &       $0.31$  &       $0.32$  &       $0.39$  &       $9.24$\%        &       $47.70$\%       &       $37.09$\%       &       $68.84$\%       \\
MSA     &       1980    &       $1.82$  &       $0.33$  &       $0.35$  &       $0.37$  &       $8.12$\%        &       $20.42$\%       &       $43.80$\%       &       $42.74$\%       \\
MSA     &       1960    &       $1.47$  &       $0.24$  &       $0.27$  &       $0.30$  &       $20.12$\%       &       $2.73$\%        &       $3.62$\%        &       $48.74$\%       \\
CZ-50 Occ.      &       1980    &       $0.11$  &       $0.05$  &       $0.06$  &       $0.06$  &       $20.22$\%       &       $12.46$\%       &       $8.57$\%        &       $4.51$\%        \\
CZ-7 Occ.       &       1980    &       $0.47$  &       $0.16$  &       $0.19$  &       $0.18$  &       $63.04$\%       &       $44.33$\%       &       $41.78$\%       &       $45.89$\%       \\
CZ-Educ.        &       1980    &       $0.65$  &       $0.15$  &       ---       &       ---       &       $67.22$\%       &       $66.12$\%       &       ---       &       ---       \\ [2pt]

\multicolumn{9}{l}{\textbf{Panel C\@: State-clustered standard error}} \\ [1pt]

CZ      &       1980    &       $0.70$  &       $0.26$  &       $0.27$  &       $0.34$  &       $23.76$\%       &       $30.73$\%       &       $5.44$\%        &       $62.41$\%       \\
CZ      &       1960    &       $1.03$  &       $0.46$  &       $0.48$  &       $0.60$  &       $2.01$\%        &       $26.87$\%       &       $12.68$\%       &       $51.41$\%       \\
MSA     &       1980    &       $2.11$  &       $0.29$  &       $0.32$  &       $0.43$  &       $3.42$\%        &       $24.52$\%       &       $49.14$\%       &       $36.30$\%       \\
MSA     &       1960    &       $1.62$  &       $0.20$  &       $0.23$  &       $0.35$  &       $17.72$\%       &       $7.59$\%        &       $12.52$\%       &       $41.45$\%       \\
CZ-50 Occ.      &       1980    &       $0.14$  &       $0.07$  &       $0.06$  &       $0.06$  &       $9.21$\%        &       $5.24$\%        &       $5.01$\%        &       $1.64$\%        \\
CZ-7 Occ.       &       1980    &       $0.64$  &       $0.23$  &       $0.26$  &       $0.25$  &       $49.70$\%       &       $27.94$\%       &       $27.90$\%       &       $29.90$\%       \\
CZ-Educ.        &       1980    &       $0.98$  &       $0.23$  &       ---       &       ---       &       $50.31$\%       &       $50.35$\%       &       ---       &       ---       \\ [2pt]

\multicolumn{9}{l}{\textbf{Panel D\@: AKM standard error}} \\ [1pt]

CZ      &       1980    &       $0.91$  &       $0.40$  &       $0.24$  &       $0.95$  &       $12.01$\%       &       $23.27$\%       &       $5.40$\%        &       $30.31$\%       \\
CZ      &       1960    &       $0.73$  &       $0.67$  &       $0.62$  &       $1.37$  &       $11.27$\%       &       $15.35$\%       &       $9.30$\%        &       $20.06$\%       \\
MSA     &       1980    &       $1.60$  &       $0.39$  &       $0.62$  &       $0.73$  &       $25.50$\%       &       $15.31$\%       &       $22.26$\%       &       $17.92$\%       \\
MSA     &       1960    &       $1.39$  &       $0.20$  &       $0.26$  &       $0.55$  &       $22.88$\%       &       $11.71$\%       &       $7.66$\%        &       $23.57$\%       \\
CZ-50 Occ.      &       1980    &       $0.15$  &       $0.06$  &       $0.06$  &       $0.05$  &       $11.63$\%       &       $10.77$\%       &       $8.16$\%        &       $10.02$\%       \\
CZ-7 Occ.       &       1980    &       $1.49$  &       $0.35$  &       $0.40$  &       $0.37$  &       $19.94$\%       &       $14.79$\%       &       $12.98$\%       &       $20.63$\%       \\
CZ-Educ.        &       1980    &       $2.40$  &       $0.49$  &       ---       &       ---       &       $17.46$\%       &       $25.09$\%       &       ---       &       ---       \\ [2pt]

\multicolumn{9}{l}{\textbf{Panel E\@: AKM0 standard error}} \\ [1pt]

CZ      &       1980    &       $\infty$        &       $\infty$        &       $\infty$        &       $\infty$        &       $2.74$\%        &       $1.63$\%        &       $3.07$\%        &       $1.00$\%        \\
CZ      &       1960    &       $3.04$  &       $3.13$  &       $2.67$  &       $6.31$  &       $3.10$\%        &       $3.16$\%        &       $5.04$\%        &       $2.26$\%        \\
MSA     &       1980    &       $30.16$ &       $7.92$  &       $17.82$ &       $12.70$ &       $1.93$\%        &       $1.25$\%        &       $0.80$\%        &       $1.41$\%        \\
MSA     &       1960    &       $\infty$        &       $\infty$        &       $\infty$        &       $\infty$        &       $2.85$\%        &       $2.84$\%        &       $2.30$\%        &       $0.52$\%        \\
CZ-50 Occ.      &       1980    &       $0.25$  &       $0.09$  &       $0.10$  &       $0.08$  &       $4.43$\%        &       $4.13$\%        &       $3.63$\%        &       $4.35$\%        \\
CZ-7 Occ.       &       1980    &       $2.89$  &       $0.66$  &       $0.76$  &       $0.72$  &       $4.67$\%        &       $3.72$\%        &       $4.08$\%        &       $4.08$\%        \\
CZ-Educ.        &       1980    &       $8.10$  &       $1.75$  &       ---       &       ---       &       $4.65$\%        &       $3.17$\%        &       ---       &       ---       \\ [2pt]
\bottomrule
\multicolumn{10}{p{6.3in}}{\scriptsize{Notes: $\Delta \log E_i$ denotes log change in native employment; $\Delta \log w_i$ denotes log change in average weekly wages of native workers. Whenever we use 1980 weights, we use observations for the time periods 1980--1990, 1990--2000, 2000--2010. Whenever we use 1960 weights, we use observations for the time period 1970--1980. We use information on 722 CZs when combined with both 1960 and 1980 weights, on 257 MSAs when combined with 1980 weights, and 217 MSAs when combined with 1960 weights. Models are weighted by the start-of-period share of national population. \textit{Robust} is the Eicker-Huber-White standard error; \textit{Cluster} is the standard error that clusters of CZs in the same state; \textit{AKM} is the standard error in \begin{NoHyper}\Cref{remark:akm-construction}\end{NoHyper}; and \textit{AKM0} is the confidence interval in \begin{NoHyper}\Cref{remark:akm0-construction}\end{NoHyper}. The median effective standard error is equal to the median length of the corresponding 95\% confidence interval divided by $2\times1.96$. Results are based on 30,000 placebo samples.}}
\end{tabular}
\end{table}

\begin{table}[!tp]
\centering
\caption{Reduced-form placebo with all origin countries}\label{tab:imm_all_placebo}
\footnotesize
\begin{tabular}{lcccccrrrr}
\toprule
                && \multicolumn{4}{c}{Median eff.\ s.e.} & \multicolumn{4}{c}{Rejection rate for $H_{0}\colon \beta=0$ at 5\%}                  \\
 \cmidrule(rl){3-6} \cmidrule(rl){7-10}
 && \multicolumn{1}{c}{{$\Delta \log E_i$}} & \multicolumn{3}{c}{{$\Delta \log w_i$}} & \multicolumn{1}{c}{{$\Delta \log E_i$}} & \multicolumn{3}{c}{{$\Delta \log w_i$}}         \\
 \cmidrule(rl){3-3} \cmidrule(rl){4-6} \cmidrule(rl){7-7} \cmidrule(rl){8-10}
 && All & All  & High-Skill & Low-Skill & \multicolumn{1}{c}{All} & \multicolumn{1}{c}{All}  & \multicolumn{1}{c}{High-Skill} & \multicolumn{1}{c}{Low-Skill} \\
Unit obs. & Weights & (1) & (2) & (3) & (4) & \multicolumn{1}{c}{(5)} & \multicolumn{1}{c}{(6)} & \multicolumn{1}{c}{(7)} & \multicolumn{1}{c}{(8)} \\
 \midrule

\multicolumn{9}{l}{\textbf{Panel A\@: Standard deviation of placebo estimate}} \\ [1pt]

CZ      &       1980    &       $0.21$  &       $0.18$  &       $0.15$  &       $0.25$  &               &               &               &               \\
CZ      &       1960    &       $0.27$  &       $0.20$  &       $0.18$  &       $0.32$  &               &               &               &               \\
MSA     &       1980    &       $0.40$  &       $0.06$  &       $0.06$  &       $0.13$  &               &               &               &               \\
MSA     &       1960    &       $0.51$  &       $0.12$  &       $0.10$  &       $0.26$  &               &               &               &               \\
CZ-50 Occ.      &       1980    &       $0.11$  &       $0.04$  &       $0.02$  &       $0.04$  &               &               &               &               \\
CZ-7 Occ.       &       1980    &       $0.51$  &       $0.15$  &       $0.2$   &       $0.22$  &               &               &               &               \\
CZ-Educ.        &       1980    &       $0.60$  &       $0.15$  &       ---       &       ---       &               &               &               &               \\ [2pt]

\multicolumn{9}{l}{\textbf{Panel B\@: Robust standard error}} \\ [1pt]

CZ      &       1980    &       $0.12$  &       $0.05$  &       $0.04$  &       $0.06$  &       $3.57$\%        &       $73.01$\%       &       $81.63$\%       &       $38.91$\%       \\
CZ      &       1960    &       $0.13$  &       $0.05$  &       $0.05$  &       $0.06$  &       $15.86$\%       &       $55.42$\%       &       $57.59$\%       &       $60.02$\%       \\
MSA     &       1980    &       $0.20$  &       $0.03$  &       $0.03$  &       $0.04$  &       $25.90$\%       &       $18.07$\%       &       $9.97$\%        &       $53.39$\%       \\
MSA     &       1960    &       $0.34$  &       $0.08$  &       $0.08$  &       $0.09$  &       $9.21$\%        &       $9.52$\%        &       $2.18$\%        &       $41.43$\%       \\
CZ-50 Occ.      &       1980    &       $0.04$  &       $0.02$  &       $0.02$  &       $0.02$  &       $62.52$\%       &       $48.73$\%       &       $9.07$\%        &       $38.04$\%       \\
CZ-7 Occ.       &       1980    &       $0.15$  &       $0.05$  &       $0.04$  &       $0.07$  &       $75.86$\%       &       $75.24$\%       &       $86.16$\%       &       $80.99$\%       \\
CZ-Educ.        &       1980    &       $0.13$  &       $0.04$  &       ---       &       ---       &       $82.08$\%       &       $49.75$\%       &       ---       &       ---       \\ [2pt]

\multicolumn{9}{l}{\textbf{Panel C\@: State-clustered standard error}} \\ [1pt]

CZ      &       1980    &       $0.09$  &       $0.07$  &       $0.05$  &       $0.10$  &       $3.46$\%        &       $46.56$\%       &       $68.91$\%       &       $13.85$\%       \\
CZ      &       1960    &       $0.16$  &       $0.09$  &       $0.07$  &       $0.12$  &       $8.50$\%        &       $20.66$\%       &       $43.84$\%       &       $29.18$\%       \\
MSA     &       1980    &       $0.28$  &       $0.04$  &       $0.03$  &       $0.07$  &       $10.98$\%       &       $5.77$\%        &       $7.30$\%        &       $21.38$\%       \\
MSA     &       1960    &       $0.39$  &       $0.09$  &       $0.09$  &       $0.13$  &       $2.80$\%        &       $7.02$\%        &       $1.29$\%        &       $33.24$\%       \\
CZ-50 Occ.      &       1980    &       $0.06$  &       $0.02$  &       $0.02$  &       $0.02$  &       $48.40$\%       &       $34.88$\%       &       $5.15$\%        &       $30.04$\%       \\
CZ-7 Occ.       &       1980    &       $0.22$  &       $0.06$  &       $0.06$  &       $0.06$  &       $30.90$\%       &       $60.28$\%       &       $74.47$\%       &       $75.69$\%       \\
CZ-Educ.        &       1980    &       $0.18$  &       $0.08$  &       ---       &       ---       &       $73.77$\%       &       $9.82$\%        &       ---       &       ---       \\ [2pt]

\multicolumn{9}{l}{\textbf{Panel D\@: AKM standard error}} \\ [1pt]

CZ      &       1980    &       $0.09$  &       $0.06$  &       $0.04$  &       $0.12$  &       $4.44$\%        &       $56.32$\%       &       $66.26$\%       &       $10.19$\%       \\
CZ      &       1960    &       $0.14$  &       $0.11$  &       $0.08$  &       $0.19$  &       $17.27$\%       &       $21.85$\%       &       $37.77$\%       &       $13.12$\%       \\
MSA     &       1980    &       $0.20$  &       $0.03$  &       $0.04$  &       $0.07$  &       $40.31$\%       &       $30.36$\%       &       $3.86$\%        &       $31.97$\%       \\
MSA     &       1960    &       $0.33$  &       $0.08$  &       $0.07$  &       $0.20$  &       $23.68$\%       &       $11.11$\%       &       $1.43$\%        &       $13.19$\%       \\
CZ-50 Occ.      &       1980    &       $0.07$  &       $0.03$  &       $0.02$  &       $0.03$  &       $32.38$\%       &       $32.88$\%       &       $25.60$\%       &       $28.50$\%       \\
CZ-7 Occ.       &       1980    &       $0.21$  &       $0.06$  &       $0.07$  &       $0.08$  &       $54.43$\%       &       $59.39$\%       &       $61.74$\%       &       $62.08$\%       \\
CZ-Educ.        &       1980    &       $0.23$  &       $0.07$  &       ---       &       ---       &       $53.00$\%       &       $39.53$\%       &       ---       &       ---       \\ [2pt]

\multicolumn{9}{l}{\textbf{Panel E\@: AKM0 standard error}} \\ [1pt]

CZ      &       1980    &       $\infty$        &       $\infty$        &       $\infty$        &       $\infty$        &       $2.52$\%        &       $0.27$\%        &       $0.33$\%        &       $0.44$\%        \\
CZ      &       1960    &       $\infty$        &       $\infty$        &       $\infty$        &       $\infty$        &       $2.52$\%        &       $0.32$\%        &       $2.03$\%        &       $0.30$\%        \\
MSA     &       1980    &       $\infty$        &       $\infty$        &       $\infty$        &       $\infty$        &       $0.54$\%        &       $0.97$\%        &       $2.47$\%        &       $0.08$\%        \\
MSA     &       1960    &       $\infty$        &       $\infty$        &       $\infty$        &       $\infty$        &       $1.06$\%        &       $0.12$\%        &       $1.58$\%        &       $0.06$\%        \\
CZ-50 Occ.      &       1980    &       $0.42$  &       $0.15$  &       $0.1$   &       $0.14$  &       $1.73$\%        &       $2.49$\%        &       $3.17$\%        &       $2.72$\%        \\
CZ-7 Occ.       &       1980    &       $\infty$        &       $\infty$        &       $\infty$        &       $\infty$        &       $1.33$\%        &       $0.73$\%        &       $0.83$\%        &       $0.56$\%        \\
CZ-Educ.        &       1980    &       $\infty$        &       $\infty$        &       ---       &       ---       &       $1.16$\%        &       $1.20$\%        &       ---       &       ---       \\ [2pt]

\bottomrule

\multicolumn{10}{p{6.3in}}{\scriptsize{Notes: $\Delta \log E_i$ denotes log change in native employment; $\Delta \log w_i$ denotes log change in average weekly wages of native workers. Whenever we use 1980 weights, we use observations for the time periods 1980--1990, 1990--2000, 2000--2010. Whenever we use 1960 weights, we use observations for the time period 1970--1980. We use information on 722 CZs when combined with both 1960 and 1980 weights, on 257 MSAs when combined with 1980 weights, and 217 MSAs when combined with 1960 weights. Models are weighted by the start-of-period share of national population. \textit{Robust} is the Eicker-Huber-White standard error; \textit{Cluster} is the standard error that clusters of CZs in the same state; \textit{AKM} is the standard error in \begin{NoHyper}\Cref{remark:akm-construction}\end{NoHyper}; and \textit{AKM0} is the confidence interval in \begin{NoHyper}\Cref{remark:akm0-construction}\end{NoHyper}. The median effective standard error is equal to the median length of the corresponding 95\% confidence interval divided by $2\times1.96$. Results are based on 30,000 placebo samples.}}
\end{tabular}
\end{table}

\begin{landscape}
\begin{table}[!tp]
\centering
\caption{Effect of immigration: analysis by CZ-Occupations and CZ-Education groups (excluding large origin countries)}\label{tab:immigration_3percent}
\tabcolsep=0.10cm
\small
\begin{tabular}{@{}lcccccccccc@{}}
\toprule
Outcome: & \multicolumn{1}{c}{{$\Delta \log E_i$}} & \multicolumn{3}{c}{{$\Delta \log w_i$}} & \multicolumn{1}{c}{{$\Delta \log E_i$}} & \multicolumn{1}{c}{{$\Delta \log w_i$}} & \multicolumn{1}{c}{{$\Delta \log E_i$}} & \multicolumn{3}{c}{{$\Delta \log w_i$}}         \\
 \cmidrule(rl){2-2} \cmidrule(rl){3-5} \cmidrule(rl){6-6} \cmidrule(rl){7-7} \cmidrule(rl){8-8} \cmidrule(rl){9-11}
Workers: & All & All  & High-Skill & Low-Skill & All & All & All & All  & High-Skill & Low-Skill \\
 \cmidrule(rl){2-2} \cmidrule(rl){3-5} \cmidrule(rl){6-6} \cmidrule(rl){7-7} \cmidrule(rl){8-8} \cmidrule(rl){9-11}
 & (1) & (2) & (3) & (4) & (5) & (6) & (7) & (8) & (9) & (10)  \\
 \midrule

& \multicolumn{4}{c}{\textit{CZ-50 Occ. (1980 weights)}} & \multicolumn{2}{c}{\textit{CZ-Educ. (1980 weights)}} & \multicolumn{4}{c}{\textit{CZ-7 Occ. (1980 weights)}} \\ [2pt]
\cmidrule(rl){2-5} \cmidrule(rl){6-7} \cmidrule(rl){8-11} \\[-7pt]
\multicolumn{5}{@{}l}{\textbf{Panel A\@: 2SLS}} \\ [2pt]

 $\hat{\beta}$  &       $-1.19$ &       $0.05$  &       $0.26$  &       $-0.14$ &       $-1.49$ &       $0.18$  &       $-1.39$ &       $0.08$  &       $0.24$  &       $-0.14$ \\
Robust  &       $[-1.55,-0.83]$ &       $[-0.09,0.20]$  &       $[0.11,0.41]$   &       $[-0.32,0.03]$  &       $[-2.09,-0.90]$ &       $[-0.20,0.56]$  &       $[-1.89,-0.90]$ &       $[-0.17,0.33]$  &       $[0.00,0.47]$   &       $[-0.43,0.15]$  \\
 Cluster        &       $[-1.89,-0.49]$ &       $[-0.35,0.46]$  &       $[-0.14,0.67]$  &       $[-0.69,0.40]$  &       $[-2.14,-0.85]$ &       $[-0.04,0.39]$  &       $[-1.87,-0.92]$ &       $[-0.11,0.27]$  &       $[0.10,0.38]$   &       $[-0.27,-0.01]$ \\
 AKM    &       $[-1.55,-0.83]$ &       $[-0.36,0.47]$  &       $[-0.17,0.69]$  &       $[-0.64,0.35]$  &       $[-2.31,-0.68]$ &       $[-0.53,0.88]$  &       $[-2.02,-0.76]$ &       $[-0.51,0.67]$  &       $[-0.31,0.79]$  &       $[-0.81,0.53]$  \\
 AKM${0}$       &       $[-1.66,-0.72]$ &       $[-0.53,0.54]$  &       $[-0.32,0.81]$  &       $[-0.92,0.39]$  &       $[-3.00,-0.14]$ &       $[-0.90,1.60]$  &       $[-2.35,-0.54]$ &       $[-0.75,0.94]$  &       $[-0.51,1.07]$  &       $[-1.12,0.80]$  \\ [6pt]

\multicolumn{5}{@{}l}{\textbf{Panel B\@: Reduced-Form}} \\ [2pt]

 $\hat{\beta}$  &       $-0.89$ &       $0.04$  &       $0.2$   &       $-0.11$ &       $-1.29$ &       $0.15$  &       $-1.05$ &       $0.06$  &       $0.18$  &       $-0.11$ \\
Robust  &       $[-1.17,-0.61]$ &       $[-0.07,0.15]$  &       $[0.06,0.33]$   &       $[-0.23,0.02]$  &       $[-1.86,-0.73]$ &       $[-0.19,0.50]$  &       $[-1.38,-0.73]$ &       $[-0.13,0.25]$  &       $[-0.01,0.37]$  &       $[-0.32,0.11]$  \\
 Cluster        &       $[-1.37,-0.41]$ &       $[-0.27,0.35]$  &       $[-0.16,0.55]$  &       $[-0.47,0.25]$  &       $[-1.69,-0.90]$ &       $[-0.02,0.33]$  &       $[-1.29,-0.82]$ &       $[-0.08,0.21]$  &       $[0.07,0.29]$   &       $[-0.20,-0.01]$ \\
 AKM    &       $[-1.35,-0.43]$ &       $[-0.28,0.36]$  &       $[-0.18,0.57]$  &       $[-0.44,0.23]$  &       $[-1.99,-0.60]$ &       $[-0.48,0.79]$  &       $[-1.54,-0.57]$ &       $[-0.39,0.52]$  &       $[-0.26,0.62]$  &       $[-0.59,0.38]$  \\
 AKM${0}$       &       $[-1.55,-0.39]$ &       $[-0.27,0.55]$  &       $[-0.16,0.79]$  &       $[-0.44,0.42]$  &       $[-2.25,-0.11]$ &       $[-0.53,1.56]$  &       $[-1.62,-0.36]$ &       $[-0.42,0.78]$  &       $[-0.29,0.86]$  &       $[-0.62,0.67]$  \\ [6pt]

\multicolumn{5}{@{}l}{\textbf{Panel C\@: First-Stage}} \\ [2pt]

 $\hat{\beta}$  &       \multicolumn{4}{c}{     $0.75$  }                                       &       \multicolumn{2}{c}{     $0.87$  }       &       \multicolumn{4}{c}{     $0.76$  }                                       \\
Robust  &       \multicolumn{4}{c}{     $[0.56,0.93]$   }                                       &       \multicolumn{2}{c}{     $[0.55,1.18]$   }       &       \multicolumn{4}{c}{     $[0.56,0.95]$   }                                       \\
 Cluster        &       \multicolumn{4}{c}{     $[0.62,0.88]$   }                                       &       \multicolumn{2}{c}{     $[0.66,1.08]$   }       &       \multicolumn{4}{c}{     $[0.60,0.91]$   }                                       \\
 AKM    &       \multicolumn{4}{c}{     $[0.36,1.13]$   }                                       &       \multicolumn{2}{c}{     $[0.44,1.30]$   }       &       \multicolumn{4}{c}{     $[0.42,1.09]$   }                                       \\
 AKM${0}$       &       \multicolumn{4}{c}{     $[0.38,1.37]$   }                                       &       \multicolumn{2}{c}{     $[0.34,1.70]$   }       &       \multicolumn{4}{c}{     $[0.38,1.24]$   }                                       \\ [2pt]

\bottomrule

  \multicolumn{11}{p{9.2in}}{\scriptsize{Notes: $\Delta \log E_i$ denotes log
  change in native employment; $\Delta \log w_i$ denotes log change in average
  weekly wages of native workers. The specifications \textit{CZ-50 Occupations},
  \textit{CZ-2 Education Groups}, and \textit{CZ-7 Occupations} differ in the
  definition of the unit of observation. In all three specifications, we use
  1980 weights and three time periods, 1980--1990, 1990--2000, and 2000--2010.
  Thus, $N = 108,300$ (722 CZs $\times$ 50 occupations $\times$ 3 time periods)
  for the \textit{CZ-50 Occupations} specification; $N = 4,332$ (722 CZs
  $\times$ 2 education groups $\times$ 3 time periods) for the \textit{CZ-2
  Education Groups} specification; and $N = 15,162$ (722 CZs $\times$ 7
  occupations $\times$ 3 time periods) for the \textit{CZ-7 Occupations}
  specification. Models are weighted by start-of-period occupation-region (or
  education group-region) share of national population. All regressions include
  occupation (or education group) and period dummies. 95\% confidence intervals
  in square brackets. \textit{Robust} is the Eicker-Huber-White standard error;
  \textit{Cluster} is the standard error that clusters of CZs in the same state;
  \textit{AKM} is the standard error in
  \begin{NoHyper}\Cref{remark:akm-construction}\end{NoHyper}; and \textit{AKM0} is the confidence interval in
  \begin{NoHyper}\Cref{remark:akm0-construction}\end{NoHyper}. We exclude from the analysis those countries of origin whose immigrant share in 1980 is larger than 3\%; i.e. $\sum_{i}ImmShare_{igt_{0}}/\sum_{i}\sum_{g'}ImmShare_{ig't_{0}}>0.03$. See
  \begin{NoHyper}\Cref{tab:origin_country_1980}\end{NoHyper} for a list of the origin countries included in the analysis.}}
\end{tabular}
\end{table}
\end{landscape}

\begin{landscape}
\begin{table}[!tp]
\centering
\caption{Effect of immigration: analysis by CZ (excluding large origin countries)}\label{tab:immigration_3percent_addres_CZ}
\tabcolsep=0.10cm
\small
\begin{tabular}{@{}lcccccccc}
\toprule
Outcome: & \multicolumn{1}{c}{{$\Delta \log E_i$}} & \multicolumn{3}{c}{{$\Delta \log w_i$}} & \multicolumn{1}{c}{{$\Delta \log E_i$}} & \multicolumn{3}{c}{{$\Delta \log w_i$}}         \\
 \cmidrule(rl){2-2} \cmidrule(rl){3-5} \cmidrule(rl){6-6} \cmidrule(rl){7-9}
Workers: & All & All  & High-Skill & Low-Skill & All & All  & High-Skill & Low-Skill \\
 \cmidrule(rl){2-2} \cmidrule(rl){3-5} \cmidrule(rl){6-6} \cmidrule(rl){7-9}
 & (1) & (2) & (3) & (4) & (5) & (6) & (7) & (8)  \\
 \midrule

& \multicolumn{4}{c}{\textit{Commuting Zone (1980 weights)}} & \multicolumn{4}{c}{\textit{Commuting Zone (1960 weights)}}  \\ [2pt]
\cmidrule(rl){2-5} \cmidrule(rl){6-9}
\multicolumn{9}{@{}l}{\textbf{Panel A\@: 2SLS Regression}} \\ [2pt]

 $\hat{\beta}$  &       $-0.89$ &       $0.42$  &       $0.56$  &       $-0.10$  &       $-1.29$ &       $-0.71$ &       $-0.34$ &       $-1.14$ \\
Robust  &       $[-1.48,-0.30]$ &       $[-0.06,0.90]$  &       $[0.16,0.95]$   &       $[-0.68,0.47]$  &       $[-3.69,1.10]$  &       $[-1.10,-0.33]$ &       $[-1.01,0.33]$  &       $[-1.66,-0.61]$ \\
 Cluster        &       $[-1.33,-0.45]$ &       $[0.15,0.69]$   &       $[0.39,0.72]$   &       $[-0.31,0.11]$  &       $[-3.96,1.37]$  &       $[-1.17,-0.26]$ &       $[-1.12,0.44]$  &       $[-1.71,-0.56]$ \\
 AKM    &       $[-1.85,0.08]$  &       $[-0.44,1.28]$  &       $[-0.12,1.24]$  &       $[-1.17,0.97]$  &       $[-3.83,1.24]$  &       $[-1.31,-0.12]$ &       $[-1.16,0.48]$  &       $[-1.89,-0.38]$ \\
 \multirow{2}{*}{AKM${0}$}       &       $[-\infty,-285.90]$       &       $[-\infty,-230.52]$       &       $[-\infty,-226.52]$       &       $[-\infty,-158.12]$       & \multirow{2}{*}{$[-\infty,\infty]$}    &       \multirow{2}{*}{$[-\infty,\infty]$}      &       \multirow{2}{*}{$[-\infty,\infty]$}      &   \multirow{2}{*}{$[-\infty,\infty]$}      \\
       &       $\cup\text{ }[-4.12,\infty]$       &       $\cup\text{ }[-2.73,\infty]$       &       $\cup\text{ }[-1.46,\infty]$       &       $\cup\text{ }[-7.24,\infty]$       &     &             &             &             \\ [6pt]

\multicolumn{9}{@{}l}{\textbf{Panel B\@: Reduced-Form Regression}} \\ [2pt]

 $\hat{\beta}$  &       $-0.75$ &       $0.35$  &       $0.47$  &       $-0.08$ &       $-1.31$ &       $-0.72$ &       $-0.34$ &       $-1.15$ \\
Robust  &       $[-1.16,-0.33]$ &       $[-0.10,0.81]$  &       $[0.07,0.86]$   &       $[-0.55,0.38]$  &       $[-3.39,0.78]$  &       $[-1.06,-0.38]$ &       $[-0.94,0.25]$  &       $[-1.61,-0.68]$ \\
 Cluster        &       $[-1.02,-0.47]$ &       $[0.12,0.59]$   &       $[0.31,0.62]$   &       $[-0.26,0.09]$  &       $[-3.57,0.96]$  &       $[-1.10,-0.35]$ &       $[-1.02,0.33]$  &       $[-1.56,-0.73]$ \\
 AKM    &       $[-1.47,-0.02]$ &       $[-0.44,1.15]$  &       $[-0.21,1.14]$  &       $[-0.96,0.79]$  &       $[-3.55,0.94]$  &       $[-1.21,-0.23]$ &       $[-1.09,0.40]$  &       $[-1.73,-0.57]$ \\
 AKM${0}$       &       $[-1.71,6.41]$  &       $[-0.73,8.03]$  &       $[-0.44,6.99]$  &       $[-1.29,8.26]$  &       $[-\infty,\infty]$      &       $[-3.23,-1.51]$ &       $[-3.37,-1.87]$ &       $[-4.01,-2.12]$ \\ [6pt]

\multicolumn{9}{@{}l}{\textbf{Panel C\@: 2SLS First-Stage}} \\ [2pt]

 $\hat{\beta}$  &       \multicolumn{4}{c}{     $0.84$  }       &                               \multicolumn{4}{c}{     $1.01$  }                                               \\
Robust  &       \multicolumn{4}{c}{     $[0.51,1.18]$   }       &                               \multicolumn{4}{c}{     $[0.46,1.56]$   }                                               \\
 Cluster        &       \multicolumn{4}{c}{     $[0.66,1.03]$   }       &                               \multicolumn{4}{c}{     $[0.65,1.37]$   }                                               \\
 AKM    &       \multicolumn{4}{c}{     $[0.37,1.31]$   }       &                               \multicolumn{4}{c}{     $[0.70,1.32]$   }                                               \\
 AKM${0}$       &       \multicolumn{4}{c}{     $[0.00,4.32]$   }       &                               \multicolumn{4}{c}{     $[-\infty,\infty]$      }                                               \\

 \bottomrule

\multicolumn{9}{p{8.5in}}{\scriptsize{Notes: $\Delta \log E_i$ denotes log change in native employment; $\Delta \log w_i$ denotes log change in average weekly wages of native workers. In the specification \textit{CZ (1980 weights)}, we use information on 722 CZs, 1980 weights and three time periods, 1980--1990, 1990--2000, 2000--2010; thus, $N=2,166$ (722 CZs $\times$ 3 time periods). In the specification \textit{CZ (1960 weights)}, we use information on 722 CZs, 1960 weights and one time period, 1970--1980; thus, $N=722$ (722 CZs $\times$ 1 time period). Models are weighted by start-of-period CZ share of national population. All regressions include period dummies. 95\% confidence intervals in square brackets. \textit{Robust} is the Eicker-Huber-White standard error; \textit{Cluster} is the standard error that clusters of CZs in the same state; \textit{AKM} is the standard error in \begin{NoHyper}\Cref{remark:akm-construction}\end{NoHyper}; and \textit{AKM0} is the confidence interval in \begin{NoHyper}\Cref{remark:akm0-construction}\end{NoHyper}. We exclude from the analysis those countries of origin whose immigrant share in year $t_{0}$ is larger than 3\%; i.e.\ $\sum_{i=1}^{N}ImmShare_{igt_{0}}/\sum_{i=1}^{N}\sum_{g'=1}^{G}ImmShare_{ig't_{0}}>
0.03$. See \Cref{tab:origin_country_1980,tab:origin_country_1960} for a list of the origin countries included in the analysis.}}
\end{tabular}
\end{table}
\end{landscape}

\begin{landscape}
\begin{table}[!tp]
\centering
\caption{Effect of immigration: analysis by MSA (excluding large origin countries)}\label{tab:immigration_3percent_addres_MSA}
\tabcolsep=0.10cm
\small
\begin{tabular}{@{}lcccccccc}
\toprule
Outcome: & \multicolumn{1}{c}{{$\Delta \log E_i$}} & \multicolumn{3}{c}{{$\Delta \log w_i$}} & \multicolumn{1}{c}{{$\Delta \log E_i$}} & \multicolumn{3}{c}{{$\Delta \log w_i$}}         \\
 \cmidrule(rl){2-2} \cmidrule(rl){3-5} \cmidrule(rl){6-6} \cmidrule(rl){7-9}
Workers: & All & All  & High-Skill & Low-Skill & All & All  & High-Skill & Low-Skill \\
 \cmidrule(rl){2-2} \cmidrule(rl){3-5} \cmidrule(rl){6-6} \cmidrule(rl){7-9}
 & (1) & (2) & (3) & (4) & (5) & (6) & (7) & (8)  \\
 \midrule

& \multicolumn{4}{c}{\textit{MSA (1980 weights)}} & \multicolumn{4}{c}{\textit{MSA (1960 weights)}}  \\ [2pt]
\cmidrule(rl){2-5} \cmidrule(rl){6-9}
\multicolumn{9}{@{}l}{\textbf{Panel A\@: 2SLS Regression}} \\ [2pt]

 $\hat{\beta}$  &       $-1.87$ &       $0.50$   &       $0.63$  &       $-0.10$  &       $-1.55$ &       $-0.32$ &       $-0.17$ &       $-0.57$ \\
Robust  &       $[-3.17,-0.58]$ &       $[-0.06,1.06]$  &       $[0.17,1.09]$   &       $[-0.93,0.73]$  &       $[-3.68,0.57]$  &       $[-0.72,0.08]$  &       $[-0.83,0.49]$  &       $[-1.08,-0.06]$ \\
 Cluster        &       $[-2.97,-0.78]$ &       $[0.23,0.77]$   &       $[0.48,0.77]$   &       $[-0.47,0.27]$  &       $[-3.94,0.84]$  &       $[-0.74,0.09]$  &       $[-0.88,0.53]$  &       $[-1.06,-0.08]$ \\
 AKM    &       $[-3.77,0.03]$  &       $[-0.44,1.44]$  &       $[-0.15,1.40]$   &       $[-1.49,1.29]$  &       $[-5.61,2.51]$  &       $[-0.90,0.25]$  &       $[-1.07,0.73]$  &       $[-1.26,0.12]$  \\
 AKM${0}$       &       $[-\infty,\infty]$      &       $[-\infty,\infty]$      &       $[-\infty,\infty]$      &       $[-\infty,\infty]$

    &       $[-\infty,\infty]$      &       $[-\infty,\infty]$      &       $[-\infty,\infty]$      &       $[-\infty,\infty]$      \\ [6pt]

\multicolumn{9}{@{}l}{\textbf{Panel B\@: Reduced-Form Regression}} \\ [2pt]

 $\hat{\beta}$  &       $-1.47$ &       $0.39$  &       $0.49$  &       $-0.08$ &       $-1.48$ &       $-0.31$ &       $-0.16$ &       $-0.54$ \\
Robust  &       $[-2.06,-0.87]$ &       $[-0.17,0.95]$  &       $[0.00,0.98]$   &       $[-0.70,0.54]$  &       $[-3.29,0.34]$  &       $[-0.61,-0.01]$ &       $[-0.75,0.42]$  &       $[-0.90,-0.19]$ \\
 Cluster        &       $[-1.94,-0.99]$ &       $[0.09,0.69]$   &       $[0.27,0.71]$   &       $[-0.35,0.19]$  &       $[-3.45,0.49]$  &       $[-0.63,0.01]$  &       $[-0.79,0.46]$  &       $[-0.86,-0.22]$ \\
 AKM    &       $[-2.43,-0.50]$ &       $[-0.50,1.28]$  &       $[-0.28,1.26]$  &       $[-1.13,0.98]$  &       $[-4.79,1.83]$  &       $[-0.73,0.11]$  &       $[-0.96,0.63]$  &       $[-0.98,-0.11]$ \\
 AKM${0}$       &       $[-2.70,8.87]$  &       $[-0.79,9.41]$  &       $[-0.53,8.41]$  &       $[-1.51,10.46]$ &       $[-\infty,\infty]$      &       $[-\infty,\infty]$      &       $[-\infty,\infty]$      &       $[-\infty,\infty]$      \\ [6pt]

\multicolumn{9}{@{}l}{\textbf{Panel C\@: 2SLS First-Stage}} \\ [2pt]

 $\hat{\beta}$  &       \multicolumn{4}{c}{     $0.78$  }       &                               \multicolumn{4}{c}{     $0.95$  }                                               \\
Robust  &       \multicolumn{4}{c}{     $[0.29,1.28]$   }       &                               \multicolumn{4}{c}{     $[0.46,1.45]$   }                                               \\
 Cluster        &       \multicolumn{4}{c}{     $[0.51,1.05]$   }       &                               \multicolumn{4}{c}{     $[0.64,1.27]$   }                                               \\
 AKM    &       \multicolumn{4}{c}{     $[0.10,1.47]$   }       &                               \multicolumn{4}{c}{     $[0.49,1.42]$   }                                               \\
 AKM${0}$       &       \multicolumn{4}{c}{     $[-0.61,5.35]$  }       &                               \multicolumn{4}{c}{     $[-\infty,\infty]$      }                                               \\ [2pt]

 \bottomrule

\multicolumn{9}{p{7.5in}}{\scriptsize{Notes: $\Delta \log E_i$ denotes log change in native employment; $\Delta \log w_i$ denotes log change in average weekly wages of native workers. In the specification \textit{MSA (1980 weights)}, we use information on 257 MSAs, 1980 weights and three time periods, 1980--1990, 1990--2000, 2000--2010; thus, $N=651$ (257 MSAs $\times$ 3 time periods). In the specification \textit{MSA (1960 weights)}, we use information on 217 MSAs, 1960 weights and one time period, 1970--1980; thus, $N=217$ (217 CZs $\times$ 1 time period). Models are weighted by start-of-period MSA share of national population. All regressions include period dummies. 95\% confidence intervals in square brackets. \textit{Robust} is the Eicker-Huber-White standard error; \textit{Cluster} is the standard error that clusters of CZs in the same state; \textit{AKM} is the standard error in \begin{NoHyper}\Cref{remark:akm-construction}\end{NoHyper}; and \textit{AKM0} is the confidence interval in \begin{NoHyper}\Cref{remark:akm0-construction}\end{NoHyper}. We exclude from the analysis those countries of origin whose immigrant share in year $t_{0}$ is larger than 3\%; i.e.\ $\sum_{i=1}^{N}ImmShare_{igt_{0}}/\sum_{i=1}^{N}\sum_{g'=1}^{G}ImmShare_{ig't_{0}}>
0.03$. See \Cref{tab:origin_country_1980,tab:origin_country_1960} for a list of the origin countries included in the analysis.}}
\end{tabular}
\end{table}
\end{landscape}

\begin{table}[!tp]
\centering
\caption{Immigration: p-values by CZ-Occ.\ and CZ-Educ. (excluding large origin countries)}\label{tab:immigration_3percent_pvalue}
\tabcolsep=0.08cm
\begin{tabular}{@{}lcccccccccc}
\toprule
 & \multicolumn{1}{c}{{$\Delta \log E_i$}} & \multicolumn{3}{c}{{$\Delta \log w_i$}} & \multicolumn{1}{c}{{$\Delta \log E_i$}} & \multicolumn{1}{c}{{$\Delta \log w_i$}} & \multicolumn{1}{c}{{$\Delta \log E_i$}} & \multicolumn{3}{c}{{$\Delta \log w_i$}}         \\
 \cmidrule(rl){2-2} \cmidrule(rl){3-5} \cmidrule(rl){6-6} \cmidrule(rl){7-7} \cmidrule(rl){8-8} \cmidrule(rl){9-11}
 & All & All  & High-Skill & Low-Skill & All & All & All & All  & High-Skill & Low-Skill \\
 \cmidrule(rl){2-2} \cmidrule(rl){3-5} \cmidrule(rl){6-6} \cmidrule(rl){7-7} \cmidrule(rl){8-8} \cmidrule(rl){9-11}
 & (1) & (2) & (3) & (4) & (5) & (6) & (7) & (8) & (9) & (10)  \\
 \midrule
& \multicolumn{4}{c}{\textit{CZ-50 Occupations}} & \multicolumn{2}{c}{\textit{CZ- Educ.}} & \multicolumn{4}{c}{\textit{CZ-7 Occupations}} \\ [2pt]
\cmidrule(rl){2-5} \cmidrule(rl){6-7} \cmidrule(rl){8-11} \\[-7pt]

\multicolumn{11}{@{}l}{\textbf{Panel A\@: 2SLS Regression}} \\ [2pt]

Robust  &       $0.000$ &       $0.459$ &       $0.001$ &       $0.116$ &       $0.000$ &       $0.357$ &       $0.000$ &       $0.519$ &       $0.046$ &       $0.348$ \\
 Cluster        &       $0.001$ &       $0.790$ &       $0.205$ &       $0.611$ &       $0.000$ &       $0.102$ &       $0.000$ &       $0.412$ &       $0.001$ &       $0.042$ \\
 AKM    &       $0.000$ &       $0.793$ &       $0.235$ &       $0.576$ &       $0.000$ &       $0.620$ &       $0.000$ &       $0.788$ &       $0.395$ &       $0.684$ \\
 AKM${0}$       &       $0.006$ &       $0.796$ &       $0.273$ &       $0.566$ &       $0.043$ &       $0.621$ &       $0.018$ &       $0.788$ &       $0.410$ &       $0.687$ \\ [6pt]

\multicolumn{11}{@{}l}{\textbf{Panel B\@: Reduced-Form Regression}} \\ [2pt]

Robust  &       $0.000$ &       $0.473$ &       $0.004$ &       $0.091$ &       $0.000$ &       $0.379$ &       $0.000$ &       $0.528$ &       $0.057$ &       $0.330$ \\
 Cluster        &       $0.000$ &       $0.795$ &       $0.275$ &       $0.565$ &       $0.000$ &       $0.089$ &       $0.000$ &       $0.409$ &       $0.001$ &       $0.036$ \\
 AKM    &       $0.000$ &       $0.801$ &       $0.304$ &       $0.538$ &       $0.000$ &       $0.635$ &       $0.000$ &       $0.792$ &       $0.422$ &       $0.674$ \\
 AKM${0}$       &       $0.006$ &       $0.796$ &       $0.273$ &       $0.566$ &       $0.043$ &       $0.621$ &       $0.018$ &       $0.788$ &       $0.410$ &       $0.687$ \\ [6pt]

\multicolumn{11}{@{}l}{\textbf{Panel C\@: First-Stage}} \\ [2pt]

Robust  &       \multicolumn{4}{c}{     $0.000$ }       &                               \multicolumn{2}{c}{     $0.000$ }       &       \multicolumn{4}{c}{     $0.000$ }                                               \\
 Cluster        &       \multicolumn{4}{c}{     $0.000$ }       &                               \multicolumn{2}{c}{     $0.000$ }       &       \multicolumn{4}{c}{     $0.000$ }                                               \\
 AKM    &       \multicolumn{4}{c}{     $0.000$ }       &                               \multicolumn{2}{c}{     $0.000$ }       &       \multicolumn{4}{c}{     $0.000$ }                                               \\
 AKM${0}$       &       \multicolumn{4}{c}{     $0.002$ }       &                               \multicolumn{2}{c}{     $0.017$ }       &       \multicolumn{4}{c}{     $0.006$ }                                               \\ [2pt]

\bottomrule

\multicolumn{11}{p{6.5in}}{\scriptsize{Notes: $\Delta \log E_i$ denotes log change in native employment; $\Delta \log w_i$ denotes log change in average weekly wages of native workers. The specifications \textit{CZ-50 Occupations}, \textit{CZ-Educ.}, and \textit{CZ-7 Occupations} differ in the definition of the unit of observation. In all three specifications, we use 1980 weights and three time periods, 1980--1990, 1990--2000, 2000--2010. Thus, $N = 108,300$ (722 CZs $\times$ 50 occupations $\times$ 3 time periods) for the \textit{CZ-50 Occupations} specification; $N = 4,332$ (722 CZs $\times$ 2 education groups $\times$ 3 time periods) for the \textit{CZ-Educ.} specification; and $N = 15,162$ (722 CZs $\times$ 7 occupations $\times$ 3 time periods) for the \textit{CZ-7 Occupations} specification. Models are weighted by start-of-period occupation-region (or education group-region) share of national population. All regressions include occupation (or education group) and period dummies. 95\% confidence intervals in square brackets. \textit{Robust} is the Eicker-Huber-White standard error; \textit{Cluster} is the standard error that clusters of CZs in the same state; \textit{AKM} is the standard error in \begin{NoHyper}\Cref{remark:akm-construction}\end{NoHyper}; and \textit{AKM0} is the confidence interval in \begin{NoHyper}\Cref{remark:akm0-construction}\end{NoHyper}. We exclude from the analysis those countries of origin whose immigrant share in year 1980 is larger than 3\%; i.e.\ $\sum_{i=1}^{N}ImmShare_{igt_{0}}/\sum_{i=1}^{N}\sum_{g'=1}^{G}ImmShare_{ig't_{0}}>
0.03$. See \Cref{tab:origin_country_1980} for a list of the origin countries included in the analysis.}}
\end{tabular}
\end{table}

\begin{table}[!tp]
\centering
\caption{Effect of immigration: p-values by CZ (excluding large origin countries)}\label{tab:immigration_3percent_pvalues_addres_CZ}
\tabcolsep=0.15cm
\begin{tabular}{@{}lcccccccc}
\toprule
Outcome: & \multicolumn{1}{c}{{$\Delta \log E_i$}} & \multicolumn{3}{c}{{$\Delta \log w_i$}} & \multicolumn{1}{c}{{$\Delta \log E_i$}} & \multicolumn{3}{c}{{$\Delta \log w_i$}}         \\
 \cmidrule(rl){2-2} \cmidrule(rl){3-5} \cmidrule(rl){6-6} \cmidrule(rl){7-9}
Workers: & All & All  & High-Skill & Low-Skill & All & All  & High-Skill & Low-Skill \\
 \cmidrule(rl){2-2} \cmidrule(rl){3-5} \cmidrule(rl){6-6} \cmidrule(rl){7-9}
 & (1) & (2) & (3) & (4) & (5) & (6) & (7) & (8)  \\
 \midrule

 & \multicolumn{4}{c}{\textit{Commuting Zone (1980 weights)}} & \multicolumn{4}{c}{\textit{Commuting Zone (1960 weights)}}  \\ [2pt]
\cmidrule(rl){2-5} \cmidrule(rl){6-9}

\multicolumn{9}{@{}l}{\textbf{Panel A\@: 2SLS Regression}} \\ [2pt]

Robust  &       $0.003$ &       $0.083$ &       $0.005$ &       $0.731$ &       $0.291$ &       $0.000$ &       $0.317$ &       $0.000$ \\
 Cluster        &       $0.000$ &       $0.002$ &       $0.000$ &       $0.338$ &       $0.342$ &       $0.002$ &       $0.390$ &       $0.000$ \\
 AKM    &       $0.071$ &       $0.335$ &       $0.109$ &       $0.853$ &       $0.318$ &       $0.018$ &       $0.416$ &       $0.003$ \\
 AKM${0}$       &       $0.216$ &       $0.360$ &       $0.174$ &       $0.855$ &       $0.393$ &       $0.179$ &       $0.472$ &       $0.142$ \\ [6pt]

\multicolumn{9}{@{}l}{\textbf{Panel B\@: Reduced-Form Regression}} \\ [2pt]

Robust  &       $0.000$ &       $0.129$ &       $0.020$ &       $0.722$ &       $0.220$ &       $0.000$ &       $0.254$ &       $0.000$ \\
 Cluster        &       $0.000$ &       $0.003$ &       $0.000$ &       $0.330$ &       $0.258$ &       $0.000$ &       $0.316$ &       $0.000$ \\
 AKM    &       $0.044$ &       $0.384$ &       $0.174$ &       $0.850$ &       $0.254$ &       $0.004$ &       $0.366$ &       $0.000$ \\
 AKM${0}$       &       $0.216$ &       $0.360$ &       $0.174$ &       $0.855$ &       $0.393$ &       $0.179$ &       $0.472$ &       $0.142$ \\ [6pt]

\multicolumn{9}{@{}l}{\textbf{Panel C\@: First-Stage}} \\ [2pt]

Robust  &       \multicolumn{4}{c}{     $0.000$ }       &                               \multicolumn{4}{c}{     $0.000$ }                                               \\
 Cluster        &       \multicolumn{4}{c}{     $0.000$ }       &                               \multicolumn{4}{c}{     $0.000$ }                                               \\
 AKM    &       \multicolumn{4}{c}{     $0.001$ }       &                               \multicolumn{4}{c}{     $0.000$ }                                               \\
 AKM${0}$       &       \multicolumn{4}{c}{     $0.050$ }       &                               \multicolumn{4}{c}{     $0.062$ }                                               \\ [4pt]

 \bottomrule

\multicolumn{9}{p{5.8in}}{\scriptsize{Notes: $\Delta \log E_i$ denotes log change in native employment; $\Delta \log w_i$ denotes log change in average weekly wages of native workers. In the specification \textit{CZ (1980 weights)}, we use information on 722 CZs, 1980 weights and three time periods, 1980--1990, 1990--2000, 2000--2010; thus, $N=2,166$ (722 CZs $\times$ 3 time periods). In the specification \textit{CZ (1960 weights)}, we use information on 722 CZs, 1960 weights and one time period, 1970--1980; thus, $N=722$ (722 CZs $\times$ 1 time period). Models are weighted by start-of-period CZ share of national population. All regressions include period dummies. 95\% confidence intervals in square brackets. \textit{Robust} is the Eicker-Huber-White standard error; \textit{Cluster} is the standard error that clusters of CZs in the same state; \textit{AKM} is the standard error in \begin{NoHyper}\Cref{remark:akm-construction}\end{NoHyper}; and \textit{AKM0} is the confidence interval in \begin{NoHyper}\Cref{remark:akm0-construction}\end{NoHyper}. We exclude from the analysis those countries of origin whose immigrant share in year $t_{0}$ is larger than 3\%; i.e.\ $\sum_{i=1}^{N}ImmShare_{igt_{0}}/\sum_{i=1}^{N}\sum_{g'=1}^{G}ImmShare_{ig't_{0}}>
0.03$. See \Cref{tab:origin_country_1980,tab:origin_country_1960} for a list of the origin countries included in the analysis.}}
\end{tabular}
\end{table}

\begin{table}[!tp]
\centering
\caption{Effect of immigration: p-values by MSA (excluding large origin countries)}\label{tab:immigration_3percent_pvalues_addres_MSA}
\tabcolsep=0.15cm
\small
\begin{tabular}{@{}lcccccccc}
\toprule
Outcome: & \multicolumn{1}{c}{{$\Delta \log E_i$}} & \multicolumn{3}{c}{{$\Delta \log w_i$}} & \multicolumn{1}{c}{{$\Delta \log E_i$}} & \multicolumn{3}{c}{{$\Delta \log w_i$}}         \\
 \cmidrule(rl){2-2} \cmidrule(rl){3-5} \cmidrule(rl){6-6} \cmidrule(rl){7-9}
Workers: & All & All  & High-Skill & Low-Skill & All & All  & High-Skill & Low-Skill \\
 \cmidrule(rl){2-2} \cmidrule(rl){3-5} \cmidrule(rl){6-6} \cmidrule(rl){7-9}
 & (1) & (2) & (3) & (4) & (5) & (6) & (7) & (8)  \\
 \midrule

 & \multicolumn{4}{c}{\textit{MSA (1980 weights)}} & \multicolumn{4}{c}{\textit{MSA (1960 weights)}}  \\ [2pt]
\cmidrule(rl){2-5} \cmidrule(rl){6-9}

\multicolumn{9}{@{}l}{\textbf{Panel A\@: 2SLS Regression}} \\ [2pt]

Robust  &       $0.005$ &       $0.080$ &       $0.008$ &       $0.812$ &       $0.152$ &       $0.115$ &       $0.610$ &       $0.028$ \\
 Cluster        &       $0.001$ &       $0.000$ &       $0.000$ &       $0.594$ &       $0.203$ &       $0.128$ &       $0.631$ &       $0.022$ \\
 AKM    &       $0.053$ &       $0.300$ &       $0.112$ &       $0.887$ &       $0.453$ &       $0.272$ &       $0.708$ &       $0.106$ \\
 AKM${0}$       &       $0.146$ &       $0.361$ &       $0.203$ &       $0.887$ &       $0.532$ &       $0.376$ &       $0.732$ &       $0.284$ \\ [6pt]

\multicolumn{9}{@{}l}{\textbf{Panel B\@: Reduced-Form Regression}} \\ [2pt]

Robust  &       $0.000$ &       $0.173$ &       $0.048$ &       $0.803$ &       $0.110$ &       $0.043$ &       $0.585$ &       $0.003$ \\
 Cluster        &       $0.000$ &       $0.012$ &       $0.000$ &       $0.564$ &       $0.141$ &       $0.059$ &       $0.607$ &       $0.001$ \\
 AKM    &       $0.003$ &       $0.388$ &       $0.213$ &       $0.884$ &       $0.381$ &       $0.150$ &       $0.687$ &       $0.015$ \\
 AKM${0}$       &       $0.146$ &       $0.361$ &       $0.203$ &       $0.887$ &       $0.532$ &       $0.376$ &       $0.732$ &       $0.284$ \\ [6pt]

\multicolumn{9}{@{}l}{\textbf{Panel C\@: First-Stage}} \\ [2pt]

Robust  &       \multicolumn{4}{c}{     $0.002$ }       &                               \multicolumn{4}{c}{     $0.000$ }                                               \\
 Cluster        &       \multicolumn{4}{c}{     $0.000$ }       &                               \multicolumn{4}{c}{     $0.000$ }                                               \\
 AKM    &       \multicolumn{4}{c}{     $0.025$ }       &                               \multicolumn{4}{c}{     $0.000$ }                                               \\
 AKM${0}$       &       \multicolumn{4}{c}{     $0.096$ }       &                               \multicolumn{4}{c}{     $0.126$ }                                               \\ [4pt]

 \bottomrule

\multicolumn{9}{p{5.5in}}{\scriptsize{Notes: $\Delta \log E_i$ denotes log change in native employment; $\Delta \log w_i$ denotes log change in average weekly wages of native workers. In the specification \textit{MSA (1980 weights)}, we use information on 257 MSAs, 1980 weights and three time periods, 1980--1990, 1990--2000, 2000--2010; thus, $N=651$ (257 MSAs $\times$ 3 time periods). In the specification \textit{MSA (1960 weights)}, we use information on 217 MSAs, 1960 weights and one time period, 1970--1980; thus, $N=217$ (217 CZs $\times$ 1 time period). Models are weighted by start-of-period MSA share of national population. All regressions include period dummies. 95\% confidence intervals in square brackets. \textit{Robust} is the Eicker-Huber-White standard error; \textit{Cluster} is the standard error that clusters of CZs in the same state; \textit{AKM} is the standard error in \begin{NoHyper}\Cref{remark:akm-construction}\end{NoHyper}; and \textit{AKM0} is the confidence interval in \begin{NoHyper}\Cref{remark:akm0-construction}\end{NoHyper}. We exclude from the analysis those countries of origin whose immigrant share in year $t_{0}$ is larger than 3\%; i.e.\ $\sum_{i=1}^{N}ImmShare_{igt_{0}}/\sum_{i=1}^{N}\sum_{g'=1}^{G}ImmShare_{ig't_{0}}>
0.03$. See \Cref{tab:origin_country_1980,tab:origin_country_1960} for a list of the origin countries included in the analysis.}}
\end{tabular}
\end{table}

\begin{landscape}
\begin{table}[!tp]
\centering
\caption{Effect of immigration: analysis by CZ-Occupations and CZ-Education groups (including all origin countries)}\label{tab:immigration_all}
\tabcolsep=3pt
\footnotesize
\begin{tabular}{@{}lcccccccccc}
\toprule
Outcome: & \multicolumn{1}{c}{{$\Delta \log E_i$}} & \multicolumn{3}{c}{{$\Delta \log w_i$}} & \multicolumn{1}{c}{{$\Delta \log E_i$}} & \multicolumn{1}{c}{{$\Delta \log w_i$}} & \multicolumn{1}{c}{{$\Delta \log E_i$}} & \multicolumn{3}{c}{{$\Delta \log w_i$}}         \\
 \cmidrule(rl){2-2} \cmidrule(rl){3-5} \cmidrule(rl){6-6} \cmidrule(rl){7-7} \cmidrule(rl){8-8} \cmidrule(rl){9-11}
Workers: & All & All  & High-Skill & Low-Skill & All & All & All & All  & High-Skill & Low-Skill \\
 \cmidrule(rl){2-2} \cmidrule(rl){3-5} \cmidrule(rl){6-6} \cmidrule(rl){7-7} \cmidrule(rl){8-8} \cmidrule(rl){9-11}
 & (1) & (2) & (3) & (4) & (5) & (6) & (7) & (8) & (9) & (10)  \\
 \midrule
& \multicolumn{4}{c}{\textit{CZ-50 Occupations}} & \multicolumn{2}{c}{\textit{CZ- 2 Education Groups}} & \multicolumn{4}{c}{\textit{CZ-7 Occupations}} \\ [2pt]
\cmidrule(rl){2-5} \cmidrule(rl){6-7} \cmidrule(rl){8-11} \\[-7pt]

\multicolumn{11}{@{}l}{\textbf{Panel A\@: 2SLS Regression}} \\ [2pt]

 $\hat{\beta}$  &       $-0.73$ &       $-0.07$ &       $0.15$  &       $-0.24$ &       $-0.53$ &       $-0.01$ &       $-0.79$ &       $-0.08$ &       $0.08$  &       $-0.27$ \\
Robust  &       $[-1.04,-0.42]$ &       $[-0.22,0.09]$  &       $[0.01,0.29]$   &       $[-0.42,-0.06]$ &       $[-1.03,-0.02]$ &       $[-0.45,0.44]$  &       $[-1.25,-0.33]$ &       $[-0.39,0.22]$  &       $[-0.17,0.33]$  &       $[-0.64,0.09]$  \\
 Cluster        &       $[-1.15,-0.31]$ &       $[-0.49,0.36]$  &       $[-0.23,0.52]$  &       $[-0.80,0.32]$  &       $[-1.03,-0.02]$ &       $[-0.24,0.23]$  &       $[-1.34,-0.25]$ &       $[-0.33,0.16]$  &       $[-0.09,0.25]$  &       $[-0.47,-0.08]$ \\
 AKM    &       $[-1.22,-0.24]$ &       $[-0.42,0.29]$  &       $[-0.18,0.47]$  &       $[-0.66,0.18]$  &       $[-1.85,0.79]$  &       $[-0.81,0.80]$  &       $[-1.67,0.09]$  &       $[-0.74,0.57]$  &       $[-0.52,0.67]$  &       $[-0.99,0.44]$  \\
 AKM${0}$       &       $[-1.61,0.24]$  &       $[-0.52,0.94]$  &       $[-0.24,1.11]$  &       $[-0.97,0.68]$  &       $[-\infty,\infty]$      &       $[-\infty,\infty]$      &       $[-\infty,\infty]$      &       $[-\infty,\infty]$      &       $[-\infty,\infty]$      &       $[-\infty,\infty]$      \\ [6pt]

\multicolumn{11}{@{}l}{\textbf{Panel B\@: Reduced-Form Regression}} \\ [2pt]

 $\hat{\beta}$  &       $-0.19$ &       $-0.02$ &       $0.04$  &       $-0.06$ &       $-0.19$ &       $0.00$  &       $-0.25$ &       $-0.03$ &       $0.02$  &       $-0.09$ \\
Robust  &       $[-0.27,-0.11]$ &       $[-0.05,0.02]$  &       $[0.00,0.08]$   &       $[-0.10,-0.02]$ &       $[-0.41,0.04]$  &       $[-0.16,0.15]$  &       $[-0.41,-0.10]$ &       $[-0.12,0.06]$  &       $[-0.06,0.11]$  &       $[-0.18,0.01]$  \\
 Cluster        &       $[-0.32,-0.06]$ &       $[-0.12,0.09]$  &       $[-0.08,0.15]$  &       $[-0.17,0.05]$  &       $[-0.38,0.01]$  &       $[-0.08,0.08]$  &       $[-0.42,-0.09]$ &       $[-0.10,0.04]$  &       $[-0.04,0.09]$  &       $[-0.13,-0.05]$ \\
 AKM    &       $[-0.38,0.01]$  &       $[-0.10,0.07]$  &       $[-0.06,0.13]$  &       $[-0.15,0.02]$  &       $[-0.74,0.37]$  &       $[-0.29,0.28]$  &       $[-0.62,0.11]$  &       $[-0.23,0.17]$  &       $[-0.17,0.22]$  &       $[-0.29,0.11]$  \\
 AKM${0}$       &       $[-1.01,0.03]$  &       $[-0.08,0.51]$  &       $[-0.04,0.60]$  &       $[-0.14,0.42]$  &       $[-\infty,\infty]$      &       $[-\infty,\infty]$      &       $[-\infty,\infty]$      &       $[-1.28,-0.26]$ &       $[-1.18,-0.20]$ &       $[-1.13,-0.37]$ \\ [6pt]

\multicolumn{11}{@{}l}{\textbf{Panel C\@: First-Stage}} \\ [2pt]

 $\hat{\beta}$  &       \multicolumn{4}{c}{     $0.26$  }       &                               \multicolumn{2}{c}{     $0.35$  }       &       \multicolumn{4}{c}{     $0.32$  }                                               \\
Robust  &       \multicolumn{4}{c}{     $[0.19,0.32]$   }       &                               \multicolumn{2}{c}{     $[0.17,0.53]$   }       &       \multicolumn{4}{c}{     $[0.20,0.44]$   }                                               \\
 Cluster        &       \multicolumn{4}{c}{     $[0.17,0.35]$   }       &                               \multicolumn{2}{c}{     $[0.24,0.46]$   }       &       \multicolumn{4}{c}{     $[0.21,0.43]$   }                                               \\
 AKM    &       \multicolumn{4}{c}{     $[0.12,0.39]$   }       &                               \multicolumn{2}{c}{     $[0.10,0.60]$   }       &       \multicolumn{4}{c}{     $[0.12,0.52]$   }                                               \\
 AKM${0}$       &       \multicolumn{4}{c}{     $[0.11,0.84]$   }       &                               \multicolumn{2}{c}{     $[-\infty,\infty]$      }       &       \multicolumn{4}{c}{     $[-\infty,\infty]$      }                                               \\ [4pt]

\bottomrule

\multicolumn{11}{p{9.2in}}{\scriptsize{Notes: $\Delta \log E_i$ denotes log change in native employment; $\Delta \log w_i$ denotes log change in average weekly wages of native workers. The specifications \textit{CZ-50 Occupations}, \textit{CZ-2 Education Groups}, and \textit{CZ-7 Occupations} differ in the definition of the unit of observation. In all three specifications, we use 1980 weights and three time periods, 1980--1990, 1990--2000, 2000--2010. Thus, $N = 108,300$ (722 CZs $\times$ 50 occupations $\times$ 3 time periods) for the \textit{CZ-50 Occupations} specification; $N = 4,332$ (722 CZs $\times$ 2 education groups $\times$ 3 time periods) for the \textit{CZ-2 Education Groups} specification; and $N = 15,162$ (722 CZs $\times$ 7 occupations $\times$ 3 time periods) for the \textit{CZ-7 Occupations} specification. Models are weighted by start-of-period occupation-region (or education group-region) share of national population. All regressions include occupation (or education group) and period dummies. 95\% confidence intervals in square brackets. \textit{Robust} is the Eicker-Huber-White standard error; \textit{Cluster} is the standard error that clusters of CZs in the same state; \textit{AKM} is the standard error in \begin{NoHyper}\Cref{remark:akm-construction}\end{NoHyper}; and \textit{AKM0} is the confidence interval in \begin{NoHyper}\Cref{remark:akm0-construction}\end{NoHyper}. We include all countries of origin in the analysis. See \Cref{tab:origin_country_1980} for a list of the origin countries included in the analysis.}}
\end{tabular}
\end{table}
\end{landscape}

\begin{landscape}
\begin{table}[!tp]
\centering
\caption{Effect of immigration: analysis by CZ (including all origin countries)}\label{tab:immigration_all_addres_CZ}
\tabcolsep=0.18cm
\begin{tabular}{@{}lcccccccc}
\toprule
Outcome: & \multicolumn{1}{c}{{$\Delta \log E_i$}} & \multicolumn{3}{c}{{$\Delta \log w_i$}} & \multicolumn{1}{c}{{$\Delta \log E_i$}} & \multicolumn{3}{c}{{$\Delta \log w_i$}}         \\
 \cmidrule(rl){2-2} \cmidrule(rl){3-5} \cmidrule(rl){6-6} \cmidrule(rl){7-9}
Workers: & All & All  & High-Skill & Low-Skill & All & All  & High-Skill & Low-Skill \\
 \cmidrule(rl){2-2} \cmidrule(rl){3-5} \cmidrule(rl){6-6} \cmidrule(rl){7-9}
 & (1) & (2) & (3) & (4) & (5) & (6) & (7) & (8)  \\
 \midrule

& \multicolumn{4}{c}{\textit{Commuting Zone (1980 weights)}} & \multicolumn{4}{c}{\textit{Commuting Zone (1960 weights)}}  \\ [2pt]
\cmidrule(rl){2-5} \cmidrule(rl){6-9}

\multicolumn{9}{@{}l}{\textbf{Panel A\@: 2SLS Regression}} \\ [2pt]

 $\hat{\beta}$  &       $-0.49$ &       $0.13$  &       $0.27$  &       $-0.2$  &       $0.05$  &       $-0.25$ &       $0.09$  &       $-0.52$ \\
Robust  &       $[-1.12,0.14]$  &       $[-0.37,0.63]$  &       $[-0.09,0.64]$  &       $[-0.85,0.44]$  &       $[-0.96,1.07]$  &       $[-0.52,0.02]$  &       $[-0.16,0.35]$  &       $[-0.86,-0.18]$ \\
 Cluster        &       $[-0.98,0.01]$  &       $[-0.15,0.41]$  &       $[0.08,0.47]$   &       $[-0.49,0.08]$  &       $[-0.93,1.03]$  &       $[-0.59,0.09]$  &       $[-0.16,0.34]$  &       $[-0.92,-0.11]$ \\
 AKM    &       $[-1.74,0.77]$  &       $[-0.88,1.14]$  &       $[-0.53,1.08]$  &       $[-1.42,1.01]$  &       $[-2.39,2.50]$  &       $[-1.14,0.64]$  &       $[-0.76,0.95]$  &       $[-1.64,0.61]$  \\
 AKM${0}$       &       $[-\infty,\infty]$      &       $[-\infty,\infty]$      &       $[-\infty,\infty]$      &       $[-\infty,\infty]$      &       $[-\infty,\infty]$      &       $[-\infty,\infty]$      &       $[-\infty,\infty]$      &       $[-\infty,\infty]$      \\ [6pt]

\multicolumn{9}{@{}l}{\textbf{Panel B\@: Reduced-Form Regression}} \\ [2pt]

 $\hat{\beta}$  &       $-0.19$ &       $0.05$  &       $0.11$  &       $-0.08$ &       $0.04$  &       $-0.17$ &       $0.06$  &       $-0.36$ \\
Robust  &       $[-0.39,0.02]$  &       $[-0.16,0.26]$  &       $[-0.07,0.28]$  &       $[-0.30,0.14]$  &       $[-0.66,0.73]$  &       $[-0.40,0.05]$  &       $[-0.12,0.24]$  &       $[-0.67,-0.04]$ \\
 Cluster        &       $[-0.37,0.00]$  &       $[-0.07,0.17]$  &       $[0.01,0.20]$   &       $[-0.17,0.01]$  &       $[-0.63,0.71]$  &       $[-0.48,0.14]$  &       $[-0.11,0.24]$  &       $[-0.79,0.08]$  \\
 AKM    &       $[-0.71,0.33]$  &       $[-0.36,0.46]$  &       $[-0.24,0.45]$  &       $[-0.52,0.37]$  &       $[-1.64,1.72]$  &       $[-0.82,0.47]$  &       $[-0.51,0.64]$  &       $[-1.20,0.49]$  \\
 AKM${0}$       &       $[-\infty,\infty]$      &       $[-\infty,\infty]$      &       $[-\infty,\infty]$      &       $[-\infty,\infty]$      &       $[-\infty,\infty]$      &       $[-\infty,\infty]$      &       $[-\infty,\infty]$      &       $[-\infty,\infty]$      \\ [6pt]

\multicolumn{9}{@{}l}{\textbf{Panel C\@: 2SLS First-Stage}} \\ [2pt]

 $\hat{\beta}$  &       \multicolumn{4}{c}{     $0.38$  }       &                               \multicolumn{4}{c}{     $0.69$  }                                               \\
Robust  &       \multicolumn{4}{c}{     $[0.20,0.57]$   }       &                               \multicolumn{4}{c}{     $[0.41,0.97]$   }                                               \\
 Cluster        &       \multicolumn{4}{c}{     $[0.27,0.49]$   }       &                               \multicolumn{4}{c}{     $[0.35,1.03]$   }                                               \\
 AKM    &       \multicolumn{4}{c}{     $[0.10,0.67]$   }       &                               \multicolumn{4}{c}{     $[0.53,0.85]$   }                                               \\
 AKM${0}$       &       \multicolumn{4}{c}{     $[-\infty,\infty]$      }       &                               \multicolumn{4}{c}{     $[-\infty,\infty]$      }                                               \\ [4pt]
 \bottomrule

\multicolumn{9}{p{8.5in}}{\scriptsize{Notes: $\Delta \log E_i$ denotes log change in native employment; $\Delta \log w_i$ denotes log change in average weekly wages of native workers. In the specification \textit{CZ (1980 weights)}, we use information on 722 CZs, 1980 weights and three time periods, 1980--1990, 1990--2000, 2000--2010; thus, $N=2,166$ (722 CZs $\times$ 3 time periods). In the specification \textit{CZ (1960 weights)}, we use information on 722 CZs, 1960 weights and one time period, 1970--1980; thus, $N=722$ (722 CZs $\times$ 1 time period). Models are weighted by start-of-period CZ share of national population. All regressions include period dummies. 95\% confidence intervals in square brackets. \textit{Robust} is the Eicker-Huber-White standard error; \textit{Cluster} is the standard error that clusters of CZs in the same state; \textit{AKM} is the standard error in \begin{NoHyper}\Cref{remark:akm-construction}\end{NoHyper}; and \textit{AKM0} is the confidence interval in \begin{NoHyper}\Cref{remark:akm0-construction}\end{NoHyper}. We include all countries of origin in the analysis. See \Cref{tab:origin_country_1980,tab:origin_country_1960} for a list of the origin countries included in the analysis.}}
\end{tabular}
\end{table}
\end{landscape}

\begin{landscape}
\begin{table}[!tp]
\centering
\caption{Effect of immigration: analysis by MSA (including all origin countries)}\label{tab:immigration_all_addres_MSA}
\tabcolsep=0.18cm
\begin{tabular}{@{}lcccccccc}
\toprule
Outcome: & \multicolumn{1}{c}{{$\Delta \log E_i$}} & \multicolumn{3}{c}{{$\Delta \log w_i$}} & \multicolumn{1}{c}{{$\Delta \log E_i$}} & \multicolumn{3}{c}{{$\Delta \log w_i$}}         \\
 \cmidrule(rl){2-2} \cmidrule(rl){3-5} \cmidrule(rl){6-6} \cmidrule(rl){7-9}
Workers: & All & All  & High-Skill & Low-Skill & All & All  & High-Skill & Low-Skill \\
 \cmidrule(rl){2-2} \cmidrule(rl){3-5} \cmidrule(rl){6-6} \cmidrule(rl){7-9}
 & (1) & (2) & (3) & (4) & (5) & (6) & (7) & (8)  \\
 \midrule

& \multicolumn{4}{c}{\textit{MSA (1980 weights)}} & \multicolumn{4}{c}{\textit{MSA (1960 weights)}}  \\ [2pt]
\cmidrule(rl){2-5} \cmidrule(rl){6-9}

\multicolumn{9}{@{}l}{\textbf{Panel A\@: 2SLS Regression}} \\ [2pt]

 $\hat{\beta}$  &       $-1.41$ &       $0.16$  &       $0.28$  &       $-0.21$ &       $-0.18$ &       $-0.14$ &       $0.18$  &       $-0.35$ \\
Robust  &       $[-2.62,-0.21]$ &       $[-0.38,0.71]$  &       $[-0.11,0.68]$  &       $[-1.06,0.63]$  &       $[-1.11,0.75]$  &       $[-0.30,0.02]$  &       $[-0.05,0.42]$  &       $[-0.56,-0.14]$ \\
 Cluster        &       $[-2.55,-0.28]$ &       $[-0.14,0.46]$  &       $[0.06,0.51]$   &       $[-0.62,0.20]$  &       $[-1.21,0.86]$  &       $[-0.31,0.04]$  &       $[-0.08,0.44]$  &       $[-0.52,-0.17]$ \\
 AKM    &       $[-3.24,0.41]$  &       $[-0.91,1.23]$  &       $[-0.61,1.18]$  &       $[-1.64,1.21]$  &       $[-5.04,4.69]$  &       $[-0.92,0.64]$  &       $[-0.90,1.26]$  &       $[-1.26,0.56]$  \\
 AKM${0}$       &       $[-\infty,\infty]$      &       $[-\infty,\infty]$      &       $[-\infty,\infty]$      &       $[-\infty,\infty]$      &       $[-\infty,\infty]$      &       $[-\infty,\infty]$      &       $[-\infty,\infty]$      &       $[-\infty,\infty]$      \\ [6pt]

\multicolumn{9}{@{}l}{\textbf{Panel B\@: Reduced-Form Regression}} \\ [2pt]

 $\hat{\beta}$  &       $-0.41$ &       $0.05$  &       $0.08$  &       $-0.06$ &       $-0.12$ &       $-0.10$ &       $0.13$  &       $-0.24$ \\
Robust  &       $[-0.61,-0.20]$ &       $[-0.13,0.23]$  &       $[-0.07,0.24]$  &       $[-0.27,0.15]$  &       $[-0.78,0.54]$  &       $[-0.22,0.03]$  &       $[-0.05,0.31]$  &       $[-0.43,-0.05]$ \\
 Cluster        &       $[-0.62,-0.19]$ &       $[-0.06,0.15]$  &       $[0.00,0.17]$   &       $[-0.16,0.03]$  &       $[-0.83,0.59]$  &       $[-0.23,0.04]$  &       $[-0.08,0.33]$  &       $[-0.42,-0.06]$ \\
 AKM    &       $[-0.91,0.10]$  &       $[-0.28,0.37]$  &       $[-0.21,0.37]$  &       $[-0.45,0.33]$  &       $[-3.47,3.23]$  &       $[-0.64,0.45]$  &       $[-0.60,0.85]$  &       $[-0.89,0.41]$  \\
 AKM${0}$       &       $[-\infty,\infty]$      &       $[-\infty,\infty]$      &       $[-\infty,\infty]$      &       $[-\infty,\infty]$      &       $[-\infty,\infty]$      &       $[-\infty,\infty]$      &       $[-\infty,\infty]$      &       $[-\infty,\infty]$      \\ [6pt]

\multicolumn{9}{@{}l}{\textbf{Panel C\@: First-Stage}} \\ [2pt]

 $\hat{\beta}$  &       \multicolumn{4}{c}{     $0.29$  }       &                               \multicolumn{4}{c}{     $0.69$  }                                               \\
Robust  &       \multicolumn{4}{c}{     $[0.10,0.48]$   }       &                               \multicolumn{4}{c}{     $[0.35,1.02]$   }                                               \\
 Cluster        &       \multicolumn{4}{c}{     $[0.16,0.41]$   }       &                               \multicolumn{4}{c}{     $[0.34,1.03]$   }                                               \\
 AKM    &       \multicolumn{4}{c}{     $[0.03,0.54]$   }       &                               \multicolumn{4}{c}{     $[0.44,0.93]$   }                                               \\
 AKM${0}$       &       \multicolumn{4}{c}{     $[-\infty,\infty]$      }       &                               \multicolumn{4}{c}{     $[-\infty,\infty]$      }                                               \\
 \bottomrule

\multicolumn{9}{p{8.5in}}{\scriptsize{Notes: $\Delta \log E_i$ denotes log
  change in native employment; $\Delta \log w_i$ denotes log change in average
  weekly wages of native workers. In the specification \textit{MSA (1980
  weights)}, we use information on 257 MSAs, 1980 weights and three time
  periods, 1980--1990, 1990--2000, 2000--2010; thus, $N=651$ (257 MSAs $\times$ 3
  time periods). In the specification \textit{MSA (1960 weights)}, we use
  information on 217 MSAs, 1960 weights and one time period, 1970--1980; thus,
  $N=217$ (217 CZs $\times$ 1 time period). Models are weighted by
  start-of-period MSA share of national population. All regressions include
  period dummies. 95\% confidence intervals in square brackets. \textit{Robust}
  is the Eicker-Huber-White standard error; \textit{Cluster} is the standard
  error that clusters of CZs in the same state; \textit{AKM} is the standard
  error in \begin{NoHyper}\Cref{remark:akm-construction}\end{NoHyper}; and
  \textit{AKM0} is the confidence interval
  in \begin{NoHyper}\Cref{remark:akm0-construction}\end{NoHyper}. We include all
  countries of origin in the analysis. See
  \Cref{tab:origin_country_1980,tab:origin_country_1960} for a list of the
  origin countries included in the analysis.}}
\end{tabular}
\end{table}
\end{landscape}

\begin{table}[!tp]
\centering
\caption{Immigration: p-values by CZ-Occ.\ and CZ-Educ. (including all origin countries)}\label{tab:immigration_all_pvalue}
\tabcolsep=0.08cm
\begin{tabular}{@{}lcccccccccc}
\toprule
 & \multicolumn{1}{c}{{$\Delta \log E_i$}} & \multicolumn{3}{c}{{$\Delta \log w_i$}} & \multicolumn{1}{c}{{$\Delta \log E_i$}} & \multicolumn{1}{c}{{$\Delta \log w_i$}} & \multicolumn{1}{c}{{$\Delta \log E_i$}} & \multicolumn{3}{c}{{$\Delta \log w_i$}}         \\
 \cmidrule(rl){2-2} \cmidrule(rl){3-5} \cmidrule(rl){6-6} \cmidrule(rl){7-7} \cmidrule(rl){8-8} \cmidrule(rl){9-11}
 & All & All  & High-Skill & Low-Skill & All & All & All & All  & High-Skill & Low-Skill \\
 \cmidrule(rl){2-2} \cmidrule(rl){3-5} \cmidrule(rl){6-6} \cmidrule(rl){7-7} \cmidrule(rl){8-8} \cmidrule(rl){9-11}
 & (1) & (2) & (3) & (4) & (5) & (6) & (7) & (8) & (9) & (10)  \\
 \midrule
& \multicolumn{4}{c}{\textit{CZ-50 Occupations}} & \multicolumn{2}{c}{\textit{CZ- Educ.}} & \multicolumn{4}{c}{\textit{CZ-7 Occupations}} \\ [2pt]
\cmidrule(rl){2-5} \cmidrule(rl){6-7} \cmidrule(rl){8-11} \\[-7pt]

\multicolumn{9}{@{}l}{\textbf{Panel A\@: 2SLS Regression}} \\ [2pt]

Robust  &       $0.000$ &       $0.402$ &       $0.042$ &       $0.011$ &       $0.040$ &       $0.978$ &       $0.001$ &       $0.586$ &       $0.543$ &       $0.139$ \\
 Cluster        &       $0.001$ &       $0.764$ &       $0.445$ &       $0.403$ &       $0.040$ &       $0.958$ &       $0.004$ &       $0.496$ &       $0.363$ &       $0.005$ \\
 AKM    &       $0.004$ &       $0.721$ &       $0.380$ &       $0.267$ &       $0.434$ &       $0.988$ &       $0.077$ &       $0.801$ &       $0.798$ &       $0.454$ \\
 AKM${0}$       &       $0.074$ &       $0.734$ &       $0.359$ &       $0.330$ &       $0.532$ &       $0.988$ &       $0.274$ &       $0.808$ &       $0.793$ &       $0.504$ \\ [6pt]

\multicolumn{9}{@{}l}{\textbf{Panel B\@: Reduced-Form Regression}} \\ [2pt]

Robust  &       $0.000$ &       $0.372$ &       $0.072$ &       $0.002$ &       $0.106$ &       $0.978$ &       $0.001$ &       $0.560$ &       $0.567$ &       $0.074$ \\
 Cluster        &       $0.005$ &       $0.752$ &       $0.515$ &       $0.282$ &       $0.068$ &       $0.957$ &       $0.002$ &       $0.447$ &       $0.422$ &       $0.000$ \\
 AKM    &       $0.059$ &       $0.704$ &       $0.442$ &       $0.161$ &       $0.513$ &       $0.988$ &       $0.170$ &       $0.793$ &       $0.805$ &       $0.395$ \\
 AKM${0}$       &       $0.074$ &       $0.734$ &       $0.359$ &       $0.330$ &       $0.532$ &       $0.988$ &       $0.274$ &       $0.808$ &       $0.793$ &       $0.504$ \\ [6pt]

\multicolumn{9}{@{}l}{\textbf{Panel C\@: First-Stage}} \\ [2pt]

Robust  &       \multicolumn{4}{c}{     $0.000$ }       &                               \multicolumn{2}{c}{     $0.000$ }       &       \multicolumn{4}{c}{     $0.000$ }                                               \\
 Cluster        &       \multicolumn{4}{c}{     $0.000$ }       &                               \multicolumn{2}{c}{     $0.000$ }       &       \multicolumn{4}{c}{     $0.000$ }                                               \\
 AKM    &       \multicolumn{4}{c}{     $0.000$ }       &                               \multicolumn{2}{c}{     $0.005$ }       &       \multicolumn{4}{c}{     $0.001$ }                                               \\
 AKM${0}$       &       \multicolumn{4}{c}{     $0.021$ }       &                               \multicolumn{2}{c}{     $0.134$ }       &       \multicolumn{4}{c}{     $0.067$ }                                               \\ [4pt]

\bottomrule

\multicolumn{11}{p{6.5in}}{\scriptsize{Notes: $\Delta \log E_i$ denotes log change in native employment; $\Delta \log w_i$ denotes log change in average weekly wages of native workers. The specifications \textit{CZ-50 Occupations}, \textit{CZ-Educ.}, and \textit{CZ-7 Occupations} differ in the definition of the unit of observation. In all three specifications, we use 1980 weights and three time periods, 1980--1990, 1990--2000, 2000--2010. Thus, $N = 108,300$ (722 CZs $\times$ 50 occupations $\times$ 3 time periods) for the \textit{CZ-50 Occupations} specification; $N = 4,332$ (722 CZs $\times$ 2 education groups $\times$ 3 time periods) for the \textit{CZ-Educ.} specification; and $N = 15,162$ (722 CZs $\times$ 7 occupations $\times$ 3 time periods) for the \textit{CZ-7 Occupations} specification. Models are weighted by start-of-period occupation-region (or education group-region) share of national population. All regressions include occupation (or education group) and period dummies. 95\% confidence intervals in square brackets. \textit{Robust} is the Eicker-Huber-White standard error; \textit{Cluster} is the standard error that clusters of CZs in the same state; \textit{AKM} is the standard error in \begin{NoHyper}\Cref{remark:akm-construction}\end{NoHyper}; and \textit{AKM0} is the confidence interval in \begin{NoHyper}\Cref{remark:akm0-construction}\end{NoHyper}. We all origin countries in the analysis. See \Cref{tab:origin_country_1980} for a list of the origin countries included in the analysis.}}
\end{tabular}
\end{table}

\begin{table}[!tp]
\centering
\caption{Effect of immigration: p-values by CZ (including all origin countries)}\label{tab:immigration_all_pvalues_addres_CZ}
\tabcolsep=0.15cm
\begin{tabular}{@{}lcccccccc}
\toprule
Outcome: & \multicolumn{1}{c}{{$\Delta \log E_i$}} & \multicolumn{3}{c}{{$\Delta \log w_i$}} & \multicolumn{1}{c}{{$\Delta \log E_i$}} & \multicolumn{3}{c}{{$\Delta \log w_i$}}         \\
 \cmidrule(rl){2-2} \cmidrule(rl){3-5} \cmidrule(rl){6-6} \cmidrule(rl){7-9}
Workers: & All & All  & High-Skill & Low-Skill & All & All  & High-Skill & Low-Skill \\
 \cmidrule(rl){2-2} \cmidrule(rl){3-5} \cmidrule(rl){6-6} \cmidrule(rl){7-9}
 & (1) & (2) & (3) & (4) & (5) & (6) & (7) & (8)  \\
 \midrule

 & \multicolumn{4}{c}{\textit{Commuting Zone (1980 weights)}} & \multicolumn{4}{c}{\textit{Commuting Zone (1960 weights)}}  \\ [2pt]
\cmidrule(rl){2-5} \cmidrule(rl){6-9}

\multicolumn{9}{@{}l}{\textbf{Panel A\@: 2SLS Regression}} \\ [2pt]

Robust  &       $0.130$ &       $0.607$ &       $0.142$ &       $0.536$ &       $0.917$ &       $0.072$ &       $0.478$ &       $0.003$ \\
 Cluster        &       $0.055$ &       $0.353$ &       $0.005$ &       $0.154$ &       $0.914$ &       $0.152$ &       $0.464$ &       $0.013$ \\
 AKM    &       $0.449$ &       $0.799$ &       $0.504$ &       $0.741$ &       $0.965$ &       $0.581$ &       $0.830$ &       $0.367$ \\
 AKM${0}$       &       $0.578$ &       $0.797$ &       $0.522$ &       $0.759$ &       $0.965$ &       $0.619$ &       $0.830$ &       $0.463$ \\ [6pt]

\multicolumn{9}{@{}l}{\textbf{Panel B\@: Reduced-Form Regression}} \\ [2pt]

Robust  &       $0.075$ &       $0.640$ &       $0.252$ &       $0.485$ &       $0.916$ &       $0.136$ &       $0.483$ &       $0.025$ \\
 Cluster        &       $0.047$ &       $0.404$ &       $0.032$ &       $0.091$ &       $0.913$ &       $0.274$ &       $0.473$ &       $0.108$ \\
 AKM    &       $0.481$ &       $0.808$ &       $0.553$ &       $0.730$ &       $0.965$ &       $0.600$ &       $0.827$ &       $0.408$ \\
 AKM${0}$       &       $0.578$ &       $0.797$ &       $0.522$ &       $0.759$ &       $0.965$ &       $0.619$ &       $0.830$ &       $0.463$ \\ [6pt]

\multicolumn{9}{@{}l}{\textbf{Panel C\@: 2SLS First-Stage}} \\ [2pt]

Robust  &       \multicolumn{4}{c}{     $0.000$ }       &                               \multicolumn{4}{c}{     $0.000$ }                                               \\
 Cluster        &       \multicolumn{4}{c}{     $0.000$ }       &                               \multicolumn{4}{c}{     $0.000$ }                                               \\
 AKM    &       \multicolumn{4}{c}{     $0.008$ }       &                               \multicolumn{4}{c}{     $0.000$ }                                               \\
 AKM${0}$       &       \multicolumn{4}{c}{     $0.179$ }       &                               \multicolumn{4}{c}{     $0.156$ }                                               \\ [4pt]

 \bottomrule

\multicolumn{9}{p{5.8in}}{\scriptsize{Notes: $\Delta \log E_i$ denotes log change in native employment; $\Delta \log w_i$ denotes log change in average weekly wages of native workers. In the specification \textit{CZ (1980 weights)}, we use information on 722 CZs, 1980 weights and three time periods, 1980--1990, 1990--2000, 2000--2010; thus, $N=2,166$ (722 CZs $\times$ 3 time periods). In the specification \textit{CZ (1960 weights)}, we use information on 722 CZs, 1960 weights and one time period, 1970--1980; thus, $N=722$ (722 CZs $\times$ 1 time period). Models are weighted by start-of-period CZ share of national population. All regressions include period dummies. 95\% confidence intervals in square brackets. \textit{Robust} is the Eicker-Huber-White standard error; \textit{Cluster} is the standard error that clusters of CZs in the same state; \textit{AKM} is the standard error in \begin{NoHyper}\Cref{remark:akm-construction}\end{NoHyper}; and \textit{AKM0} is the confidence interval in \begin{NoHyper}\Cref{remark:akm0-construction}\end{NoHyper}. We include all origin countries in the analysis. See \Cref{tab:origin_country_1980,tab:origin_country_1960} for a list of the origin countries included in the analysis.}}
\end{tabular}
\end{table}

\begin{table}[!tp]
\centering
\caption{Effect of immigration: p-values by MSA (including all origin countries)}\label{tab:immigration_all_pvalues_addres_MSA}
\tabcolsep=0.15cm
\begin{tabular}{@{}lcccccccc}
\toprule
Outcome: & \multicolumn{1}{c}{{$\Delta \log E_i$}} & \multicolumn{3}{c}{{$\Delta \log w_i$}} & \multicolumn{1}{c}{{$\Delta \log E_i$}} & \multicolumn{3}{c}{{$\Delta \log w_i$}}         \\
 \cmidrule(rl){2-2} \cmidrule(rl){3-5} \cmidrule(rl){6-6} \cmidrule(rl){7-9}
Workers: & All & All  & High-Skill & Low-Skill & All & All  & High-Skill & Low-Skill \\
 \cmidrule(rl){2-2} \cmidrule(rl){3-5} \cmidrule(rl){6-6} \cmidrule(rl){7-9}
 & (1) & (2) & (3) & (4) & (5) & (6) & (7) & (8)  \\
 \midrule

 & \multicolumn{4}{c}{\textit{Commuting Zone (1980 weights)}} & \multicolumn{4}{c}{\textit{Commuting Zone (1960 weights)}}  \\ [2pt]
\cmidrule(rl){2-5} \cmidrule(rl){6-9}

\multicolumn{9}{@{}l}{\textbf{Panel A\@: 2SLS Regression}} \\ [2pt]

Robust  &       $0.021$ &       $0.562$ &       $0.159$ &       $0.621$ &       $0.709$ &       $0.096$ &       $0.131$ &       $0.001$ \\
 Cluster        &       $0.015$ &       $0.294$ &       $0.014$ &       $0.306$ &       $0.737$ &       $0.118$ &       $0.168$ &       $0.000$ \\
 AKM    &       $0.130$ &       $0.768$ &       $0.534$ &       $0.769$ &       $0.943$ &       $0.727$ &       $0.739$ &       $0.450$ \\
 AKM${0}$       &       $0.343$ &       $0.771$ &       $0.554$ &       $0.778$ &       $0.944$ &       $0.753$ &       $0.730$ &       $0.562$ \\ [6pt]

\multicolumn{9}{@{}l}{\textbf{Panel B\@: Reduced-Form Regression}} \\ [2pt]

Robust  &       $0.000$ &       $0.616$ &       $0.302$ &       $0.571$ &       $0.717$ &       $0.127$ &       $0.172$ &       $0.012$ \\
 Cluster        &       $0.000$ &       $0.375$ &       $0.061$ &       $0.203$ &       $0.737$ &       $0.168$ &       $0.237$ &       $0.010$ \\
 AKM    &       $0.115$ &       $0.782$ &       $0.584$ &       $0.757$ &       $0.943$ &       $0.731$ &       $0.735$ &       $0.467$ \\
 AKM${0}$       &       $0.343$ &       $0.771$ &       $0.554$ &       $0.778$ &       $0.944$ &       $0.753$ &       $0.730$ &       $0.562$ \\ [6pt]

\multicolumn{9}{@{}l}{\textbf{Panel C\@: First-Stage}} \\ [2pt]

Robust  &       \multicolumn{4}{c}{     $0.003$ }       &                               \multicolumn{4}{c}{     $0.000$ }                                               \\
 Cluster        &       \multicolumn{4}{c}{     $0.000$ }       &                               \multicolumn{4}{c}{     $0.000$ }                                               \\
 AKM    &       \multicolumn{4}{c}{     $0.029$ }       &                               \multicolumn{4}{c}{     $0.000$ }                                               \\
 AKM${0}$       &       \multicolumn{4}{c}{     $0.165$ }       &                               \multicolumn{4}{c}{     $0.200$ }                                               \\ [4pt]

 \bottomrule

\multicolumn{9}{p{5.8in}}{\scriptsize{Notes: $\Delta \log E_i$ denotes log change in native employment; $\Delta \log w_i$ denotes log change in average weekly wages of native workers. In the specification \textit{CZ (1980 weights)}, we use information on 722 CZs, 1980 weights and three time periods, 1980--1990, 1990--2000, 2000--2010; thus, $N=2,166$ (722 CZs $\times$ 3 time periods). In the specification \textit{CZ (1960 weights)}, we use information on 722 CZs, 1960 weights and one time period, 1970--1980; thus, $N=722$ (722 CZs $\times$ 1 time period). Models are weighted by start-of-period CZ share of national population. All regressions include period dummies. 95\% confidence intervals in square brackets. \textit{Robust} is the Eicker-Huber-White standard error; \textit{Cluster} is the standard error that clusters of CZs in the same state; \textit{AKM} is the standard error in \begin{NoHyper}\Cref{remark:akm-construction}\end{NoHyper}; and \textit{AKM0} is the confidence interval in \begin{NoHyper}\Cref{remark:akm0-construction}\end{NoHyper}. We include all origin countries in the analysis. See \Cref{tab:origin_country_1980,tab:origin_country_1960} for a list of the origin countries included in the analysis.}}
\end{tabular}
\end{table}

\end{appendices}

\clearpage
\bibliographystyle{aer}
\bibliography{Refs}